\documentclass[twoside,12pt]{article}
\usepackage{epsfig}
\usepackage{graphics}
\usepackage{graphicx}
\usepackage{amssymb}
\usepackage{amsmath}
\usepackage{multicol}

\newcommand{\pol}[1]{\mathaccent"017E{#1}}
\newcommand{\half}{\mbox{${\textstyle \frac{1}{2}}$}}
\newcommand{\fmn}[2]{\mbox{${\textstyle \frac{#1}{#2}}$}}

\newcommand{\CT}{\cos\theta_{\eta}^{*}}

\begin{document}

\title{Production of $\eta$ and $\eta^{\prime}$ Mesons on Nucleons and Nuclei}

\author{B.~Krusche$^1$ and C.~Wilkin$^2$\\ \\
$^1$ Department of Physics, University of Basel, CH-4056 Basel, Switzerland\\
$^2$ Physics and Astronomy Department, UCL, Gower Street, London WC1E 6BT, UK}

\maketitle
\begin{abstract}
The production of $\eta$ and $\eta^{\prime}$ mesons in photon- and
hadron-induced reactions on free and quasi-free nucleons and on
nuclei is reviewed. The extensive database on $\gamma N \to \eta N$, for both
proton and neutron targets, is described in detail and its implications for
the search for $N^{\star}$ resonances much heavier than the dominant
$S_{11}(1535)$ discussed. Though less is currently known about the production
of the $\eta^{\prime}$ or of $\eta\pi$ pairs, these also offer tantalizing
prospects in the search for the missing isobars. The more limited data
available on pion-induced production are still necessary ingredients in the
partial wave analysis discussed.

The production of the $\eta$-meson in $pp$ and $pn$ collisions shows once
again the strong influence of the $S_{11}(1535)$ isobar, which is in contrast
to the relatively weak behaviour seen near threshold for $\eta^{\prime}$
production. This difference is reflected in the important final state
interaction effects of the $\eta$ in nuclei that may even lead to this meson
being ``bound'' in some systems. The evidence for this is reviewed for both
$\gamma A$ and $p A$ collisions. The inclusive photoproduction of $\eta$,
$\eta^{\prime}$, and $\eta\pi$ pairs from nuclei provides further information
regarding the production mechanism and the interaction of the $\eta$ and
$\eta^{\prime}$ with nuclei and the $\eta\pi$ pairs may even allow access to
low mass $\eta A$ systems that are forbidden in direct single-meson
photoproduction.
\end{abstract}

%
%
\section{Introduction}
\label{Introduction}\setcounter{equation}{0}

Although the isovector $\pi$ and the isoscalar $\eta$ and $\eta^{\prime}$
represent the non-strange members of the fundamental pseudoscalar meson
nonet, their interactions with nucleons are dramatically different. The
$s$-wave $\pi N$ interactions are very weak and at only a few MeV above
threshold the $p$-wave, driven by the $\Delta(1232)$ isobar, starts to
dominate. In complete contrast, the $S_{11}$ $N^*(1535)$ resonance sits very
close to the $\eta N$ threshold, which means that the $s$-wave $\eta N$
interaction is extremely strong and its effects are felt in many of the
phenomena discussed in this review. Finally, there is far less unambiguous
evidence for isobars that have large decay widths into the $\eta^{\prime}N$
channel so that one would expect the near-threshold behaviour of
$\eta^{\prime}$ production to be very different from that of the $\eta$, and
this is precisely what one finds.

This review is concerned with the production of $\eta$ and $\eta^{\prime}$
mesons with both photons and hadronic probes. Earlier articles have surveyed
various aspects of the
field~\cite{Faldt_02,Moskal_02,Kelkar_13,Machner_14,Krusche_11}, but possibly
with not such a wide remit. Our primary thrust will be to discuss physics
phenomena rather than the details of theoretical models. The first part of
this survey is concerned with \emph{elementary reactions}, by which we mean
production reactions that involve only a single nucleon. It is immediately
apparent here that the sad lack of secondary pion beams in the world compared
to the richness of possibilities offered at electron machines means that the
vast majority of the new data in this field were obtained with photon (or
electron) beams. This imbalance has, of course, to be reflected in our
review.

Because of its large mass and isoscalar nature, the $\eta N$
system has a very important role to play in the search for the
\emph{missing} $N^{\star}$ ($I=\frac{1}{2}$) isobars. This
search would be simpler if one had a complete set of
measurements of the $\pi^-p\to\eta n$ reaction, because there
is here only one isospin channel. In photoproduction there are
contributions from both isoscalar and isovector photons (plus
interference terms) so that an experimental programme involving
(quasi-free) neutrons is completely essential to resolve some
of the ambiguities. Nevertheless, the rich collection of
photoproduction data, both current and under analysis,
involving both polarized photons and nucleons, means that these
results largely dominate the partial wave analyses. Though
there are data also on the photoproduction of both the
$\eta^{\prime}$ and $\eta\pi$ pairs from nucleons, these have
yet to have their full impact on the search for heavy nucleon
isobars.

Once the $\eta$ meson has been produced, it interacts very
strongly with nucleons and this is seen in its photoproduction
off light nuclei but possibly even more clearly in
nucleon-nucleon or nucleon-nucleus reactions. The interaction
is so strong, and attractive that it might even lead to the
formation of quasi-bound states of an $\eta$ meson with a
nucleus. The evidence for this is reviewed in both
electromagnetic and hadronic reactions. The searches above
threshold suggest that such a state might exist for a light
nucleus such as $_{\eta}^3$He, though no evidence for this is
found from other decay channels looking in the bound state
region.

Though the beam intensities can be very strong, and one is not hindered by
the suppression associated with the fine structure constant, proton-induced
reactions have a serious problem with the enormous total hadronic background.
This is especially troublesome if one cannot trigger on a clean $\eta$ signal
and there have been few experiments with heavier nuclei. On the other hand,
the selection rules in coherent $(\gamma,\eta)$ reactions strongly hinder
most direct searches for mesic nuclei but this will not be the case for the
$(\gamma,\eta\pi)$ reactions that can be measured on nuclei as well as
nucleons.

Much less is known about the production and interaction of the
$\eta^{\prime}$ meson. Photon-induced transparency measurements suggest that
the interaction with nucleons is weakly attractive but data from $pp\to
pp\eta^{\prime}$ near threshold find little indication for any significant
$\eta^{\prime}p$ interaction. This point is very relevant in the discussion
of possible quasi-bound states of an $\eta^{\prime}$ with a nucleus.

%
%

%
%

\section{Elementary reactions}
\label{sec:elementary}\setcounter{equation}{0}%

In this section we summarize the basic properties of the
production of $\eta$ and $\eta^{\prime}$ mesons from free and
quasi-free nucleons using photon and pion beams. Production
reactions off the proton can be studied with the free protons
contained in hydrogen targets. But, in the case of
photon-induced reactions, the disentanglement of the isospin
composition of the reaction amplitudes requires also
measurements off neutrons. This is only possible in quasi-free
kinematics with neutrons bound in light nuclei, in particular
the deuteron. Therefore we discuss here reactions with both
free protons and also quasi-free nucleons from light nuclei,
when the aim is the study of the elementary cross section for
neutrons.

\subsection{Production in $\boldsymbol{\gamma N}$ reactions}%
\label{s2ec:N_gamma}%
Photon-induced reactions have cross sections that are typical for
electromagnetic interactions and are thus much smaller than those for
reactions induced by hadrons via the strong force. However, over the last two
decades a large effort has been made to study meson production reactions with
real (tagged photons) or virtual (electron scattering) photon beams with
experiments at modern electron-accelerator facilities. Experiments such as
Crystal Barrel/TAPS \cite{Aker_92,Gabler_94} at ELSA \cite{Hillert_06} in
Bonn (Germany), the CLAS detector at Jlab in Newport News \cite{Mecking_03}
(USA), the GRAAL facility at ESRF in Grenoble \cite{Bartalini_05} (France),
the Crystal Ball/TAPS setup \cite{Gabler_94,Starostin_01} at MAMI
\cite{Kaiser_08} in Mainz (Germany), the LEPS facility at SPring-8 in Osaka
\cite{Sumihama_06} (Japan), and the GeV-$\gamma$ experiment at LNS at Tohoku
University \cite{Yamazaki_05} in Sendai (Japan) have dominated this research.
A short overview of these experiments is given in \cite{Krusche_11}.

As a consequence of these efforts, for many final states the
precision of the results from electromagnetic production
reactions may now be as good or even better than those from
hadron beams. This is particularly true for the production of
$\eta$ and $\eta^{\prime}$ mesons. As a side remark, for these
two mesons photon-induced production off protons (which is a
very `clean' production reaction) even becomes important for
the investigation of (rare) meson decays. Recent results from
this programme at MAMI for $\eta$ mesons include measurements
of the slope parameter for the Dalitz plot of the
$\eta\rightarrow 3\pi^0$ decay \cite{Unverzagt_09,Prakhov_09},
of the $\eta$ transition form factor from the $\eta\rightarrow
e^+e^-\gamma$ Dalitz decay \cite{Arguar_14}, and of the rare
$\eta\rightarrow\pi^0\gamma\gamma$ decay \cite{Nefkens_14}.
Programmes for the study of rare $\eta^{\prime}$ decays are
under way at ELSA and at MAMI.

In the meantime, the GRAAL experiment has been shut-down and the CLAS
detector has stopped data taking and is under preparation for the Jlab 12-GeV
upgrade (though many observables measured by CLAS for different reactions are
still under analysis and will only become available over the next few years).
Programmes for the measurement of single and double polarization observables
are still under way at the ELSA and MAMI facilities, where the emphasis is
gradually shifting towards the neutron target. These two experiments will
fill the gap in the data base for photoproduction reactions with more than
one neutral particle (nucleon or meson) in the final state.

The experimental approaches at these laboratories were in general different
(and complementary), especially for the reactions discussed in this review
($\gamma N\rightarrow N\eta$, $\gamma N\rightarrow N\eta^{\prime}$, $\gamma
N\rightarrow N\eta\pi$). The main features of the facilities that are most
important for the experiments discussed here are:
\begin{itemize}
\item{
The CEBAF Large Acceptance spectrometer (CLAS) \cite{Mecking_03} was
housed in Hall B of the Thomas Jefferson National Accelerator Facility in
Newport News.
This experiment was optimized for the detection of charged particles
(protons, charged pions,...) using a magnetic field. The resolution for
the momenta of charged particles was thus superior to the other
experiments. Neutral particles, such as neutrons or mesons decaying into
photons, could only be identified via missing-mass analysis (only a small
fraction of the total solid angle at forward angles was equipped with an
electromagnetic calorimeter and this provided only a very restricted
acceptance for photons). This experiment therefore achieved excellent
performance for final states with not more than one neutral particle but
could not measure final states with neutral meson production off neutrons
or pairs of neutral mesons (e.g. $\pi^0\pi^0$ or $\pi^0\eta$). The
maximum electron beam energy was 6 GeV; linearly and circularly polarized
beams and longitudinally and transversely polarized targets have been
used.}
\end{itemize}
The other experiments avoided magnetic spectroscopy and employed
electromagnetic calorimeters with (almost) $4\pi$ coverage for the detection
of photons, charged hadrons (protons, charged pions,...), and neutrons. For
these experiments the reaction identification is based on combined invariant
and missing-mass analyses (kinematic fitting is often used for the proton
target).
\begin{itemize}
\item{Crystal Barrel/TAPS at ELSA: the combination of the Crystal Barrel
    (CBB, CsI(Tl) scintillators) \cite{Aker_92} detector with the TAPS
    calorimeter (BaF$_2$ scintillators) \cite{Gabler_94} provides a
    device that can detect and identify photons and recoil nucleons over
    almost the full solid angle. However, in the configuration used up to
    now, only a small part of the calorimeter (at laboratory polar angles
    below 30$^{\circ}$) provided trigger information, because only this
    part (TAPS forward detector and CBB forward plug) was equipped with
    photomultipliers. The main part of the CBB was read out by
    photodiodes, which did not deliver timing and trigger information.
    Additional trigger information was derived from a detector for
    charged particle identification that surrounded the target. Thus, for
    measurements with a free proton target, the recoil proton provided
    the trigger. Measurements off quasi-free neutrons were so far only
    possible for high-multiplicity final states (such as $\eta\rightarrow
    3\pi^0\rightarrow 6\gamma$ or $\eta^{\prime}\rightarrow
    \pi^0\pi^0\eta\rightarrow 6\gamma$), for which the probability of
    some photons hitting the calorimeter at small angles is sufficiently
    large. Currently the CBB is undergoing an upgrade (readout with
    Avalanche Photodiodes) which will eliminate this restriction. The
    maximum electron beam energy is 3.5 GeV; linearly and circularly
    polarized beams are available and transversely and longitudinally
    polarized targets have been used.}
\item{Crystal Ball/TAPS at MAMI: The detector setup is very similar to
    that at ELSA, also covering almost $4\pi$ with the Crystal Ball (CBM,
    NaI scintillators) \cite{Starostin_01} and using TAPS as a forward
    wall (the TAPS calorimeter has been split into two parts, allowing
    simultaneous experiments at ELSA and MAMI). Since the Crystal Ball
    modules are equipped with photomultipliers, this detector can trigger
    on photons over the full solid angle covered. It is thus capable of
    also measuring reactions with low photon multiplicities off
    quasi-free neutrons such as $\gamma n\rightarrow n\eta\rightarrow
    n\gamma\gamma$. The maximum electron beam energy is 1.5 GeV;
    polarized beams and polarized targets are available.}
\item{The GRAAL experiment: The central detector component
    \cite{Bartalini_05} was also an electromagnetic calorimeter
    constructed from BGO modules (covering polar angles from 25$^{\circ}$
    to 155$^{\circ}$) combined with forward detectors (wire chambers,
    time-of-flight hodoscopes from plastic scintillators,
    lead-scintillator sandwich) used mainly for recoil nucleon detection.
    In contrast to the previous experiments, which all used
    bremsstrahlung beams, GRAAL located at the ESRF in Grenoble, used
    laser backscattering. The maximum photon energy was 1.5 GeV; linearly
    and circularly polarized photon beams were available, but no
    polarized targets were used.}
\item{The BGO-OD experiment at ELSA: This experiment \cite{Bantes_14} is
    still in the setup phase. It is mentioned here because experiments to
    search for $\eta^{\prime}$ mesic states have been suggested for this
    unique facility \cite{Nanova_12p}. The main components are the BGO
    calorimeter, previously used in the GRAAL experiment, and a dipole
    magnet (and tracking detectors) at forward angles. It will thus
    combine an almost $4\pi$ coverage for photons with magnetic
    spectrometry for forward-going charged particles. These features
    allow an efficient detection of neutral mesons, using invariant-mass
    analysis for their photon decays, together with high-resolution
    momentum spectrometry for recoil protons.}
\end{itemize}

The main motivation for the study of photon-induced meson production off
nucleons is the investigation of the excitation spectrum of the nucleon,
which is one of the most important testing grounds for Quantum Chromodynamics
(QCD) in the non-perturbative regime (see \cite{Crede_13} for a recent
summary). In the past, most information about excited nucleon states was
gained from pion scattering; even today most states listed in the summary
tables of the Review of Particle Physics (PDG) \cite{PDG_Rev} were
`discovered' in this reaction. Only in the last update of the PDG
\cite{Beringer_12} a few states also appear that were first reported from
photon-induced reactions. This situation will change drastically in the near
future when the large body of data from photon-induced reactions become
available, not only for the $N\pi$ channel, but also for many other final
states.

Photoproduction of $\eta$ and $\eta^{\prime}$ mesons is
complementary to the $N\pi$ final state in several aspects. The
isoscalar $\eta$, $\eta^{\prime}$ are only emitted in the decay
of isospin $I=$1/2 $N^{\star}$ resonances. Furthermore, due to
their large mass, the number of contributing partial waves at a
given incident photon energy is more limited than for pions.
This not only simplifies the model analyses, but it makes these
mesons particularly promising tools for the study of low-spin
$N^{\star}$ states at reasonably high excitation energies. Such
states are at the core of the `missing resonance' problem: many
more states have been predicted by nucleon models than have
been so far observed experimentally~\cite{PDG_Rev}. For most
quantum numbers only the lowest lying state has so far been
observed in experiment though models predict many higher lying
states.

The photoproduction of $\pi\eta$ pairs, like double pion
production, aims at the study of excited nucleon states that
have only a small branching ratio for direct decay to the
nucleon ground state but decay in a cascade via an intermediate
excited state. In such a cascade decay, isospin conservation
allows $\eta$ emission for $N^{\star}\rightarrow N^{(\star)}$
and $\Delta^{\star}\rightarrow \Delta^{(\star)}$ transitions,
while all combinations are possible for the pion emission. As
discussed later, this reaction offers interesting possibilities
for the search for $\eta$-mesic states for nuclei for which
coherent production of single $\eta$ mesons is forbidden or
suppressed.

Photoproduction reactions off protons can be studied with hydrogen targets
(typically liquid H$_2$). However, since the electromagnetic interaction is
charge dependent, measurements off the neutron are also required. The isospin
decomposition for photoproduction of isoscalar mesons like the $\eta$ and
$\eta^{\prime}$ is simple; the amplitudes for the reactions off protons and
neutrons are:
\begin{eqnarray}
\label{eq:iso}
A(\gamma p\rightarrow \eta p) & = & (A^{IS}+A^{AV})\\
A(\gamma n\rightarrow \eta n) & = & (A^{IS}-A^{AV})\nonumber
\end{eqnarray}
with the (photon) isoscalar and isovector amplitudes $A^{IS}$ and $A^{IV}$.
The data base for reactions with neutrons in the initial state is so far much
more limited than for the corresponding reactions off protons. Since free
neutrons are not available, it is unavoidable to extract the information
about the elementary $\gamma n\rightarrow N x$ reactions ($N$: nucleon, $x$:
any meson or mesons) from experiments using nuclear targets. Such
measurements are technically more challenging and, for the interpretation of
the results, several complications must be taken into account.

The experimental issues arise from the necessary coincident
detection of the recoil nucleons. Experiments based on magnetic
momentum analysis of the produced particles are therefore
limited to final states with charged mesons (e.g. $\gamma
n\rightarrow p\pi^-$). Electromagnetic calorimeters that cover
large solid angles are much better suited for this purpose.
However, even in this case, neutron detection (with typical
efficiencies on the order of 10-30\%) reduces the statistical
precision of the results compared to the corresponding
measurements off proton targets. The neutron detection
efficiency, and the possible contamination of the neutron
sample with misidentified protons, introduce additional
systematic uncertainties. The statistical quality of the data
achievable has been much enhanced over the last few years,
mainly due to the ever increasing speed of the data acquisition
systems (this is in particularly true for experiments such as
Crystal Ball/TAPS at MAMI, where the primary beam intensity can
be increased to whatever values the detection system can cope
with). Even measurements of double polarization observables off
neutrons are thus feasible.

Systematic effects from the detection of the recoil neutrons
are most important for cross section measurements (they cancel
at least partially in asymmetries). Although the Monte Carlo
codes (mostly Geant4 \cite{Geant4}) used for the simulation of
detection efficiencies include special program packages for
neutrons, these are not as precise as for protons and by no
means as precise as for electromagnetic showers. Precision can
be increased by comparing the simulated detection efficiencies
to measurements using, for example, reactions like $\gamma
p\rightarrow p\pi^0$, $\gamma p\rightarrow p\eta$ (for recoil
protons) or $\gamma p\rightarrow n\pi^+$, $\gamma p\rightarrow
n\pi^0\pi^+$ (for recoil neutrons) with free protons targets,
where the detection efficiency as function of the nucleon
momentum and polar angle can be directly determined from the
ratios of events with and without coincident nucleons.

Most experiments with light nuclear targets (deuterons, $^3$He) discussed
below used, in addition, the internal consistency of the data. For this
purpose the cross sections of the two exclusive reactions $\gamma
d\rightarrow xp(n)$ ($\sigma_p$) and $\gamma d\rightarrow xn(p)$ ($\sigma_n$)
($x$: any meson or system of mesons, $p$,$n$ detected participant nucleons,
($n$),($p$) non-detected spectator nucleons) are compared to the cross
section $\sigma_{\rm incl}$ of the inclusive reaction $\gamma d\rightarrow x
X$ ($X$: candidate for recoil proton or recoil neutron or not present). The
inclusive cross section accepts all events, independent of whether a
candidate for a recoil nucleon was detected or not. It thus includes, for
example, the roughly 70 - 90\% of events with recoil neutrons that escape
detection. For most reactions discussed below the cross section of the
coherent process $\gamma d\rightarrow dx$ is negligible (or can be determined
and included in the comparison) so that $\sigma_{\rm incl} = \sigma_p +
\sigma_n$ must hold. Since $\sigma_{\rm incl}$ is independent of recoil
nucleon detection efficiencies this comparison provides a powerful cross
check.

The interpretation of the results obtained in quasi-free kinematics is
complicated by two factors. All results depend on the momentum distribution
assumed for the bound nucleons and final state interaction (FSI) processes
may obscure the properties of the elementary reactions off free nucleons.

The importance of Fermi motion effects can vary significantly. For example,
cross section measurements at energies away from the production threshold of
the investigated process will be only somewhat smeared out with respect to
the free reactions. In regions where the cross section does not show sharp
variations or narrow structures, such effects can easily be accounted for by
folding model results with the assumed nucleon momentum distributions. On the
other hand, fast variations of the cross sections can be strongly suppressed
and asymmetries can be significantly modified by the Fermi motion.

For differential cross sections and asymmetries it is important
to extract the meson polar and azimuthal angles in the true
photon-nucleon centre-of-momentum (c.m.) frame, taking into
account the incident nucleon momenta, so that the $z$-axis is
no longer along the photon beam axis. Fortunately, most
experiments allow a complete kinematic reconstruction of the
Fermi momenta. For a typical $\gamma d\rightarrow x N_1(N_2)$
reaction the kinematics of the initial state (photon with
energy $E_{\gamma}$ moving along beam axis and hitting a
deuteron at rest) is completely determined. In the final state
the three-momentum $\vec{p}_{x}$ and the mass $m_{x}$ of the
meson (or system of mesons) $x$ are known. For the detected
participant nucleon $N_{1}$ all angles and its mass are known.
This means that the direction $\vec{p}_{N_1}/|\vec{p}_{N_1}|$
of its momentum is measured. In the case of recoil protons,
their kinetic energy can often also be extracted from the
energy deposition in the calorimeter, but this is not true for
recoil neutrons. Finally, for the undetected spectator nucleon
($N_2$) only its mass is known. This means that four kinematic
quantities in the final state - the kinetic energy of the
participant nucleon and the three-momentum of the spectator -
are missing. These four quantities can be extracted from the
four equations that follow from energy-momentum conservation.
The four-momenta of all three final state particles can thus be
reconstructed and used to determine the effective c.m.\ system
and the effective total energy $W$ in that system in plane-wave
approximation. The only disadvantage of this method is that the
extraction of $W$ depends on the experimental resolution
(angles, deposited energies) of the production detector while,
for tagged photon measurements off free protons, it depends
only on the momentum resolution for the scattered electrons in
the tagging device, which is usually much better (for example
for Crystal Ball/TAPS at MAMI a few MeV from tagging compared
to a few 10 MeV from kinematic reconstruction).

Final state interactions between the two recoil nucleons, a
nucleon and a meson, or even three-particle effects, show two
completely different faces for the research reviewed in this
paper. For the measurement of elementary reactions off neutrons
discussed in this section, they present difficulties that must
be understood and corrected for as far as possible. On the
other hand, FSI effects offer the only access to the
interaction of short lived mesons (which cannot be prepared as
secondary beams) with nucleons. The experiments aiming at
meson-nucleus bound states discussed later in this review are
entirely based on meson-nucleon (meson-nucleus) FSI.

The influence of FSI on the elementary reactions is so far not very well
understood. It depends strongly for example on the reaction channels and its
effects are also probably rather different for different observables for the
same reaction channel. Experimentally one can, of course, get some
information about FSI effects from a comparison of the results for a
quasi-free $\gamma d\rightarrow xp (n)$ reaction with free
$\gamma p\rightarrow xp$ data. It might even be tempting to use the deviations
between these two reactions as a correction to approximate the observables of
the $\gamma n\rightarrow xn$ reaction by the results measured for the
quasi-free $\gamma d\rightarrow xn (p)$ meson production, but this
approximation is certainly not always valid.

Examples for strong FSI effects reported in the literature are, for example,
those for pion photoproduction off quasi-free nucleons
\cite{Krusche_99,Darwish_03,Tarasov_11,Dieterle_14}. However, for the same
reactions (e.g., $\gamma d\rightarrow (n)p\pi^0$) effects for polarization
observables seem to
be less important \cite{Krusche_14}. As it turns out (see results discussed
below) for single $\eta$ and $\eta^{\prime}$ meson photoproduction FSI
effects are not even important for cross sections but play a substantial role
for the photoproduction of $\eta\pi$ pairs.

\subsubsection{\it Photoproduction of $\eta$-mesons off nucleons}
\label{s3ec:N_gamma_eta_free}

\begin{figure}[h!]
\centerline{
\epsfig{file=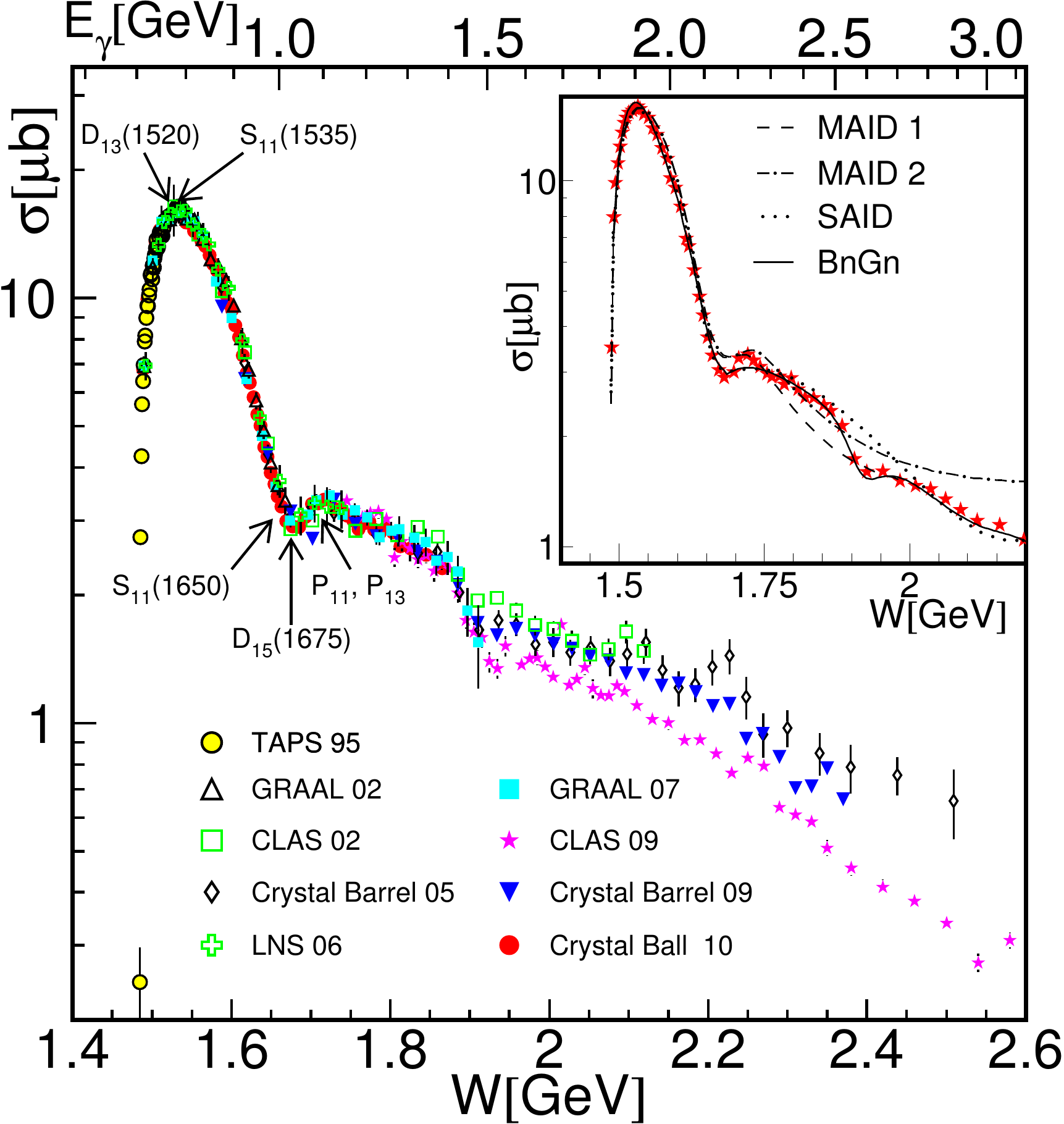,scale=0.5}
\epsfig{file=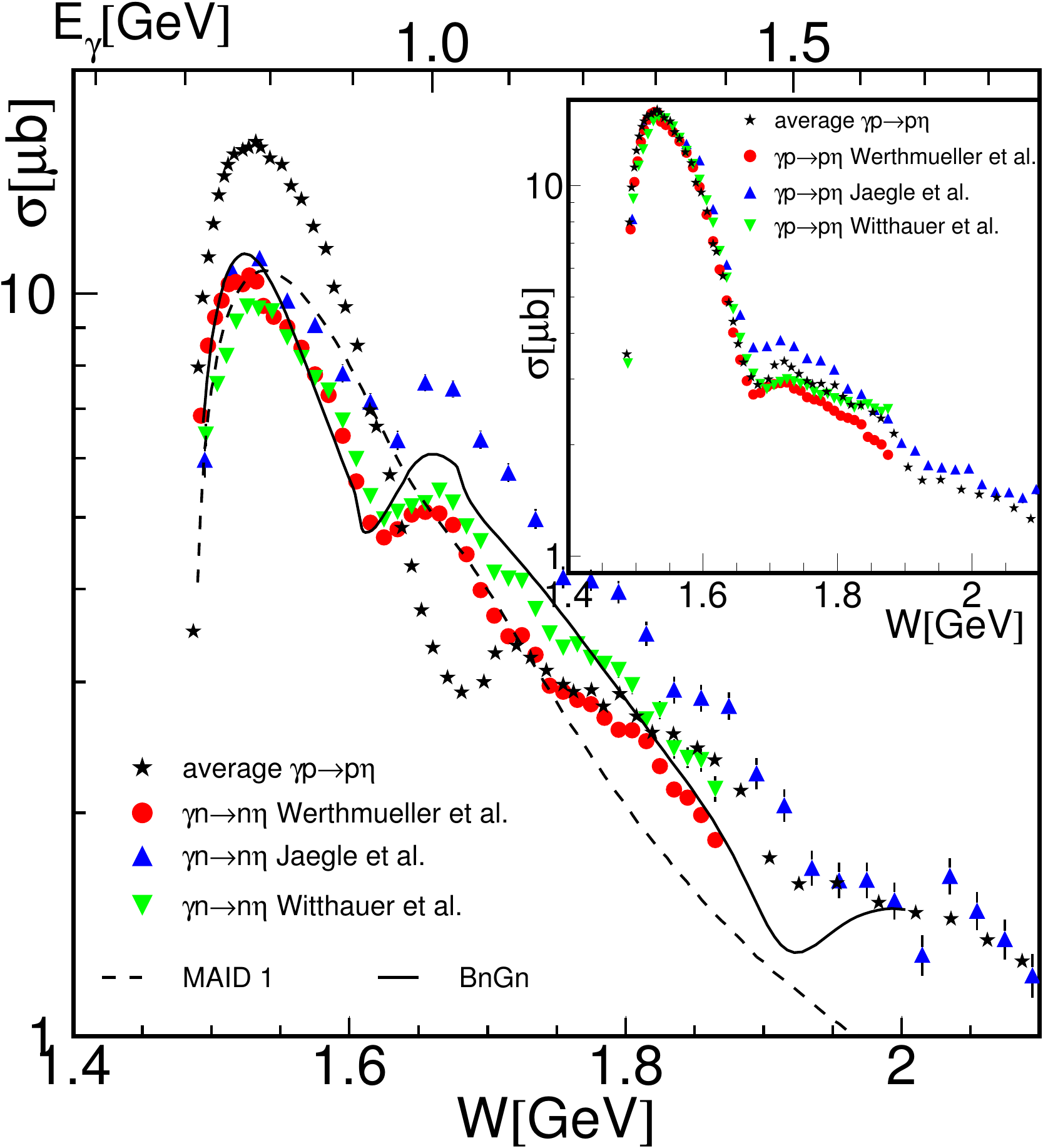,scale=0.5}
}
\vspace*{0.75cm}
\epsfig{file=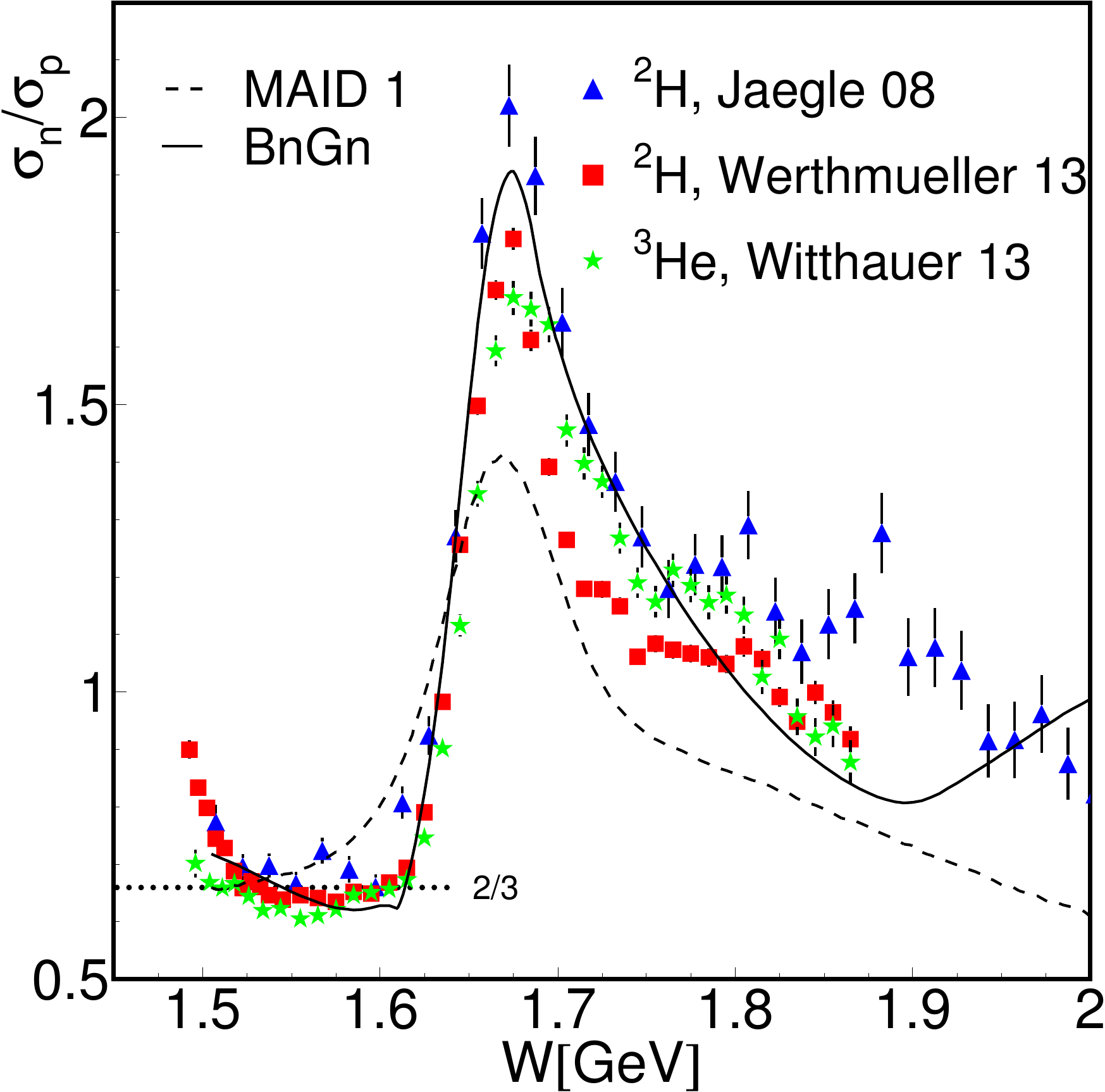,scale=0.5}

\vspace*{-9.2cm} \hspace*{9.5cm}\begin{minipage}{8.5cm}
\caption{Upper left corner: Total cross section for $\gamma
p\rightarrow p\eta$ for free protons. Main figure shows data
from \cite{Krusche_95} (TAPS 95), \cite{Renard_02} (GRAAL 02),
\cite{Dugger_02} (CLAS 02), \cite{Crede_05} (Crystal Barrel
05), \cite{Nakabayashi_06} (LNS 06), \cite{Bartalini_07} (GRAAL
07), \cite{Williams_09} (CLAS 09), \cite{Crede_09} (Crystal
Barrel 09), and \cite{McNicoll_10} (Crystal Ball 10). The
insert shows the average of the data, together with model
results from \cite{Chiang_02} (MAID~1), \cite{Chiang_03} (MAID~2),
\cite{McNicoll_10} (SAID), and
\cite{Anisovich_12a,Anisovich_12} (BnGa). Upper right corner:
Total cross section for $\gamma n\rightarrow n\eta$ for
quasi-free photoproduction off neutrons bound in the deuteron
or $^3$He nuclei (see text) from
\cite{Jaegle_11,Werthmueller_13,Witthauer_13}. Model curves
from MAID 1 (solid) \cite{Chiang_02} and BnGa
\cite{Anisovich_13}. Insert: cross sections for quasi-free
reactions off protons from same references compared to average
free-proton cross section. All quasi-free $^3$He data scaled
up by a factor of 1.4. Bottom left: ratios of quasi-free
neutron and proton cross sections for data and models shown in
the earlier panels. \label{fig:eta_tot}}
\end{minipage}
\end{figure}

The photoproduction of $\eta$-mesons off free protons was studied at all the
major tagged photon facilities, sometimes even with repeated and improved
experiments. Quasi-free $\eta$ production off neutrons has also recently
attracted much interest. We first summarize the available data base. The
total cross section data are shown in Fig.~\ref{fig:eta_tot}. For
publications that quoted only values for the differential cross sections (as
long as they covered a reasonable range in the polar angle) the total cross
section was extracted from fits to the angular distributions (see below).

\begin{figure}[t]
\begin{center}
\epsfig{file=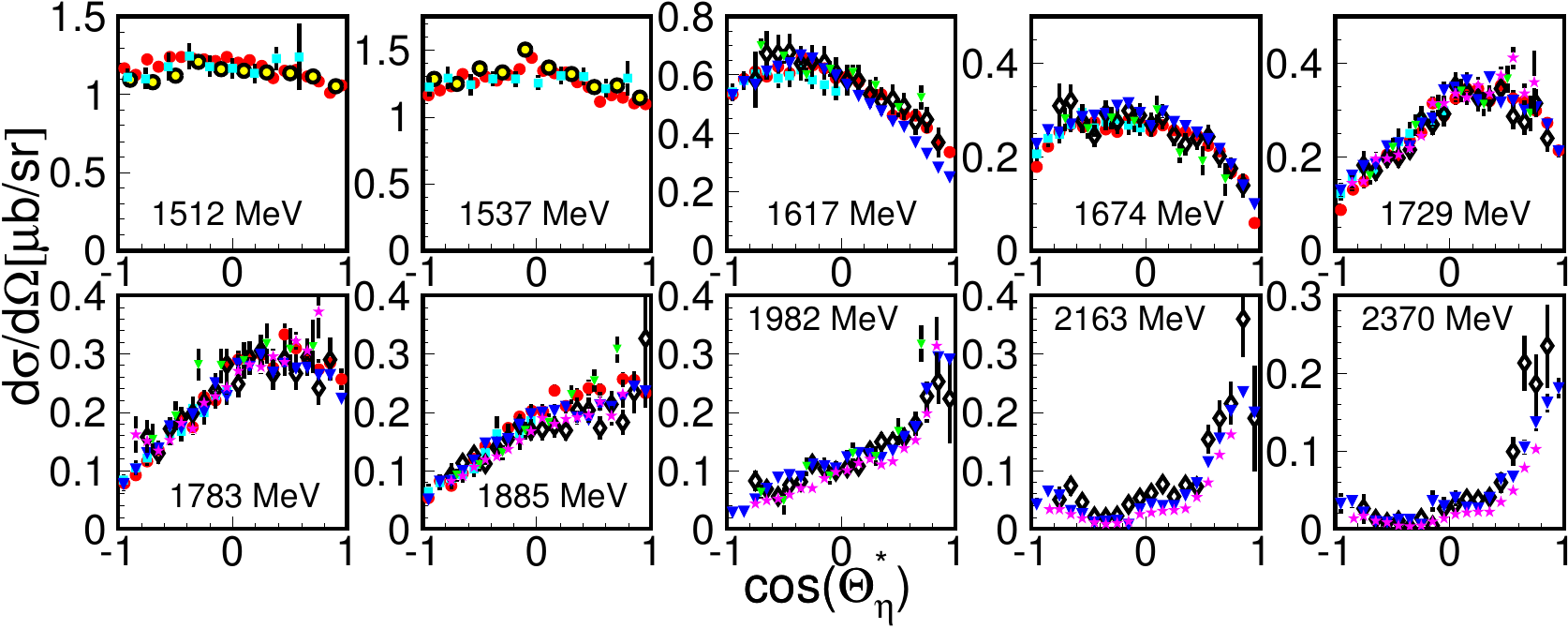,scale=1.0}
\begin{minipage}[t]{16.5 cm}
\caption{Comparison of angular distributions from different measurements for
some typical c.m.\ energies $W$ for $\gamma p\rightarrow p\eta$. Symbols are as
in Fig.~\ref{fig:eta_tot}. \label{fig:eta_p_diff}}
\end{minipage}
\end{center}
\end{figure}

\newpage
Angular distributions for the $\gamma p\rightarrow p\eta$
reaction have been reported in
\cite{Krusche_95,Renard_02,Dugger_02,Crede_05,Nakabayashi_06,Bartalini_07,Bartholomy_07,Crede_09,Williams_09,Sumihama_09,McNicoll_10}	
(we have omitted some older references that can be found in
\cite{Krusche_03}). Typical examples that demonstrate the level
of agreement between the experiments are shown in
Fig.~\ref{fig:eta_p_diff}. Several different experimental
approaches have been used so that the sources of systematic
error are different. The first CLAS experiment \cite{Dugger_02}
identified the $\eta$ mesons through a missing-mass analysis of
the recoil protons and determined the absolute normalization of
the cross sections by comparing the data for single $\pi^0$
production (analyzed in the same way) with the results of the
SAID partial wave analysis. This CLAS experiment included all
the $\eta$ decay channels. The second CLAS experiment
\cite{Williams_09} used the $\eta\rightarrow \pi^+\pi^-\pi^0$
decay channel, detected the two charged pions and the recoil
proton, reconstructed the $\pi^0$ in a kinematic fit, and then
studied the invariant mass of the three pions. The cross
sections were absolutely normalized.

\begin{figure}[h!]
\begin{center}
\epsfig{file=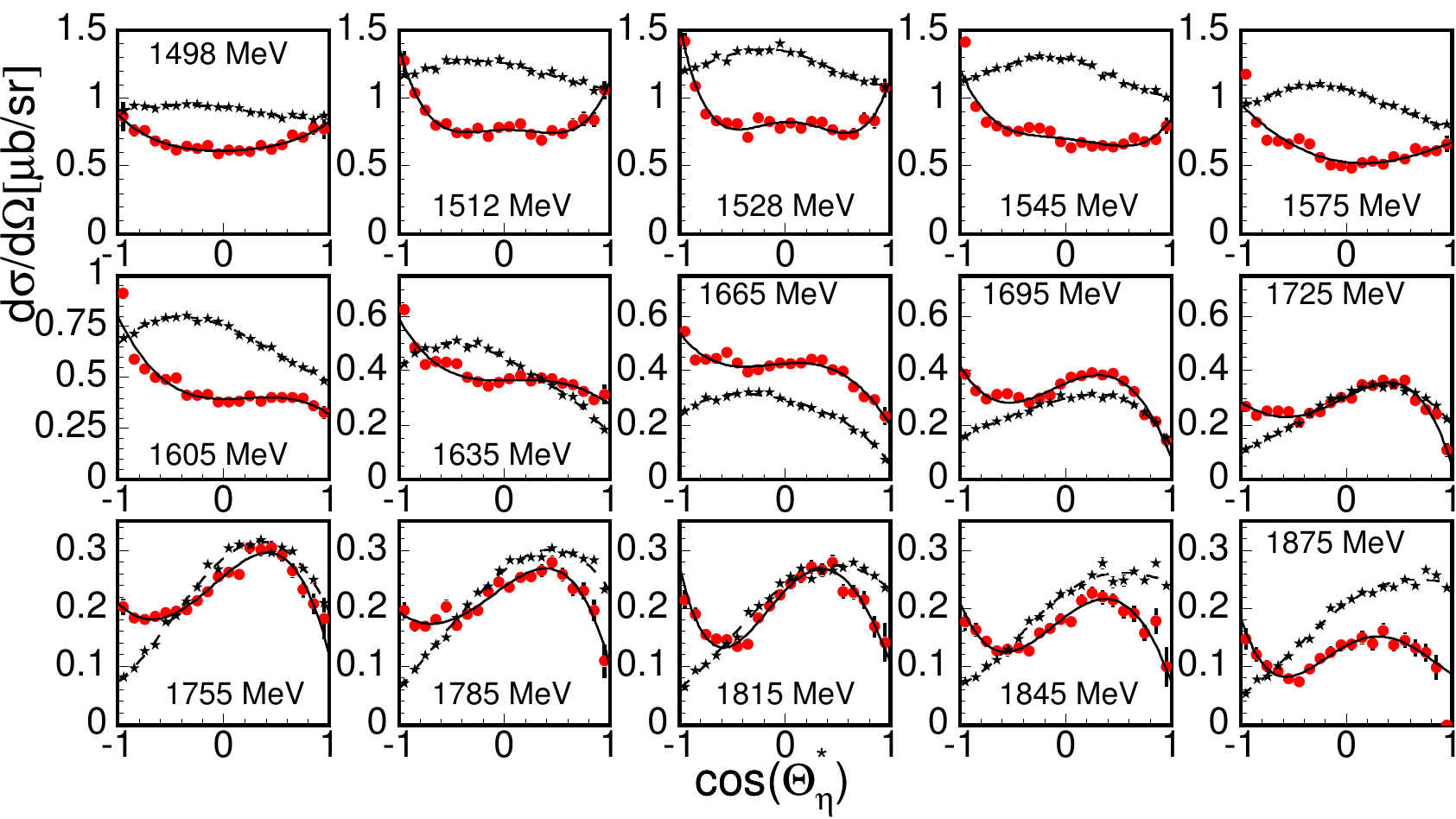,scale=1.0}
\vspace*{-0.5cm}
\begin{minipage}[t]{16.5 cm}
\caption{Angular distributions for $\gamma n\rightarrow n\eta$
\cite{Werthmueller_13} (red dots) compared to the results for $\gamma
p\rightarrow p\eta$ from \cite{McNicoll_10} (black stars). Solid and dashed
curves are fits to the data with Eq.~(\ref{eq:diff_legendre}).
\label{fig:eta_n_diff}}
\end{minipage}
\end{center}
\end{figure}

\begin{figure}[htb]
\centerline{
\epsfig{file=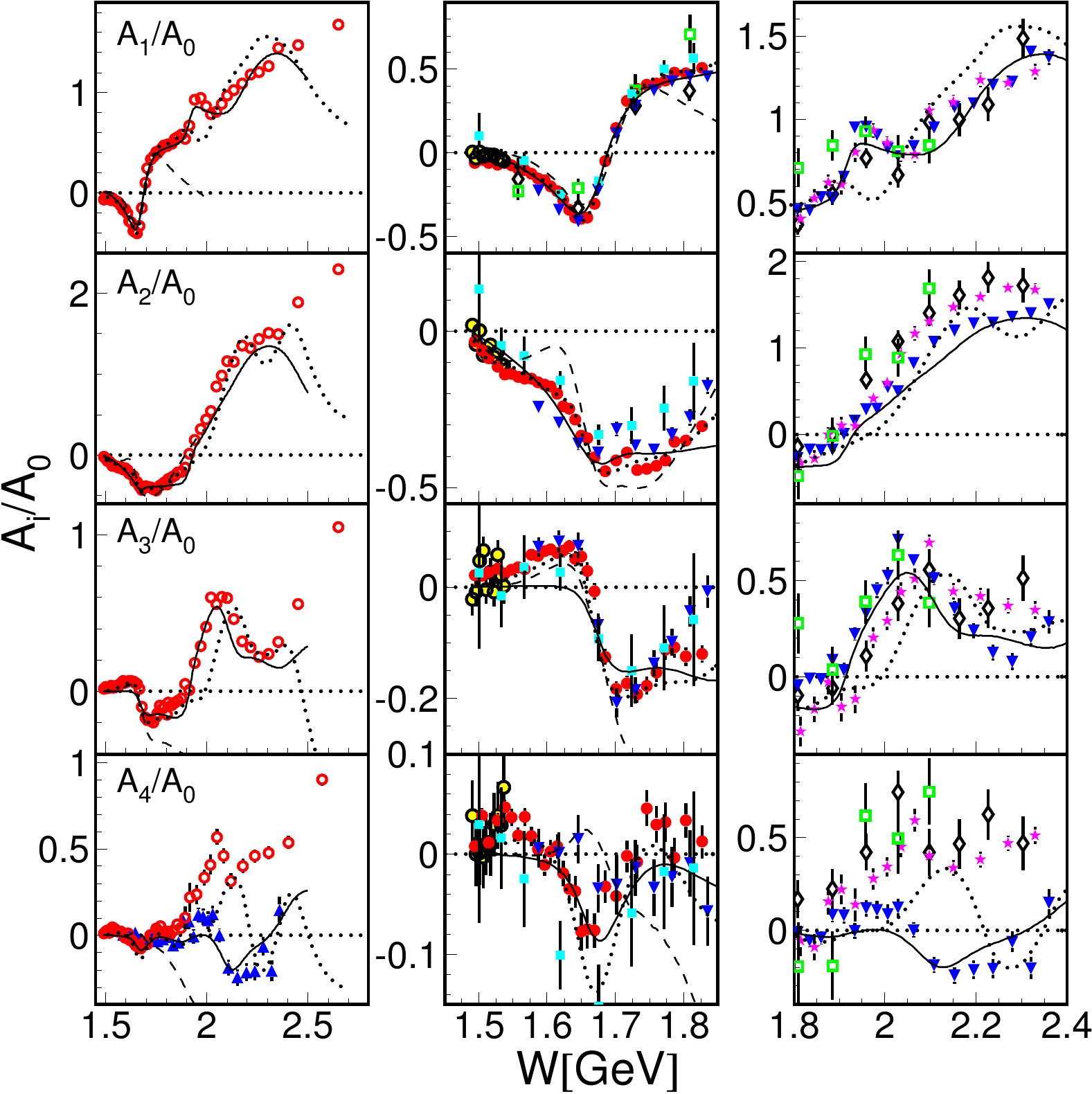,scale=0.88}
}
\centerline{
\begin{minipage}[t]{16.5 cm}
\caption{Legendre coefficients obtained from fits to the $\gamma p\rightarrow
p\eta$ angular distributions using Eq.~(\ref{eq:diff_legendre}). Left column:
average of all experimental data compared to MAID 1 \cite{Chiang_02}, SAID
\cite{McNicoll_10}, and BnGa \cite{Anisovich_12a,Anisovich_12} (notation as
in Fig.~\ref{fig:eta_tot}). The $A_4/A_0$ data from Ref.~\cite{Crede_09} is
not included in the average but shown separately. Central column: individual
data sets for low energy data (symbols as in Fig.~\ref{fig:eta_tot}). Right
column: same for high energy data. \label{fig:p_legendre}}
\end{minipage}
}
\end{figure}

The first Crystal Barrel experiment
\cite{Crede_05,Bartholomy_07} used the $\eta\rightarrow
2\gamma$ and $\eta\rightarrow 3\pi^0\rightarrow 6\gamma$ decay
channels of the $\eta$ meson (analyzed by invariant mass after
kinematic fitting) but triggered on the recoil proton. This
experiment also used the SAID $\pi^0$ results for absolute
normalization. The second experiment \cite{Crede_09} was
similar, but used an absolute measurement of the photon flux
and target density for normalization. Also in the GRAAL
experiment \cite{Bartalini_07} the two- and six-photon decay
channels of the $\eta$-meson were analyzed and an absolute
normalization was made. All the experiments discussed above
detected the coincident recoil protons (which is useful to
ensure overdetermined kinematics but also introduces a further
source of systematic uncertainty). The two MAMI experiments used
the two-photon decay of the $\eta$ \cite{Krusche_95} or the six-photon
decay \cite{McNicoll_10} and did not require detection of the
recoil proton. Both experiments were absolutely normalized.

\begin{figure}[tb]
\begin{center}
\epsfig{file=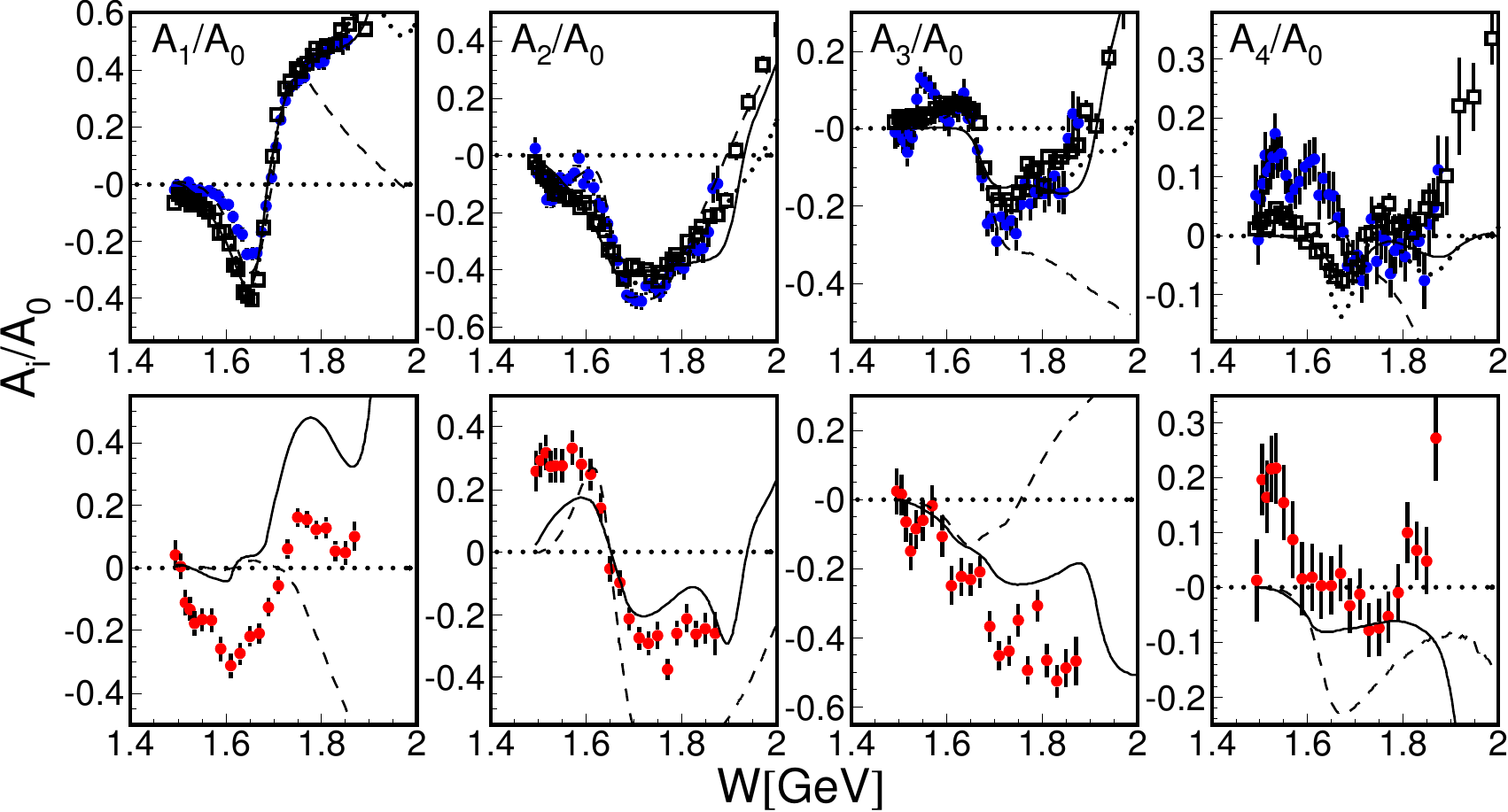,scale=1.0}
\begin{minipage}[t]{16.5 cm}
\caption{Upper row: Coefficients of the Legendre series of
Eq.~(\ref{eq:diff_legendre}) for $\gamma p\rightarrow p\eta$ for a free
proton target (averaged over all experimental data, as in the left column of
Fig.~\ref{fig:p_legendre}, black open squares) compared to the results for
quasi-free $\gamma d\rightarrow \eta p(n)$
\cite{Werthmueller_13,Werthmueller_14} (blue dots). Model curves are from
MAID 1 \cite{Chiang_02} (dashed), BnGa \cite{Anisovich_12a,Anisovich_12}
(solid), and SAID \cite{McNicoll_10} (dotted). Lower row: coefficients for
quasi-free $\gamma d\rightarrow \eta p(n)$
\cite{Werthmueller_13,Werthmueller_14} compared to model results for MAID 1
\cite{Chiang_02} (dashed) and BnGa \cite{Anisovich_13} (solid).
\label{fig:eta_n_diff_coeff}}
\end{minipage}
\end{center}
\end{figure}

The angular distributions were fitted in a series of Legendre
polynomials up to fourth order:
\begin{equation}
\frac{d\sigma}{d\Omega}(W, \CT) = \frac{q_{\eta}^{*}(W)}{k_{\gamma}^{*}(W)}
\sum_{i=0}^{N} A_i(W) P_i(\CT)\,,
\label{eq:diff_legendre}
\end{equation}
where the coefficient $A_{0}$ is related to the total cross
section $\sigma(W)$ by $A_{0}(W) = 4\pi
(k_{\gamma}^{*}(W)/q_{\eta}^{*}(W)) \sigma (W)$. Here
$q_{\eta}^{*}$ and $k_{\gamma}^{*}$ are c.m.\ momenta of the
$\eta$ and $\gamma$, respectively. The coefficients for fits
with $N$=4 normalized to $A_{0}$ are summarized in
Figs.~\ref{fig:p_legendre} and \ref{fig:eta_n_diff_coeff}.

The measurement of angular distributions and total cross
sections for the quasi-free $\gamma n\rightarrow n\eta$
reaction has recently made much progress. Early experiments
measured only the inclusive reaction $\gamma d\rightarrow \eta
(np)$ (without the detection of recoil nucleons)
\cite{Krusche_95b} or detected only the recoil nucleons using
missing-mass analysis \cite{Hoffmann_97}. The first fully
exclusive measurement, which already used a complete
reconstruction of the final state kinematics to remove the
Fermi motion effects, was done in the threshold region
\cite{Weiss_03}. This reaction generated a lot of interest when
the first exclusive measurements \cite{Kuznetsov_07} at higher
incident photon energies carried out at the GRAAL facility
revealed a pronounced, narrow structure at incident photon
energies around 1 GeV, which is absent for the $\gamma
p\rightarrow p\eta$ reaction. Subsequently, the quasi-free
production of $\eta$ mesons off neutrons was studied in detail
with deuteron targets at LNS in Tohoku \cite{Miyahara_07}, at
ELSA in Bonn \cite{Jaegle_08,Jaegle_11}, and at MAMI in Mainz
\cite{Werthmueller_13,Werthmueller_14}. At MAMI it was also
studied for neutrons bound in $^3$He nuclei
\cite{Witthauer_13}. The experimental strategies were also
different. The Tohoku experiment measured the inclusive
quasi-free reaction $\gamma d\rightarrow \eta (np)$ and
subtracted the cross section of the $\gamma p\rightarrow p\eta$
reaction, after folding the latter with the momentum
distribution of the bound nucleons. The experiments at GRAAL,
ELSA, and MAMI measured the $\eta$-mesons in coincidence with
the recoil neutrons and extracted the cross sections as
function of the incident photon energy (obtaining Fermi-smeared
results) and as function of kinematically reconstructed total
c.m.\ energy $W$, thus eliminating Fermi motion effects.
Results for the latter type of analysis are summarized in
Figs.~\ref{fig:eta_tot}, \ref{fig:eta_n_diff}, and
\ref{fig:eta_n_diff_coeff}.

Before we discuss the experimental results, we first give a
short summary of existing modelling approaches, which are quite
diverse. For the reaction off free protons, they include the
SAID partial wave analysis \cite{McNicoll_10} (and references
therein), effective Lagrangian models of different types such
as the MAID isobar model (MAID 1) \cite{Chiang_02} and the
Reggeized MAID isobar model (MAID 2) \cite{Chiang_03},
different coupled channel models such as the Giessen model
\cite{Shklyar_13}, the Bonn-Gatchina (BnGa) model
\cite{Anisovich_12a,Anisovich_12}, and even an approach based
on the constituent quark model \cite{Saghai_01}. The $\gamma
n\rightarrow n\eta$ reaction was also analyzed in the framework
of the MAID models \cite{Chiang_02,Chiang_03} and with the BnGa
coupled-channel analysis \cite{Anisovich_13}.

The experimental results for differential cross sections off
free protons are in good overall agreement. The largest
systematic discrepancy occurs in the fits of the angular
distributions for the $A_4$ coefficient (see
Fig.~\ref{fig:p_legendre}) where above $W\approx$ 1.8 GeV
the results from Ref.~\cite{Crede_09} disagree with all other
measurements. Here one should note, however, that this
measurement is the only one that reports results for extreme
forward angles with small statistical uncertainties. The $A_4$
coefficient is very sensitive to these extreme angles, so that
the deviation could also be a reflection of the other
measurements not having sufficient coverage in this region.

The agreement between the measurements of the quasi-free
production off nucleons bound in light nuclei is in part less
good. The Fermi-smeared cross sections as function of the
incident photon energy from the ELSA \cite{Jaegle_11} and the
MAMI \cite{Werthmueller_13,Werthmueller_14} experiments are in
good (for participant protons) and in reasonable (for neutrons)
agreement (see Fig.~\ref{fig:d_compa}, left hand side in
Appendix A). However, the cross sections extracted with the
reconstructed kinematics from the ELSA and the MAMI experiments
agree only in the range of the $S_{11}$ resonance peak, for
quasi-free protons as well as for neutrons, but they deviate at
higher energies, where the ELSA cross sections are larger. The
effect is similar for protons and neutrons. The
$\sigma_n/\sigma_p$ cross section ratios, also shown in
Fig.~\ref{fig:eta_tot}, are not much different and certainly
agree within experimental systematic uncertainties. The problem
seems to be related to an energy-dependent normalization issue
for the kinematically reconstructed ELSA data~ \cite{Jaegle_11}
(see Appendix A for a detailed discussion). Here one should
note that these data included only four points per angular
distribution compared to the 20 points for
\cite{Werthmueller_13,Werthmueller_14}. Detailed angular
distributions for the ELSA experiment \cite{Jaegle_11} were
only given for the Fermi-smeared $E_{\gamma}$ data, which agree
with the MAMI data within systematic uncertainties.

No significant FSI effects have been noted, within experimental
uncertainties, from the comparison of free and quasi-free $\gamma
p\rightarrow p\eta$ data measured with a deuterium target. The quasi-free
total cross section from \cite{Werthmueller_13} is close to the free proton
results and also the Legendre coefficients from the fits of the angular
distributions shown in Fig.~\ref{fig:eta_n_diff_coeff} are very similar.
Deviations occur only for the high order $A_4$ coefficient at low energies,
where systematic uncertainties due to the detection of the recoil protons are
difficult to control for certain angular regions, as discussed in
\cite{Werthmueller_14}. Also for the ELSA deuterium measurement
\cite{Jaegle_11} there is agreement between the measured quasi-free proton
cross section and free proton data after these have been folded with the
deuteron Fermi motion. This is not the case for the $^3$He target for which
quasi-free cross sections for protons as well as for neutrons are suppressed
by roughly a factor of 1.4 compared to free proton data (or quasi-free proton
and neutron data from the deuteron target). Some possible reasons for this
are discussed in
\cite{Witthauer_13}. Fermi motion effects are much stronger for the $^3$He
target and a kinematic reconstruction is only possible under the assumption
that there is no relative momentum between the two spectator nucleons. FSI
effects may be significantly larger than for the deuteron but there are so
far no model calculations for such effects. These data have only been used to
demonstrate that the general behaviour of the quasi-free neutron cross
section, in particular the narrow structure around incident photon energies
of 1 GeV, does not depend much on the target nucleus. The $^3$He results have
been scaled up by a factor of 1.4 in Fig.~\ref{fig:eta_tot}.

We discuss the experimental results for three different ranges
of incident photon energy, i.e., for different total c.m.\
energies $W$. These are the threshold region that extends
through the $S_{11}$(1535) peak, the following region up to
$W\approx 2$~GeV, and higher energies. For the proton, as well
as for the neutron target, the threshold region up to incident
photon energies of 0.9 GeV ($W\approx 1.6$~GeV) is quite well
understood. The reaction is dominated by the excitation of the
$S_{11}$(1535) resonance \cite{Krusche_97} via the $E_{0+}$
amplitude, leading to the strong peak at $\approx 1.53$~GeV.
This resonance overlaps with the $\eta$-production threshold at
$W=1486$~MeV ($E_{\gamma}^{\rm thr}= 707.8$~MeV) and has an
unusually large branching ratio of around 50\% for the emission
of $\eta$ mesons. It is responsible for the strong $\eta$ -
nucleon interaction in the threshold region (on which the
discussion about possible $\eta$ - nucleus bound states is
based). The properties of the $E_{0+}$ amplitude are reflected
in the energy dependence of the total cross section (almost
perfectly $\propto (E_{\gamma}-E_{\gamma}^{\rm thr})^{1/2}$ in
the threshold region \cite{Krusche_95c,Nikolaev_14}) and in the
shape of the angular distributions (almost isotropic). The
dominance of this state has also allowed a detailed
investigation of the dependence of its electromagnetic coupling
$A_{1/2}$ on the four momentum transfer $Q^2$ in
electroproduction, which shows a surprisingly slow falloff
\cite{Thompson_01}. This feature is important for understanding
the quark model structure of this isobar. The deviation from
isotropy in the angular distributions close to threshold is due
to an interference between the leading $E_{0+}$ multipole and
the $E_{2-}$ and $M_{2-}$ multipoles related to the excitation
of the nearby $D_{13}$(1520) resonance. The interference term
is proportional to ${\rm
Re}[E_{0+}^{\star}(E_{2-}-3M_{2-})]\cos^2\theta_{\eta}^{\star}$
\cite{Krusche_95} and is thus reflected in the $A_2$
coefficient of the Legendre series of
Eq.~(\ref{eq:diff_legendre}). The branching ratio for the
$D_{13}(1520)\rightarrow N\eta$ decay is very small; the
Particle Data Group quotes (0.23$\pm$0.04)\% \cite{Beringer_12}
and this is mainly determined by the beam asymmetries of the
$\gamma p\rightarrow p\eta$ reaction (see below).

The measured ratio of neutron $\sigma_n$ and proton $\sigma_p$
cross sections in this energy range is
$\sigma_n/\sigma_p\approx 2/3$ (see Fig.~\ref{fig:eta_tot}).
The ratio increases towards threshold but this is an artefact
of the plane-wave impulse approximation interpretation. Very
close to threshold, energy and momentum conservation force the
`participant' and `spectator' nucleons to have almost identical
momenta and the kinematics become similar to the coherent
reaction. This means that the simple participant - spectator
picture breaks down and the measured cross section ratio will
approach unity.

The $\sigma_n/\sigma_p$ ratio can be translated into one for
the magnitudes of the electromagnetic couplings of the $S_{11}$
state to protons and neutrons (see \cite{Krusche_03} for
details): $|A_{1/2}^n|/|A_{1/2}^p|\approx 0.8$. This means,
according to Eq.~(\ref{eq:iso}), that either the isoscalar or
the isovector part of the coupling must be small. The relative
sign of the electromagnetic helicity couplings can be fixed in
two ways. The cross section for the coherent reaction $\gamma
d\rightarrow d\eta$ will be much larger when the coupling is
dominantly isoscalar. The experimental results
\cite{Hoffmann_97,Weiss_01} clearly favour a dominant isovector
coupling and thus a negative relative sign. An alternative
access to the sign comes from the interference term between the
$S_{11}$ and $D_{13}$ states. It was shown in
Ref.~\cite{Weiss_03} that, with a few simple approximations,
the interference term is proportional for protons and neutrons
to the products of their helicity couplings $A^N_{1/2}(S_{11})$
and $A^N_{1/2}(D_{13})$. Since the helicity couplings of the
$D_{13}$ state are well known from pion photoproduction and
have the same sign for protons and neutrons, a negative
relative sign for the $S_{11}$ state will result in angular
distributions for $\eta$ production with opposite curvature for
protons and neutrons. The precise angular distributions shown
in Fig.~\ref{fig:eta_n_diff} have this behaviour; the
corresponding $A_2$ coefficients in
Fig.~\ref{fig:eta_n_diff_coeff} have opposite sign in the
relevant energy region. The interference effect is larger for
the neutron because the $D_{13}$ neutron coupling is a factor
of $\approx 2.5$ larger than its proton coupling
\cite{Beringer_12}.

The same arguments can be made for the beam asymmetry $\Sigma$
discussed below. The interference effect is even more
pronounced in this observable \cite{Fantini_08}, but the
relevant term here is $\propto {\rm
Re}[E_{0+}^{\star}(E_{2-}+M_{2-})]$, which involves the product
of the $A_{1/2}$ coupling of the $S_{11}$ with the $A_{3/2}$
coupling of the $D_{13}$. Since both couplings have a relative
negative sign between protons and neutrons, this interference
term has the same sign for the proton and the neutron. All the
more advanced analyses (e.g. MAID \cite{Chiang_02}, BnGa
\cite{Anisovich_13}) come to the same conclusions and indicate
also that contributions from the non-resonant background are
small in this energy range. There is only one experimental
result that does not fit into this picture. This is the target
asymmetry measured by the PHOENICS experiment at ELSA at the
end of the nineties \cite{Bock_98}. As discussed, e.g., in
Ref.~\cite{Krusche_03}, none of the existing models could
reproduce the observed nodal structure of this observable for
incident photon energies below 800~MeV. Tiator and
collaborators \cite{Tiator_99} showed, with a truncated
multipole analysis, that the measured target asymmetry forces a
large and strongly varying phase between the $s$-wave $E_{0+}$
multipole and the $d$-wave $E_{2-}$, $M_{2-}$ multipoles. Since
the first is dominated by the $S_{11}$(1535) excitation and the
latter by the $D_{13}$(1520) state, this would indicate a
strongly varying phase between two resonances with almost
identical excitation energies and thus mean that at least one
of them would show an unusual behaviour. This problem persisted
until very recently when new and statistically much more
precise data for the target asymmetry were obtained at MAMI and
ELSA. The results from MAMI have been published in the meantime
\cite{Akondi_14} (see discussion at the end of this chapter)
and did not confirm the critical nodal structure. The ELSA
results are sill under analysis, but preliminary results seem
to support this finding.

In view of the experiments to search for $\eta$ - nucleus bound states, which
are discussed later, one should note that $\eta$ photoproduction in the
threshold region is completely dominated by the isovector component of the
$E_{0+}$ multipole. This has strong consequences for coherent production
processes of the $\gamma A\rightarrow A\eta$ type, which are the most natural
entrance channel for the formation of bound states. Since $E_{0+}$ is a
spin-flip multipole the final state $s$-wave is forbidden for $J=0$ and,
because it is dominantly isovector, it is strongly suppressed for $I=0$.

The energy region above the $S_{11}$ peak up to $W$ values around 2 GeV is
complicated and far from being understood in terms of nucleon resonance
physics. Above $W\approx$ 1.65 GeV ($E_{\gamma}\approx$1~GeV) the
results from model analyses differ already for the total cross section of the
$\gamma p\rightarrow p\eta$ reaction (see Fig.~\ref{fig:eta_tot}). The
measured angular distributions change rapidly in this energy range. The
Legendre coefficients $A_{1},...,A_4$ (see Fig.~\ref{fig:p_legendre}, central
column) show a strong energy dependence around $W$= 1.65 GeV. The large
and fast varying values of the $A_1$ coefficient point to an
interference between the strong $S_{11}$ contribution and $p$-wave states
($S_{11}$ - $P_{11}$ interference is $\propto \cos\theta_{\eta}^{\star}$).
This effect was also observed in $\eta$ electroproduction \cite{Denizli_07}
and found to be almost independent of the four momentum transfer $Q^2$, which
would mean that the relevant multipoles - $E_{0^+}$ and $M_{1-}$ - have a
similar $Q^2$-dependence. At these energies contributions can be expected
from the $S_{11}$(1650), $D_{15}$(1675), $D_{13}$(1700), $P_{11}$(1710), and
$P_{13}$(1720) resonances. Branching ratio estimates given by PDG
\cite{Beringer_12} are 5 - 15\% for the $S_{11}$(1650), 10--30\% for the
$P_{11}$, (4$\pm$1)\% for the $P_{13}$, and (0$\pm$1)\% for the two $d$-wave
states. The values extracted from different analyses have a significant
spread for some resonances and have recently changed, due to the results from
the analysis of the new photoproduction data in the framework of the BnGa
model \cite{Anisovich_12a,Anisovich_12}.

\begin{figure}[thb]
\begin{center}
\epsfig{file=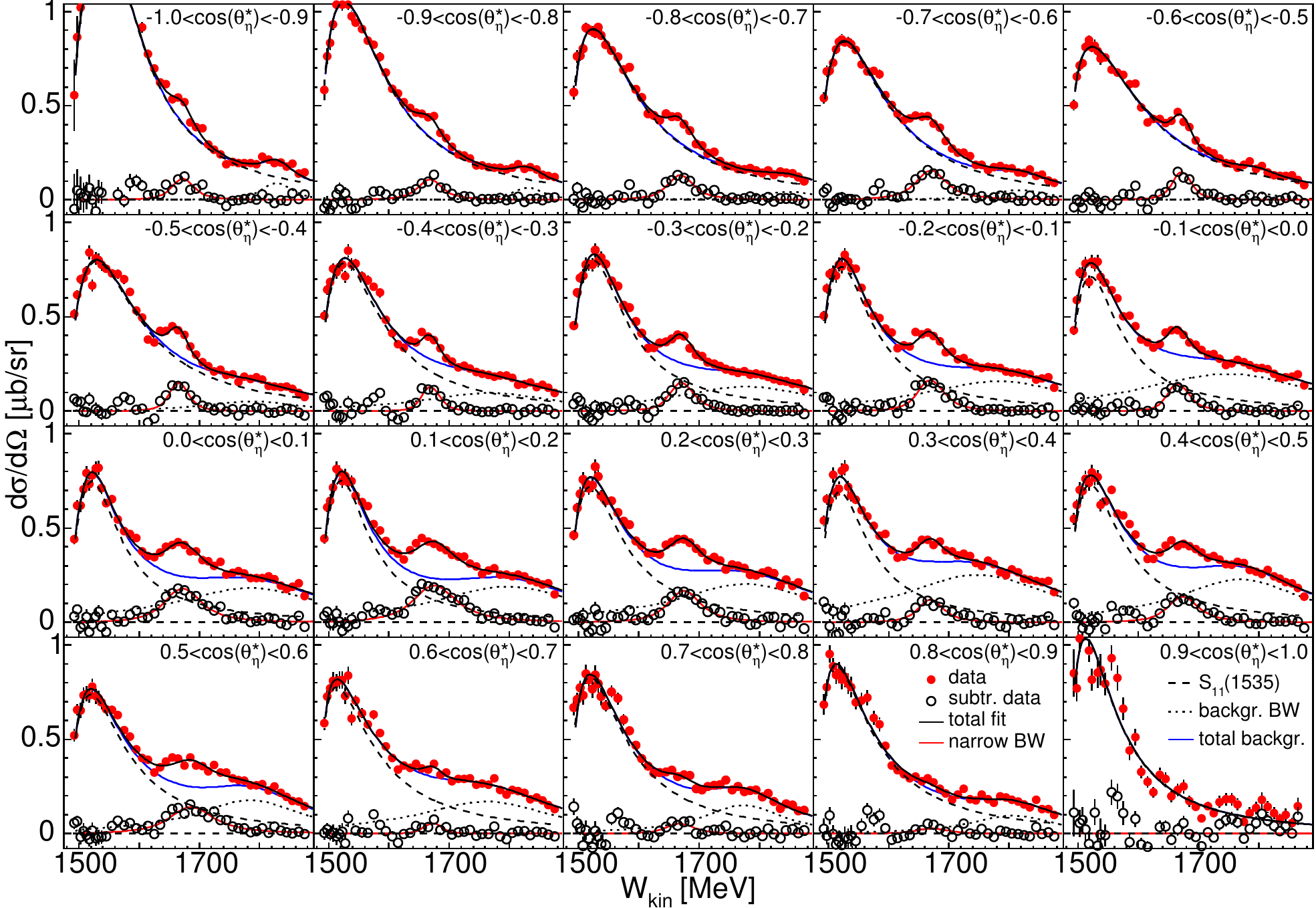,scale=0.85}
\begin{minipage}[t]{16.5 cm}
\caption{Differential cross sections for the $\gamma n\rightarrow n\eta$
reaction \cite{Werthmueller_13} for different bins of polar angle as function
of the $n\eta$ c.m.\ energy $W$. The curves are a phenomenological fit with a
Breit-Wigner resonance for the $S_{11}$ state, a second BW for the narrow
structure, and a third BW for the broad background underneath the narrow
structure (see text). The open symbols represent data with $S_{11}$ and
background fit contributions subtracted. \label{fig:n_bump}}
\end{minipage}
\end{center}
\end{figure}

The picture becomes more complicated by the results for the $\gamma
n\rightarrow n\eta$ reaction. The most striking feature is the extremely
steep rise of the $\sigma_n/\sigma_p$ cross section ratio above $W$ = 1.6
GeV. All experiments that measured total cross sections reported this effect;
the results from \cite{Jaegle_11,Werthmueller_13,Witthauer_13} are compared
in Fig.~\ref{fig:eta_tot}. Interestingly, a strong rise of this ratio was
predicted by the MAID model \cite{Chiang_02}, though not as prominent and
steep as was later observed. This prediction was even among the motivations
for the experimental activities but, as it turns out now, it was made for the
wrong reasons. The MAID fit of the proton data produced an $N\eta$ decay
branching ratio of the $D_{15}$(1675) resonance ($\approx$~17\%) much
larger than any other
analysis; PDG~\cite{Beringer_12} has it compatible with zero. Since this
state is Moorhouse-suppressed \cite{Moorhouse_66} in the electromagnetic
excitation off the proton, it couples more strongly to the neutron. This
state thus made a large contribution to the $\gamma n\rightarrow n\eta$
reaction in MAID. However, a comparison (see Fig.~\ref{fig:eta_tot})
of the total cross section predicted by MAID already shows that the model
misses completely the peak-like structure in the neutron excitation function.
The peak in the cross section ratio results entirely from the `dip' in the
proton excitation function at the same energy. The `dip' is reproduced
because the model was fitted to the proton data, but it is then of course
questionable whether it reflects the correct physics. The description of
the angular distributions (see, e.g., Fig.~\ref{fig:eta_n_diff_coeff})
and the beam asymmetries discussed below (see Fig.~\ref{fig:sigma_coeff})
by the MAID fit, in particular for the neutron, is also not good above
$W$=1.6~GeV so that the extraction of resonance contributions in this
energy range is almost certainly not realistic. A refit of the model
to the data now available would be desirable.

The agreement between data and the BnGa fit is better for all observables,
but this model was also fitted to all the available results. For the $\gamma
n\rightarrow n\eta$ reaction, the BnGa fit included the Fermi-smeared cross
section data from \cite{Jaegle_08} and the beam asymmetries from GRAAL
\cite{Fantini_08}. This explains why the total cross section without Fermi
smearing as function of $W$ is closer to the new MAMI data from
\cite{Werthmueller_13} (which were not yet included in the fit) than to the
older ELSA data from \cite{Jaegle_11}. The fit to the neutron data might be
improved by a refit of the new differential cross section data (this is
already under way). Typical examples for the evolution of the fit results,
due to the continuing improvements of the experimental data base, are the
$P_{11}$(1710) and $P_{13}$(1720) states. The MAID model \cite{Chiang_02},
and also the Giessen coupled-channel model \cite{Shklyar_07}, found a
significant contribution to $\gamma p\rightarrow p\eta$ from the $P_{11}$ and
almost none from the $P_{13}$ resonance. The early fits with the BnGa
coupled-channel model, on the other hand, found no signature for the $P_{11}$
but a large branching ratio of 30\% \cite{Crede_05} for the $P_{13}$. In the
meantime, the situation in the BnGa fit has reversed; in the most recent
version \cite{Anisovich_12a} the branching ratio for the $P_{11}$ has risen
to (17$\pm$10)\% and the one for the $P_{13}$ has dropped to (3$\pm$2)\%.

Some comments should be made about the pronounced narrow
structure in the excitation functions around
$E_{\gamma}\approx1$~GeV. More than ten years ago there were
predictions for such a structure related to the conjectured
baryon anti-decuplet \cite{Diakonov_97} of pentaquarks. Taking
together, the results from
\cite{Diakonov_97,Polyakov_03,Arndt_04,Azimov_05} suggested
that the non-strange $P_{11}$-like member of the anti-decuplet
should be electromagnetically excited more strongly on the
neutron, should have a large decay branching ratio to N$\eta$,
an invariant mass around 1.7 GeV/$c^2$, a width of a few tens
of MeV/$c^2$, and a radiative coupling to the neutron
corresponding to $A_{1/2}^n=15\times 10^{-3}$~GeV$^{-1/2}$.
The structure that was subsequently observed in the $\gamma
n\rightarrow n\eta$ reaction fitted exactly to these
properties. It was seen in {\it all} experiments that looked
for it (even using the in-principle unfavourable $^3$He target
nucleus) and the statistical significance, in particular in the
most recent data \cite{Werthmueller_13}, is beyond any doubt.
In this sense it is very different from the experimental claims
for the observation of the exotic pentaquark, which could not
be reproduced by subsequent experiments. Although the existence
of this structure is unambiguous, its nature is not yet
understood. Phenomenological analyses of the excitation
function found for the position and widths of the structure
values of (we cite the most recent results from
\cite{Werthmueller_13,Werthmueller_14}; other experiments found
comparable values) $W = (1670\pm 5)$ MeV with an intrinsic
width of $\Gamma\approx 30$~MeV. This would be a very unusual
value for a nucleon resonance (which in this energy range
rather have widths above 150~MeV), but a bump in an excitation
function need not necessarily correspond to a nucleon
resonance.

It was already argued in \cite{Jaegle_11} that the `bump' in
$\gamma n\rightarrow n\eta$ and the `dip' (although less
prominent) in $\gamma p\rightarrow p\eta$ are probably related
structures. The structure in the proton excitation function has
been recently investigated in detail with the BnGa coupled
channel analysis \cite{Anisovich_13a}. The main result was that
the narrow `dip' cannot be reasonably well explained by broad
resonances and standard background amplitudes. The fit could be
improved by either introducing a narrow $P_{11}$ state or by a
threshold effect resulting from a strong coupling of the
$\gamma p\rightarrow\omega p$ reaction to the $S_{11}$ partial
wave (however, most recent analyses by the Bonn-Gatchina group
including experimental data for $\omega$ production do not
favour a strong coupling). Various scenarios have been
suggested in the literature for the narrow structure in the
neutron excitation function. They range from different
coupled-channel effects of known nucleon resonances
\cite{Shklyar_07,Shyam_08}, interference effects in the
$S_{11}$ partial wave \cite{Anisovich_09, Anisovich_13},
effects from the opening of strangeness thresholds
\cite{Doering_10}, to intrinsically narrow states
\cite{Arndt_04,Anisovich_09,Choi_06,Fix_07,Shrestha_12}. The
available data are still insufficient for an unambiguous
analysis. However, the precise angular distributions shown in
Fig.~\ref{fig:n_bump} have added some new clues. They do not
favour a scenario where a narrow $P_{11}$ resonance interferes
with the broad $S_{11}$ structures. An interference term
between the $M_{1-}$ multipole from a $P_{11}$ excitation and
the $E_{0+}$ multipole from $S_{11}$ excitations has an angular
dependence $\propto \cos\theta_{\eta}^{\star}$. This means that
the narrow structure should have a maximum contribution either
at extreme forward or at extreme backward angles, a linear
dependence across the angular distribution, and then a minimum
at the opposite extreme angle. Depending on the sign, one would
then expect a maximum bump at forward angle and a dip at
backward angle (or vice versa). But the excitation functions
for different bins of polar angle (see Fig.~\ref{fig:n_bump})
show the most pronounced bump-like structure at intermediate
angles and almost no effect at forward and backward angles.

\begin{figure}[t!]
\begin{center}
\epsfig{file=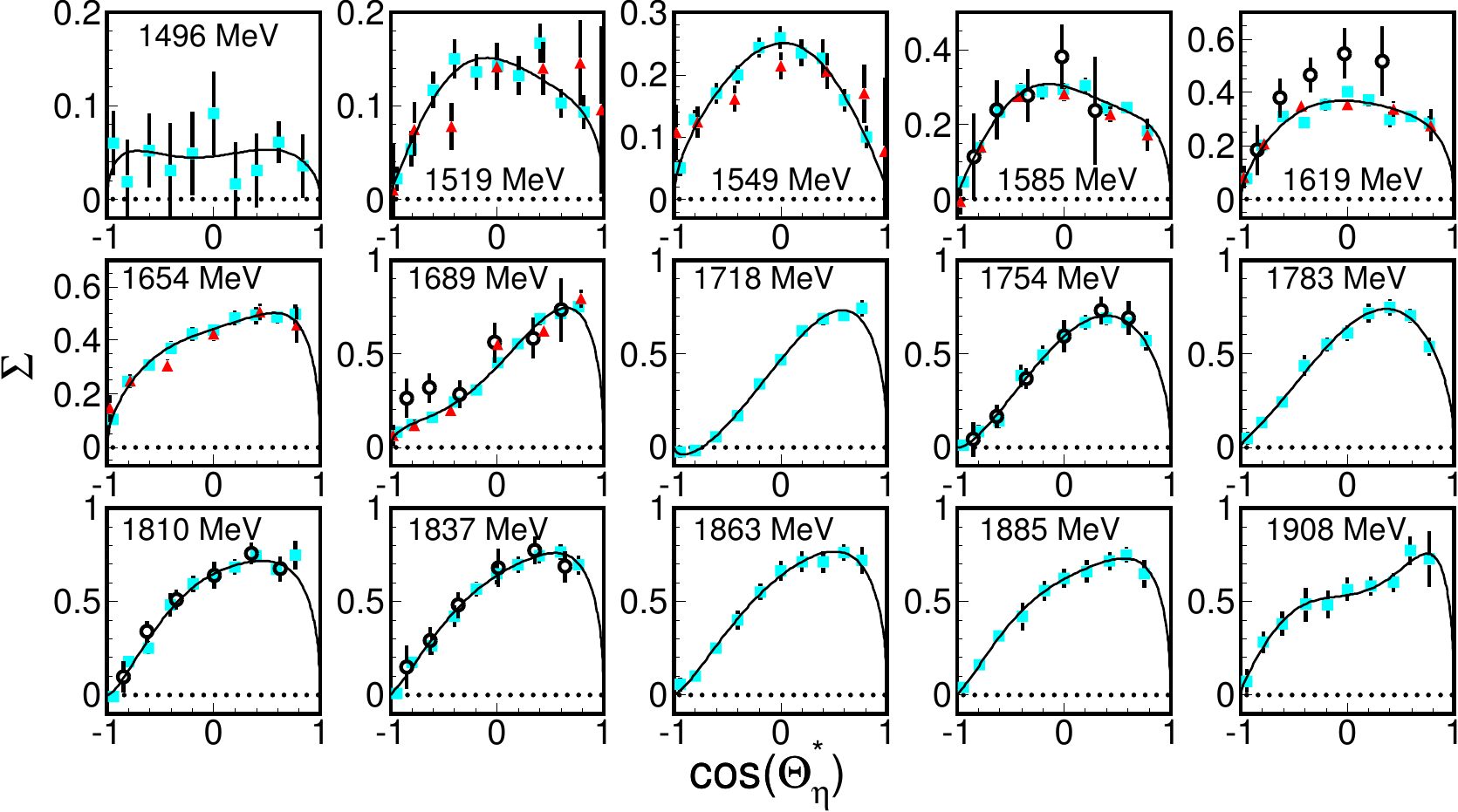,scale=1.}
\epsfig{file=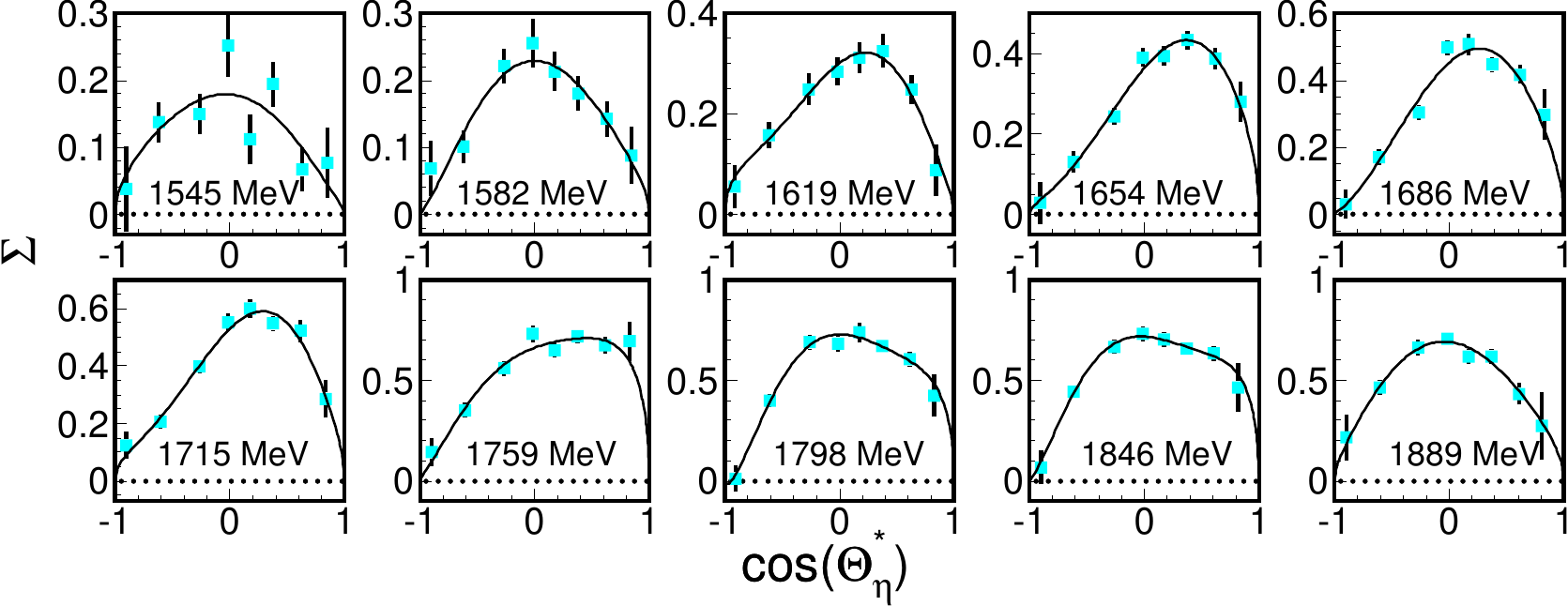,scale=1.}
\begin{minipage}[t]{16.5 cm}
\caption{Top row: Beam asymmetries $\Sigma$ for the free $\gamma p\rightarrow
p\eta$ reaction. Light blue dots \cite{Bartalini_07} and red triangles
\cite{Ajaka_98} from GRAAL experiment, open circles \cite{Elsner_07} from
CBELSA/TAPS. Solid lines: fits with Eq.~(\ref{eq:sigma_legendre}) to data
from \cite{Bartalini_07}. Results are given for the different values of $W$
indicated in the panels. Bottom: Beam asymmetries $\Sigma$ for the quasi-free
$\gamma n\rightarrow n\eta$ reaction from the GRAAL experiment
\cite{Fantini_08}; the solid lines are fits with
Eq.~(\ref{eq:sigma_legendre}). \label{fig:eta_sigma}}
\end{minipage}
\end{center}
\end{figure}

A fit of the BnGa model to the previous ELSA data \cite{Jaegle_08} gave
comparable quality for a scenario with a narrow $P_{11}$ state or
interference effects in the $S_{11}$ channel \cite{Anisovich_09}. A refit
\cite{Anisovich_14} to the more precise angular distributions from the new
MAMI data \cite{Werthmueller_13} preferred the solution without a narrow
$P_{11}$ state, for which the `bump' in the total cross section arises
entirely from subtle interference effects in the $S_{11}$ channel.  This
solution requires, however, a change of sign of the electromagnetic coupling
of the $S_{11}$(1650) resonance for the neutron. Most analyses (MAID, SAID)
previously found a negative sign for the $A^n_{1/2}$ (in the following all
values are in units of $10^{-3}$ GeV$^{-1/2}$) neutron helicity coupling of
this four star state (PDG: $-$15$\pm$21) and thus a destructive interference
between the two $S_{11}$ states for the neutron (as for the proton). A
negative sign was also preferred by quark models (e.g., Capstick
\cite{Capstick_92}: $-35$; Burkert et al. \cite{Burkert_03}: $-$31$\pm$3).
However, the most recent analyses of the Bonn-Gatchina group
\cite{Anisovich_13} (25$\pm$20) and Shresta and Manley \cite{Shrestha_12}
(11$\pm$2) find positive values corresponding to a constructive interference.
This solution might be appealing, in so far as it could explain the
structures in the proton and neutron excitation function with the same
mechanism. However, it requires significant fine-tuning in the $S_{11}$
sector and it remains to be seen whether the results for polarization
observables, which will soon be available, fit this interpretation.

For $W\gtrsim 2$~GeV, $t$-channel contributions from vector mesons exchange
become important and the angular distributions start to peak at forward
angles. These contributions have been analyzed for the proton target in Ref.
\cite{Bartholomy_07}. Data for the neutron target in this energy range have
only been reported from the ELSA experiment for the Fermi-smeared data
\cite{Jaegle_08,Jaegle_11} up to incident photon energies of 2.5~GeV
(corresponding to $W\approx 2.36$~GeV). They also show the pronounced peaking
at forward angles. The large diffractive contributions could obscure the
small signals from the nucleon resonances that have been suggested by the
model fits. However, the interpretation of the data is even more tentative
than at lower photon energies. The first analysis of the ELSA data for
$\gamma p\rightarrow p\eta$ in the framework of the BnGa model
\cite{Crede_05} claimed, for example, contributions from $F_{15}$(2000),
$D_{15}$(2070), $D_{13}$(2080), and $P_{13}$(2200) states. In particular, the
contribution of the $D_{15}$(2070) (which was introduced by this analysis as
a previously unknown nucleon resonance) was strong. In the more recent BnGa
fit \cite{Anisovich_12a, Anisovich_12}, only two states with significant
coupling to $N\eta$ are reported above 2~GeV, viz.\ the $F_{15}$(2000) and
the $D_{15}$(2060), the latter now with a branching ratio of (4$\pm$2)\%.

It is clear that there are still many open questions about the resonance
contributions to $\gamma N\rightarrow N\eta$ above the $S_{11}$(1535) region.
Cross section data alone, even with improved precision, can certainly not
solve this problem since a unique description of photoproduction of a single
pseudoscalar meson off nucleons requires the measurement of at least eight
carefully selected observables \cite{Chiang_97}.

\begin{figure}[tb]
\begin{center}
\epsfig{file=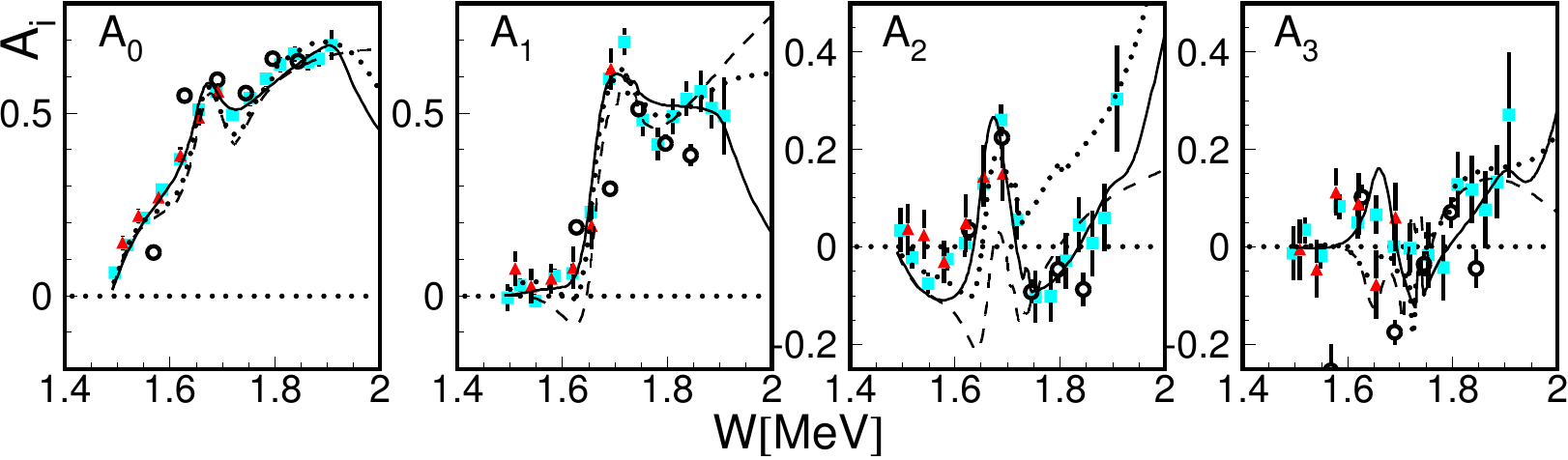,scale=1.0}

\vspace*{0.3cm}
\epsfig{file=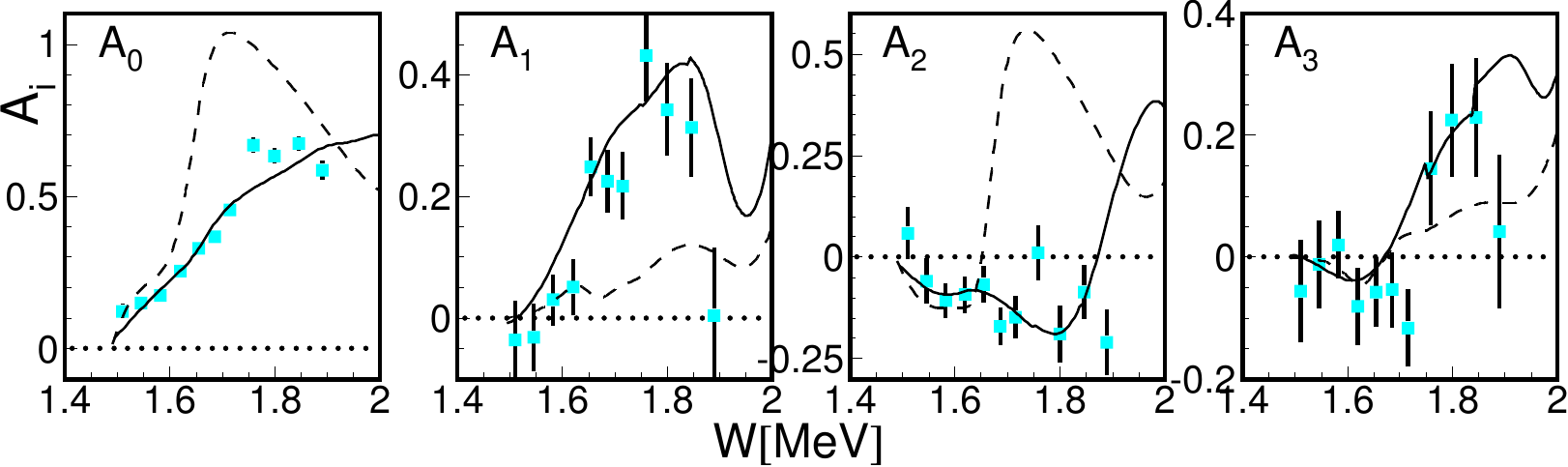,scale=1.0}
\begin{minipage}[t]{16.5 cm}
\caption{Coefficients of the Legendre series (\ref{eq:sigma_legendre}) fitted to the beam asymmetries
$\Sigma(W, \CT)$ as function of $W$. Upper part: $\gamma p\rightarrow p\eta$,
Bottom: $\gamma n\rightarrow n\eta$. Symbols for data are as in Fig.~\ref{fig:eta_sigma}.
Solid curves: results of BnGa fit \cite{Anisovich_12,Anisovich_12a} for proton and \cite{Anisovich_13}
for neutron target, dashed: MAID model \cite{Chiang_02}, dotted: SAID partial wave analysis \cite{McNicoll_10}
(only for proton target).
\label{fig:sigma_coeff}}
\end{minipage}
\end{center}
\end{figure}

Apart from angular distributions the only other reasonably
precise results that are available so far (very recent new
polarization data \cite{Akondi_14} are shortly discussed at the
end of this chapter) are for the photon beam asymmetry $\Sigma$
(linearly polarized photon beam, unpolarized target) for free
protons and quasi-free neutrons. These data have gone into the
model analyses discussed above. They have been measured mainly
at the GRAAL facility \cite{Bartalini_07,Fantini_08,Ajaka_98}
and at ELSA \cite{Elsner_07}. The primary results are
summarized in Fig.~\ref{fig:eta_sigma} and have been fitted
with the Legendre series
\begin{equation}
\Sigma(W, \CT) = \sin^2\theta_{\eta}^{*}\sum_{i=0}^{3} A_i(W) P_i(\CT)\,,
\label{eq:sigma_legendre}
\end{equation}
with the fitted coefficients being shown in
Fig.~\ref{fig:sigma_coeff}. All the experimental results for
the $\gamma p\rightarrow p\eta$ reaction
\cite{Bartalini_07,Ajaka_98,Elsner_07} are in reasonable
agreement; for $\gamma n\rightarrow n\eta$ only results from
the GRAAL experiment \cite{Fantini_08} are so far available. We
have already discussed above the strong contribution of the
$S_{11}$(1535) - $D_{13}$(1520) interference to the beam
asymmetry, which is responsible for the fast rise of the
$A_{0}$ coefficient from threshold. In a similar behaviour to
that seen for the angular $\gamma p\rightarrow p\eta$
distributions, the corresponding beam asymmetry $\Sigma$ also
shows a strong energy dependence for $W$ between 1.6 and
1.8~GeV. The $A_{1}$ coefficient rises sharply and $A_{2}$ (and
perhaps also $A_{3}$) shows a narrow structure around 1.65~GeV.
The MAID \cite{Chiang_02}, SAID \cite{McNicoll_10}, and BnGa
\cite{Anisovich_12,Anisovich_12a} analyses reproduce this
behaviour reasonably well. However, a word of caution is needed
here. The fact that the fits (with many free parameters) can
describe the data does not of course guarantee that the right
physics is included in the models. An instructive example of
this is given in \cite{Elsner_07}.

\begin{figure}[tb]
\begin{center}
\epsfig{file=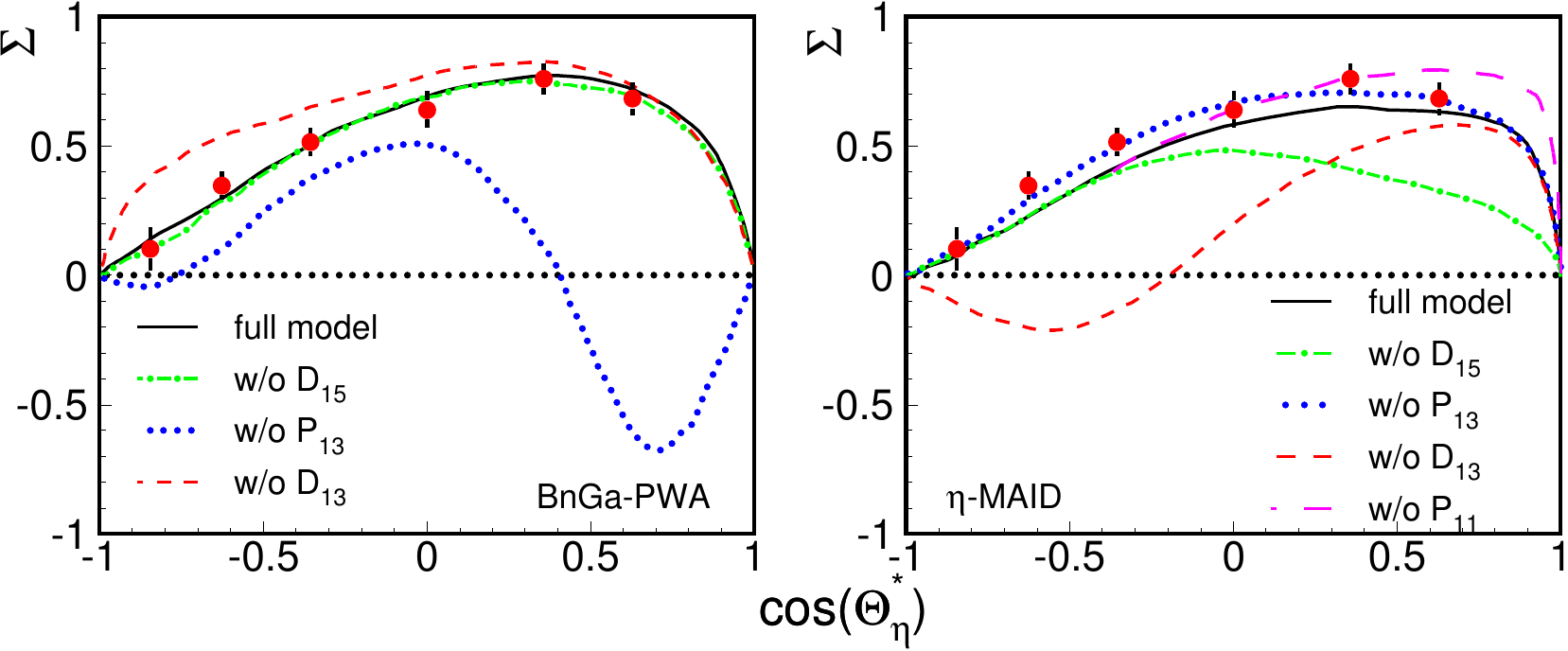,scale=0.95}
\begin{minipage}[t]{16.5 cm}
\caption{Beam asymmetry data from \cite{Elsner_07} for $\gamma p\rightarrow
p\eta$ for incident photon energies around 1.25~GeV ($W\approx1.8$~GeV)
compared to results from the MAID model \cite{Chiang_02} (right hand side)
and the BnGa fit \cite{Anisovich_05} (left hand side). Solid (black) curves:
full fits, dotted (blue): without the $P_{13}$(1720) state,
dashed (red): without $D_{13}$(1520), long dashed (magenta): without
$P_{11}$(1710) (no difference to full model for BnGa model),
dash-dotted (green): without $D_{15}$(1675). \label{fig:elsner}}
\end{minipage}
\end{center}
\end{figure}

The measured beam asymmetries around $W\approx1.8 $~GeV are
compared in Fig.~\ref{fig:elsner} to results from the MAID
\cite{Chiang_02} and BnGa models \cite{Anisovich_05} (note that
these are not the most recent results from BnGa
\cite{Anisovich_12,Anisovich_12a}). Both models described the
beam asymmetry comparably well (see figure) and were also in
reasonable agreement with the angular distributions.
Nevertheless, as the figure shows, the physics contents of the
fits were quite different, which is demonstrated by
`switching-off' certain resonance contributions. While for the
BnGa model the $P_{13}$ state was very important, it played
almost no role in the MAID model which, on the other hand,
included large contributions from the $D_{13}$ and $D_{15}$
$d$-wave states. The model results start to diverge quickly
above the region where experimental $\Sigma$ values are
available (see Fig.~\ref{fig:sigma_coeff}, top row).

Little structure is observed for the $\gamma n\rightarrow
n\eta$ reaction around $W\approx$1.7~GeV, although also in this
case the $A_0$ coefficient shows an abrupt rise around this
energy. However, resolution effects might play a role here.
Although the c.m.\ angles were reconstructed from the final
state kinematics taking into account Fermi motion, the results
are given in terms of the incident photon energy without
correction for Fermi-smearing effects. Nevertheless, the
effects do not seem to be very large since the quasi-free
proton results are close to those of the free proton data. The
BnGa model agrees reasonably well with these data (because it
has been fitted to them). The MAID model largely overestimates
the asymmetries between 1.6 - 1.8 GeV, mainly because of the
large contribution of the $D_{15}$(1675) state, which is
clearly not realistic. Agreement with the Reggeized version of
MAID \cite{Chiang_03} is better \cite{Fantini_08} because a
smaller $N\eta$ branching ratio for the $D_{15}$ is used.

Apart from the beam asymmetries, until very recently only the
previously mentioned results for the target asymmetry in the
$S_{11}$ peak \cite{Bock_98}, and a few low statistics values
for the separation of the total cross section into
$\sigma_{1/2}$ and $\sigma_{3/2}$ components (longitudinally
polarized target, photon beam circularly polarized parallel
($\sigma_{3/2}$) or antiparallel ($\sigma_{1/2}$)) right in the
maximum of the $S_{11}$ peak, have been reported
\cite{Ahrens_03}. The latter only confirm the expectation that
$\sigma_{1/2}$ is very dominant in the $S_{11}$ peak
($\sigma_{3/2}$ is compatible with zero, within large
statistical uncertainties). However, this situation has now
changed dramatically. The CLAS experiment and the experiments
at MAMI and ELSA have acquired data for the $\gamma
p\rightarrow p\eta$ reaction for the single polarization
observables $\Sigma$, $T$, $P$ ($P$ measured not by analysis of
the recoil nucleon polarization but as double polarization
observable with linearly polarized beam and transversely
polarized target), and the double polarization observables $E$,
$F$, $G$, and $H$ (not all experiments measure all observables,
but all observables are measured in at least one, most in two,
and some even in three experiments). These data are presently
under analysis (preliminary results from CLAS for the $E$
observable have, e.g., been shown at the NSTAR 2011 workshop
\cite{Morrison_12}). Data on $E$, $T$, and $F$ for the $\gamma
n\rightarrow n\eta$ reaction have been measured at MAMI and
further measurements of observables with linearly polarized
photon beams for the neutron target are planned at ELSA.
Therefore, at present it would be completely premature to draw
final conclusions on nucleon resonance contributions to
$\eta$-photoproduction, in particular at higher incident photon
energies. The large body of new data will have a strong impact
on all partial wave analyses.

\begin{figure}[tb]
\begin{center}
\epsfig{file=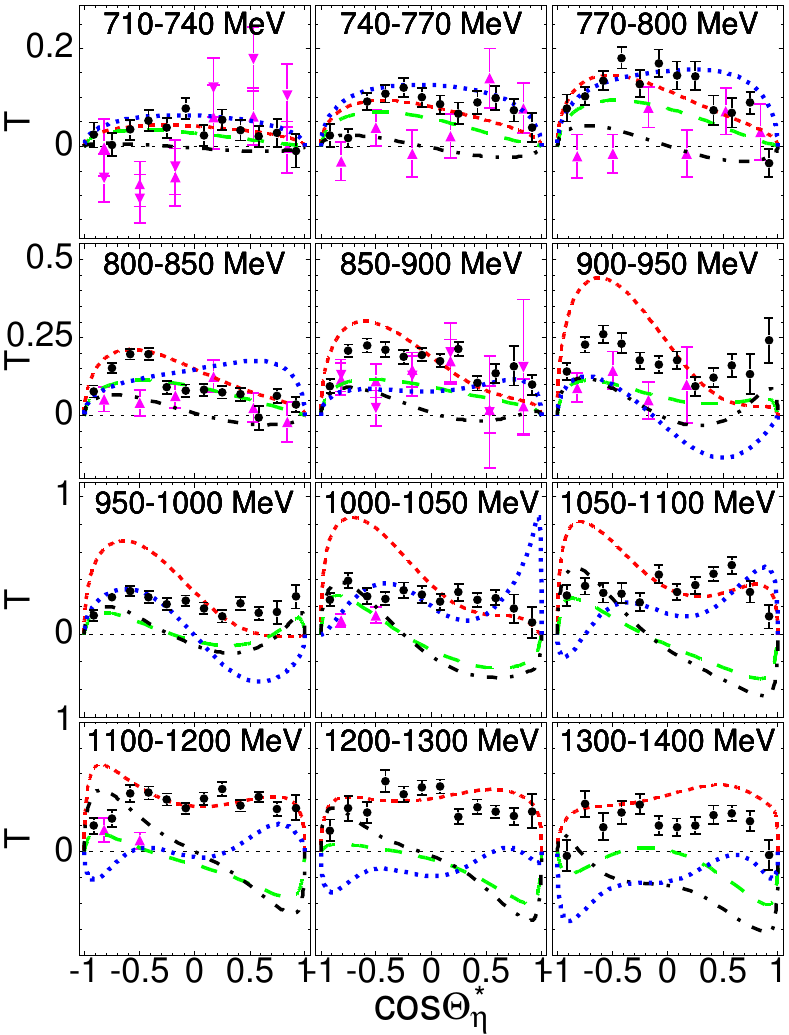,scale=1.0}
\epsfig{file=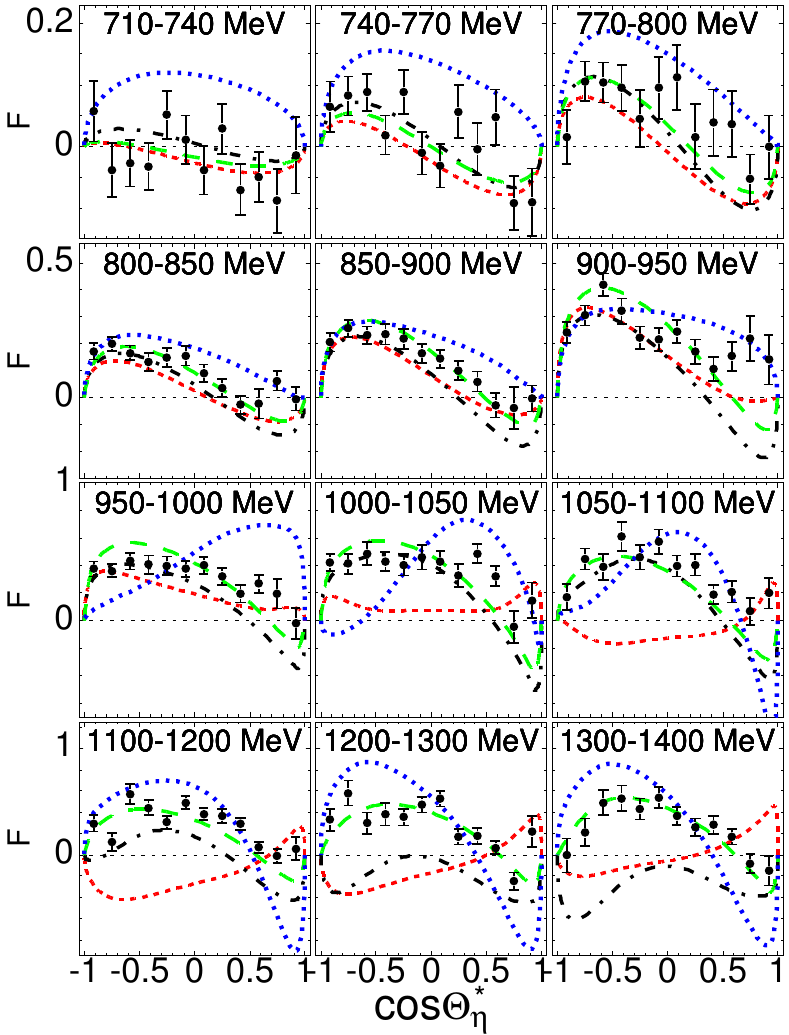,scale=1.0}
\vspace*{-0.5cm}
\begin{minipage}[t]{16.5 cm}
\vspace*{-0.5cm}
\caption{Left hand side: target asymmetry $T$ for the $\gamma
p\rightarrow p\eta$ reaction. Right hand side: double
polarization observable $F$ for the same reaction.
Ranges of incident photon energies quoted in the figure.
Experimental results (black dots) are from Crystal Ball/TAPS at MAMI
\cite{Akondi_14}. For $T$ also previous results (magenta triangles)
from PHOENICS at ELSA \cite{Bock_98} are shown (for these data
the energy bins do not correspond exactly to the present
results, see \cite{Akondi_14} for details). The curves are
predictions from the following models: dashed (red):
$\eta$-MAID \cite{Chiang_02}, long-dashed (green): Giessen
model \cite{Shklyar_13}, dash-dotted (black): BnGa
\cite{Anisovich_12}, dotted (blue) SAID \cite{McNicoll_10}.
\label{fig:akondi}}
\end{minipage}
\end{center}
\end{figure}

The first data from the new measurements of $\gamma p\rightarrow p\eta$ with polarized
beams and polarized targets have just been published \cite{Akondi_14}. They come from
a measurement with Crystal Ball/TAPS at MAMI with transversally polarized target
(solid buthanol) and circularly polarized photon beam, giving access to the target
asymmetry $T$ and the double polarization observable $F$ defined by \cite{Barker_75}
\begin{equation}
\frac{d\sigma}{d\Omega} =
\frac{d\sigma_o}{d\Omega}
\left(1+Tp_T\sin\phi+Fp_Tp_{\odot}\cos\phi\right)
\end{equation}
where $d\sigma_o$ is the unpolarized cross section, $p_T$ is
the degree of transverse target polarization in the
$y$-direction, $p_{\odot}$ the degree of right circular beam
polarization, and $\phi$ the azimuthal angle of the meson.

The much more precise new results for the threshold behaviour
of the target asymmetry clearly refute the sign change between
forward and backward polar angles reported from the only
previously available data \cite{Bock_98} and thus settle the
issue of the unnatural phase between the $S_{11}$(1535) and
$D_{13}$(1520) multipoles discussed above. The comparison of
the experimental results to different predictions from reaction
models/partial-wave analysis in Fig.~\ref{fig:akondi}
demonstrates the selective power of these new observables. None
of the models that describe differential cross sections and
beam asymmetries quite well agrees with the new data for the
asymmetries. We mention only one example; predictions for the
target asymmetries in the range between 950 - 1050 MeV, where
the peculiar structures in the $\gamma N\rightarrow N\eta$
excitation functions were observed, disagree strongly with data
for all models. Consequently, conclusions like the one drawn by
the latest Bonn-Gatchina analysis \cite{Anisovich_14}, that the
bump-like structure in the neutron excitation function can be
explained by $S_{11}$ interferences alone, will have to be
cross checked against the new data (in this case upcoming
polarization data for the neutron target will be even more
important).

In summary, all reaction models will have to be refitted against the new polarization
data (results for $E$, $P$, $H$, $G$... are still to come) and then one will have
to check whether different approaches converge in terms of nucleon resonance
contributions (which seems to happen for single $\pi^0$ production, where more
observables are already available).

\subsubsection{\it Photoproduction of $\eta^{\prime}$-mesons off nucleons}
\label{ssec:etaprime}

In the context of nucleon resonance physics, the $\eta^{\prime}$ meson can be
regarded as the heavy twin of the $\eta$, where the isoscalar nature ensures
that only $N^{\star}$ resonances contribute. The much larger mass shifts the
interesting $W$ range upwards. The production threshold at $W = 1.896$~GeV
corresponds to an incident photon energy of 1.447~GeV for a free proton
target. This means that one can still expect in an energy range above
$W\approx 1.9$~GeV, where $\eta$ production shows already a complicated
behaviour, a situation where only a few low-order partial waves contribute.
This is the $W$ range where most of the predicted, but so far unobserved,
$N^{\star}$ states should be located.

\begin{figure}[htb]
\begin{center}
\epsfig{file=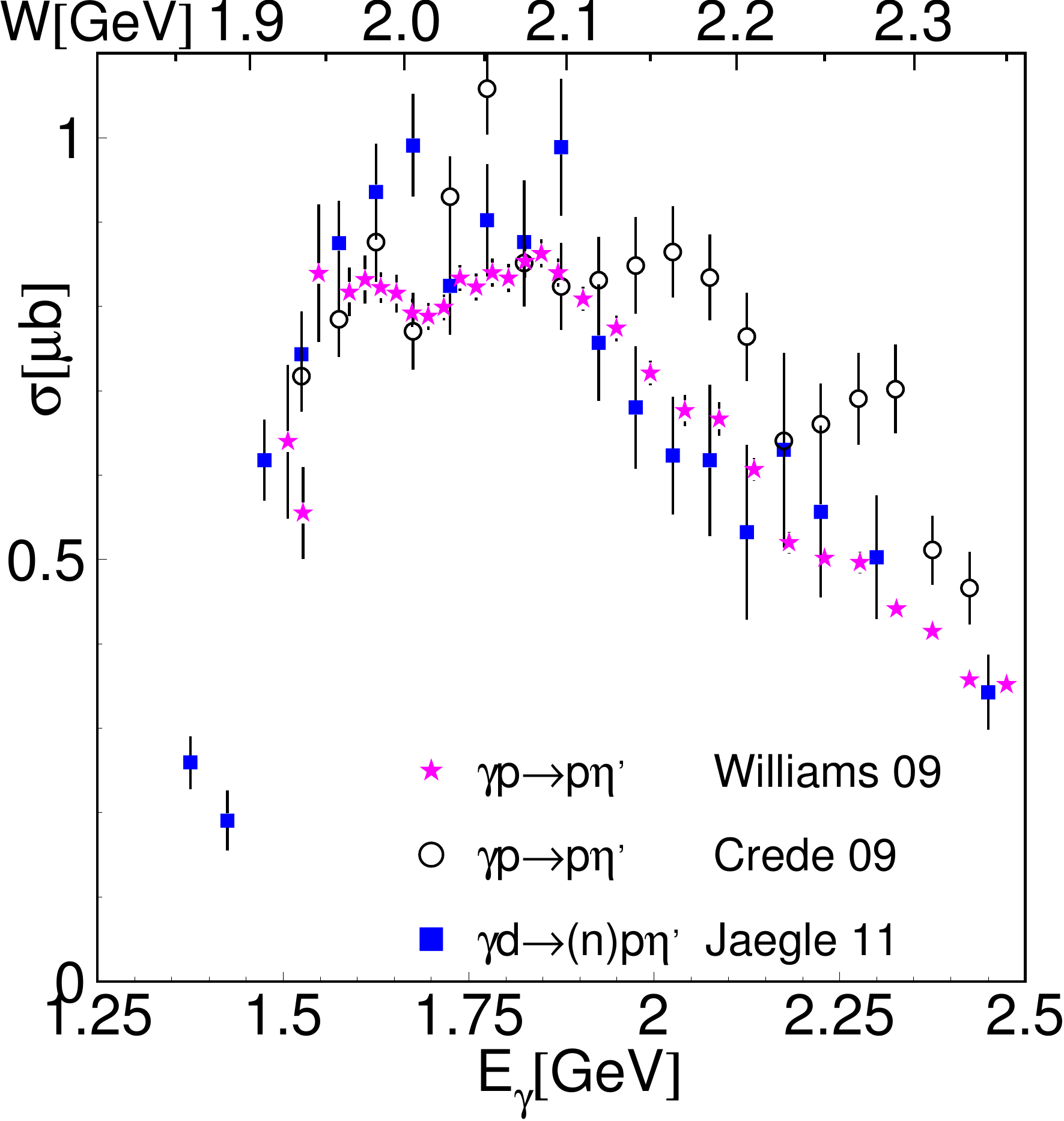,scale=0.46}
\epsfig{file=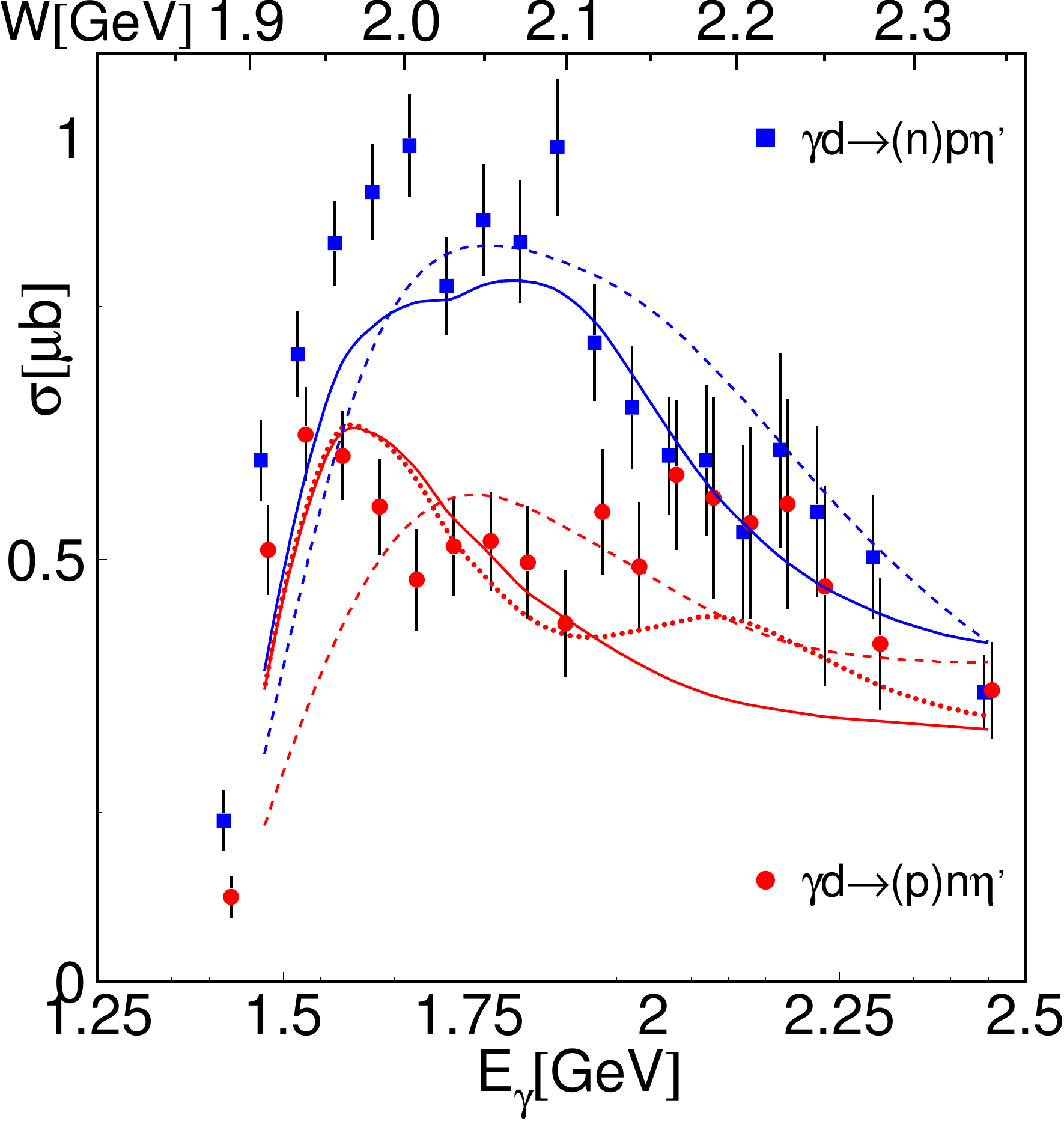,scale=0.46}
\begin{minipage}[t]{16.5 cm}
\caption{Total cross section for the $\gamma N\rightarrow N\eta^{\prime}$
reactions. Left hand side: data for $\gamma p\rightarrow p\eta^{\prime}$ from
\cite{Crede_09} (Cred\'{e} 09, open black circles), \cite{Williams_09} (Williams
09, magenta stars) and the quasi-free $\gamma d\rightarrow (n)p\eta^{\prime}$
\cite{Jaegle_11} (Jaegle 11, blue squares) reactions. Right hand side:
quasi-free data for $\gamma d\rightarrow (n)p\eta^{\prime}$ (participant
proton) and $\gamma d\rightarrow (p)n\eta^{\prime}$ (participant neutron)
from \cite{Jaegle_11}. Curves: model results from ETA$^{\prime}$-MAID
\cite{Chiang_03} (dashed blue proton target, dashed red neutron target),
solid blue NH-model \cite{Nakayama_06} for proton target, solid and dotted
red different solutions from NH model for neutron target (see discussion in
\cite{Jaegle_11}). \label{fig:etap_tot}}
\end{minipage}
\end{center}
\end{figure}

However, the available data base for $\eta^{\prime}$
photoproduction is much more scarce than for the $\eta$ meson
and, as a consequence, resonance contributions are not yet well
determined. The reason is twofold. The total cross section of
$\eta^{\prime}$ production reaches at most the 1~$\mu b$ level
($\eta$ production falls below this level only above $W\approx
2.3$~GeV). Secondly, the decay modes of the $\eta^{\prime}$ are
much less accessible for most experiments. For electromagnetic
calorimeters that cover large solid angles, both the
$\eta\rightarrow 2\gamma$ (branching ratio $\approx 39\%$) and
the $\eta\rightarrow 3\pi^0\rightarrow 6\gamma$ (branching
ratio $\approx 31\%$) \cite{Beringer_12} decay are favourable.
The two-photon decay of the $\eta^{\prime}$ meson has only a
probability of $\approx 2.2\%$ and the largest decay branching
ratio for photon final states is only $\approx 8.3\%$ for the
$\eta^{\prime}\rightarrow\eta\pi^0\pi^0\rightarrow 6\gamma$
decay \cite{Beringer_12}. Finally, few experiments reach the
necessary incident photon energies so that there are
statistically relevant results for total cross section and
angular distributions only from the CLAS and ELSA facilities.

Until the end of the 1990s the only data came from old bubble chamber
measurements \cite{ABBHHM_68,AHHM_76} at DESY. They were analyzed in 1995 by
Zhang, Mukhopadhyay, and Benmerouche \cite{Zhang_95}, preparing the ground
for the analysis of the expected results from the modern tagged photon
experiments, in the framework of an effective Lagrangian model. They reported
strong contributions from vector meson exchange ($\omega$, $\rho$ mesons) and
a leading nucleon resonance contribution from a $D_{13}$(2080) two-star
state. This, however, did not come from the multipoles ($E_{2-}$, $M_{2-}$),
where this state is resonant, but from a \emph{background} contribution of
this state to the $E_{0+}$ multipole. In the meantime, due to the results of
more recent coupled-channel analyses, this resonance has been split into two
states, the three-star $D_{13}$(1875) and the two-star $D_{13}$(2120)
\cite{Beringer_12}.

The first \emph{modern} measurement with a tagged photon beam came from the
SAPHIR experiment at ELSA \cite{Ploetzke_98}, which used the $\gamma
p\rightarrow p\eta^{\prime}\rightarrow p\pi^+\pi^-\eta\rightarrow
p\pi^+\pi^-\pi^+\pi^-\pi^0$ reaction chain. However, it was based on only 250
events so that the statistical quality was not significantly better than the
earlier bubble chamber results. The total cross section and angular
distributions from this measurement have been interpreted in completely different
ways. In the SAPHIR paper \cite{Ploetzke_98}, the main emphasis was on the
rapid rise of the total cross section from threshold and the linear behaviour
of the angular distributions (i.e., $\propto \cos\theta^*$) between the
backward and forward directions. The authors argued that the minimal model to
describe such a behaviour has to include two resonances of opposite parity
and the most natural choice is an $S_{11}$ and a $P_{11}$ state. A simple fit
gave excitation energies and widths of $W\approx 1.897$~GeV, $\Gamma\approx
0.4$~GeV for the $S_{11}$ and $W\approx 1.986$~GeV, $\Gamma\approx 0.3$~GeV
for the $P_{11}$ state.

\begin{figure}[t!]
\begin{center}
\epsfig{file=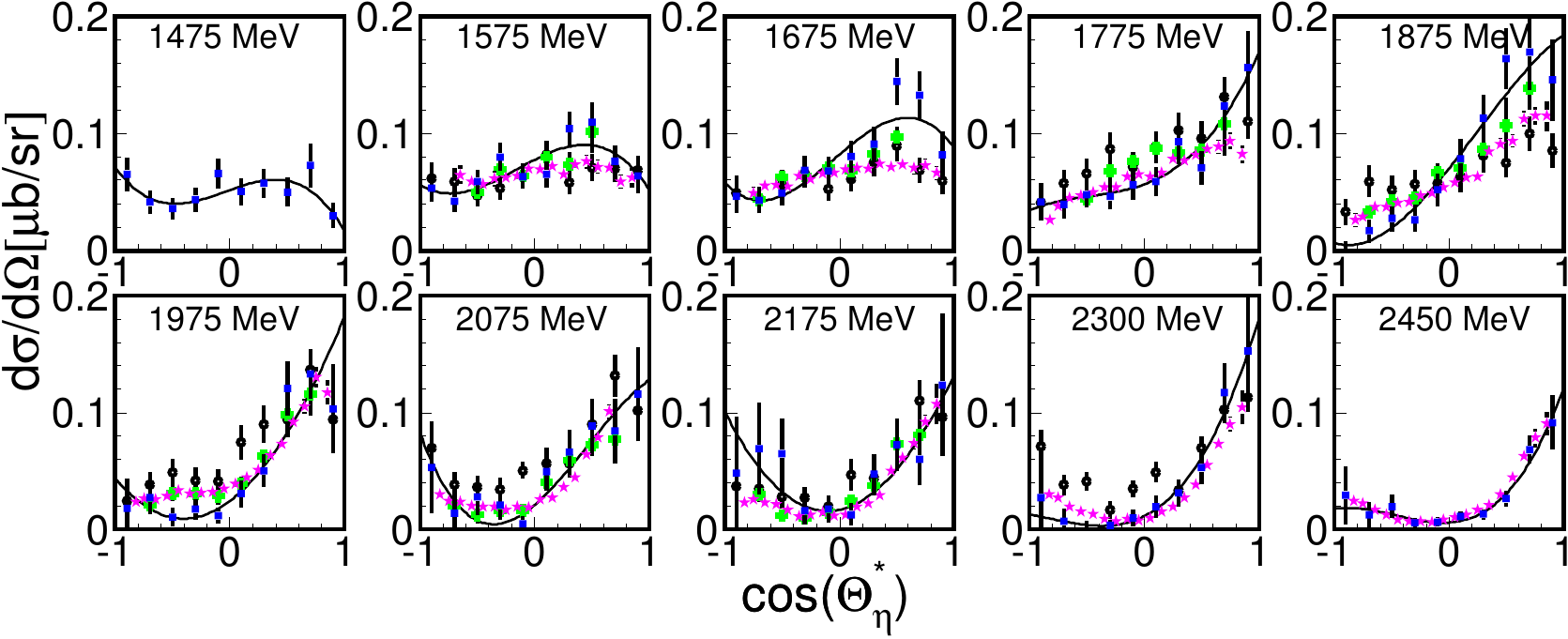,scale=1.0}
\vspace*{-0.5cm}
\begin{minipage}[t]{16.5 cm}
\caption{Selected angular distributions (mean incident photon energies quoted
in the figure) for the reaction $\gamma p\rightarrow
p\eta^{\prime}$ off free protons (magenta stars: \cite{Williams_09}, black
open circles: \cite{Crede_09}, green stars: \cite{Dugger_06}) and off
quasi-free protons (blue squares: \cite{Jaegle_11a} bound in the deuteron).
Black solid lines: fit of the quasi-free proton data with
Eq.~(\ref{eq:diff_legendre}) ($N$=3).
\label{fig:etap_diff_p}}
\end{minipage}
\end{center}
\end{figure}
\begin{figure}[h!]
\begin{center}
\epsfig{file=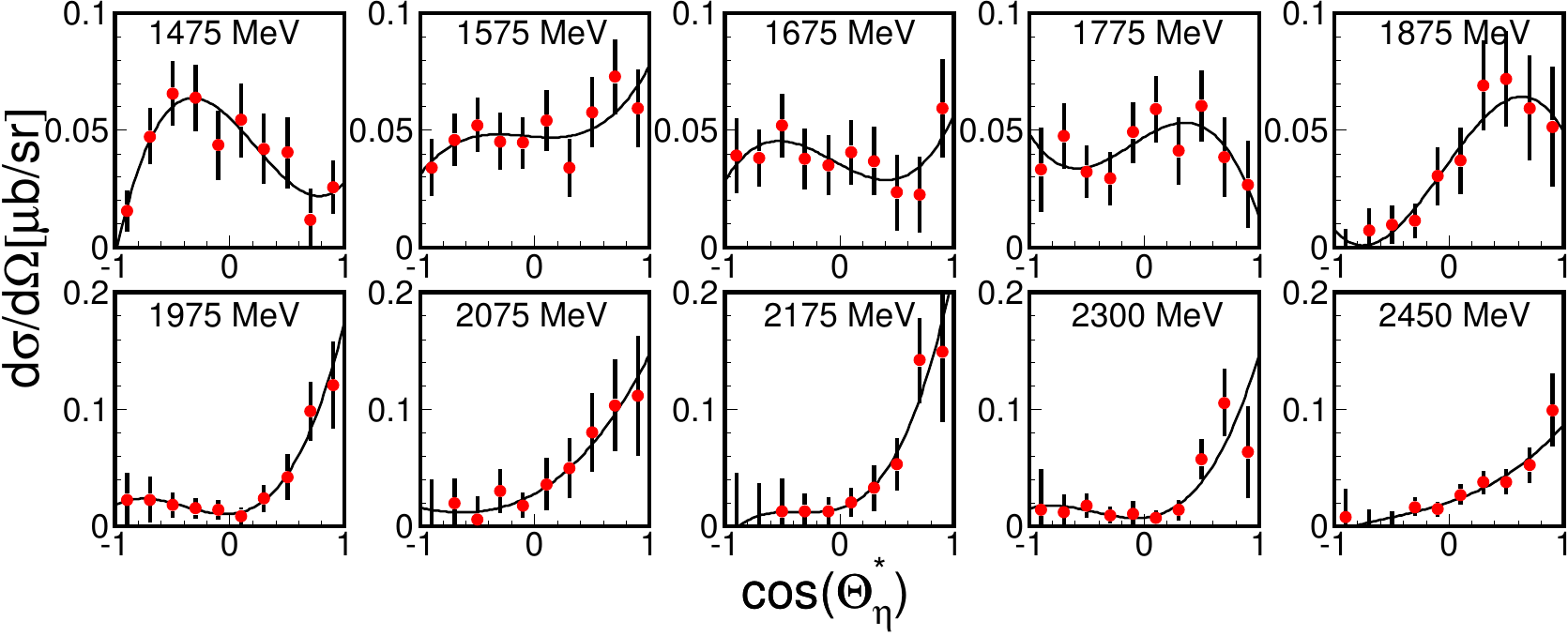,scale=1.0}
\vspace*{-0.5cm}
\begin{minipage}[t]{16.5 cm}
\caption{Selected angular distributions for different values of $E_{\gamma}$
for the reaction $\gamma n\rightarrow
n\eta^{\prime}$ off quasi-free neutrons bound in the deuteron. Solid lines:
fits with Eq.~(\ref{eq:diff_legendre}) ($N$=3). \label{fig:etap_diff_n}}
\end{minipage}
\end{center}
\end{figure}

\begin{figure}[tb]
\begin{center}
\epsfig{file=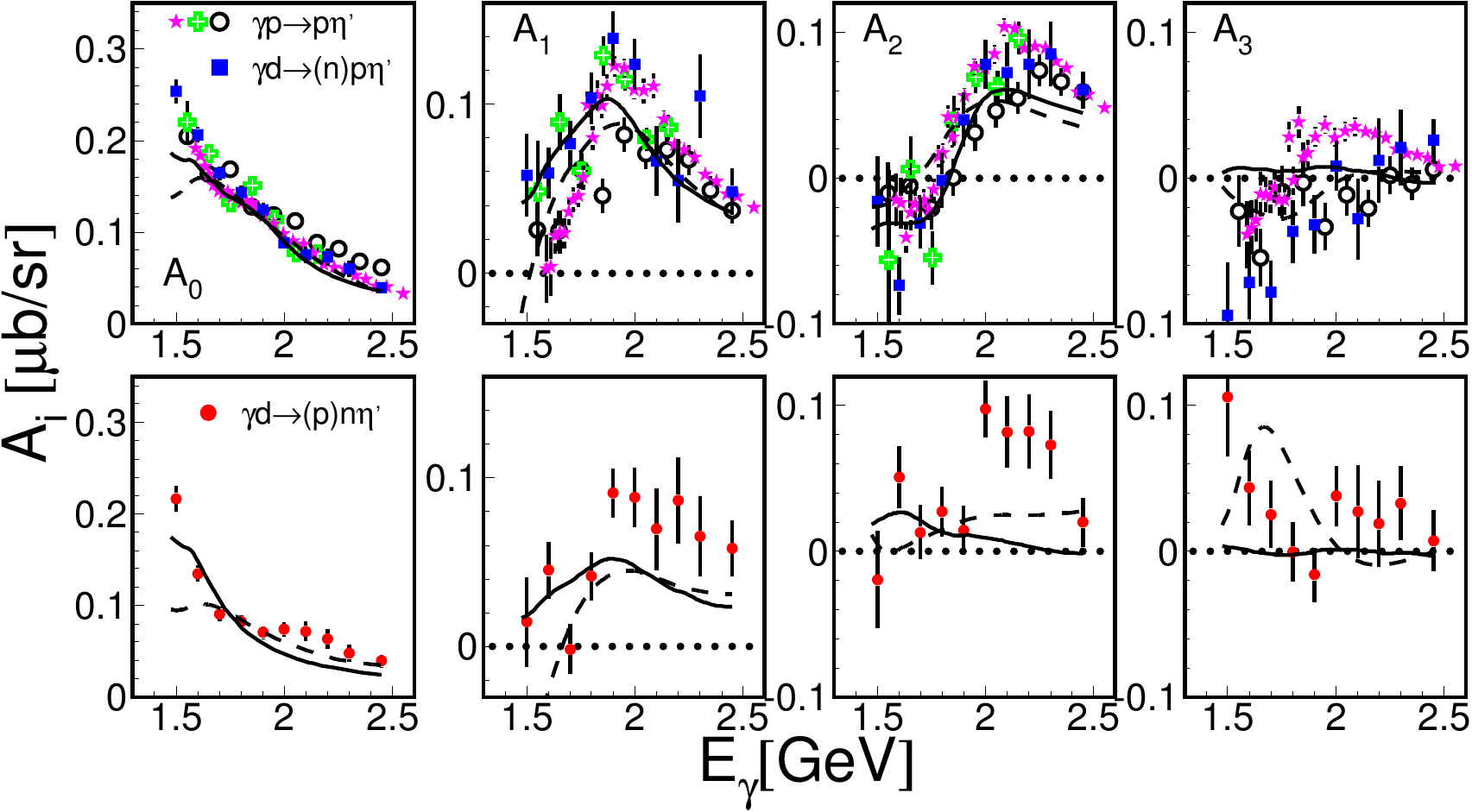,scale=1.0}
\begin{minipage}[t]{16.5 cm}
\caption{Coefficients of the Legendre series of Eq.~(\ref{eq:diff_legendre})
for $\eta^{\prime}$ photoproduction from fits with $N=3$ ($N=2$ for the data
from \cite{Dugger_06}). Top row: free proton data from
\cite{Dugger_06,Williams_09,Crede_09} and quasi-free proton data from
\cite{Jaegle_11a}. Symbols as in Fig.~\ref{fig:etap_diff_p}. Bottom row:
quasi-free neutron data from \cite{Jaegle_11a}. Curves: model results. Solid:
NH model (solution (I), see \cite{Nakayama_06},\cite{Jaegle_11a}), dashed:
ETA$^{\prime}$-MAID \cite{Chiang_03}. \label{fig:etap_coeff}}
\end{minipage}
\end{center}
\end{figure}

The same data were analyzed in the framework of a Reggeized effective
Lagrangian model (ETA$^{\prime}$-MAID) \cite{Chiang_03}. There was also some
evidence in this approach for contributions from an $S_{11}$ state in the
mass range 1.932 - 1.959~GeV, but a $P_{11}$ state was not absolutely
necessary. The shape of the angular distributions could be reproduced by an
interference of the $E_{0+}$ multipole from the $S_{11}$ resonance with the
(in this case large) contributions from the Reggeized $\rho$ and $\omega$
$t$-channel exchange. Elster and coworkers \cite{Sibirtsev_04} also found
large contributions from vector meson exchange and even argued that such
contributions plus the tail of the well-known $S_{11}$(1535) resonance could
explain the data. More refined analyses were excluded by the low statistical
quality of the cross section data and the lack of any results for
polarization observables.

In the meantime the situation for the angular distributions has improved
somewhat. The reaction $\gamma p\rightarrow p\eta^{\prime}$ has been measured
twice by the CLAS experiment at Jlab \cite{Dugger_06,Williams_09} and by the
CBELSA/TAPS experiment \cite{Crede_09}. The quasi-free $\gamma d\rightarrow
(n)p\eta^{\prime}$ and $\gamma d\rightarrow (p)n\eta^{\prime}$ reactions
(spectator nucleons in brackets) have also been measured at CBELSA/TAPS
\cite{Jaegle_11a}. Total cross sections from these measurements are compared
in Fig.~\ref{fig:etap_tot}. The two CLAS experiments published only angular
distributions and the total cross sections have been extracted from fits of
the distributions of \cite{Williams_09} using Eq.~(\ref{eq:diff_legendre})
and taking the leading Legendre polynomial coefficient. The fits of the data
from \cite{Dugger_06} are less stable, due to the smaller angular range, but
the comparison of the $A_{0}$ coefficients of all data sets in
Fig.~\ref{fig:etap_coeff} shows that the earlier CLAS data are also
compatible with the other measurements.

As was the case for $\eta$-production, these experiments used different
approaches to measure the reaction. The first CLAS measurement was based
purely on the detection and momentum analysis of the recoil protons; the
$\eta^{\prime}$ meson was then identified in a missing-mass analysis. This
technique included all the decay channels of the meson but the background
underneath the missing-mass peak was large. The proton detection efficiency
was determined empirically using the $\gamma p\rightarrow p\pi^+\pi^-$
reaction. The photon flux was measured absolutely and the method for its
extraction was tested with photoproduction of $\pi^0$ mesons, for which
precise cross section data are known (although not in the photon energy range
of interest). The second CLAS experiment \cite{Williams_09} used the
$\eta^{\prime}\rightarrow \pi^+\pi^-\eta$ decay channel (branching ratio
$\approx 43\%$ \cite{Beringer_12}), reconstructed the $\eta$ meson as a
missing particle and then analyzed the invariant mass of the $\pi^+\pi^-\eta$
system, which resulted in a much lower background level. This experiment also
had absolute normalization.

The two measurements at ELSA \cite{Crede_09,Jaegle_11a} used the
$\eta^{\prime}\rightarrow \pi^0\pi^0\eta\rightarrow 6\gamma$ decay (effective
branching ratio 8.3\%). Both were absolutely normalized; the free-proton
measurement used kinematic fitting while the quasi-free reactions from the
deuteron target were studied with a combined invariant- and missing-mass
analysis. The quasi-free measurement off the deuteron was not analyzed with
full kinematic reconstruction (which it was for the $\eta$ data), which would be
needed to eliminate the effects from Fermi motion, because the statistical
quality was not sufficient. The quasi-free $\gamma n\rightarrow
n\eta^{\prime}$ reaction was analyzed in two different ways, once with the
coincident detection of the recoil neutron and once as the difference between
the simultaneously measured inclusive cross section (without condition for
recoil nucleons) and the quasi-free proton cross section, measured in
coincidence with recoil protons. Both results agreed within statistical
uncertainties and were averaged. The comparison of the total cross sections
for the reactions off protons (see Fig.~\ref{fig:etap_tot}) show reasonable
agreement (but see discussion below). Within the level of precision reached
there are no systematic deviations between the free and quasi-free proton
data, at least not more significant than between the different data sets for
the free proton (although effects arising from Fermi motion were not
reconstructed for the quasi-free data). All data sets deviate much more
strongly in absolute scale from the earlier SAPHIR measurement (which had a
maximum cross section of almost 2~$\mu$b, although with large uncertainties)
and also in the shape of the angular distributions, where the more recent
data show a slower rise to forward angles in the near-threshold region. We
therefore ignore the SAPHIR data in the further discussion.

Typical angular distributions for the proton and neutron target are shown in
Figs.~\ref{fig:etap_diff_p} and \ref{fig:etap_diff_n}. In the near-threshold
region they are rather flat, which would be consistent with a substantial
contribution from an $s$-wave. At the highest incident photon energies the
angular distributions for protons and neutrons peak at forward angles, as
expected for non-resonant $t$-channel contributions. However, they seem to
rise also for extreme backward angles, which might be an indication for
significant $u$-channel nucleonic contributions, but could also arise from
the excitation of nucleon resonances (see discussion in \cite{Nakayama_06}).
The coefficients of the Legendre fits Eq.~(\ref{eq:diff_legendre}) are
summarized in Fig.~\ref{fig:etap_coeff}. They do not show such rapid
variations as observed for $\eta$ production.

The experimental results have been analyzed with different model approaches
\cite{Chiang_03,Nakayama_06,Zhong_11,Huang_13}, in particular in the context
of nucleon resonance physics, but the results are not yet conclusive. A
reaction model based on a relativistic meson-exchange model of hadronic
interactions developed by Nakayama and Haberzettl (NH-model)
\cite{Nakayama_06} was first employed to describe the earlier CLAS data
\cite{Dugger_06}. The authors claimed significant contributions from
$t$-channel exchange, for which descriptions in terms of ordinary
vector-meson exchange or Regge trajectories gave comparable results (with a
slight advantage for the standard $\rho$, $\omega$ exchange). Contributions
from $u$-channel nucleonic currents were found to be small but not well
determined, due to the ambiguities in the $s$-channel resonance
contributions. Different sets of $N^{\star}$ states were tested, in
particular for the $S_{11}$, $P_{11}$, $P_{13}$, and $D_{13}$ partial waves.
Some evidence was found that such states do contribute, but the solutions
were far from unique. It was pointed out that polarization observables,
especially beam and target asymmetries, would be much more sensitive to
discriminate between the different fits.

The results from the ETA$^{\prime}$-MAID model \cite{Chiang_03} and the NH
model were later updated to include also the quasi-free neutron data (see the
discussion in \cite{Jaegle_11a}). Both models found contributions from the
$S_{11}$, $P_{11}$, $P_{13}$, and $D_{13}$ partial waves (in the NH model the
latter two were below threshold). However, the resonance parameters,
including also the neutron/proton electromagnetic coupling ratios, differed
significantly between the two approaches. Total cross sections and Legendre
coefficients of these models are compared to the measured values in
Fig.~\ref{fig:etap_tot} and \ref{fig:etap_coeff}.

In a different approach, Zhong and Zhao \cite{Zhong_11} analyzed the free and
quasi-free nucleon data within a chiral quark model. They included only $s$-
and $u$-channel contributions and omitted completely the $t$-channel terms in
order to avoid double counting. In this model the leading contributions came
from the $u$-channel and the $S_{11}$(1535) resonance, actually a
constructive interference between these two terms was held responsible for
the strong rise of the total cross section from threshold. The $S_{11}$(1650)
also had a significant effect. Contributions from the higher lying
$S_{11}$(1920) were only small and the most significant resonant term above
threshold was associated with a $D_{15}$(2080) state.

Most recently, Huang, Haberzettl, and Nakayama \cite{Huang_13} have updated
the NH model to include all the available $\gamma N \to N \eta^{\prime}$
data together with the results from the $NN\rightarrow NN\eta^{\prime}$
and $\pi N\rightarrow N\eta^{\prime}$ hadronic reactions. Apart from
$t$-channel exchange, they find now mainly contributions from
$P_{13}$(1720), $P_{13}$(2050), $S_{11}$(1925), and $P_{11}$(2130) states
for the photon-induced reactions.
However, the parameters of these resonances depend strongly on the different
data sets used. In particular, the systematic differences between the most
recent free-proton data from CLAS \cite{Williams_09} and CBELSA/TAPS
\cite{Crede_09} have a significant effect (compare the total cross sections
in Fig.~\ref{fig:etap_tot}, left hand side). The fit to the CBELSA data
produces a relatively narrow peak structure around an invariant mass of
2.05~GeV. This peak hinges, however, on only one data point which, at
the level of statistical uncertainties, could well be a simple fluctuation.
The fit of the CBELSA data thus produces a relatively narrow $P_{13}$(2050)
state (widths $\Gamma\approx 50$~MeV), while for the CLAS data the width of
the same state is around $\Gamma\approx 140$~MeV. The contribution of the
$P_{11}$(2130) state is also quite different for the two fits. Finally, one
should note that the structure ascribed in the NH model to the $P_{13}$(2050)
state is assigned in the chiral quark model to a $D_{15}$(2080) resonance
\cite{Zhong_11}.

In spite of the recent efforts in modelling the reaction, many
ambiguities still exist in the interpretation of the $\gamma
N\rightarrow N\eta^{\prime}$ data in terms of nucleon resonance
contributions. Up to now there are not even total cross
sections and angular distributions that are sufficiently well
determined experimentally and polarization observables are not
yet avaiable. This situation will change in the near future.
The total cross section and angular distributions right at
threshold have been already precisely measured at MAMI (data
under analysis). Polarization observables have been measured
with CLAS at Jlab and with CBELSA/TAPS. These data sets are
also still under analysis.

\begin{figure}[tb]
\begin{center}
\epsfig{file=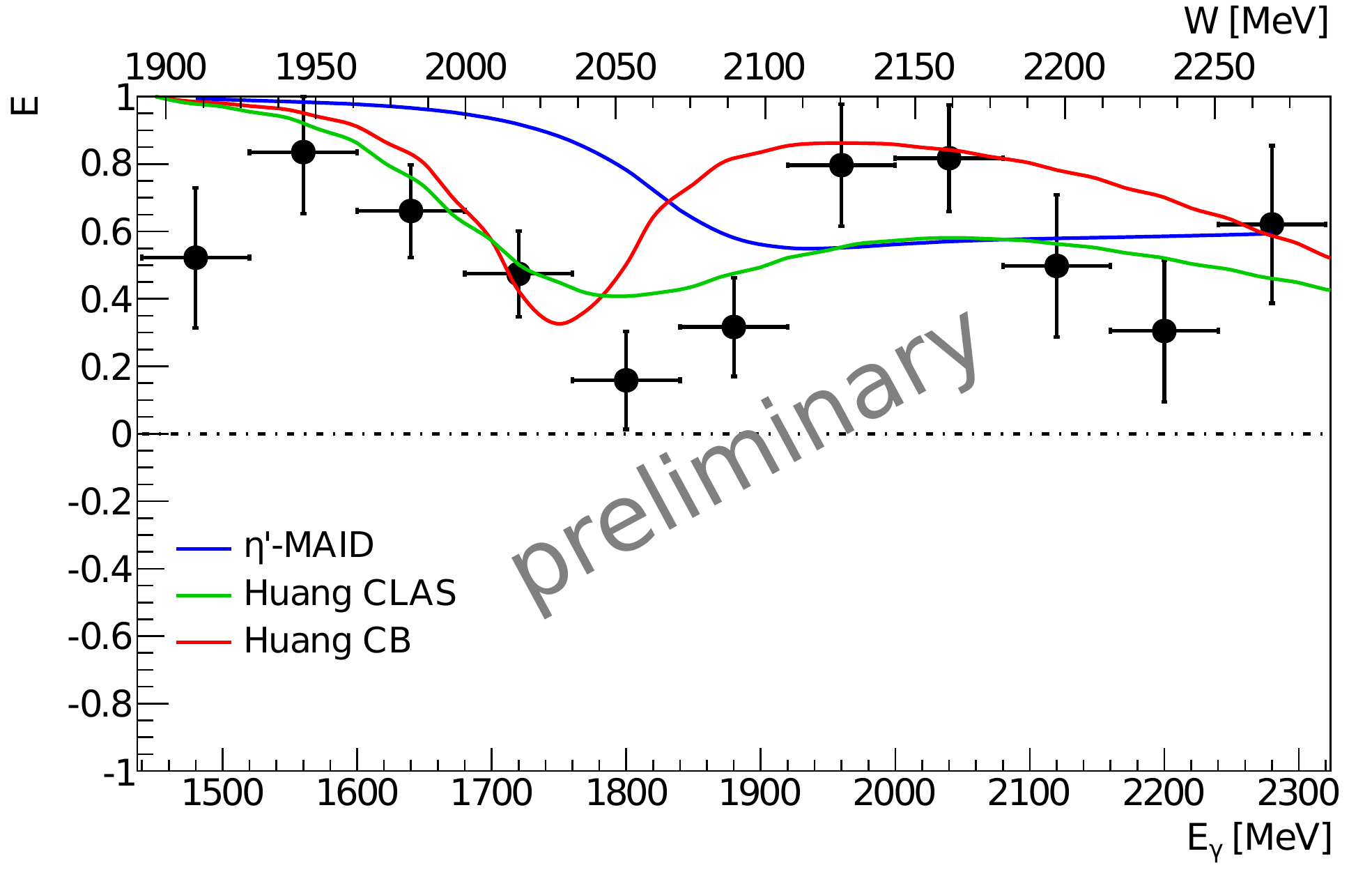,scale=0.42}
\epsfig{file=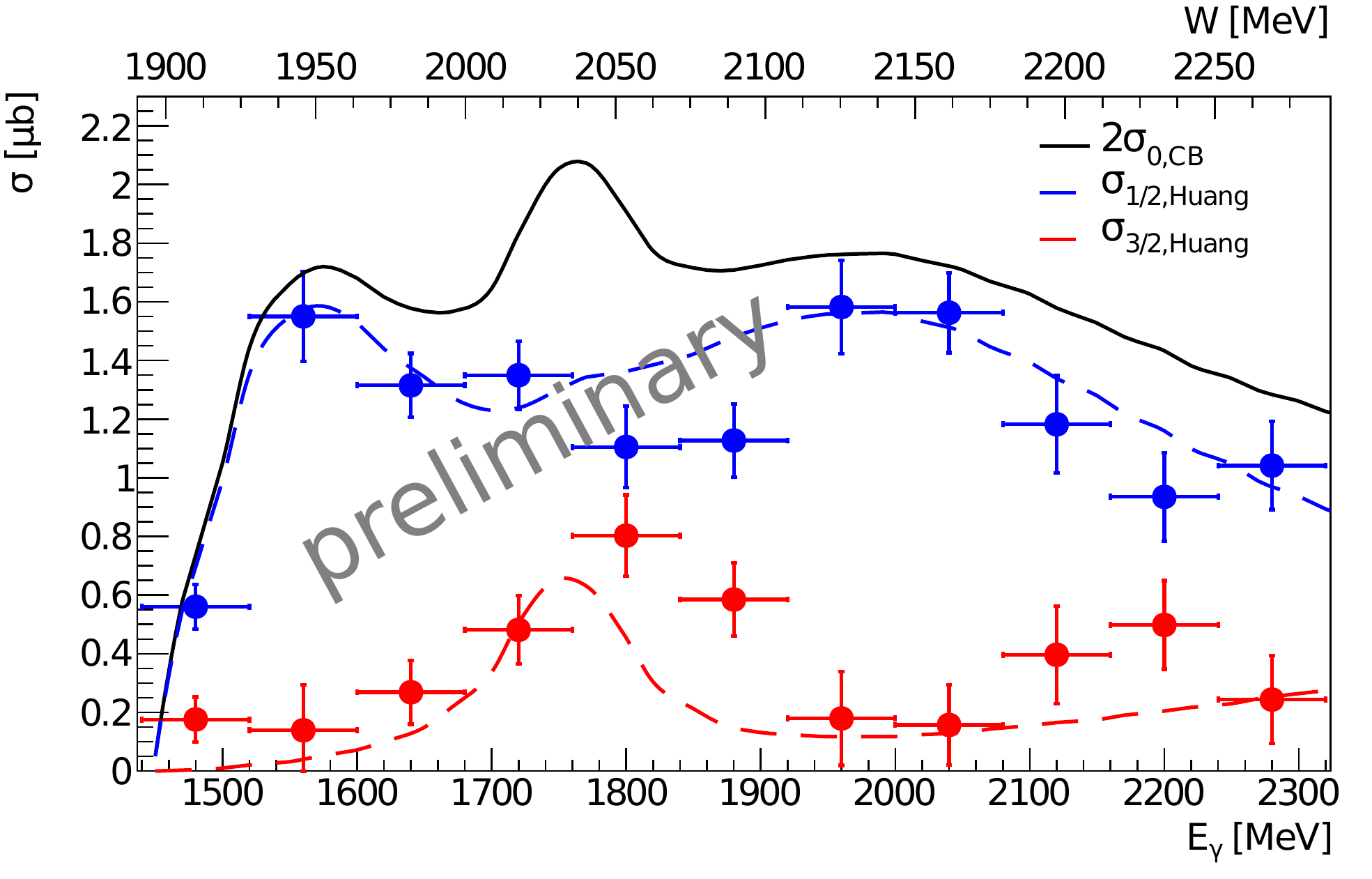,scale=0.42}
\begin{minipage}[t]{16.5 cm}
\caption{Preliminary $\gamma p \to p\eta^{\prime}$ results for the double
polarization observable $E$ (left hand side) and the split of the total cross
section (right hand side) into the $\sigma_{1/2}$ and $\sigma_{3/2}$ helicity
components from the CBELSA/TAPS experiment \cite{Afzal_14}. Curves on the
left are model predictions from ETA$^{\prime}$-MAID (blue) \cite{Chiang_03}
and the NH model \cite{Huang_13} (green: NH fit to CLAS data, red: NH fit to
CBELSA data). Curves on the right: NH fit to CBELSA data, black: total cross
section, blue: $\sigma_{1/2}$, red: $\sigma_{3/2}$. \label{fig:afzal}}
\end{minipage}
\end{center}
\end{figure}

As one example of the data to come, we show in Fig.~\ref{fig:afzal}
preliminary results for a measurement with a circularly polarized photon beam
and a longitudinally polarized target. These allow the extraction of the
double polarization observable $E$ and the division of the total cross
section into its helicity components $\sigma_{1/2}$ and $\sigma_{3/2}$. The
$E$ asymmetry is positive over the entire range of photon energies measured,
which means that $\sigma_{1/2}$ is always dominant. This is apparent in the
separation of the helicity components shown on the right hand side of the
figure. Here one should note, however, that only the values for the $E$
asymmetry were extracted directly from data; the points for the two helicity
components were then obtained by applying this asymmetry to the model fit for
the total cross section. Nevertheless, the $\sigma_{3/2}$ component seems to
show a local maximum around $W=2050$~MeV where the NH-model \cite{Huang_13}
predicts the contribution from the $P_{13}$ state and the chiral quark model
the $D_{15}$ state. Both these states could have significant $\sigma_{3/2}$
couplings so that the $E$ asymmetry cannot distinguish between them. However,
the upcoming results for observables such as beam and target asymmetries will
be more selective between different spin configurations.

First experimental data for the beam asymmetry $\Sigma$ for the
$\gamma p\rightarrow p\eta^{\prime}$ reaction at threshold
measured with a linearly polarized photon beam have very
recently been reported by the GRAAL collaboration
\cite{Sandri_14}. The results are very surprising. The
asymmetry shows a strong polar-angle dependence (positive for
forward angles, negative for backward angles with a zero
crossing around 90$^{\circ}$.), the closer to threshold the
stronger. Such a behaviour is not in agreement with dominant
contributions from $S-$ and $P-$wave states in the threshold
region, which were favoured by all previous model analyses. The
authors discuss possible $P$ - $D$ or $S$ - $F$ interferences.
It is obvious that data for $\Sigma$ with finer energy binning
and also measurements of further polarization observables will
be needed to finally identify the dominant contributions to
$\eta^{\prime}$ threshold production. In this sense, this
reaction even in the immediate threshold region is still much
less understood then $\eta$-production or the production of
$\eta\pi$-pairs discussed in the next section.

\subsubsection{\it Photoproduction of $\eta\pi$ pairs off nucleons}
\label{ssec:etapi}

The photoproduction of meson pairs plays an increasingly
important role in the investigation of nucleon resonances. High
lying states have an appreciable phase-space available for
sequential decay modes involving an intermediate excited state.
Furthermore, sometimes such decays might be preferred for
nucleon structure reasons. In some higher lying multiplets in
the quark model, both oscillators may be excited. It is then
plausible that such states will de-excite first one oscillator
in a decay to an appropriate excited nucleon state, which
subsequently decays to the nucleon ground state. It is thus
possible that direct decays to the nucleon ground state could
be strongly suppressed for an entire multiplet of states which
must therefore be searched for and studied through cascade
decays. The problem is similar to ones in nuclear physics. If
only direct decays to the ground state of nuclei had been
looked for, many excitation degrees of freedom, such as
collective rotations or vibrations, would have been missed.

The many degrees of freedom associated with the photoproduction of
pseudoscalar meson pairs complicate, of course, the interpretation of the
results. Such transitions involve eight complex amplitudes \cite{Roberts_05}
that are functions of five kinematic variables (for example, two Lorentz
invariants and three angles). The measurement of eight independent
observables would be needed just to extract the magnitudes of all the
amplitudes in a unique way (not even considering ambiguities arising from
finite statistical precision of the data). Fixing in addition the phases
would require the measurement of 15 observables. It is therefore clear that
the analysis of differential cross section data alone cannot solve the
problem. A complete measurement appears totally unrealistic but the
measurement of at least a few polarization observables will be indispensable.

It is mostly the production of pion pairs that was so far
studied (see e.g., \cite{Sarantsev_08,Zehr_12,Kashevarov_12}
and references therein), but over the last few years
$\eta\pi$-pairs have also moved into the frame. Their
production is more selective since the isoscalar $\eta$ can be
only emitted in $N^{\star}\rightarrow N^{(\star)}$ and
$\Delta^{\star}\rightarrow \Delta^{(\star)}$ decays. One may
therefore expect that two classes of nucleon resonances may
play important roles for $\eta\pi$ production, viz.\ excited
$\Delta^{\star}$-states with significant $\eta$-decays to the
$\Delta$(1232) and $N^{\star}$ and $\Delta^{\star}$ resonances
decaying to S$_{11}$(1535) via pion emission. Both types of
decays select in a unique way sub-classes of nucleon states
with special properties. At higher incident photon energies
there should also be contributions from the decay of the
$a_{0}$(980) meson, which decays dominantly to $\eta\pi$
\cite{Beringer_12}. Particularly interesting for nucleon
resonance physics is the neutral $\eta\pi^0$ decay where the
non-resonant background is suppressed. Most recent results were
obtained for this channel.

\begin{figure}[tb]
\begin{center}
\epsfig{file=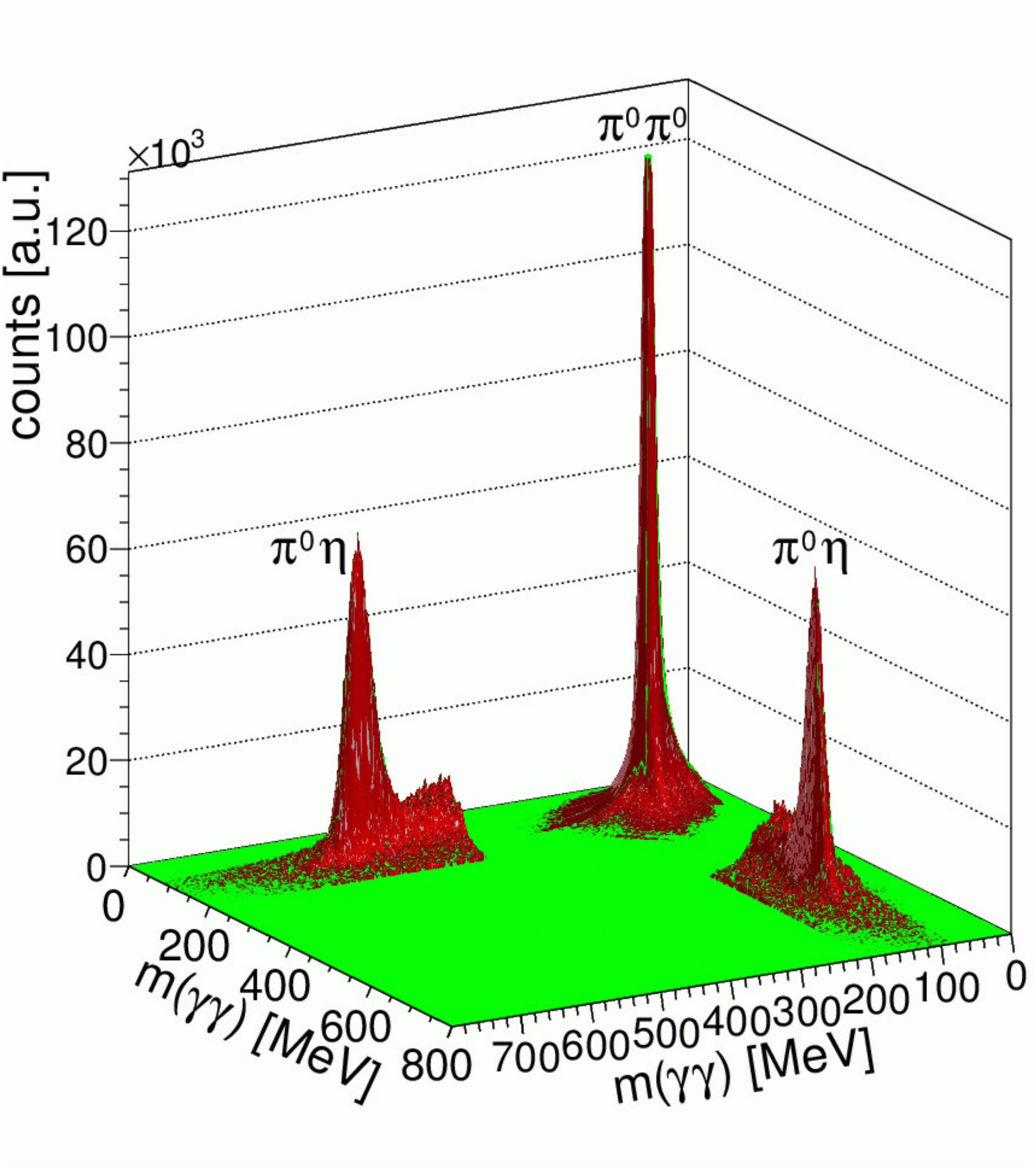,scale=0.37}
\epsfig{file=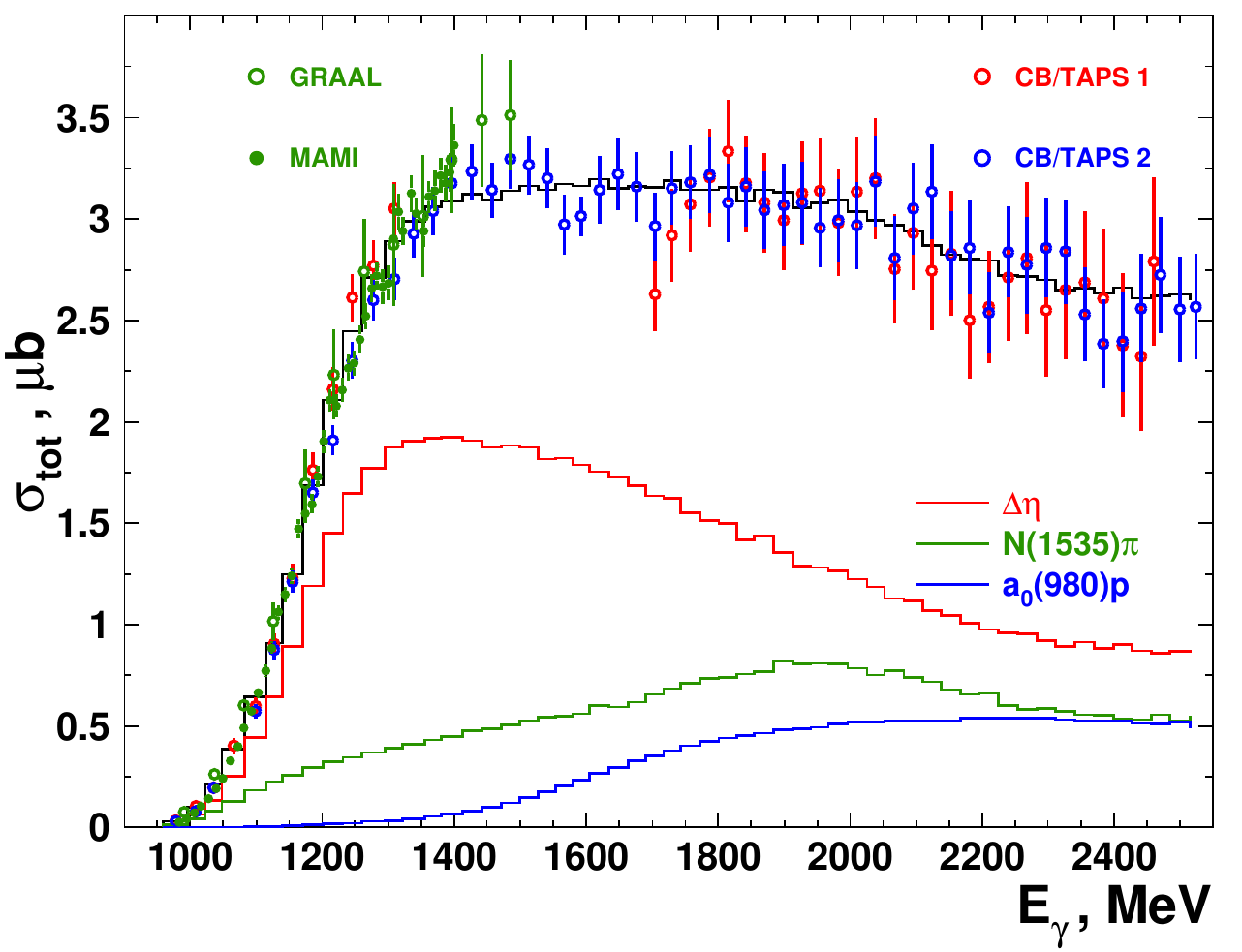,scale=0.73}
\begin{minipage}[t]{16.5 cm}
\caption{Left hand side: typical two-dimensional invariant mass spectrum for
four-photon events from $\gamma p \to pX$ with $X=2\pi^0$ and $\eta\pi^0$
peaks (data from Crystal Ball/TAPS at MAMI \cite{Keshelashvili_14},
the region around the $\eta\pi$ peak is scaled up).
Right hand side: Total cross section for $\gamma
p\rightarrow p\pi^0\eta$ from \cite{Ajaka_08} (GRAAL, green open circles),
Crystal Ball/TAPS at MAMI \cite{Kashevarov_09} (filled green circles) and two
data sets from CBELSA/TAPS \cite{Gutz_14} (CBTAPS 1, CBTAPS 2, open red and
blue circles). The red, green, and blue histograms show the results of the
BnGa analysis for the $\Delta\eta$, $S_{11}$(1535)$\pi^0$, and
$a_{0}$(980)$p$ final states; the black histogram represents the full fit.
\label{fig:etapi_tot}}
\end{minipage}
\end{center}
\end{figure}

Total cross sections, invariant mass distributions, and also
some polarization observables for the $\gamma p\rightarrow
p\pi^0\eta$ reaction have been measured at LNS
\cite{Nakabayashi_06}, GRAAL \cite{Ajaka_08}, ELSA
\cite{Gutz_14,Horn_08a,Horn_08b,Gutz_08,Gutz_10}, and at MAMI
\cite{Kashevarov_09,Kashevarov_10}. The three-body final state
cannot be identified through a missing-mass analysis of the
recoil proton and all decay modes of the $\eta\pi^0$ pair with
reasonably large branching ratios involve four or more photons
(from the $\pi^0$ decay and the $\eta\rightarrow 2\gamma$,
$\eta\rightarrow 3\pi^0\rightarrow 6\gamma$, or
$\eta\rightarrow\pi^0\pi^+\pi^-\rightarrow 2\gamma\pi^+\pi^-$
decays). Only experiments equipped with electromagnetic
calorimeters that cover almost 4$\pi$ can successfully
investigate this reaction. All the recent measurements used the
$\eta\pi^0\rightarrow 4\gamma$ decay channel (branching ratio
$\approx$38.9\%) and identified the reaction by combining invariant-
and missing-mass analyses, with or without kinematic fitting.

A typical two-dimensional invariant-mass spectrum for the reaction is shown
on the left-hand side of Fig.~\ref{fig:etapi_tot}. The results for the total
cross section are summarized on the right of the same figure. The overall
shape of the excitation function, with the rapid rise from the threshold at
$E_{\gamma}^{\rm thr}$ = 930~MeV and the flattening out up to $E_{\gamma}=
2.5$~GeV is well established, but there are still discrepancies between
different measurements in the absolute normalization. The data from the
earlier Crystal Barrel experiment \cite{Horn_08a,Horn_08b} (not shown in the
figure) reach a maximum cross section of almost $4\mu b$ (but have systematic
uncertainties on the 20\% level); the recent CBELSA/TAPS data \cite{Gutz_14}
shown in the figure have been normalized in the absolute scale to the low
energy GRAAL and MAMI data (renormalization factors on the order of 20\%).

\begin{figure}[t!]
\epsfysize=9.0cm
\centerline{
\epsfig{file=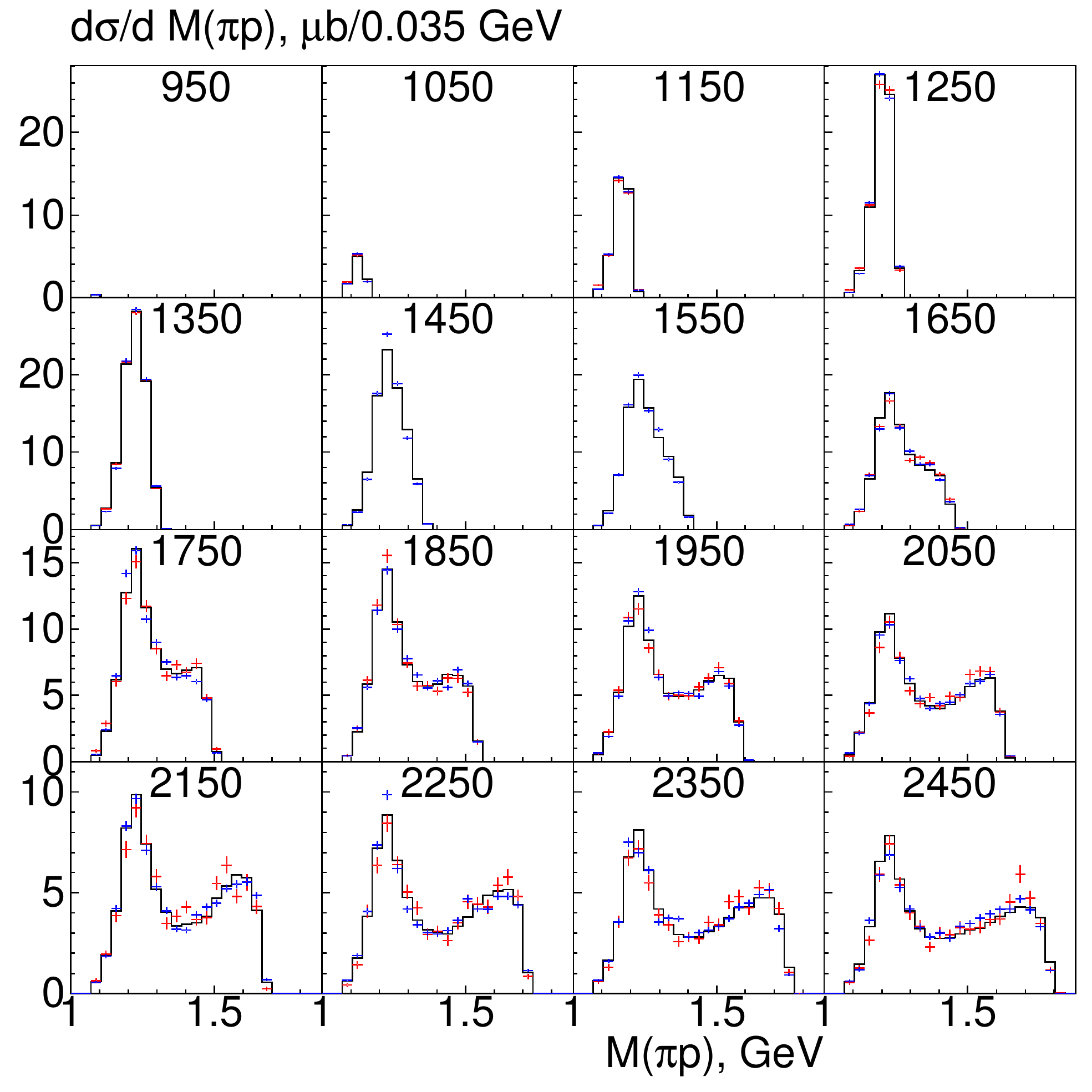,scale=0.45}
\epsfig{file=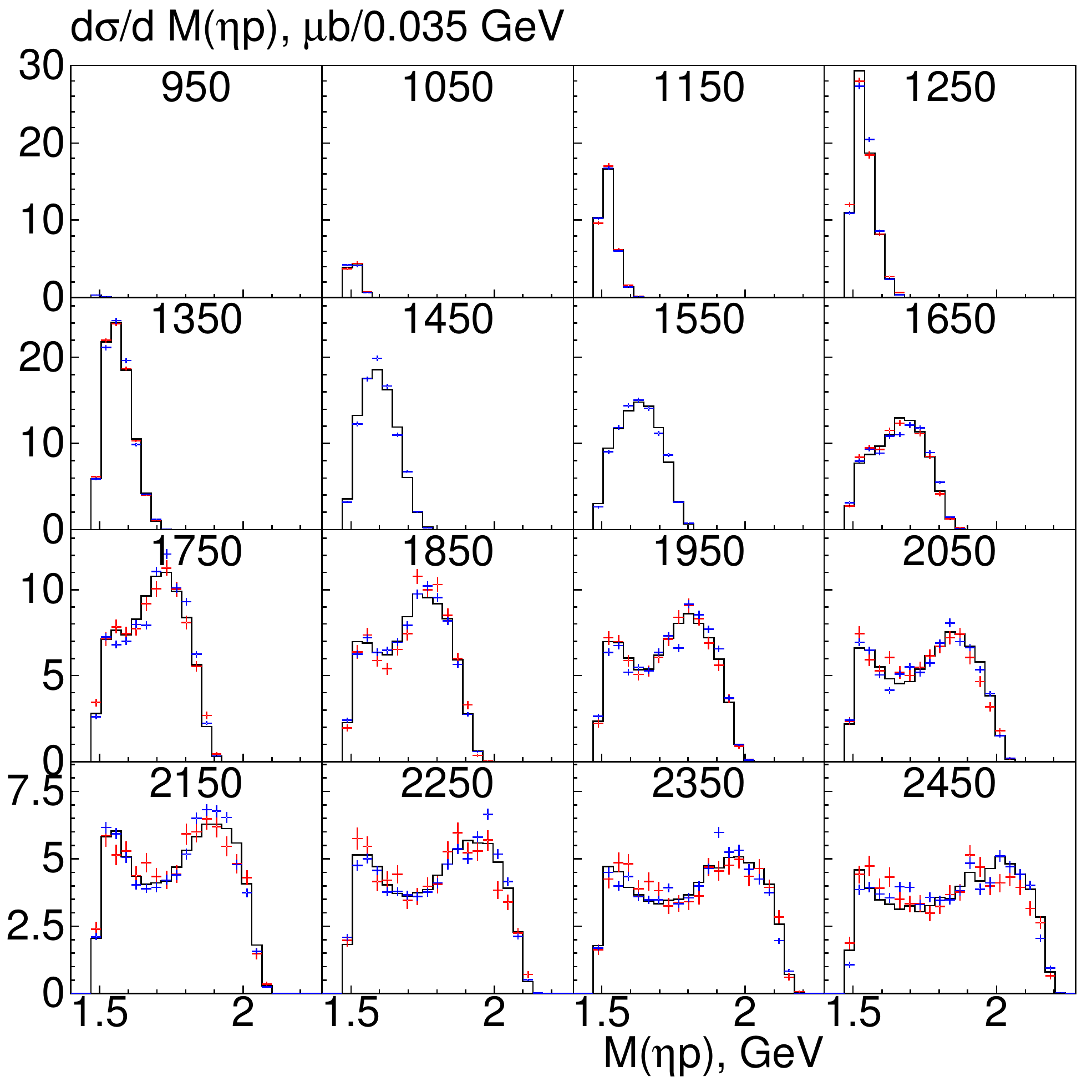,scale=0.45}}
\hspace*{0.3cm}\epsfig{file=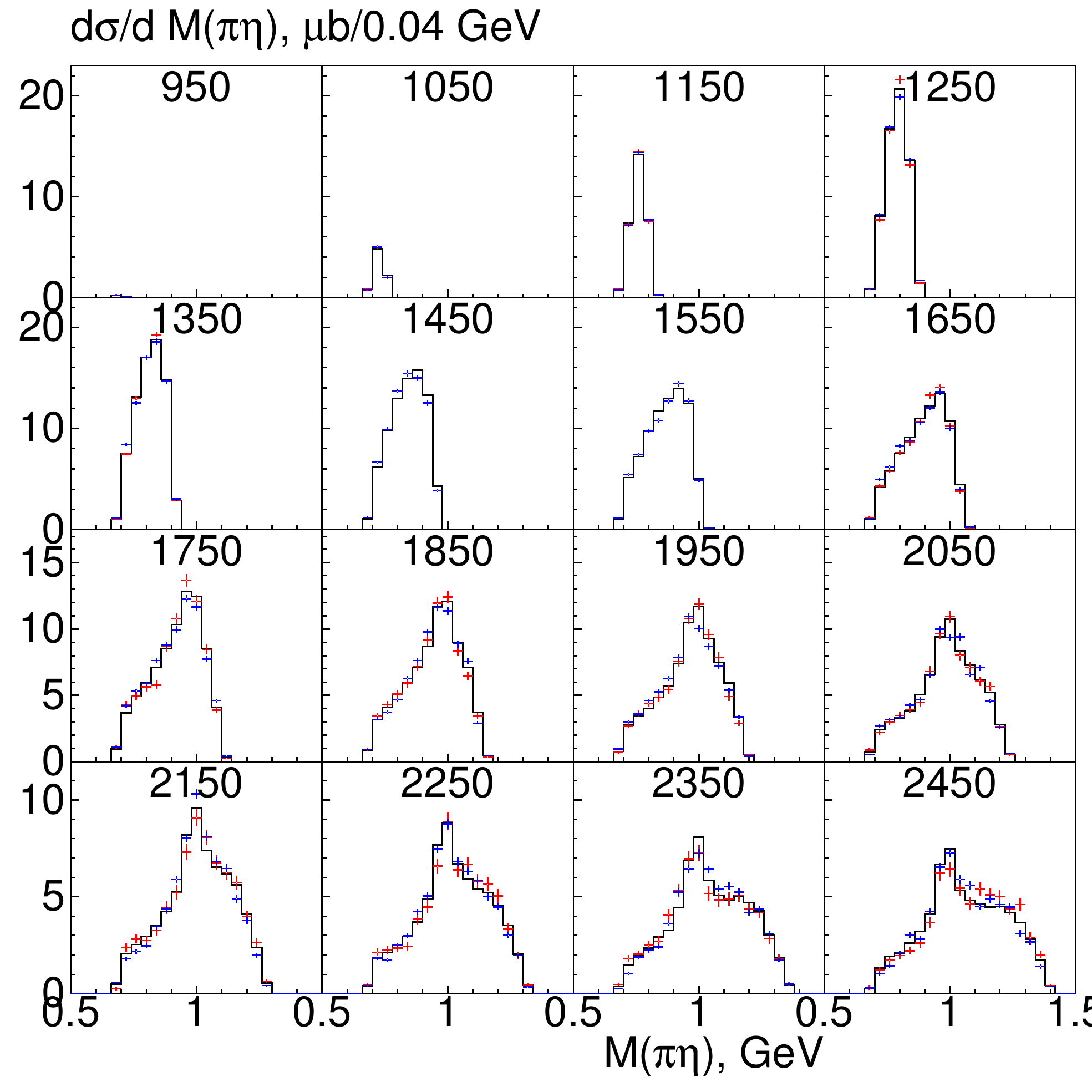,scale=0.45}

\vspace*{-8.5cm} \hspace*{9.75cm}\begin{minipage}[t]{8.5 cm}
\caption{Invariant mass distributions of $\pi^0 p$ (top, left
hand side), $\eta p$ (top, right hand side), and $\pi^0\eta$
pairs (bottom) from the $\gamma p\rightarrow p\pi^0\eta$
reaction for different values of the incident photon energy.
The data are taken from Ref.~\cite{Gutz_14} (red and blue
points from CBTAPS~1 and CBTAPS~2 data) and the BnGa model fits
shown are discussed in this reference. \label{fig:etapi_diff}}
\vspace*{4cm}
\end{minipage}
\end{figure}

For the unpolarized data angular distributions (in different frames) and
invariant mass distributions of the meson pairs and the nucleon - meson pairs
(Dalitz plots) have been measured. Typical results for the invariant mass
distributions from the most recent CBELSA/TAPS measurement are summarized in
Fig.~\ref{fig:etapi_diff}. Already without a detailed model it is obvious
from these results that the three reaction types discussed above make
significant contributions: $\Delta^{*}\rightarrow P_{33}(1232)\eta$ decays
produce a pronounced peak in the $p\pi^0$ invariant mass around
1232~MeV/$c^2$, $(N^{\star},\Delta^{\star})\rightarrow S_{11}(1535)\pi^0$
decays lead to an accumulation of strength in the $p\eta$ spectra at the
lower limit of possible invariant masses (the sum of the proton and $\eta$
mass of 1485~MeV/$c^2$ is not much below the resonance position), and the
$a_{0}$(980) meson shows up as a peak in the $\eta\pi^0$ invariant masses.

\begin{figure}[tb]
\begin{center}
\epsfig{file=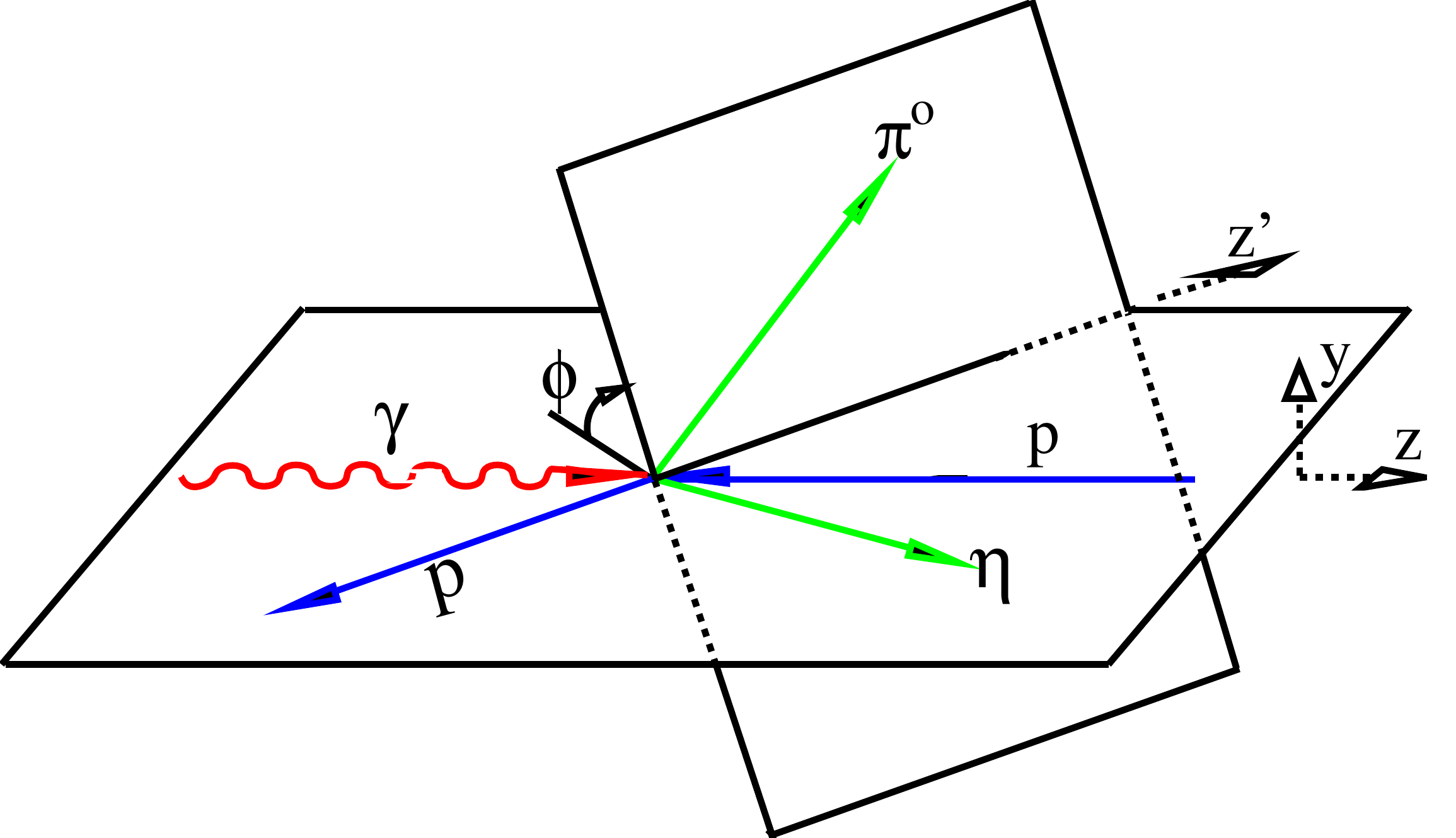,scale=0.43}
\hspace*{0.5cm}
\epsfig{file=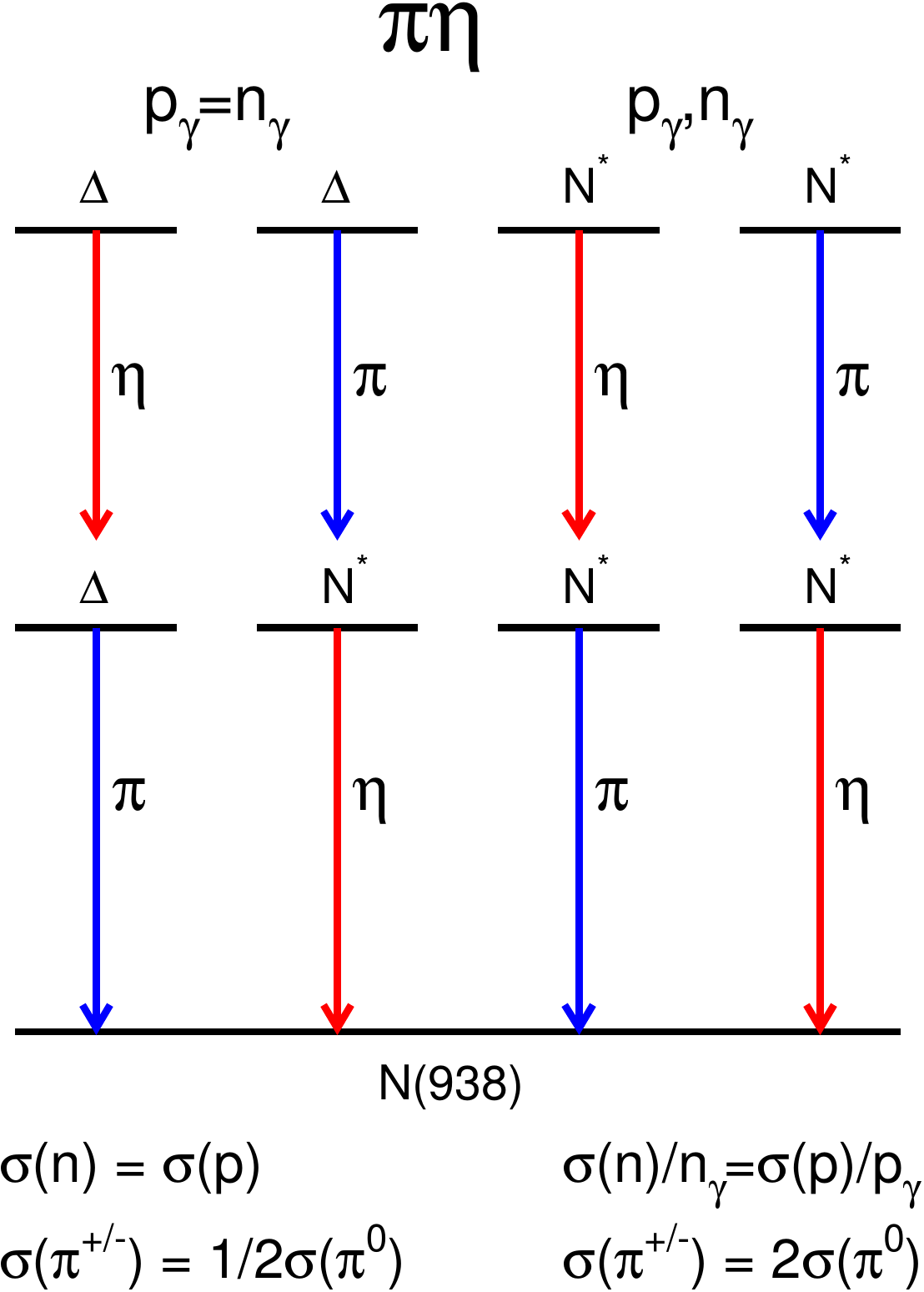,scale=0.450}
\begin{minipage}[t]{16.5 cm}
\caption{Left hand side: Vector and angle definitions in the c.m.\ system.
$\phi$ is the angle between the reaction plane (defined by $\vec{k}$ and
$\vec{p}_{p}$) and the production plane of the two mesons (defined by
$\vec{p}_{\pi^0}$ and $\vec{p}_{\eta}$). Right hand side: possible decay
chains and their cross section relations for production off protons and
neutrons, and for neutral and charged pions. \label{fig:asym_def}}
\end{minipage}
\end{center}
\end{figure}

Polarization observables have so far only been explored with polarized photon
beams (linearly polarized at GRAAL and ELSA \cite{Ajaka_08,Gutz_14,Gutz_10}
and circularly polarized at MAMI \cite{Kashevarov_10}). Such observables for
a final state with a nucleon and a pair of pseudoscalar mesons can be
analyzed in different ways (see Refs.~\cite{Roberts_05,Fix_11} for a general
discussion). For the results from the GRAAL \cite{Ajaka_08} and CBELSA/TAPS
\cite{Gutz_14} experiments, beam asymmetries were analyzed in quasi-two-body
kinematics assuming $\gamma p\rightarrow \eta X$, $\gamma p\rightarrow \pi^0
Y$, and $\gamma p\rightarrow p Z$, where $X\rightarrow p\pi^0$, $Y\rightarrow
p\eta$, and $Z\rightarrow \eta\pi^0$ were considered as intermediate state
particles. One can, however, also use a full three-body approach, which
involves additional degrees of freedom. The kinematics are depicted in the
left-hand side of Fig.~\ref{fig:asym_def}. In contrast to single meson
production, two independent planes can be defined, for example the reaction
plane spanned by the incident photon and the recoil nucleon and the
production plane spanned by the two mesons. Modulations of the differential
cross section can then be observed as function of the angle $\phi$ between
the two planes.

For a linearly polarized beam the differential cross section is then given by
\begin{equation}
\frac{d\sigma}{d\Omega}=
\left(\frac{d\sigma}{d\Omega}\right)_0
\left[1+p_l(I^c(\phi)\cos2\phi+I^s(\phi)\sin2\phi)\right],
\label{eq:linpol}
\end{equation}
where $p_l$ is the degree of linear polarization and $I^s$ and
$I^c$ are polarization observables with
\begin{eqnarray}
I^s(\phi)  =   -I^s(2\pi-\phi)  &=& \sum_{n=1}^{\infty} a_n \sin n\phi\:,\\
\label{eq:polex1}
I^c(\phi)  =  \phantom{-}I^c(2\pi-\phi)  &=& \sum_{n=0}^{\infty} b_n \cos n\phi\;.
\label{eq:polex2}
\end{eqnarray}
The constant ($n=0$) term in the expansion of $I^c$ corresponds to the
two-body beam asymmetry $\Sigma$.

For a circularly polarized photon beam (which due to parity invariance does
not produce asymmetries for two-body final states) one gets
\begin{equation}
\frac{d\sigma}{d\Omega}=
\left(\frac{d\sigma}{d\Omega}\right)_0
\left[1+p_{\odot}I^{\odot}(\phi)\right]\;,
\label{eq:circpol}
\end{equation}
where $p_{\odot}$ is the degree of circular polarization and
$I^{\odot}(\phi)$ can be also expanded in the form of
Eq.~(\ref{eq:polex1}).

The polarization observables are very sensitive to small reaction amplitudes
that enter only via interference terms. The asymmetry $I^{\odot}$, e.g., has
been recently much explored for the photoproduction of pion pairs
\cite{Strauch_05,Krambrich_09,Oberle_13,Oberle_14}, where it uncovered many
deficiencies in the existing model analyses of these reactions. The
asymmetries for the $\eta\pi^0$ final state are substantial; experimental
results for $I^s$ and $I^c$ are discussed in \cite{Gutz_14,Gutz_10} and
results for $I^{\odot}$ in \cite{Kashevarov_10}.

The experimental results have been studied within the BnGa coupled-channel
analysis (see \cite{Gutz_14,Horn_08a,Horn_08b,Gutz_10}), an isobar model
developed by Fix and collaborators (see
\cite{Kashevarov_09,Kashevarov_10,Fix_10}), and were also compared to the
results from a chiral unitary approach (see
\cite{Ajaka_08,Doering_06a,Doering_06b}). Although there are differences
between the details of the results from the models, the main features seem to
be quite stable. All models agree that close to threshold the reaction is
dominated by the excitation of the $D_{33}$(1700) resonance and its decay to
$\eta P_{33}(1232)$. Contributions from non-resonant backgrounds, such as
nucleon Born or $t$-channel exchanges, appear to be small.

The division of the total cross section into the $\eta P_{33}(1232)$,
$S_{11}(1535)\pi^0$, and $a_0(980)p$ final states in the BnGa model analysis
is shown in Fig.~\ref{fig:etapi_tot}. The dominance of the $\Delta\eta$ final
state in the threshold region is obvious. Apart from the $D_{33}(1700)$, the
BnGa analysis finds contributions from several other $\Delta$-resonances
$S_{31}(1900)$, $F_{35}(1905)$ (both weak), and from $P_{31}(1910)$,
$P_{33}(1920)$, and $D_{33}(1940)$ with decays into $\Delta\eta$ and
$S_{11}(1535)\pi^0$ (the $\Delta\eta$ decays are always stronger) and
contributions from $N^{\star}\rightarrow S_{11}(1535)\pi$ decays for
$P_{11}(1710)$, $P_{11}(1880)$, $P_{13}(1900)$, $P_{11}(2100)$, and
$D_{13}(2120)$. The analysis by Fix and coworkers \cite{Kashevarov_09,Fix_10}
does not extend over the same energy range (it used mainly the MAMI and GRAAL
data up to an incident photon energy of 1.5~GeV) and did not clearly identify
contributions from the $S_{11}(1535)\pi$ final state. For the $\Delta\eta$
final state, contributions from $D_{33}(1700)$, $P_{33}(1600)$ (weak),
$P_{31}(1750)$, $F_{35}(1905)$, $P_{33}(1920)$, and $D_{33}(1940)$ are
quoted. The two states $P_{33}(1920)$, and $D_{33}(1940)$ were seen in both
analyses. These two states were first reported from the $\gamma p\rightarrow
p\eta\pi^0$ reaction by Horn and coworkers \cite{Horn_08a} and are very
interesting for nucleon structure because they form a parity doublet.

\begin{figure}[tb]
\begin{center}
\epsfig{file=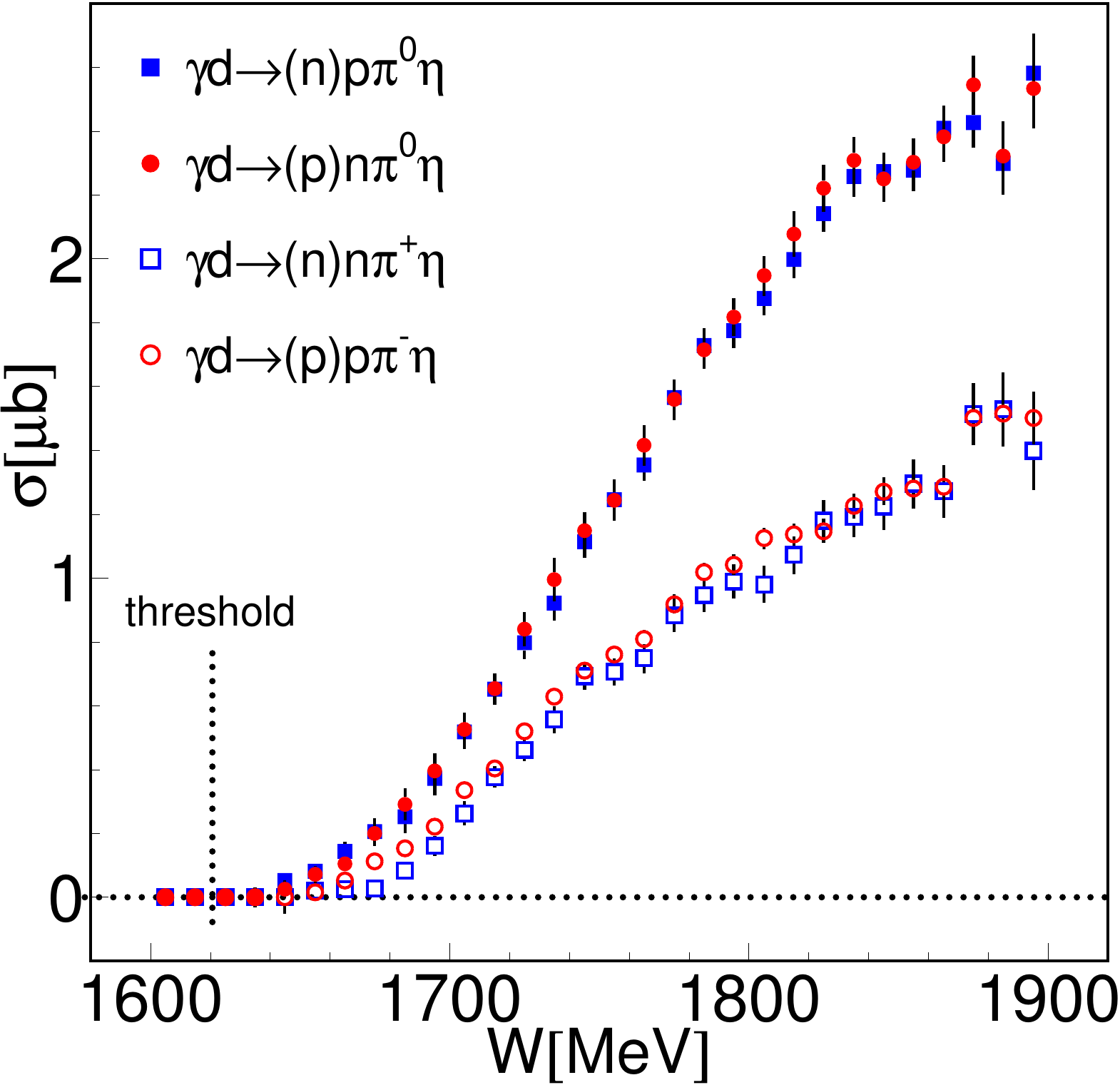,scale=0.450}
\epsfig{file=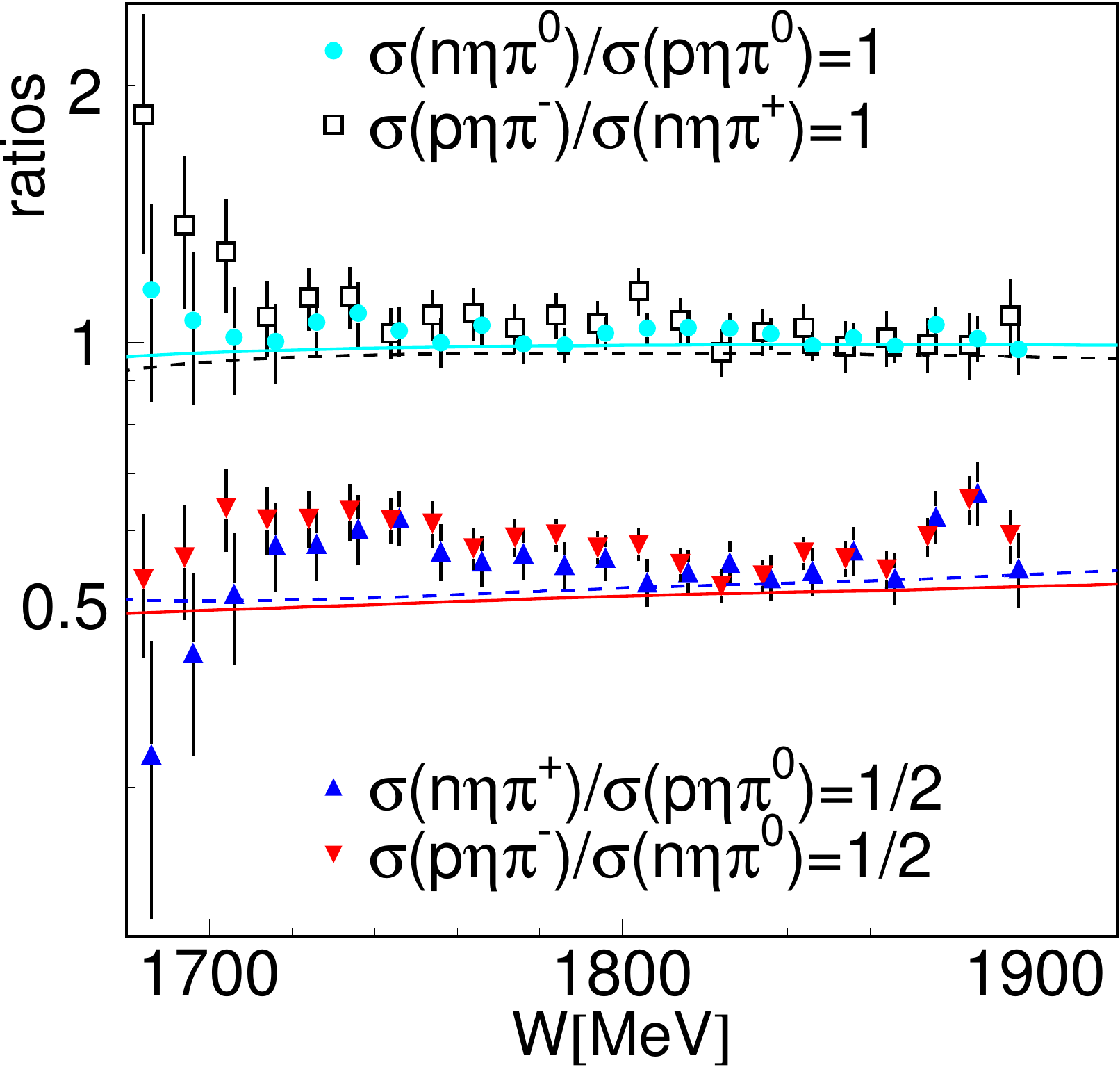,scale=0.450}
\begin{minipage}[t]{16.5 cm}
\caption{Quasi-free excitation functions for $\gamma N\rightarrow N\eta\pi$
measured with a deuteron target. Left hand side: total cross sections, right
hand side: cross section ratios. All data are preliminary. The model results
are from \cite{Kashevarov_09,Fix_13}. \label{fig:etapi_alex}}
\end{minipage}
\end{center}
\end{figure}

The $\gamma p\rightarrow p\pi^0\eta$ reaction is obviously an excellent tool
for the study of the $D_{33}(1700)$ state. D\"oring, Oset, and Strottman
\cite{Doering_06a,Doering_06b} have made predictions within a coupled-channel
chiral unitary theory for meson-baryon scattering in which this state is
dynamically generated. They predict a strong decay to the $S_{11}(1535)\pi$
final state and this is clearly reflected in the invariant mass distributions
within their model. However, it does not agree well with experiment;
invariant mass distributions with a dominant decay to $\Delta\eta$ fit much
better \cite{Ajaka_08}.

Further information about the reaction mechanism comes from the isospin
degree of freedom. The possible sequential resonance decays are summarized on
the right-hand side of Fig.~\ref{fig:asym_def}. The $\gamma N\rightarrow
\Delta^{\star}\rightarrow \eta\Delta(1232)\rightarrow \eta\pi N$ reaction
chain is characterized by the equal electromagnetic $\gamma N\Delta$
couplings for the excitation of $I=3/2$ states on neutrons and protons, the
isoscalar nature of the $\eta$, and the isovector nature of the pion. It then
follows immediately that the above reaction chain must have the same isospin
pattern as photoproduction of single pions through the $\Delta$-resonance:
\begin{equation}
 \sigma (\gamma p\rightarrow \eta\pi^0 p) = \sigma (\gamma n\rightarrow \eta\pi^0 n) =
2\sigma (\gamma p\rightarrow \eta\pi^+ n)  = 2\sigma (\gamma n\rightarrow \eta\pi^- p)\;.
\label{eq:isorel}
\end{equation}
Preliminary results from a measurement of all four possible final states at
MAMI in quasi-free kinematics from a deuteron target shown in
Fig.~\ref{fig:etapi_alex} \cite{Krusche_14b} demonstrate that, up to
invariant masses of 1.9~GeV this relation holds almost perfectly. This
means that significant contributions from $N^{\star}$ excitations are ruled
out for this energy range. The isospin argument does not help to distinguish
between the $\Delta^{\star}\rightarrow P_{33}(1232)\eta\rightarrow N\eta\pi$
and $\Delta^{\star}\rightarrow S_{11}(1532)\pi\rightarrow N\eta\pi$ decay
chains. Their contributions can be probed, e.g., by the energy distributions
of the mesons discussed in Sec.~\ref{ssec:photon_coh}.

It should be noted that important FSI effects have been
observed for $\eta\pi$ photoproduction on the deuteron. The
cross sections for the $\eta\pi^0$ final state are suppressed
for quasi-free protons by the order of 25\% with respect to
free proton targets, although one should bear in mind that
there are still issues concerning the absolute normalization of
the free proton data sets.

%
%

\subsection{$\boldsymbol{\eta}$ production in $\boldsymbol{\pi N}$ collisions} %

In contrast to the photoproduction case, most of the measurements of the
near-threshold $\pi^-p\to \eta n$ differential cross section were carried out
many years ago and some inconsistencies are apparent in the data
base~\cite{Bulos_64,Deinet_69,Richards_70,Binnie_73,Debeham_75,Feltesse_75,Brown_79,Baker_79}.
There are, however, two more recent experiments where the $\eta$ meson was
reliably identified through its $2\gamma$ or $6\gamma$ decay
mode~\cite{Prakhov_05,Bayadilov_12}.

\begin{figure}[htb]
\begin{center}
\includegraphics[width=0.45\textwidth]{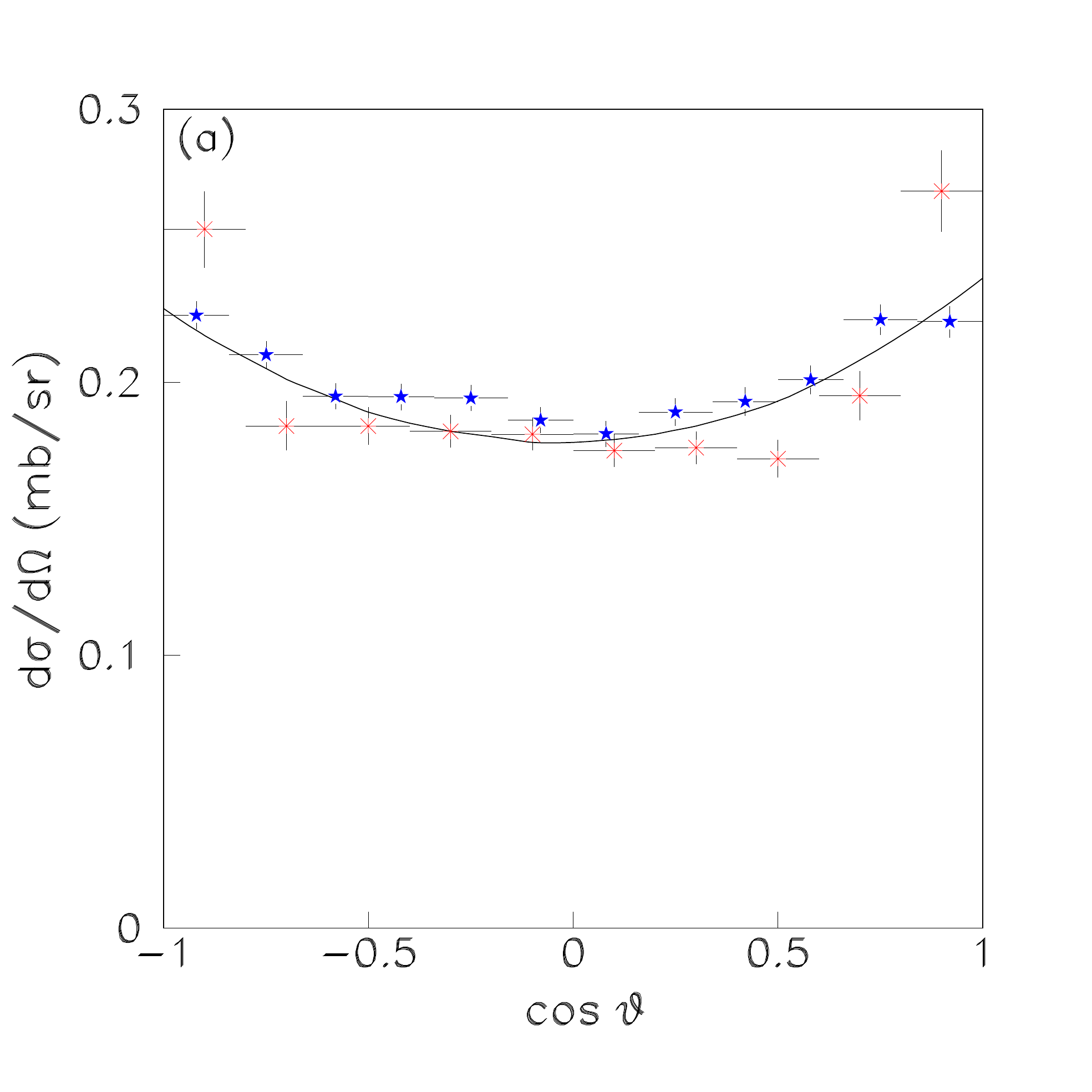}
\includegraphics[width=0.45\textwidth]{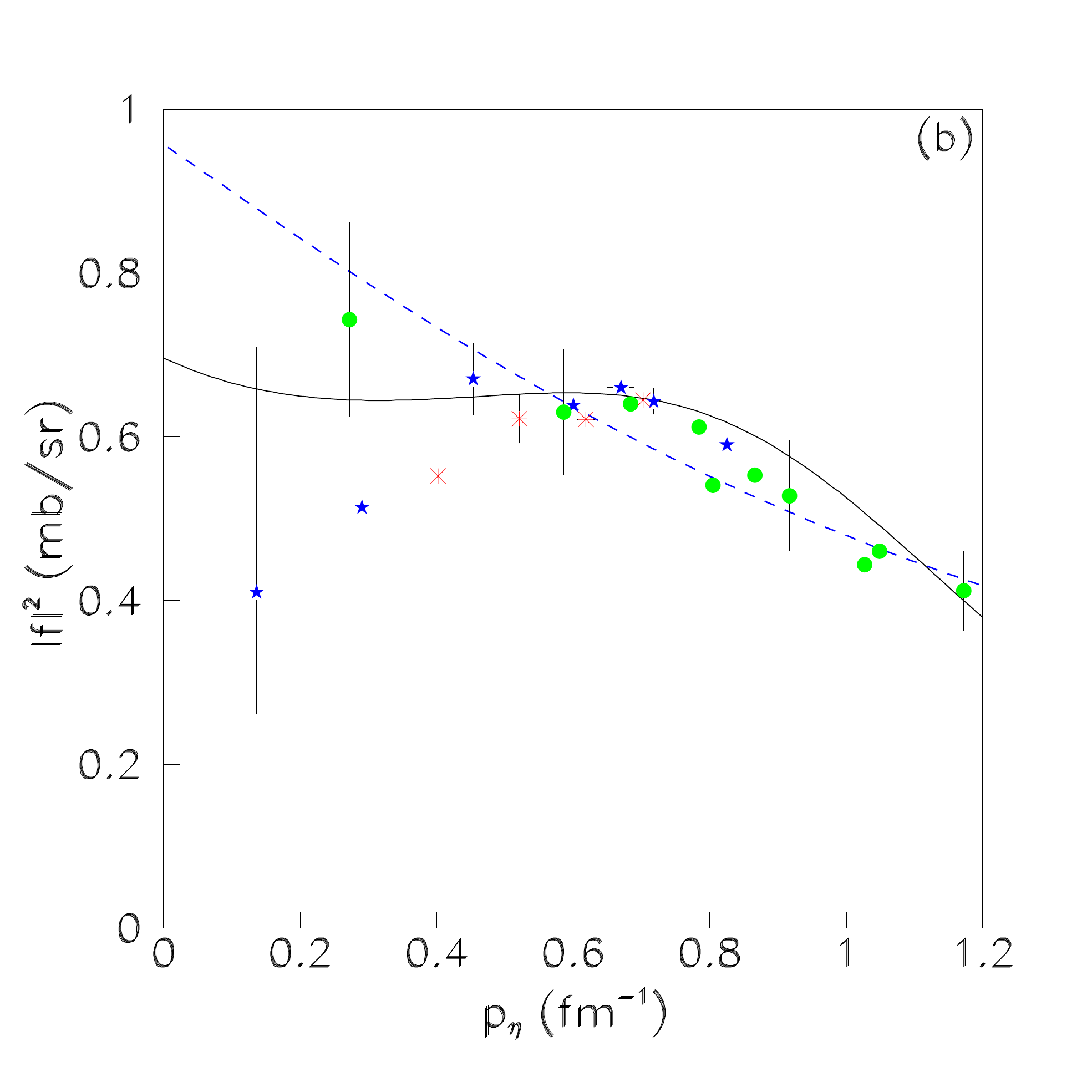}
\caption{\label{fig:Prakhov_stot} (a) The $\pi^-p\to \eta n$ differential
cross section measured at a pion laboratory momentum of 732~MeV/$c$ (blue
stars)~\cite{Prakhov_05} and 730~MeV/$c$ (red crosses)~\cite{Bayadilov_12}.
The systematic uncertainties were $\approx 6\%$ for the Brookhaven
experiment~\cite{Prakhov_05} and rather more at Gatchina~\cite{Bayadilov_12},
mainly due to the evaluation of the acceptance. The curve represents the
results of the Bonn-Gatchina partial wave analysis~\cite{Anisovich_12}. (b)
Average amplitude-squared of the $\pi^-p\to \eta n$ reaction showing the
results of Ref.~\cite{Prakhov_05} (blue stars) and \cite{Bayadilov_12} (red
crosses) as well as a selection of older
data~\cite{Bulos_64,Deinet_69,Binnie_73,Brown_79} (green circles). The solid
(black) line represents the contribution from the $N^*(1535)$ isobar with
parameters taken from Ref.~\cite{Shrestha_12}. The (blue) dashed line is an
effective-range fit~\cite{Wilkin_93} to the older data. }
\end{center}
\end{figure}

At threshold the production cross section is dominated by the
$S_{11}(1535)$ isobar but, at a little higher energy, the
differential cross section develops an anisotropy, as
illustrated by the data~\cite{Prakhov_05,Bayadilov_12} shown in
Fig.~\ref{fig:Prakhov_stot}a at a c.m.\ momentum
$p_{\eta}\approx 140$~MeV/$c$, i.e., an invariant mass of
$W\approx 1515$~MeV/$c^2$. Because the distribution is fairly
symmetric, it was suggested that the main deviations from
isotropy could arise from the interference between $s$- and
$d$-waves, where the $d$-wave amplitude could be quite
small~\cite{Prakhov_05,Bayadilov_12}. However, the partial wave
analysis curve that is also shown involves significant $p$-wave
contributions~\cite{Anisovich_12}.

In addition to the binning in the incident pion momentum, a more serious
problem in interpreting the data is the uncertainty in the knowledge of the
absolute pion momentum, which is quoted as being $\approx \pm
2.5$~MeV/$c$~\cite{Prakhov_05} or $\approx \pm
1.5$~MeV/$c$~\cite{Bayadilov_12} in the two more recent experiments. Such
uncertainty is particularly significant close to threshold, as can be seen in
the evaluation of the average amplitude-squared:
\begin{equation}
\label{ampsq}
\overline{|f(\pi^-p\to \eta n)|^2} = (p_{\pi}/p_{\eta})\sigma_{\rm tot}(\pi^-p\to \eta n)/4\pi,
\end{equation}
where $p_{\pi}$ and $p_{\eta}$ are c.m.\ momenta. Thus the biggest
contribution to the error bars in Fig.~\ref{fig:Prakhov_stot}b near threshold
comes from the uncertainty in the absolute value of the incident beam
momentum, which is a systematic effect. In all cases the $s$-$d$ interference
drops out in the evaluation of $\overline{|f|^2}$. The agreement with the
selected older data~\cite{Bulos_64,Deinet_69,Binnie_73,Brown_79} is
reasonable.

Also shown in Fig.~\ref{fig:Prakhov_stot}b is the original scattering-length
fit~\cite{Wilkin_93} to the older results. This clearly fails to reproduce
the new data close to threshold but, on the other hand, a parameterization of
the $S_{11}(1535)$ resonance~\cite{Shrestha_12} describes well the forward
dip in $\overline{|f|^2}$. However, other masses and widths for the $S_{11}$
resonance give an equally good description so that these data are not precise
enough to tie down the resonance parameters accurately. It is important to
note that the scattering lengths for both the solid and dashed lines in
Fig.~\ref{fig:Prakhov_stot}b are rather similar. The major difference lies in
the neglect of the effective range term in the early work~\cite{Wilkin_93}
and the consequent scaling to fit the data in the high $p_{\eta}$ region.

The data on $\eta^{\prime}$ production in pion-nucleon
collisions are even rarer. Apart from early bubble chamber
work~\cite{Landolt_88,Rader_72}, there were counter
measurements of the $\pi^- p \to n\eta^{\prime}$ cross sections
near threshold~\cite{Binnie_73}.
%
%
\subsection{Partial wave analysis}

The only $\pi^-p\to \eta n$ results available are unpolarized differential
cross sections so that the use of these data alone would necessarily involve
significant ambiguities. In addition, as remarked in the previous section,
the data are not very precise and their main use seems to be to determine
well the branching ratios of different nucleon resonances into the $\eta n$
channel. However, the $\eta n$ system is very strongly coupled to the
$\pi^-p$, as is seen in the energy dependence of the pion charge-exchange
reaction $\pi^-p\to\pi^0n$ in the backward direction, which has a clear cusp
at the $\eta n$ threshold~\cite{Starostin_05}.

Several coupled-channel partial wave analyses have been carried out in recent
years~\cite{Anisovich_12,Shklyar_13,Shrestha_12,Arndt_06}, generally using
also information gained from $\eta$ photoproduction from the proton. Although
the main resonance findings are qualitatively similar, there are significant
differences in the details that were discussed in the photoproduction
section.

%
%

%
%

\section{Production in $\boldsymbol{\gamma A}$ reactions}
\label{gamma-A}\setcounter{equation}{0}

In section~\ref{sec:elementary} we already discussed photoproduction of
mesons off light nuclei where the aim was the study of quasi-free production
on the neutron. However, the photoproduction of mesons off medium and heavy
nuclei is mostly used as a tool for the investigation of meson - nucleon
(meson - nucleus) interactions and in-medium properties of hadrons. The
experiments can be grouped into three different types of final states
(although this characterization is not always unique and intermediate
situations exist), where different physics topics are explored.
\begin{itemize}
\item{In breakup reactions, at least one nucleon is removed from the
    nucleus and this includes the so-called quasi-free processes. In the
    simplest picture of such a reaction, the incident photon interacts
    with one individual nucleon, the `participant' off which the meson is
    produced, while the rest of the nucleons in the nucleus act only as
    spectators. This is the approach used for the extraction of
    elementary cross sections off neutrons bound in light nuclei but, for
    heavier nuclei, FSI processes involving the `spectator' nucleons will
    always be important. Reactions of this type are explored for the
    study of meson - nucleus interactions and for the investigation of
    hadron in-medium properties. A recent review of the in-medium
    properties of scalar and vector mesons derived from photoproduction
    and other reactions is given by Leupold, Mosel, and Metag
    \cite{Leupold_10}; results for the in-medium properties of nucleon
    resonances are summarized, e.g., in \cite{Krusche_05}. }
\item{Coherent meson photoproduction is characterized by a final system
    where the target nucleus remains in its ground state. By exploiting
    the spin and isospin quantum numbers of the nucleus one can in
    principle project out specific parts of the elementary reaction
    amplitudes. This was, for example, used (as discussed in section
    \ref{ssec:photon_coh}) to unravel the isospin decomposition of the
    electromagnetic $S_{11}$(1535) excitation using the photoproduction
    of $\eta$-mesons on the deuteron. The coherent reaction is also used
    as a doorway for the search of meson-nucleus bound states, such as
    the $\eta$ or $\eta^{\prime}$ mesic nuclei. However, with few
    exceptions, the typical cross sections are small and the experiments
    are demanding (in most cases coherent reactions are difficult to
    identify in the presence of much larger contributions from breakup
    reactions). Detailed studies are so far only available for $\pi^0$
    mesons to investigate the in-medium behaviour of the $\Delta(1232)$
    resonance (see e.g., \cite{Drechsel_99,Krusche_02}) and, in a
    completely different context, for the extraction of nuclear mass form
    factors \cite{Krusche_05a,Maghrbi_13,Tarbert_14}.}
\item{In incoherent production the final-state nucleus is excited (but
    otherwise identical with the initial state nucleus) and de-excites
    typically by the emission of $\gamma$-radiation. Such processes
    provide additional selection possibilities as spin- and isospin
    filters or may be used to study transition form factors. However,
    they are still almost unexplored, due to the small reaction cross
    sections. For $\pi^0$ photoproduction there is a recent measurement
    for the 4.4 MeV excited state of $^{12}$C \cite{Tarbert_08}. }
\end{itemize}

\subsection{Inclusive production}
\label{ssec:photon_incl}

In this section we will summarize the results for reactions where only the
produced meson is used to characterize the final state. Since the cross
sections for coherent production of $\eta$ and $\eta^{\prime}$ mesons are
very small (see section \ref{ssec:photon_coh}), the inclusive reactions are
completely dominated by the breakup contributions. The interaction of mesons
with nuclei has contributed greatly to our knowledge of the strong
interaction. Many properties of meson - nucleon potentials have been studied
with elastic or inelastic reactions induced by pion or kaon beams. However,
such beams are only available for long-lived, charged mesons. The interaction
of short-lived neutral mesons, such as the $\eta$, $\eta^{\prime}$, and
$\omega$, can only be studied in indirect ways when they are produced by some
initial reaction in a nucleus and then interact within the same nucleus. Such
measurements provide data for the extraction of meson - nucleus potentials,
of meson in-medium properties like modified masses and/or lifetimes, and may
also test the in-medium properties of nucleon resonances.

\begin{figure}[thb]
\begin{center}
\epsfig{file=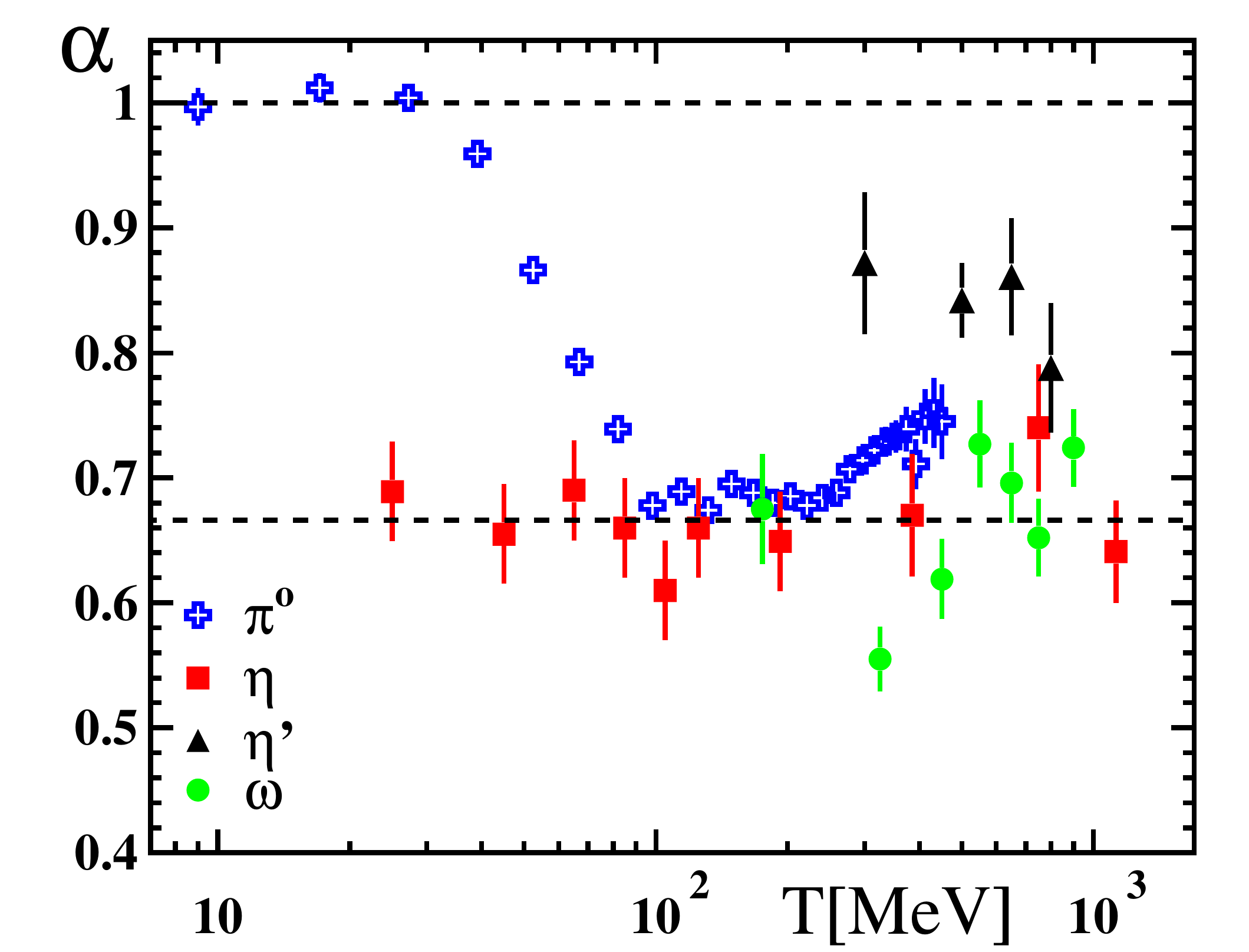,scale=0.475}
\begin{minipage}[t]{16.5 cm}
\caption{Scaling parameter $\alpha$ as a function of meson kinetic energy $T$
for $\pi^o$ \cite{Krusche_04}, $\eta$ \cite{Roebig_96,Mertens_08},
$\eta^{\prime}$ \cite{Nanova_12}, and $\omega$ mesons \cite{Kotulla_08}.
\label{fig:alpha}}
\end{minipage}
\end{center}
\end{figure}

Photoproduction is advantageous because there are almost (apart from small
shadowing corrections \cite{Bianchi_96}) no complications from initial state
interactions. Due to the small cross section of the electromagnetic reaction,
photons interact also with nucleons bound deeply in heavy nuclei, while
hadron-induced reactions explore mostly the nuclear surface. A comparison of
the cross sections for photoproduction of mesons off nuclei with different
mass numbers therefore gives an easy tool to study the absorption properties
of nuclear matter for a particular meson. Within some modelling approaches -
for example the Glauber-type approximations as in \cite{Roebig_96} -
these can be converted into a meson - nucleon absorption cross section.

A convenient technique is the study of the scaling of the meson production
cross sections with nuclear mass number $A$. This typically exhibits a
power-law behaviour
\begin{equation}
\frac{d\sigma}{dT}(T)\propto A^{\alpha(T)},
\end{equation}
where $T$ is the kinetic energy of the mesons. Qualitatively, small
absorption cross sections (\emph{transparent nucleus}) correspond to a
scaling of nuclear meson production with the volume of the nuclei, i.e., with
their mass number $A$ ($\alpha\approx 1$), while large absorption cross
sections (\emph{black nucleus}) result in a scaling corresponding to the
nuclear surface, i.e., with $\alpha\approx 2/3$. Typical results are
summarized in Fig.~\ref{fig:alpha} for $\pi^0$, $\eta$, $\eta^{\prime}$, and
$\omega$ mesons (details for $\eta$ and $\eta^{\prime}$ are discussed below).
The absorption properties for these mesons are quite different. Pions are
strongly absorbed when their kinetic energy is high enough ($>100$~MeV) to
excite the $\Delta(1232)$ resonance, but nuclei are transparent for
low-energy pions \cite{Krusche_04}.

The pion - nucleon interaction is weak at low momenta so that
pions cannot be bound in nuclei (though negative pions can be
bound with the help of the Coulomb force \cite{Geissel_02}).
The situation is completely different for $\eta$ mesons. Almost
independent of their kinetic energy, over a wide range from a
few 10~MeV to 1~GeV, the scaling coefficient is close to 2/3
\cite{Roebig_96,Mertens_08}. The absorption is large, which
means that the meson - nucleon interaction is strong. The
difference in the low-energy behaviour for pions and $\eta$
mesons is due to the properties of the $S_{11}$(1535) nucleon
resonance. An $s$-wave resonance with strong coupling to
$N\eta$ located very close to (and overlapping with) the $\eta$
production threshold generates a very efficient absorption
channel for low-energy $\eta$ mesons. The $\eta$ - nucleon
interaction at low relative momenta is strong, which is the
basis for the investigation of possible $\eta$-mesic states to
be discussed later. The absorption of $\omega$ mesons is also
strong \cite{Kotulla_08} but $\eta^{\prime}$ mesons show an
intermediate behaviour, with scaling coefficients around 0.8 -
0.9 \cite{Nanova_12}.

Inelastic meson - nucleon cross sections have been estimated from the scaling
of the cross sections, either from the scaling coefficients in Glauber-type
analyses or from an analysis of the so-called transparency ratios, discussed
below in the framework of different models. The $\eta$-nucleon absorption
cross section $\sigma_{\eta N}$ in nuclear matter was determined from such
data \cite{Roebig_96} to be around 30 mb, corresponding to a typical mean
free path of $\lambda\approx 2$ fm. For the $\omega$ meson an absorption
cross section of 70 mb was deduced \cite{Kotulla_08}, which is roughly a
factor of three larger than the accepted $\sigma_{\omega N}$ cross section
for free nucleons. A similar result had been previously reported from the
LEPS collaboration for the $\phi$ meson \cite{Ishikawa_05}. In this case an
absorption cross section of $\approx 30$ mb was deduced, which has to be
compared to the free nucleon absorption cross section of 7.7 - 8.7~mb. Both
experiments have produced strong evidence for the much discussed in-medium
modification of vector mesons. The inelastic $\eta^{\prime}N$ cross section,
on the other hand, was estimated in the range of 3 - 10~mb from similar
analyses \cite{Nanova_12}. The absorption cross sections are responsible for
an increase of the widths of the mesons in nuclear matter (due to the
absorption-related decrease of their lifetimes), which is connected to the
imaginary part of the meson - nucleus optical potential. The real part of the
potential is reflected in a modification of the meson masses. For $\omega$
mesons a very strong broadening from $\approx 8.5$~MeV in vacuum to 130 -
150~MeV in nuclear matter has been reported \cite{Kotulla_08} while, for the
$\eta^{\prime}$, in-medium widths on the order of 15 - 25~MeV have been
extracted \cite{Nanova_12} (compared to $\approx 0.2$~MeV in vacuum).

In the low-density approximation the absorption cross section and in-medium
width are related by \cite{Nanova_12,Kotulla_08}
\begin{equation}
\Gamma = \rho_o\,\sigma_{\rm inel}\,\beta\;,
\end{equation}
where $\Gamma$ is the meson width in normal nuclear matter density $\rho_o$,
$\sigma_{\rm inel}$ is the inelastic cross section, and $\beta = p_m/E_m$ the
meson velocity. If the same relation is assumed for $\eta$ mesons, the
absorption cross section of $\approx 30$~mb corresponds to an in-medium width
of 70 - 95~MeV (vacuum width 1.3~keV) for $\eta$ mesons with kinetic energies
between 200 and 1000~MeV. For mesons with lifetimes as long as those of the
$\eta$ and $\eta^{\prime}$, a measurement of the absorption cross sections is
the only way to access the in-medium width. A direct measurement of the width
is not possible since the mesons are either absorbed by nucleons or escape
from the nucleus and subsequently decay in vacuum.

\begin{figure}[thb]
\begin{center}
\epsfig{file=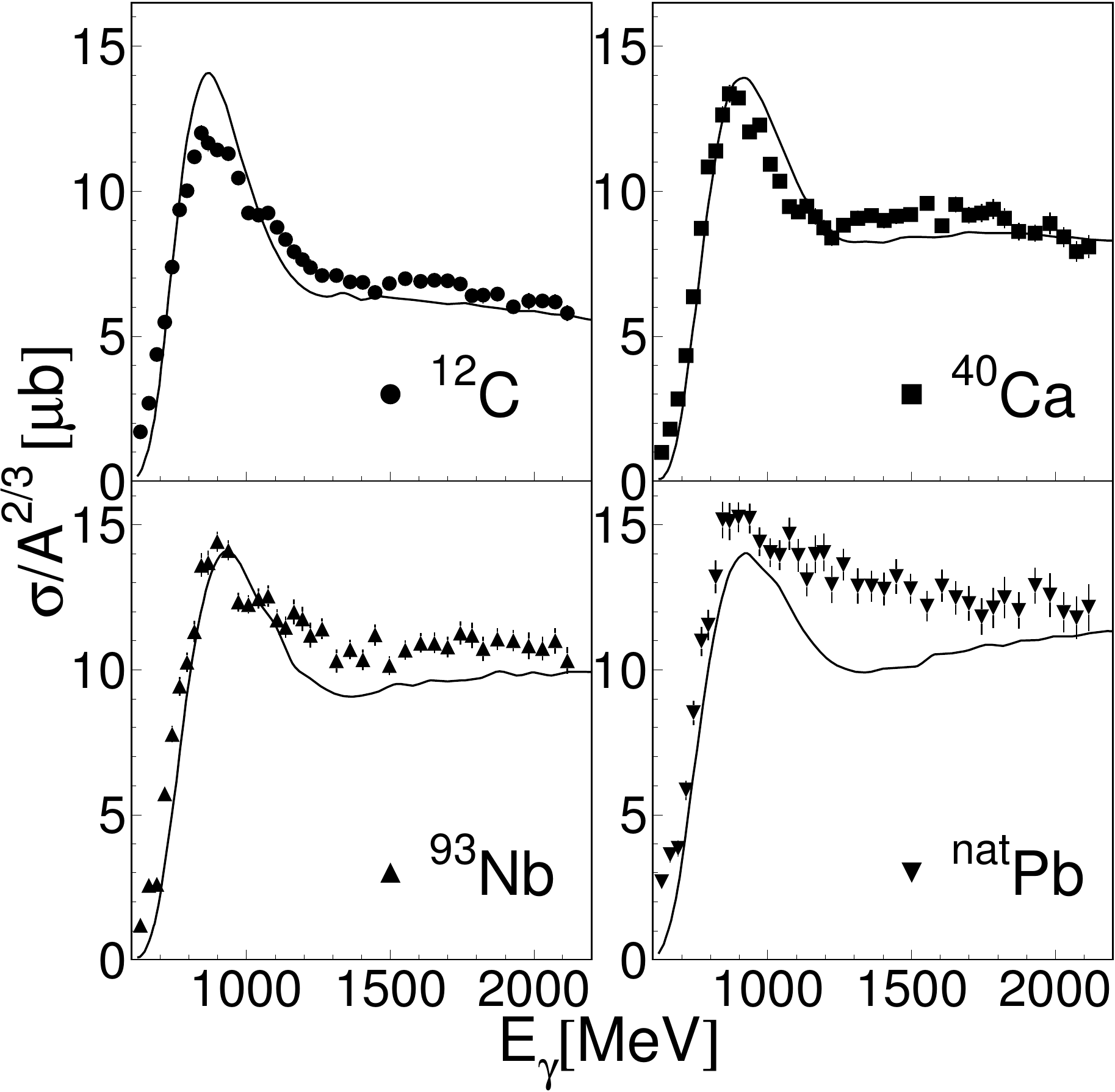,scale=0.32}
\epsfig{file=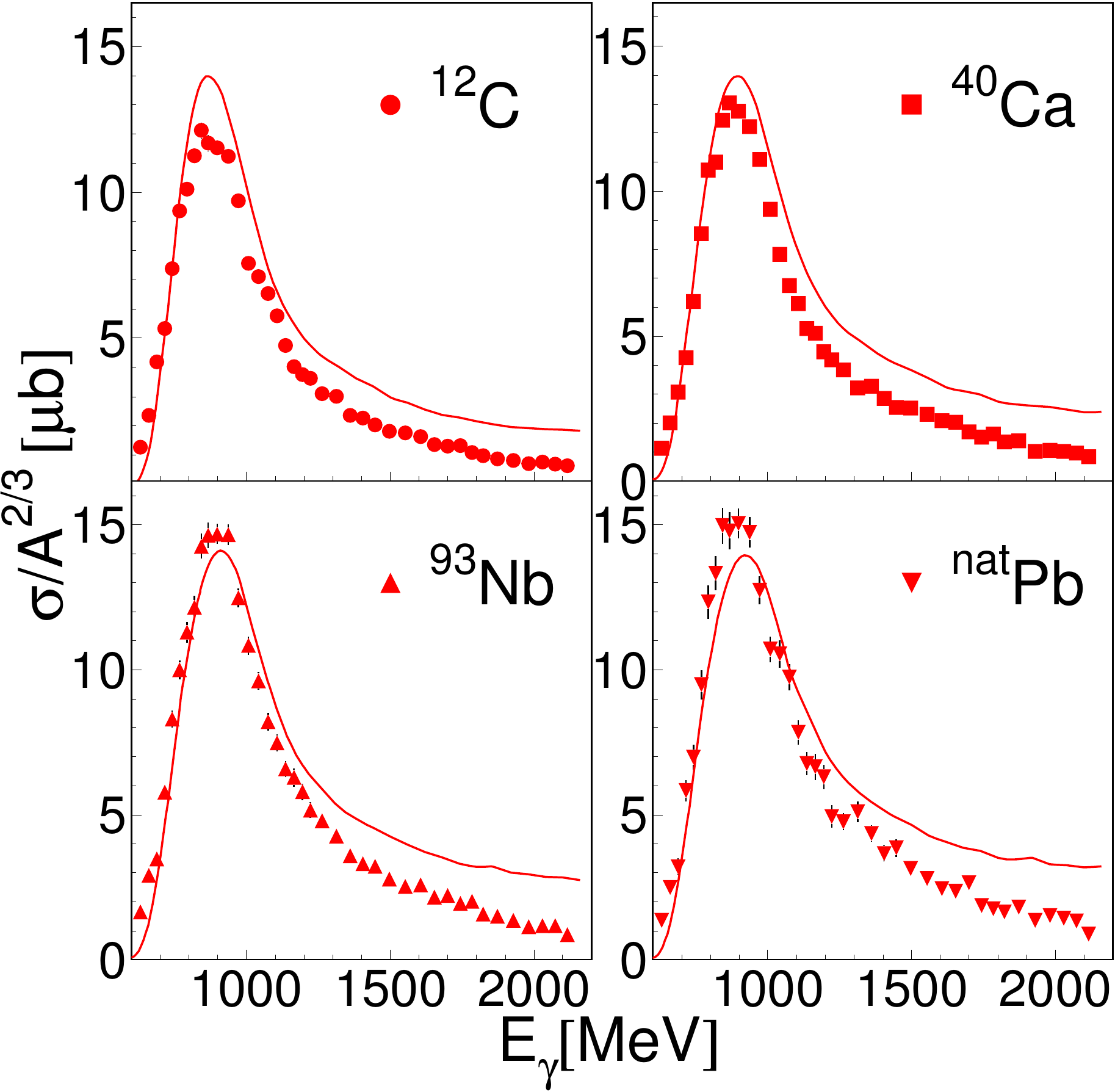,scale=0.32}
\epsfig{file=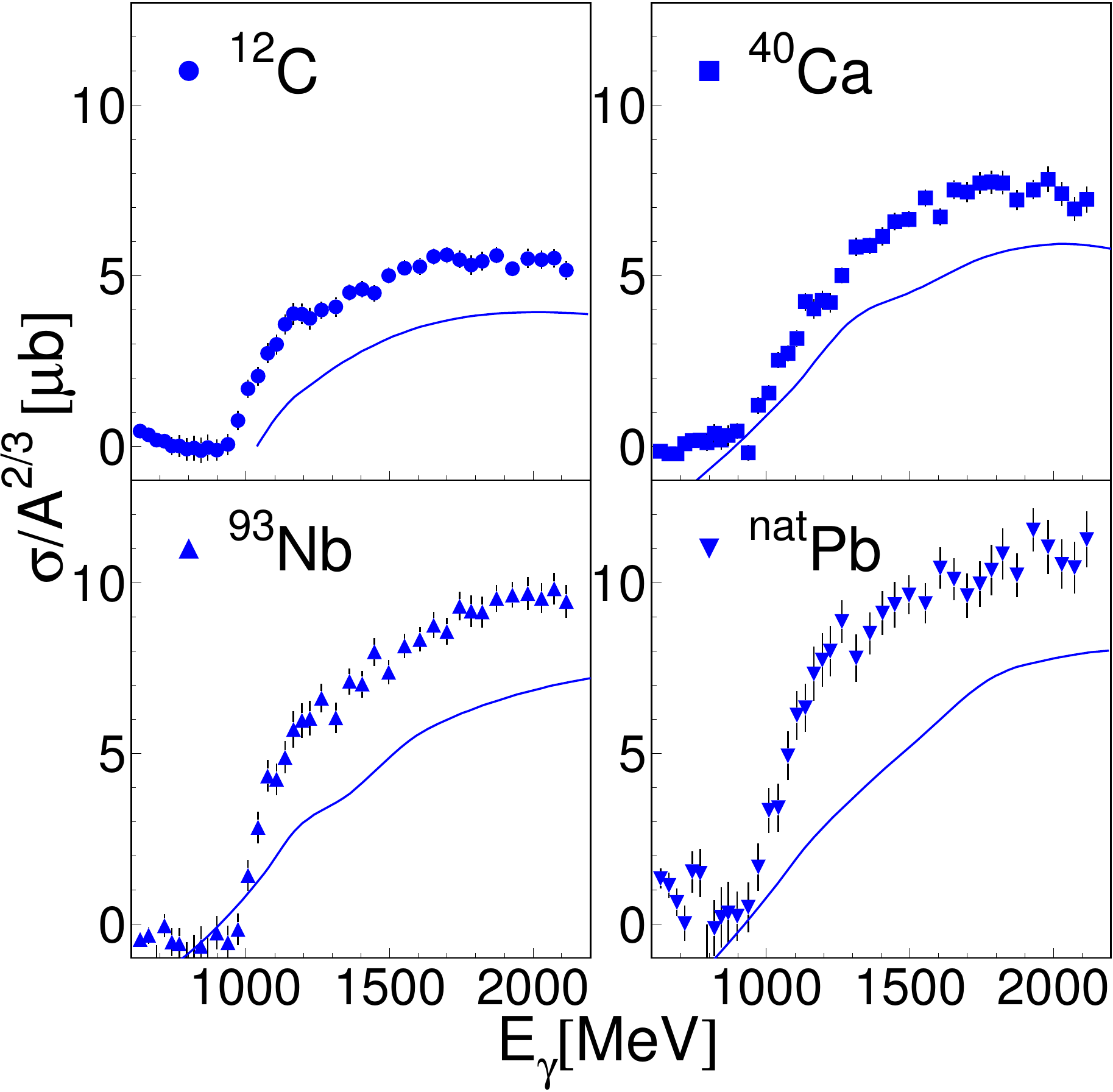,scale=0.32}
\vspace*{-1.cm}
\begin{minipage}[t]{16.5 cm}
\caption{Total cross sections for $\eta$ photoproduction off heavy nuclei
from \cite{Mertens_08}. Left hand side: fully inclusive $\eta$ production
($\gamma A\rightarrow X\eta$, no restriction on $X$, which can be a
multi-particle final state, possibly also including further mesons such as
pions). Centre: exclusive quasi-free $\eta$-production $\gamma A\rightarrow
(A-1)N\eta$ (identified by cuts on the reaction kinematics). Right hand side:
difference between the inclusive and exclusive cross section, which arise
mainly from $\gamma A\rightarrow (A-1)N\eta\pi$. Data from \cite{Mertens_08},
theory curves from GiBUU model
\cite{Hombach_95,Effenberger_97,Lehr_00,Lehr_01} taken from
\cite{Mertens_08}. \label{fig:eta_a_tot}}
\end{minipage}
\end{center}
\end{figure}

We now discuss in greater detail the results for $\eta$ and $\eta^{\prime}$
production. Inclusive photoproduction of $\eta$ mesons from nuclei has been
measured for the deuteron
\cite{Jaegle_11,Krusche_95b,Hoffmann_97,Weiss_03,Miyahara_07}, for $^{3}$He
\cite{Witthauer_13}, and for heavier nuclei such as $^{12}$C, $^{40}$Ca,
$^{63}$Cu, $^{93}$Nb, $^{\rm nat}$Pb
\cite{Roebig_96,Mertens_08,Yorita_00,Kinoshita_06} at MAMI, at KEK, in
Tohoku, and at ELSA. In general, the agreement between the different
measurements is good (see comparisons in \cite{Mertens_08}). Typical results
for total inclusive cross sections for heavy nuclei from the most recent
measurement \cite{Mertens_08}, covering the widest energy range, are
summarized in Fig.~\ref{fig:eta_a_tot} (angular and energy-dependent
differential cross sections are also given in \cite{Mertens_08}). The
experimental results are compared to calculations in the framework of the
Giessen version (see \cite{Buss_12} for a recent review) of the
Boltzmann-Uehling-Uhlenbeck (BUU) transport model, which incorporates all
relevant elementary production cross sections and then traces the space-time
evolution of an ensemble of interacting particles $N, N^{\star}, \Delta, \pi,
\eta,...$ in nuclear matter from the moment of their creation to their
absorption or their escape through the outer boundaries of the nucleus.
Effects like Fermi motion and Pauli-blocking of final states are of course
included. In short, the model incorporates all \emph{trivial} in-medium
effects (such as, e.g., collision broadening of hadrons in nuclear matter due
to the additional decay channels) and, as long as all reaction cross sections
are known precisely enough, can be used as a kind of \emph{null} hypothesis
for non-trivial quantum mechanical in-medium effects (such as, e.g., chiral
restoration effects).

The production of $\eta$-mesons from nuclei is complicated by two factors. At
incident photon energies above 930~MeV the cross section for the production
of $\eta\pi$ pairs (see Sec.~\ref{ssec:etapi}) rises rapidly and reaches
values comparable to single $\eta$ production. But these two reactions are
not easily distinguishable. Experimental difficulties arise in particular in
the case of charged pions, which might escape detection when they are too low
in energy or emitted along the beam-pipe. Furthermore, the pions may be lost
due to FSI in the production nucleus. However, one could argue that, for the
measurement of the absorption properties of $\eta$ mesons in nuclear matter,
such effects do not matter much because the absorption probability of an
$\eta$ depends only on its kinetic energy and not on its initial production
reaction. More problematic are the $\eta$ that are produced in secondary
reactions where, in the first step for example a pion is produced, which is
then reabsorbed by a nucleon to produce an $\eta$. This reaction chain
contributes significantly since the production cross section for pions is
large and the $S_{11}$(1535) resonance, which has an almost 50:50 branching
ratio to $N\pi$ and $N\eta$, offers a very efficient conversion mechanism. In
contrast to the direct processes, secondary production will not scale linearly
with the mass number so that the scaling behaviour of the final and initial
states will be mixed.

\begin{figure}[thb]
\begin{center}
\epsfig{file=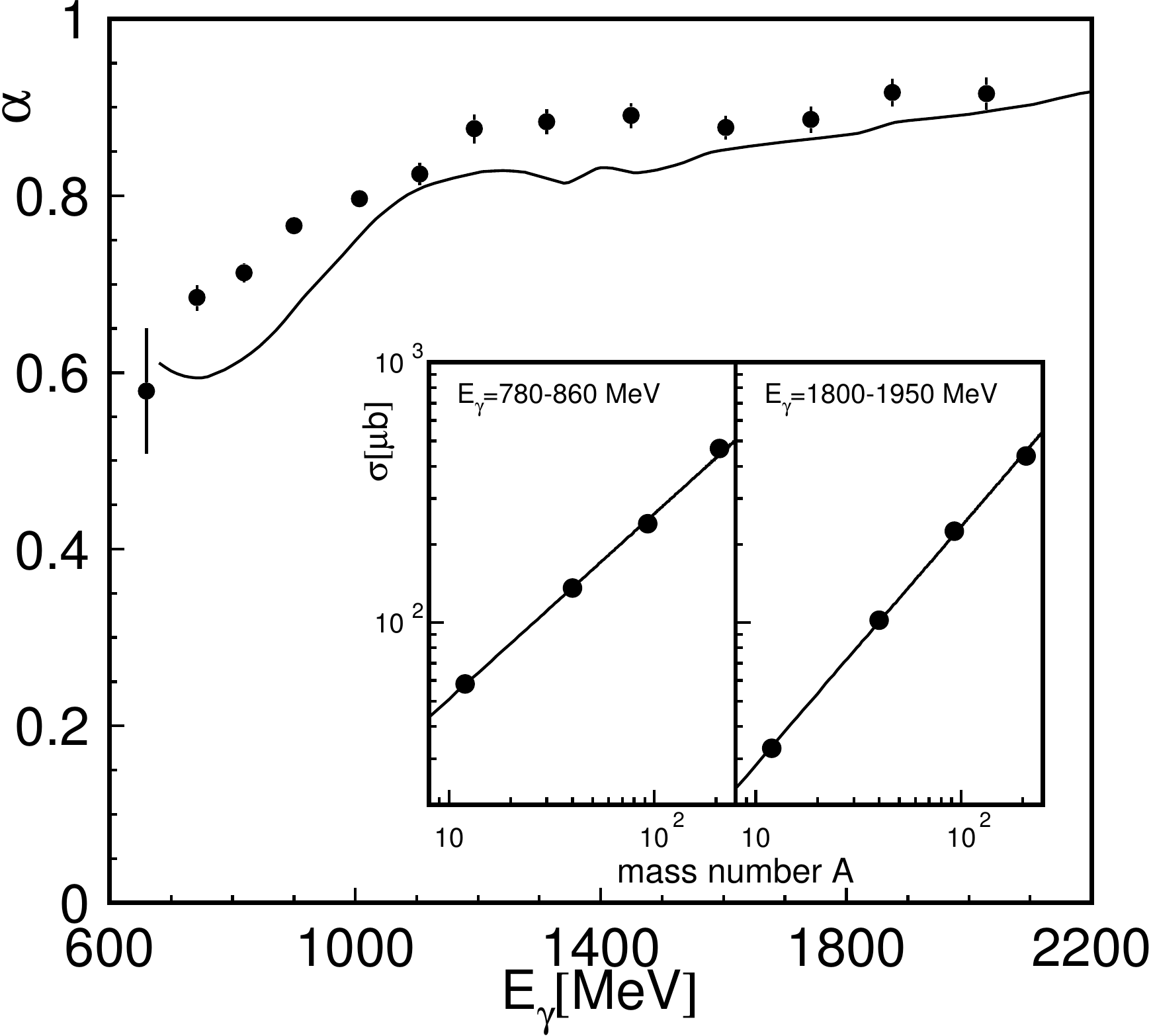,scale=0.36}
\epsfig{file=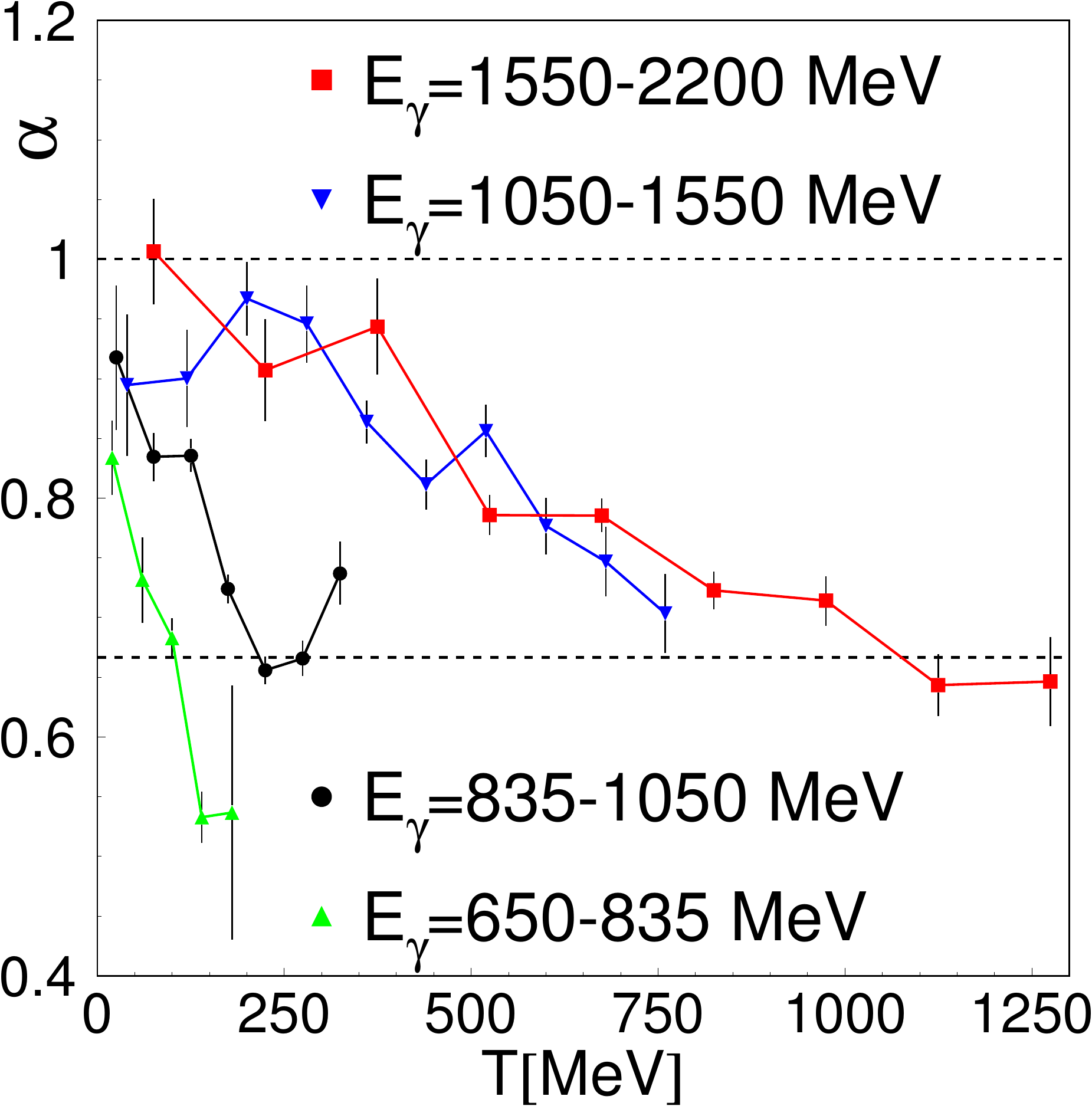,scale=0.30}
\epsfig{file=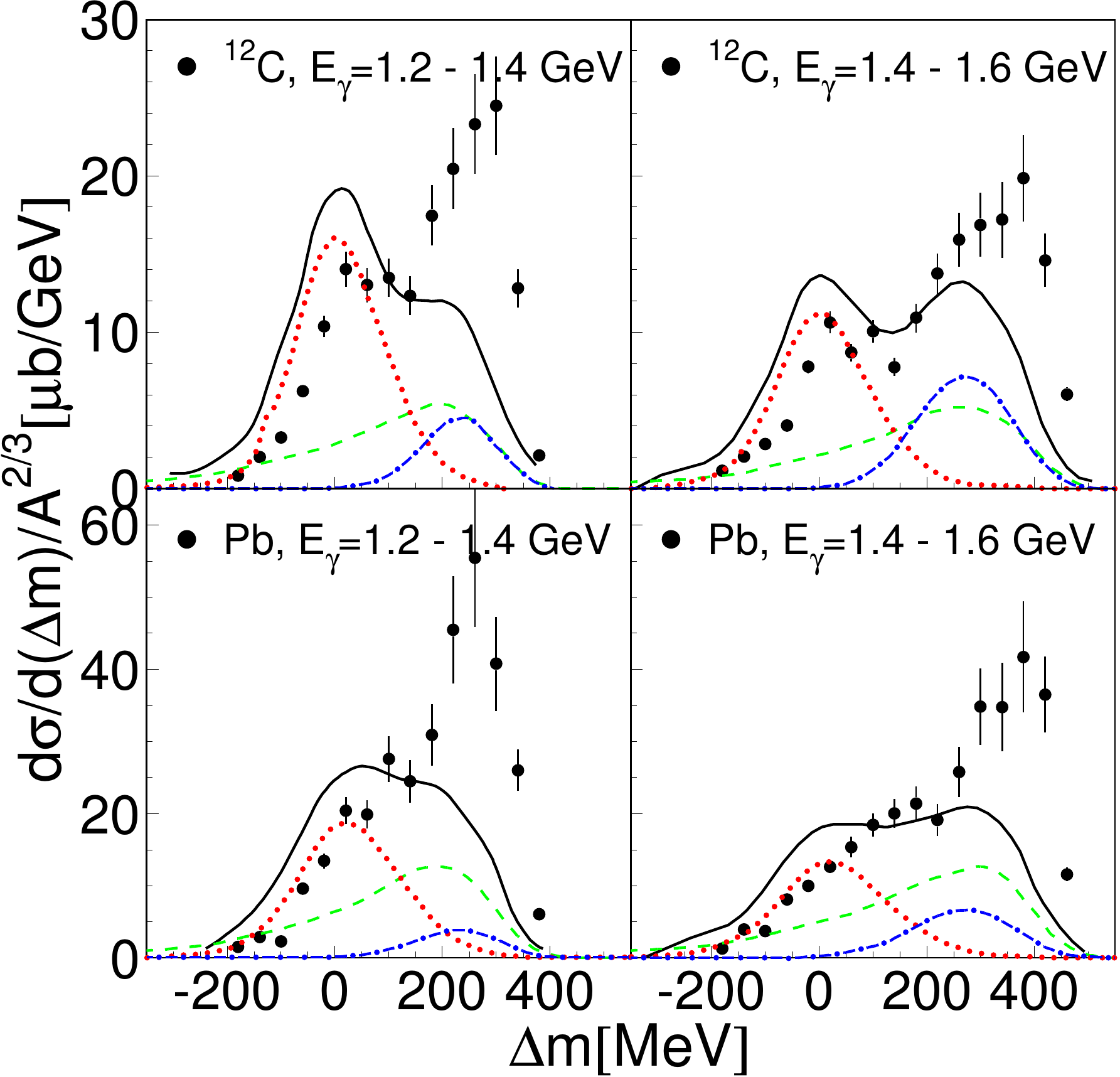,scale=0.33}
\begin{minipage}[t]{16.5 cm}
\caption{Left hand side: scaling parameter $\alpha$ as function of
$E_{\gamma}$ for fully inclusive $\eta$ production. Insert: two examples of
individual fits. Centre: scaling parameter $\alpha$ as a function of the
$\eta$ kinetic energy for different ranges of incident photon energy. Right
hand side: missing-mass distributions. All data from \cite{Mertens_08}.
Results from the GiBUU model for the missing mass \cite{Mertens_08}, solid
(black) curves: full model, dotted (red): single, quasi-free $\eta$
production, dash-dotted (blue) $\eta\pi$ production, dashed (green) $\eta$
production from secondary processes. All cross sections normalized to
$A^{2/3}$, figures from Ref.~\cite{Mertens_08}. \label{fig:alpha_eta}}
\end{minipage}
\end{center}
\end{figure}

The left hand side of Fig.~\ref{fig:eta_a_tot} shows the total inclusive
reaction cross section $\gamma A\rightarrow \eta X$ (no condition on $X$) for
different nuclei. The agreement with the BUU calculation is quite good for
carbon, calcium, and niobium, but somewhat poorer for lead. The central part
of the figure shows the best possible approximation for single, quasi-free
production of $\eta$ mesons, which has been selected by a cut on the missing
mass of the reaction (cut at $\Delta m>140$~MeV/$c^2$, see the missing-mass
spectra in Fig.~\ref{fig:alpha_eta}). This cut removes the production of
$\eta\pi$ pairs but not all the contributions from secondary reactions.
Nevertheless, the comparison of data and BUU predictions shows that the
$S_{11}$(1535) resonance does not show any unexpected in-medium
modifications. The peak shape agrees reasonably well with the BUU predictions
and the cross sections scales with $A^{2/3}$. The right hand side of the
figure shows the difference of the inclusive and exclusive reactions. This
cross section rises strongly at the $\eta\pi$ production threshold (for free
nucleons at 930~MeV) and has an energy dependence similar to the
photoproduction of $\eta\pi$ pairs. The peak cross sections are in the range
of 2.2$\mu {\rm b}/A$ for $^{12}$C and 1.7$\mu {\rm b}/A$ for Pb, which are
in reasonable
agreement with the $\approx 2\mu {\rm b}/A$ cross section average for the
$\pi^0\eta$ and $\pi^{\pm}\eta$ final states measured for the deuteron target
(see Sec.~\ref{ssec:etapi}). In this sense one may regard the data as an
indirect measurement of $\eta\pi$ production off nuclei. These data include,
of course, contributions from secondary processes. The BUU model results
suggest that this contribution is at the same level or even larger than
$\eta\pi$ production. However, the results of this model are not in good
agreement with data. They underestimate the difference cross section (see
Fig.~\ref{fig:eta_a_tot}, right hand side) and do not reproduce the shape of
the missing-mass spectra. A comparison of the data and model results
demonstrates that the excess of the measured cross section compared to the
BUU results is related to large missing masses in the region where both
$\pi\eta$ and secondary production peak, while single $\eta$ production seems
to be much better described by the model (see Fig.~\ref{fig:alpha_eta}, right
hand side).

The analysis of the scaling behaviour for $\eta$ mesons is
summarized in Fig.~\ref{fig:alpha_eta}. The left hand side
shows the scaling coefficient for the total cross section. It
rises from $\approx 2/3$ at threshold to almost unity at the
highest incident photon energies. It might be tempting to
interpret this as showing that close to threshold nuclear
matter is black for $\eta$ mesons, due to absorption into the
S$_{11}$(1535) state, but then becomes more transparent at
higher incident photon energies, where no strong coupling to
resonances exists. However, the central part of the figure
shows that this is not the case. Plotted are the scaling
coefficients as functions of the meson kinetic energy for
different ranges of incident photon energy. The expectation is
that the absorption properties of the nucleus, and thus the
scaling, should depend mainly on the kinetic energy of the
mesons, regardless of the initial photon energy, but the
opposite is the case. For all the incident photon energies
investigated, the scaling coefficients drop for constant
$E_{\gamma}$ from almost unity at small $T_{\eta}$ to $\approx
2/3$ at maximum $T_{\eta}$; for a given kinetic energy they are
very dependent on $E_{\gamma}$. Such a behaviour can arise if
the initial production cross sections (before FSI) do not scale
with the mass number, which would be the case if there were
significant contributions from secondary production processes.
The \emph{true} dependence of $\alpha$ on $T_{\eta}$ can thus
be tested best with the scaling at kinetic energies close to
the maximum values of $T_{\eta}$ for a given incident photon
energy (because secondary production processes are then
strongly suppressed by the kinematics). For this a compromise
has to be made between the suppression of the secondary
processes and the statistical precision. The results from
\cite{Mertens_08}, which are included in Fig.~\ref{fig:alpha},
were analyzed under the condition that
\begin{equation}
T_{\eta} > (E_{\gamma} - m_{\eta})/2,
\label{eq:mertens_cut}
\end{equation}
where $E_{\gamma}$ is the incident photon energy and $m_{\eta}$ the mass of
the $\eta$ meson. The results indicate strong absorption for all the $\eta$
kinetic energies investigated.

\begin{figure}[thb]
\begin{center}
\epsfig{file=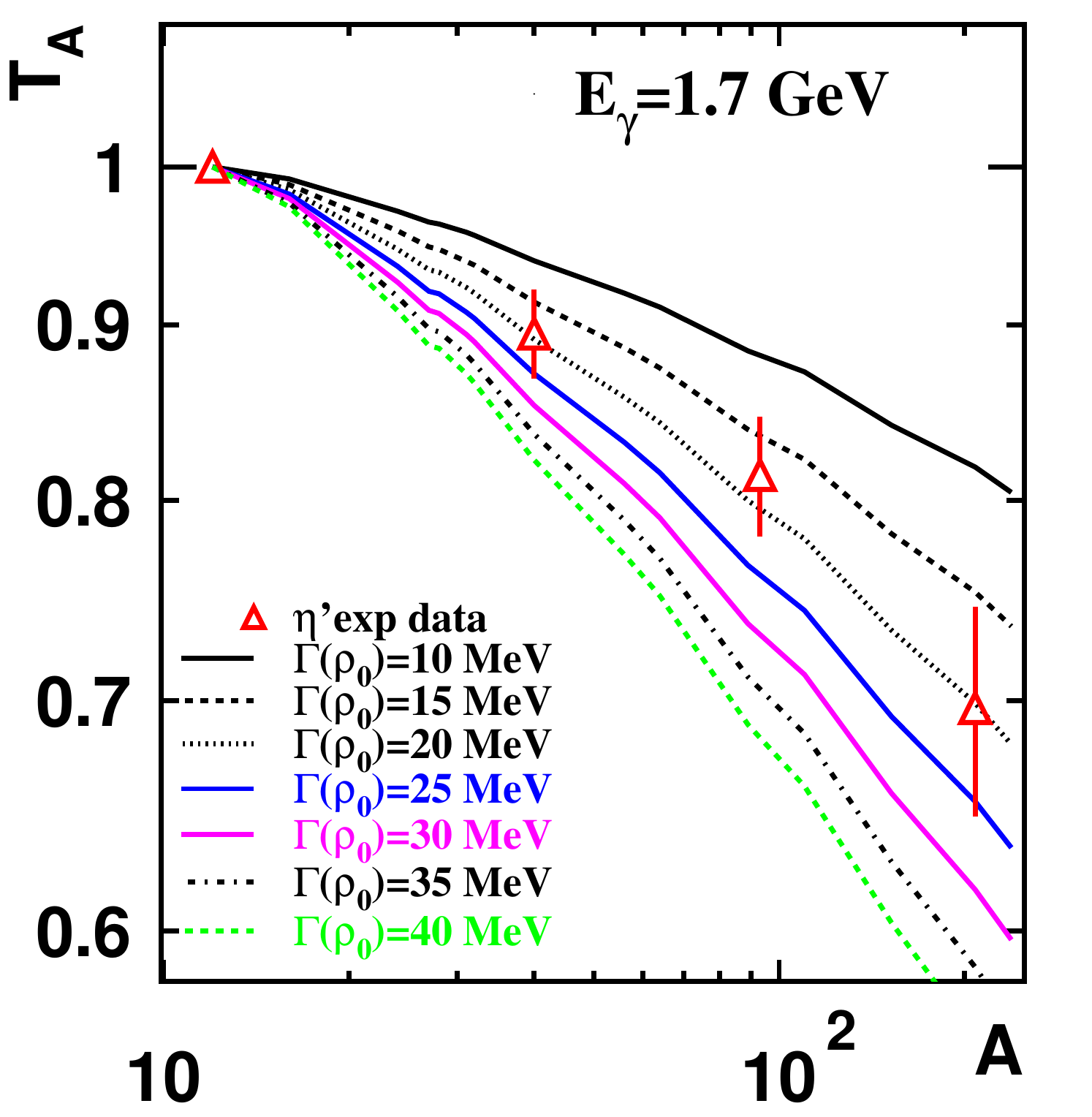,scale=0.29}
\epsfig{file=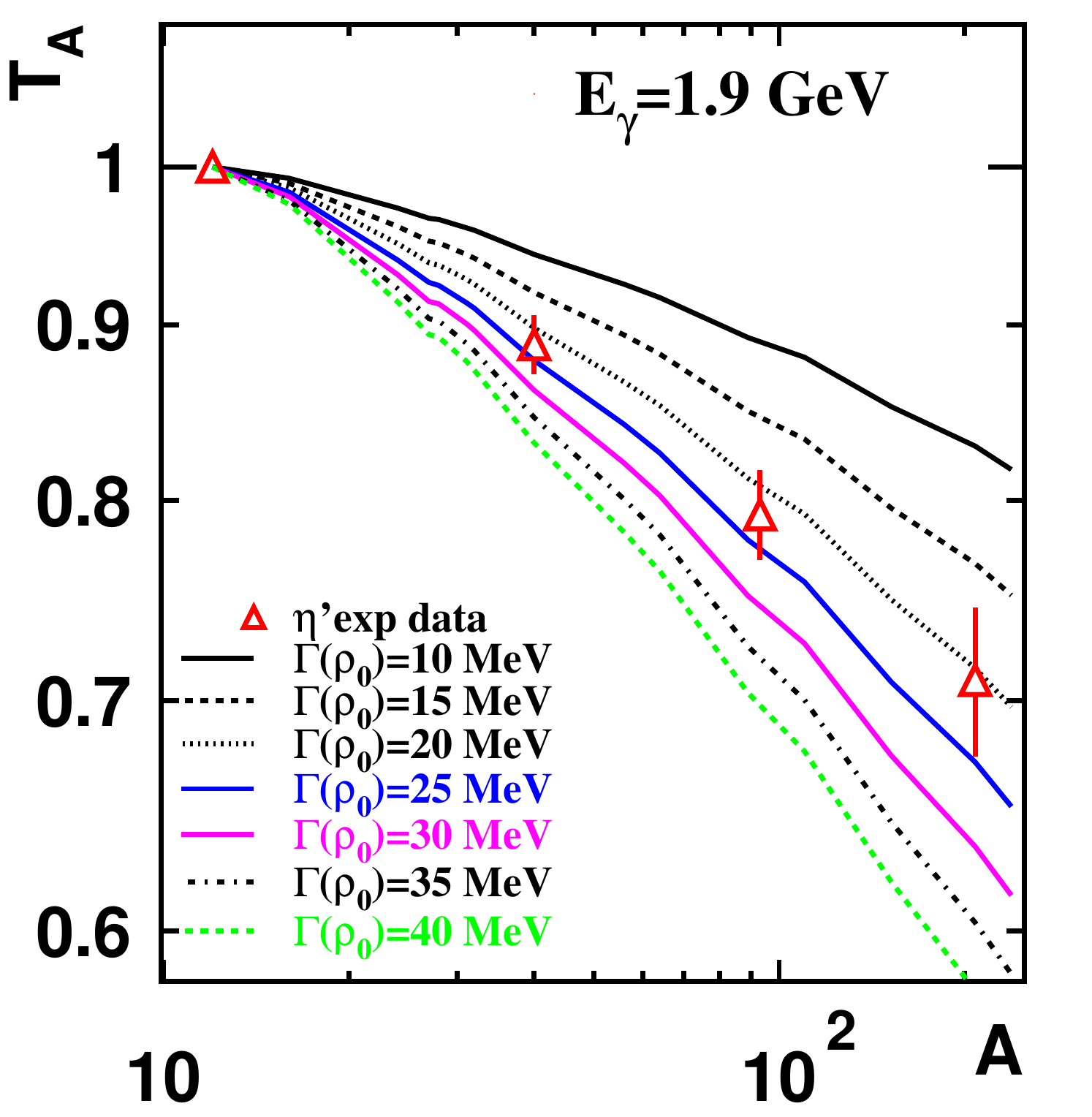,scale=0.29}
\epsfig{file=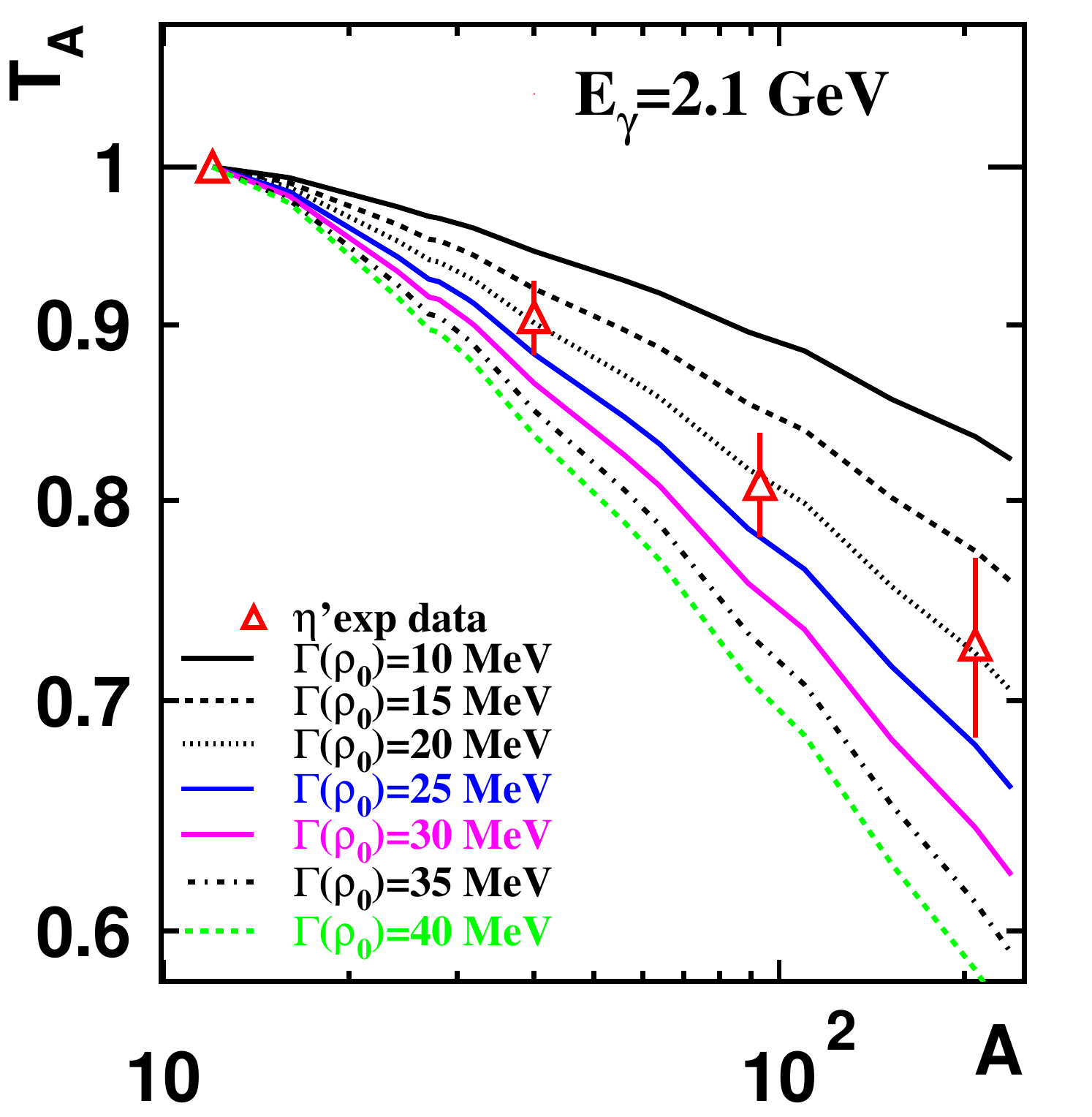,scale=0.29}
\epsfig{file=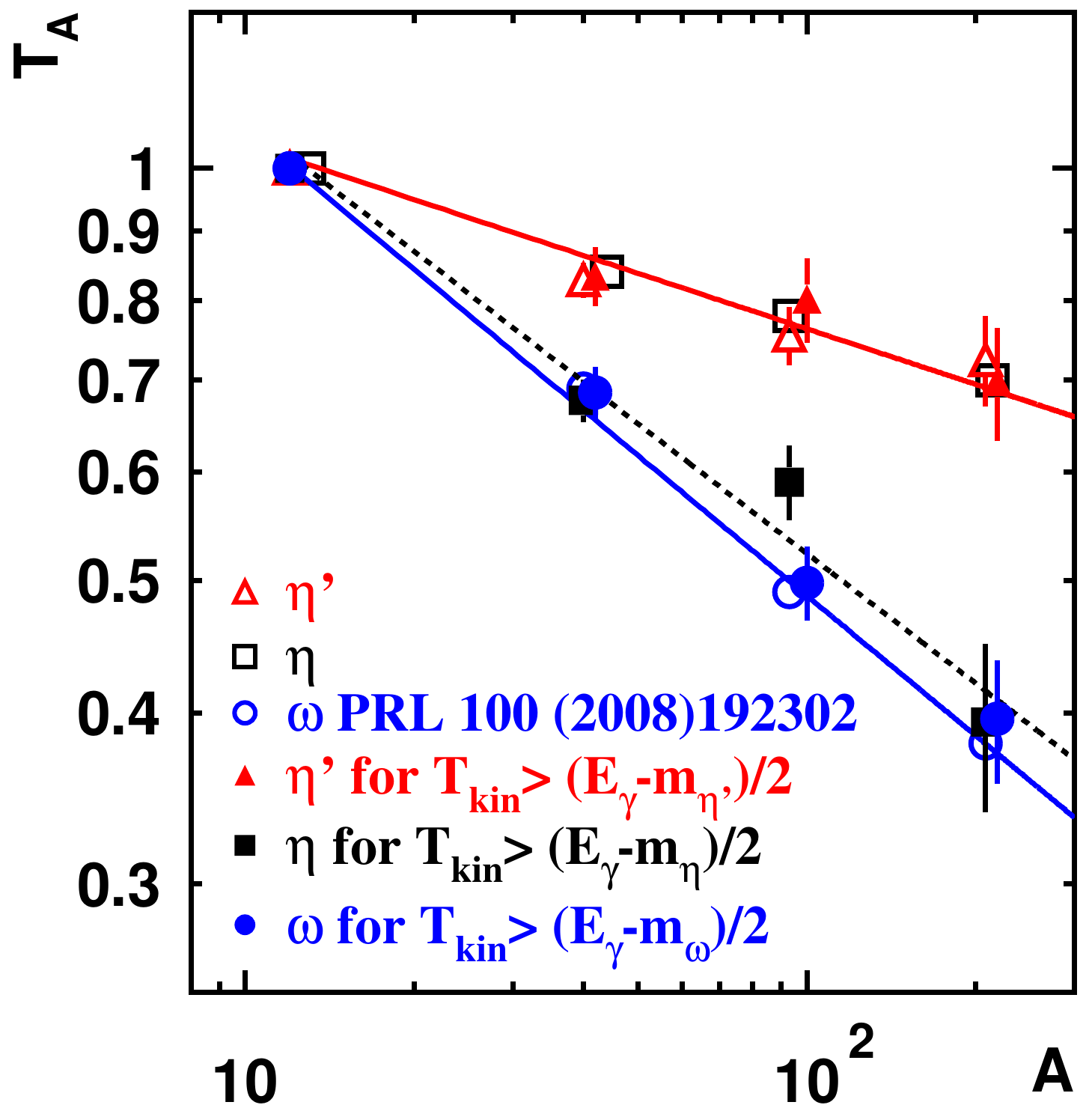,scale=0.29}
\begin{minipage}[t]{16.5 cm}
\caption{First three figures from left to right: transparency ratios as
functions of the nuclear mass number for three ranges of incident photon
energy. Data (red triangles) compared to model predictions for different
values of the $\eta^{\prime}$ in-medium width. Right hand side: transparency
ratios for $\eta$, $\eta^{\prime}$, and $\omega$  production with (closed
symbols) and without (open symbols) cut on $T_{kin} > (E_{\gamma} - m_{\rm
meson})/2$. Figures from \cite{Nanova_12}. \label{fig:trans_ratio}}
\end{minipage}
\end{center}
\end{figure}

The photoproduction of $\eta^{\prime}$ mesons was analyzed in the same way
(actually for the only experimental results available so far, the data from
the Crystal Barrel/TAPS experiment analyzed in \cite{Mertens_08} for $\eta$
production were used). The $\eta^{\prime}$ mesons were identified via an
invariant-mass analysis of their
$\eta^{\prime}\rightarrow\eta\pi^0\pi^0\rightarrow 6\gamma$ decay
\cite{Nanova_12}. The analysis of the nuclear absorption properties used the
so-called transparency ratio $T$, which compares the total production cross
section $\sigma_{\gamma A\rightarrow mX}$ from a nucleus with mass number $A$
to $A$-times the elementary production cross section $\sigma_{\gamma
N\rightarrow mX}$ on the nucleon:
\begin{equation}
T_A=\frac{12 \, \sigma_{\gamma A\rightarrow mX}}{A\;\sigma_{\gamma
^{12}{\rm C}\rightarrow m X}}
\label{eq:transp}
\end{equation}
where $m$ denotes any meson. In order to account for the cross section
difference  between the proton and neutron, and also possible secondary
production processes, it is better to normalize to an average nucleon cross
section measured for a light target nucleus with equal numbers of protons and
neutrons. Generally this is taken to be carbon, which leads to the $T_A$
defined in Eq.~(\ref{eq:transp}). This type of analysis is equivalent to the
evaluation of the scaling coefficients $\alpha$, which were also extracted
from the data.

The deviations of the cross sections from the scaling with the mass number
$A$ can be related to the in-medium width of the produced meson (the basic
idea is simple; the smaller the scaling coefficient the larger the absorption
probability, the larger the reduction of lifetime of the meson and the larger
its effective in-medium width). The formalism, which does not require a
particular model for the elementary production process, is given in
\cite{Hernandez_92,Magas_05,Kaskulov_07}; details such as corrections for
photon shadowing effects \cite{Bianchi_96} and two-body absorption processes
are discussed in \cite{Nanova_12}. Results for $T_A$ for $\eta^{\prime}$
production off nuclei are summarized in Fig.~\ref{fig:trans_ratio} and
compared to predictions assuming different values for the $\eta^{\prime}$
in-medium width \cite{Nanova_12}. The comparison reveals an in-medium width
of 15 - 25~MeV at normal nuclear matter density $\rho=\rho_o$, which
corresponds to an imaginary part of the $\eta$ - nucleus optical potential of
$-$(10$\pm$2.5)~MeV \cite{Nanova_13}.

The absorption probability and the imaginary part of the
optical potential for $\eta^{\prime}$ mesons are significantly
smaller than for other mesons for which similar analyses have
been performed. This is shown at the right hand side of
Fig.~\ref{fig:trans_ratio}, where the transparency ratios for
$\eta$, $\omega$, and $\eta^{\prime}$ mesons are compared. The
slope for the $\eta^{\prime}$ meson is much less steep. The
analysis demonstrates also that the complications from
secondary production mechanisms is only important for $\eta$
mesons. The $\eta$, $\omega$, and $\eta^{\prime}$ data have
been analyzed with and without a cut of
Eq.~(\ref{eq:mertens_cut}) on the meson kinetic energy. The
effect of the cut, which selects mesons with large kinetic
energy that are unlikely to be produced by secondary processes,
is almost negligible for $\eta^{\prime}$ and $\omega$ but is
strong for the $\eta$, due to the efficient $\pi N\rightarrow
S_{11}(1535)\rightarrow N\eta$ conversion chain. Such a
conversion mechanism is absent for the $\eta^{\prime}$ and the
$\pi N\rightarrow N\eta^{\prime}$ cross section of only
$\approx$ 0.1 mb is unusually small \cite{Landolt_88,Rader_72}.

\begin{figure}[thb]
\begin{center}
\epsfig{file=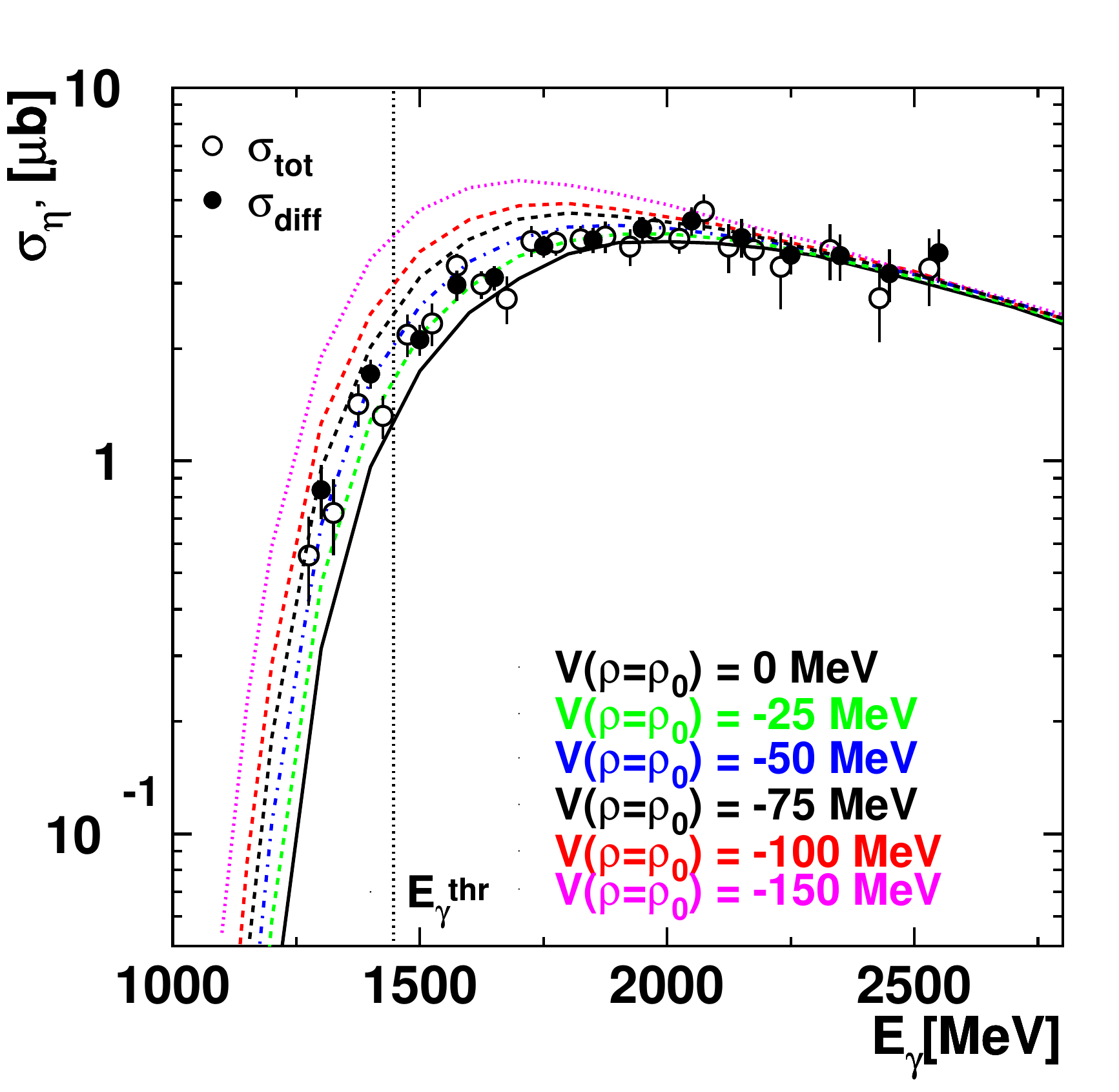,scale=0.305}
\epsfig{file=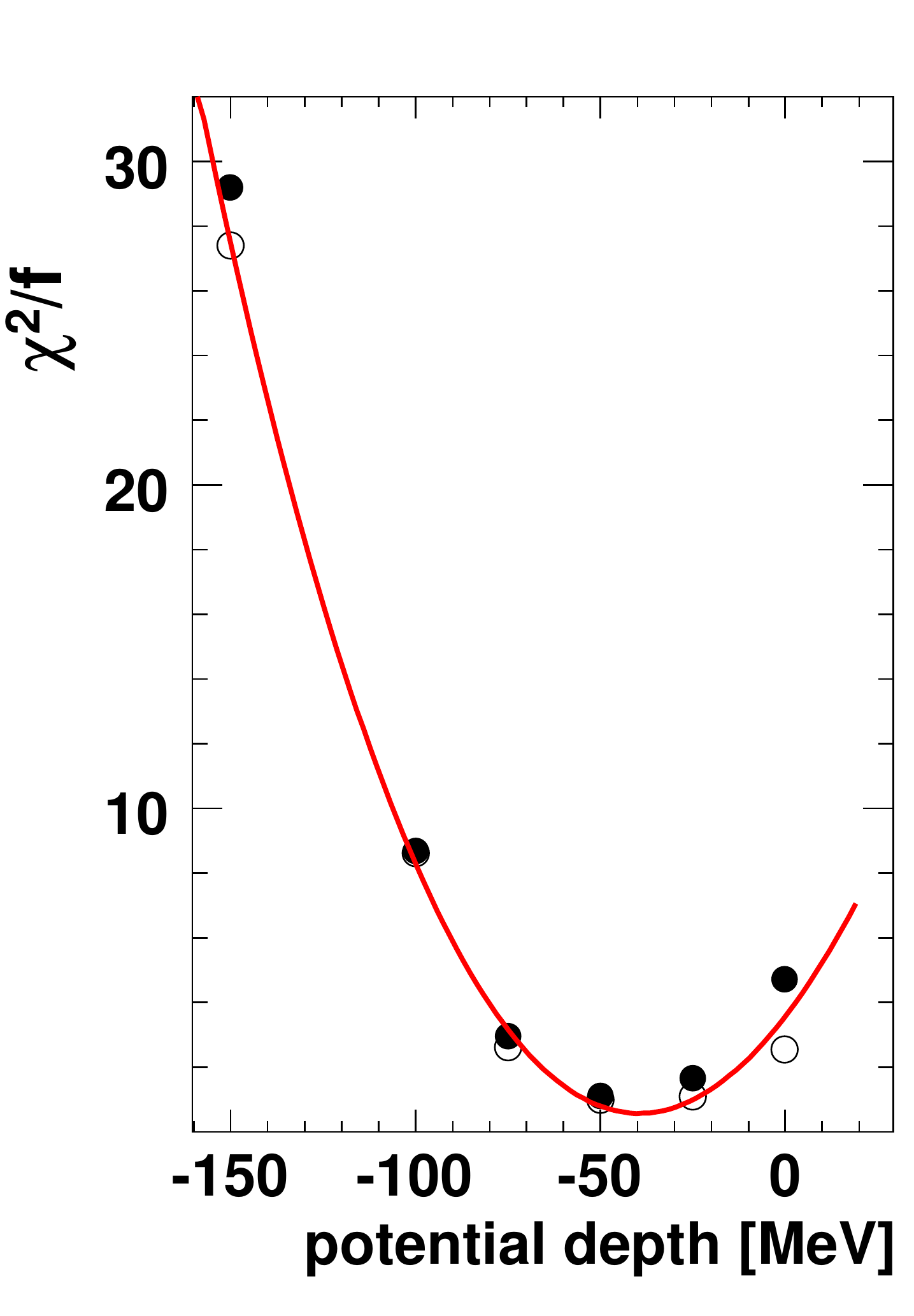,scale=0.245}
\epsfig{file=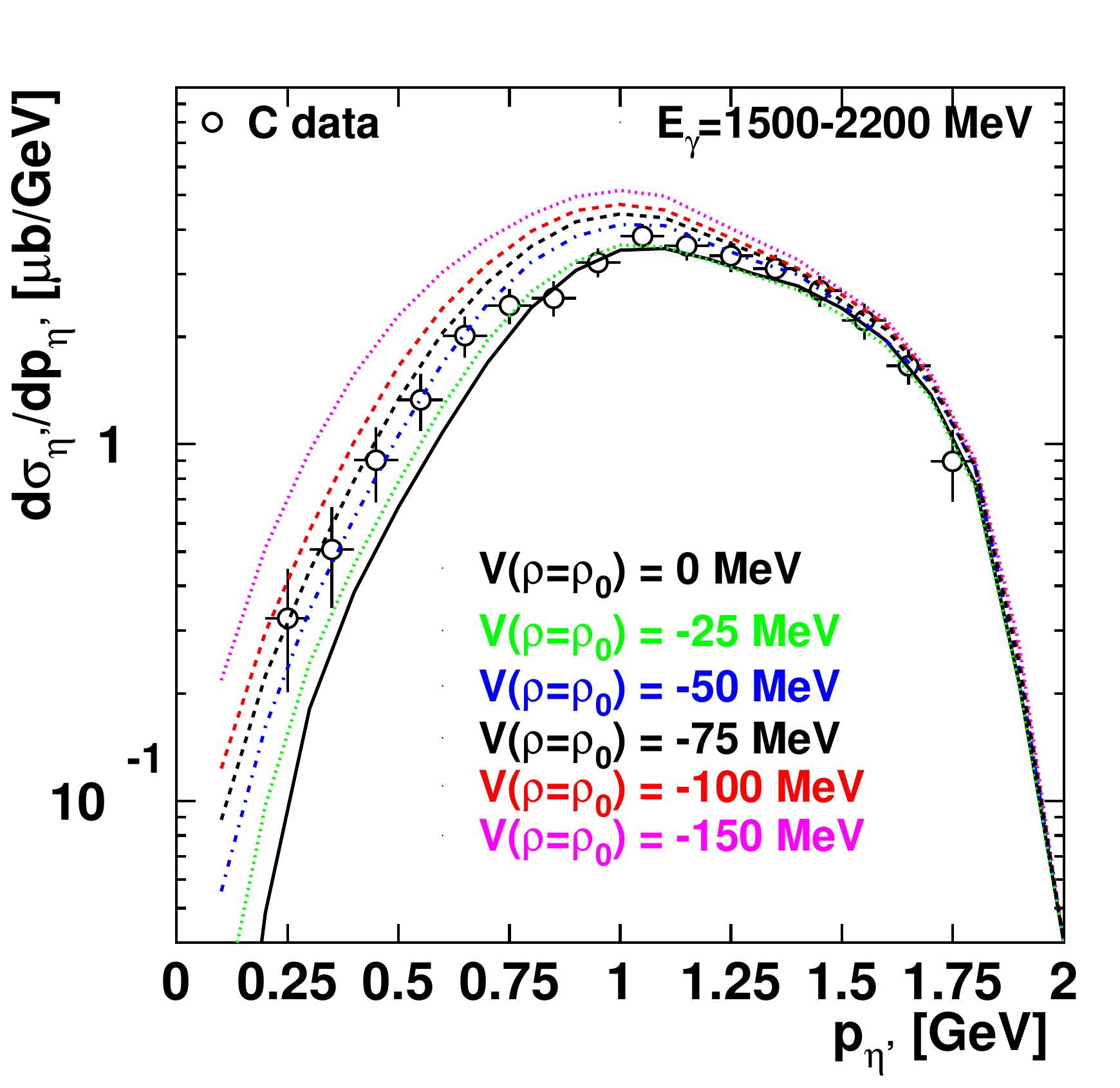,scale=0.305}
\epsfig{file=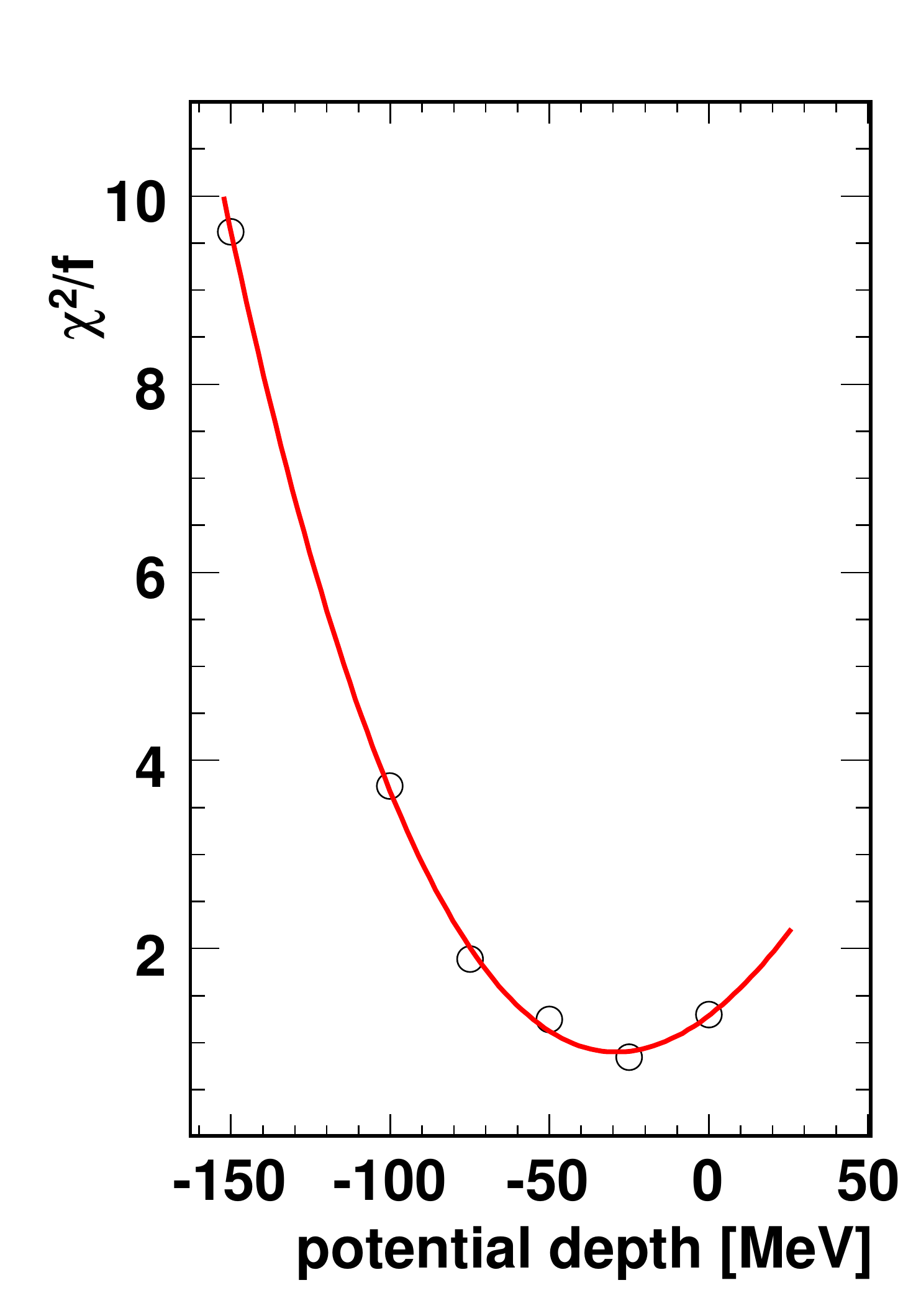,scale=0.245}
\begin{minipage}[t]{16.5 cm}
\caption{Left hand side: the threshold behaviour of $\gamma A\rightarrow X
\eta^{\prime}$ \cite{Nanova_13}. The vertical line indicates the threshold
for free nucleons. Open and closed symbols represent two different analyses
of the data, which should be equivalent (see \cite{Nanova_13} for details).
The curves represent model predictions for different assumptions on the
depths of the real potential \cite{Paryev_13}. The reduced $\chi^2$ for the
fit of the theory curves to the data is shown as function of the potential
depth. Right hand side: comparison of data \cite{Nanova_13} and model
calculations \cite{Paryev_13} for the momentum distributions of the
$\eta^{\prime}$ mesons for incident photon energies between 1500 - 2200~MeV
and reduced $\chi^2/f$ for the fit of model results to the data. All model
results are normalized by a common factor (0.75) to the total cross section
of the data for $E_{\gamma}>2.2$~GeV. Figures are taken from
\cite{Nanova_13}. \label{fig:optic}}
\end{minipage}
\end{center}
\end{figure}

The measurement of the absorption cross section gives practically the only
access to the imaginary part of the nucleus - meson optical potentials for
mesons that live long enough in vacuum to escape from nuclei unless they are
absorbed by a nucleon. The situation is different for very short-lived mesons
decaying inside the nucleus, which have been explored by direct measurements.
One example is the $\rho$ in-medium width, which was studied by the CLAS
collaboration \cite{Nasseripour_07,Wood_08} using the $\rho\rightarrow e^+
e^-$ Dalitz decay. However, already for $\omega$ mesons with much shorter
lifetimes than the $\eta$ and $\eta^{\prime}$, line-shape measurements are
challenging \cite{Thiel_13}.

A similar restriction applies to the $\eta$ and $\eta^{\prime}$ measurements
of the real part of the optical potential where a direct measurement of a
mass shift of the mesons in nuclear matter is also excluded. One may,
however, explore two indirect consequences of a meson mass shift in nuclear
matter and this has been done for the $\eta^{\prime}$ produced off $^{12}$C
nuclei by Nanova and coworkers \cite{Nanova_13} with data taken at the
CBELSA/TAPS experiment. The first method is based on the energy dependence of
the total cross section close to the production threshold. Due to the nuclear
Fermi motion, the production threshold for $\gamma A\rightarrow  mX$ ($m$ any
meson) is pushed far below the production threshold for free nucleons. Such
reactions can, in principle, occur almost down to the threshold of the coherent
reaction $\gamma A\rightarrow mA$ (only shifted slightly upward due to the
nucleon binding energy). The production threshold for
$\eta^{\prime}$ mesons off the free nucleon is at $\approx 1.447$~GeV, while
the threshold energy for the coherent reaction is only $\approx 1$~GeV. The
absolute value and the energy dependence of the cross section around and
below the free production threshold depend on the in-medium mass of the
meson.

The use of the threshold effect, is of course, quite model dependent. One
needs predictions of the nuclear cross section based on the elementary
production cross sections for protons and neutrons, the nuclear spectral
function taking into account the bound-nucleon momentum distribution, and the
influence of FSI processes. The analysis in \cite{Nanova_13} used model
predictions by Paryev \cite{Paryev_13} which had, as input, the elementary
production cross sections off protons and neutrons from \cite{Crede_09} and
\cite{Jaegle_11a} and, for final state absorption, an inelastic in-medium
$\eta^{\prime}N$ cross section of $\sigma_{\rm inel}$ that is similar to the
values extracted from the nuclear transparency measurements discussed above
($\sigma_{\rm inel}$= 3 - 10 mb). Model predictions for nuclear potential
depths of $V=0,\,-25,\,-50,\,-75,\,-100,\,-150$~MeV and normal nuclear
density $\rho_o$ are compared to the data on the right hand side of
Fig.~\ref{fig:optic}. It should, however, be noted that the model does not
reproduce the absolute magnitude of the measured cross sections. Part of the
difference may be related to the photon shadowing effect \cite{Bianchi_96},
which was not included into the model. It accounts, however, only for a 10\%
correction while the mismatch is on the 25\% level. All model results have
therefore been renormalized to the measured cross section for incident photon
energies above 2.2~GeV. This is the energy range where the model does not
show any sensitivity to the nuclear potential (the same renormalization
factor was used for all model curves). After this renormalization, the
reduced $\chi^2$ for fits of the data to the model curves were determined
(see Fig.~\ref{fig:optic}). As a result a potential depth of
$-$(40$\pm$6)~MeV was extracted. There are also, of course, systematic
uncertainties in this procedure, but strongly attractive potentials on the
order of $-100$~MeV or deeper appear improbable.

Another means to access the in-medium mass shift comes from the momentum
distribution of the observed mesons. Mesons that are produced inside the
nucleus with a reduced mass must somehow restore their on-shell value when
they emerge into free space. As a consequence of energy conservation, this
will happen at the expense of their kinetic energy. Therefore, the kinetic
energy spectra (or the momentum distributions) also carry, of course in a
model-dependent way, information about the in-medium mass. This was
demonstrated with GiBUU calculations \cite{Weil_13} and is also included in
the model predictions from Paryev \cite{Paryev_13}. The measured cross
sections and the model predictions for the momentum distributions are
compared on the right hand side of Fig.~\ref{fig:optic}. The model results
were renormalized to the data with the same common normalization constant as
for the total cross sections. The reduced $\chi^2$ of the fit to the data
results in a potential depth of $-$(32$\pm$11)~MeV. The systematic
uncertainty is somewhat larger than for the analysis of the threshold
behaviour of the cross section because the experimental resolution for the
$p_{\eta^{\prime}}$ (between 25 and 50~MeV) must also be considered
while resolution effects for the incident photon energy ($\approx 4$ MeV) are
negligible. The final result for the real part of the potential is
\cite{Nanova_13}
\begin{equation}
V_{o}(\rho=\rho_o)=-(37\pm10_{\rm stat}\pm10_{\rm sys})\; {\rm MeV}.
\end{equation}
Taken together with the results for the imaginary part ($-$10$\pm$2.5~MeV)
from the transparency measurements  \cite{Nanova_12}, this leads to an optical
potential for $\eta^{\prime}$ mesons in a carbon nucleus of:
\begin{equation}
U_{\eta^{\prime}}(r) = V(r)+iW(r) = (V_o +iW_o)\,{\rho(r)}/{\rho_o} = -(37 +i10)\, {\rm MeV}\, {\rho(r)}/{\rho_o},
\end{equation}
where $r$ is the distance of the meson from the centre of the nucleus and the
uncertainties in the numerical values of the potential parameters are given
above.

\begin{figure}[htb]
\begin{center}
\epsfig{file=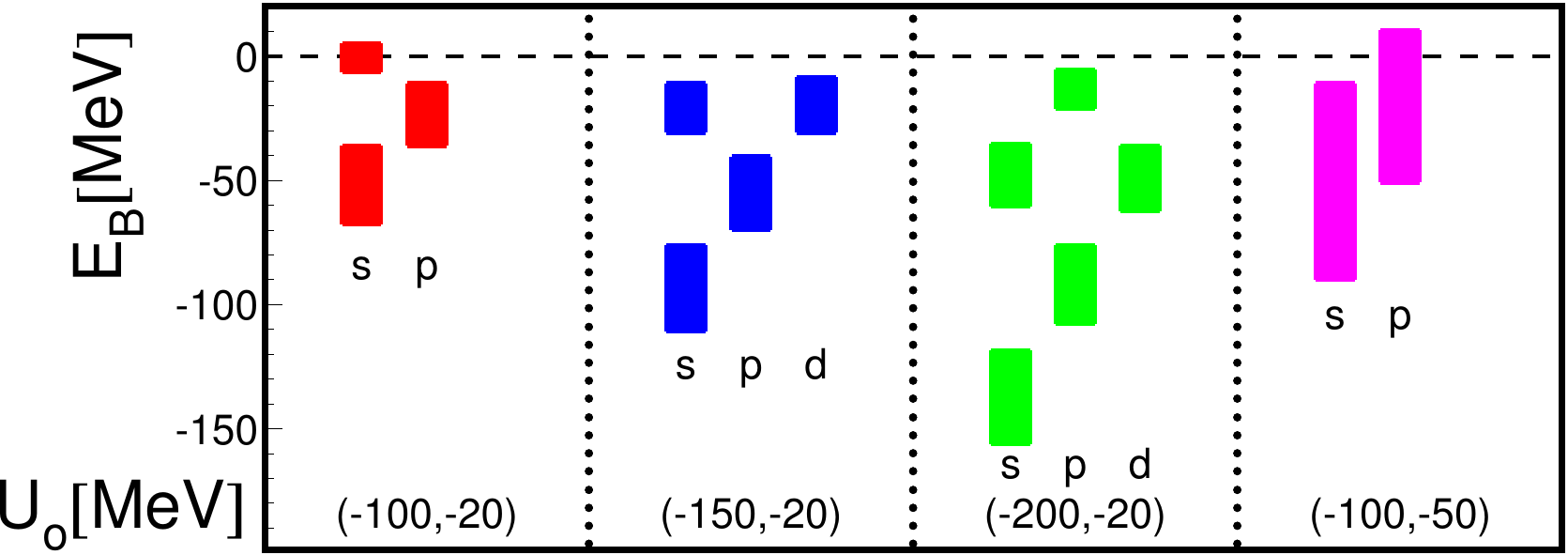,scale=0.90}
\begin{minipage}[t]{16.5 cm}
\caption{Predicted positions and widths of quasi-bound $\eta^{\prime}$ states
states in $^{12}$C for different optical potential parameters \cite{Jido_12}.
\label{fig:jido}}
\end{minipage}
\end{center}
\end{figure}

The potential parameters are important input for the question whether
quasi-bound states exist for the $\eta^{\prime}$ meson in nuclei.
Figure~\ref{fig:jido} summarizes the predictions of the positions and widths
of quasi-bound $\eta^{\prime}$ states in $^{12}$C nuclei for different
potential depths made by Jido and co-workers \cite{Jido_12}. For sufficiently
large real parts of the potential, several states with widths much smaller
than the binding energy are predicted. However, one should note that all
predictions refer to depth of the real potential of 100~MeV or larger, while
the experimental results suggest a more shallow potential with a depth not
larger than 50~MeV. A large imaginary part of the potential results in very
broad states, which would probably escape detection. The combined results from
\cite{Nanova_12,Nanova_13} suggest that the potential is attractive ($V_o <
0$), although only moderately strong. However, since $|V_o|>>|W_o|$, it is
not impossible that reasonably narrow states might be formed.

Photon-induced searches for quasi-bound $\eta^{\prime}$ states are planned
for the BGO-OD experiment at ELSA, where two types of measurements are
envisaged \cite{Nanova_12p}. Both of them are based on reactions where the
$\eta^{\prime}$ is produced in quasi-free kinematics off a proton, which is
kicked out of the target nucleus ($^{12}$C) at very forward angles so that it
takes away most of the momentum of the incident photon, leaving the
$\eta^{\prime}$ almost at rest in the residual nucleus. In an inclusive
measurement, only the momentum of the fast proton will be measured
(the magnetic spectrometer of BGO-OD will have a momentum resolution of 1 - 2\%).
The reaction identification will then be based on the missing-mass spectrometry
of the proton. Characteristic spectra from the formation of quasi-bound states
have been predicted, e.g., by Nagahiro and coworkers \cite{Nagahiro_05}.
In a semi-inclusive mode the decay of the $\eta^{\prime}$-mesic state should
be detected in coincidence with the fast forward proton. States below threshold
cannot, of course, decay via $\eta^{\prime}$ emission; they will emit lighter
mesons or nucleons. Model results from Oset and Ramos \cite{Oset_11} indicate
that the dominant decay mode should be $\eta^{\prime}N\rightarrow \eta N$,
so the planned experiment will detect $\eta$ mesons in coincidence with fast
forward protons.

However, one should keep in mind that the cross sections for these reactions
are small (predictions for the inclusive experiment are on the order of a few
nb/sr MeV \cite{Nagahiro_05} for potential parameters ($V_o,W_o$)=($-100,-5$)
MeV). The background from competing processes may be substantial and so these
proposed measurements appear quite challenging. Furthermore, not only the nuclear
ground state but also excited nuclear states may couple so that the structures
may be additionally washed out.

\subsection{Coherent production}
\label{ssec:photon_coh}
\subsubsection{Single $\eta$-production off $^{3}$He and $^{7}$Li nuclei}

The coherent photoproduction of mesons can give valuable
information on the isospin decomposition of the production
amplitudes. This has been exploited for $\eta$ production,
where measurements of the breakup reaction
\cite{Krusche_95b,Hoffmann_97,Weiss_03} revealed a
neutron/proton cross section ratio of 2/3. According to
Eq.~(\ref{eq:iso}), this indicated that either the isoscalar or
the isovector photons dominate the reaction. The small cross
sections reported for the coherent $\gamma d\rightarrow d\eta$
reaction \cite{Krusche_95b,Hoffmann_97,Weiss_01} established
the dominance of the isovector part (see \cite{Krusche_03} for
details).

For heavy mesons like the $\eta$ or $\eta^{\prime}$, due to the relatively
large momentum transfers involved, coherent production from even light nuclei
is strongly suppressed by the nuclear form factors. Only a few data sets are
therefore available and these are mostly for incident photon energies in the
immediate neighbourhood of the threshold. In the case of $\eta$
photoproduction, the situation is particularly unfavourable. For spin $J=0$
nuclei, such as $^4$He, coherent photoproduction of pseudoscalar mesons in
relative $s$-wave is forbidden due to spin/parity conservation; even very
close to the production threshold only breakup reactions contribute
\cite{Hejny_99,Hejny_02}. Furthermore, due to the quantum numbers of the
dominating amplitude, which is the isovector component of the $E_{0+}$
multipole (from the excitation of the $S_{11}$(1535) resonance), there are
large cancellations even when $J\neq 0$.

The motivation to measure the coherent reaction comes mainly from the search
for $\eta$-mesic nuclei (see discussion in Sec.~\ref{nuclei}). The basic idea
is that $\eta$-mesic states that overlap the coherent production threshold
should strongly influence the threshold behaviour of the reaction. This
concerns the energy dependence of the total cross section as well as the
angular distributions. Due to the above constraints, only nuclei with
$J,I\neq$0 are promising candidates for this approach. Among the stable light
nuclei this condition is fulfilled for $^3$He ($I=1/2,J^{\pi}=1/2^+$) and
$^7$Li ($I=1/2,J^{\pi}=3/2^-$).

The reactions have been studied at MAMI, first with the TAPS detector
\cite{Pfeiffer_04} for the $^3$He target, and later with the Crystal
Ball/TAPS setup for $^3$He \cite{Pheron_12} and $^7$Li \cite{Maghrbi_13}. The
experimental challenge is that the recoil nuclei are stopped in the liquid
helium or solid lithium targets so that a direct identification of the
coherent process is impossible (active gaseous helium targets provide much
too low luminosities). For detectors that cover large solid angles, such as
Crystal Ball/TAPS, a significant suppression of the breakup background is
achieved by vetoing events where, in addition to the meson-decay photons, a
recoil nucleon is detected. However, low energy recoil protons and recoil
protons at extreme forward angles escape detection and the detection
efficiency for the recoil neutrons is, at best, in the 30 - 40\% range.

\begin{figure}[htb]
\begin{center}
\epsfig{file=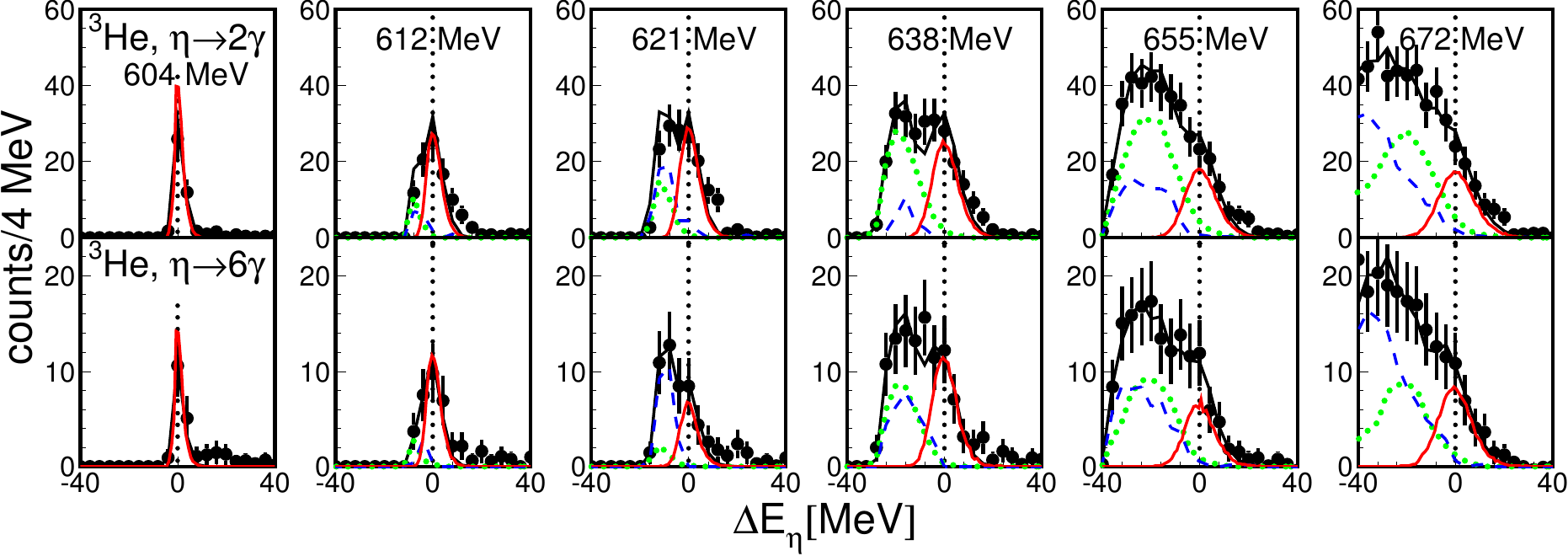,scale=0.95}
\begin{minipage}[t]{16.5 cm}
\caption{Missing energy spectra for $\eta$ production off $^3$He nuclei for
different bins of incident photon energy~\cite{Pheron_12}. Top row for the $\eta\rightarrow
2\gamma$, bottom row for the $\eta\rightarrow 3\pi^0\to 6\gamma$ decay. Black
symbols with error bars: measured data, solid (red) curves: simulated line
shapes for the coherent $\gamma ^3{\rm He}\rightarrow \eta ^3{\rm He}$
reaction, dashed (blue): breakup reaction with recoil taken by a nucleon,
dotted (green): recoil taken by a di-nucleon, solid (black): sum of all.
\label{fig:missene}}
\end{minipage}
\end{center}
\end{figure}

The final separation of breakup and coherent reactions must
rely on the reaction kinematics. This is done with a
missing-energy analysis, where the measured kinetic energy of
the meson in the c.m.\  frame is compared to that extracted
from the incident photon energy for the two-body $\eta$ -
nucleus final state. The missing-energy spectra for the $^3$He
target and both analyzed decay branches of the $\eta$-meson are
summarized in Fig.~\ref{fig:missene} \cite{Pheron_12}; results
for the $^7$Li target \cite{Maghrbi_13} are qualitatively
similar. While the identification of the $\eta$ mesons via
invariant mass analysis is clean (in particular the
$\eta\rightarrow 3\pi^0$ decay channel is almost
background-free in the near-threshold region, see
\cite{Maghrbi_13}) residual background from breakup reactions
becomes dominant in the $^3$He case already 40~MeV above
threshold. Only the fitting of the simulated line shapes for
the coherent and breakup reactions to the experimental data
allows the extraction of the coherent part. It is only in the
first energy bin, centred around 604~MeV, that a clean coherent
signal is visible because this energy is below the threshold
for a breakup reaction.

\begin{figure}[thb]
\begin{center}
\epsfig{file=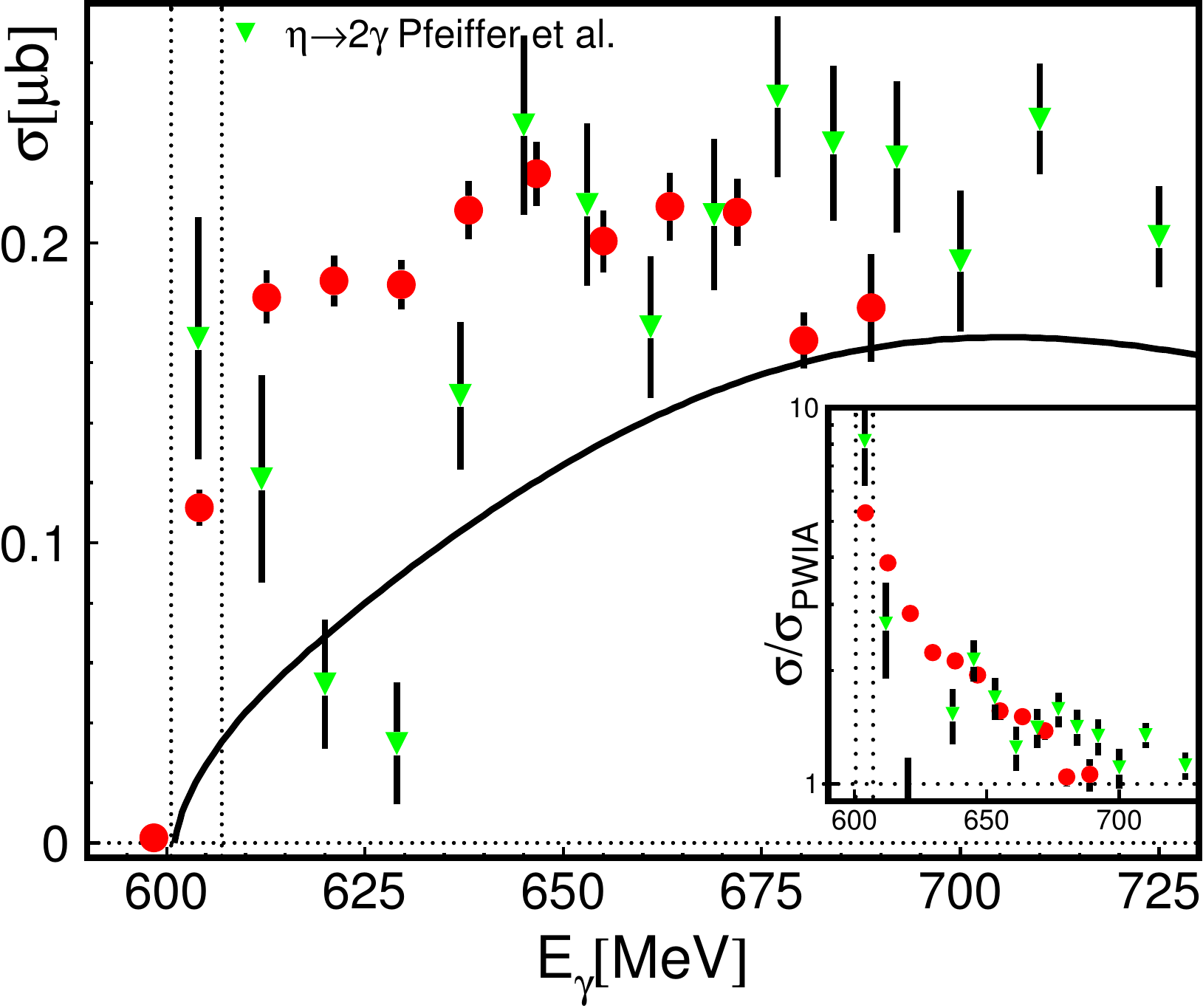,scale=0.5}
\epsfig{file=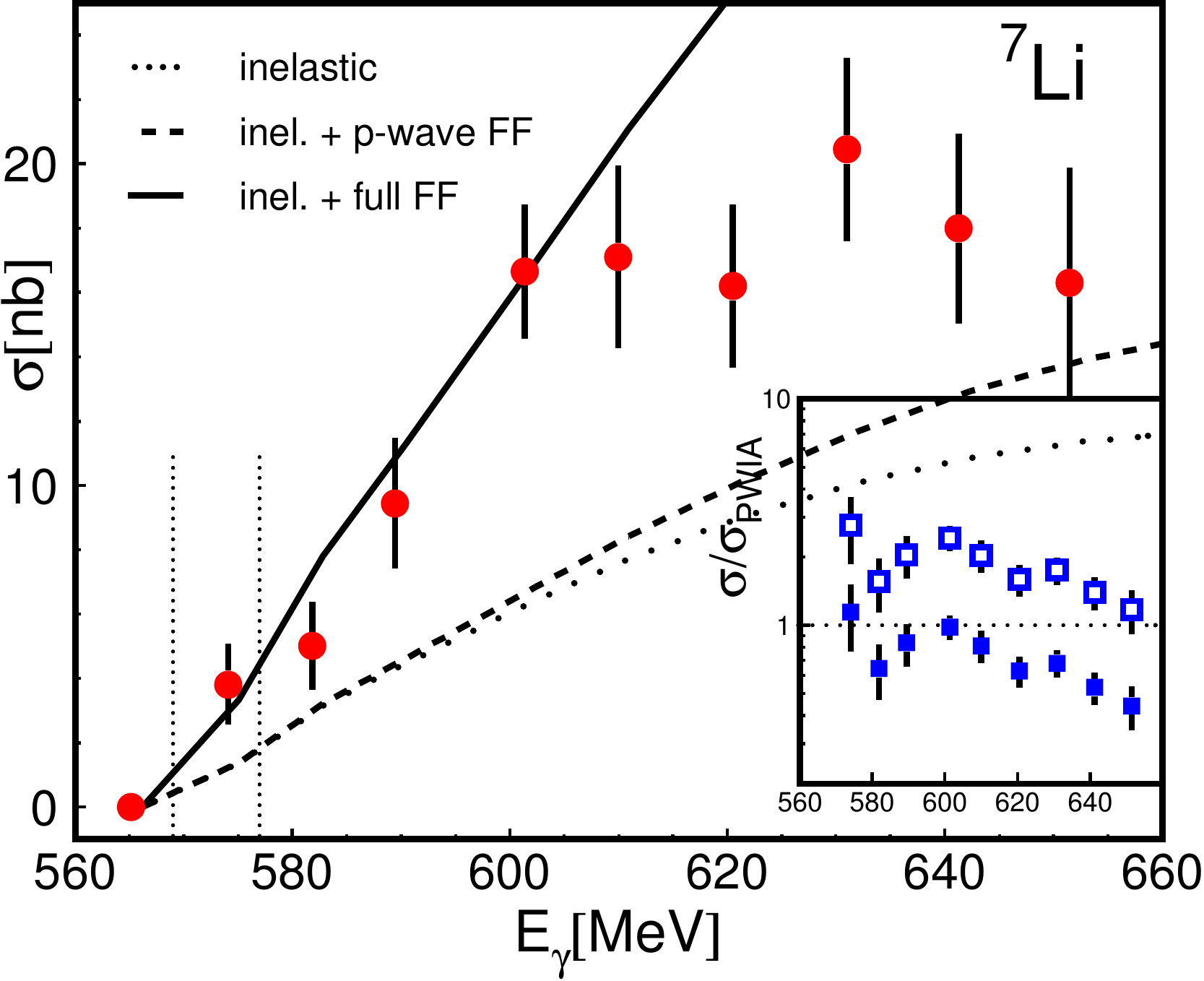,scale=0.5}
\begin{minipage}[t]{16.5 cm}
\vspace*{-0.5cm}
\caption{Total cross section for the $\gamma ^3\mbox{He}\rightarrow \eta
^3\mbox{He}$ \cite{Pheron_12} (left hand side) and $\gamma
^7\mbox{Li}\rightarrow \eta ^7\mbox{Li}$ \cite{Maghrbi_13} (right hand side)
coherent $\eta$-production reactions. The green points for $^3$He are from
\cite{Pfeiffer_04}. The curves represent PWIA modelling. For the $^7$Li
target the inelastic contribution (excitation of low lying nuclear state) is
shown separately and calculations using the full elastic form factor or only
the $p$-wave part are shown. The dotted lines indicate coherent and breakup
thresholds. The inserts show the ratio of data to the PWIA prediction; in the
$^7$Li case this is shown separately for the full and $p$-wave form factor.
\label{fig:eta_coh}}
\end{minipage}
\end{center}
\end{figure}

Total cross sections extracted from such analyses are shown in
Fig.~\ref{fig:eta_coh}. The agreement between the two MAMI experiments for
$^3$He is reasonable, but the older data \cite{Pfeiffer_04} show a dip around
625~MeV that was not confirmed in the later experiment. This is in the most
critical region as far as the separation of the coherent components and
breakup reactions is concerned; at lower photon energies the breakup
background is not so important and at higher energies it is better separated
from the coherent channel in the missing energy plots. Since the Pfeiffer et
al.\ measurement \cite{Pfeiffer_04} suffered from significantly larger
breakup background than the more recent experiment (the apparatus covered a
smaller fraction of the solid angle so that the vetoing of events with recoil
nucleons was less efficient), it seems likely that systematic uncertainty in
this region was larger. The statistical fluctuations are obviously also much
more significant. The $^7$Li data are so far the only results for coherent
production from nuclei heavier than the deuteron or $^3$He. The cross section
is smaller by one order of magnitude than that on $^3$He and the rise at
threshold is less steep (the plateau value of the cross section is reached
for $^3$He within $\approx10$~MeV, for $^7$Li only $\approx$40~MeV above
threshold).

The energy dependence of the $^3$He data is striking. The cross section rises
very steeply from the production threshold and then stays almost constant.
Already the data point between the coherent and breakup thresholds reaches
more than half the plateau value. Here, one should note that the experimental
resolution for the incident photon energy is much better than the bin size in
the figure \cite{Nikolaev_14} so that resolution effects for the slope are
negligible. A similar but even more spectacular behaviour has been observed
for the $dp\rightarrow {\rm ^3He}\;\eta$ reaction measured at COSY
\cite{Mersmann_07,Smyrski_07} (see Fig.~\ref{fig:Mersmann} and the discussion
in Sec.~\ref{ssec:mersmann}). This has been interpreted as at least a
signature of strong FSI effects and possibly some evidence for the formation
of a mesic state (the similar and unusual threshold behaviour of reactions
with different initial states is most likely due to the FSI). The behaviour
of the angular distributions supports this picture. Close to threshold they
are almost isotropic and not forward peaked as one would expect from the
influence of the nuclear form factor, which becomes significant only at
higher energies. The lithium results behave more as one would expect.

\begin{figure}[thb]
\begin{center}
\epsfig{file=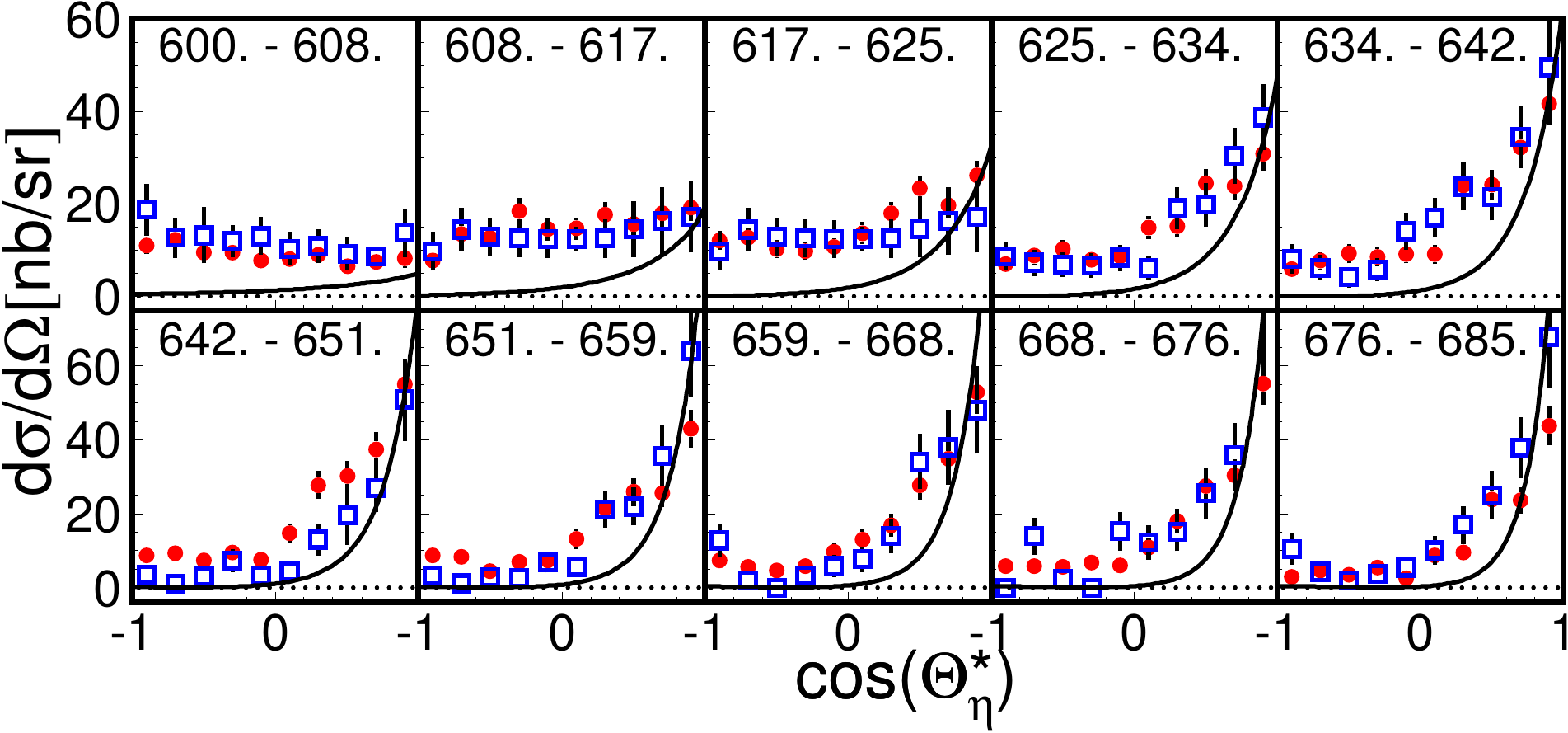,scale=0.585}
\epsfig{file=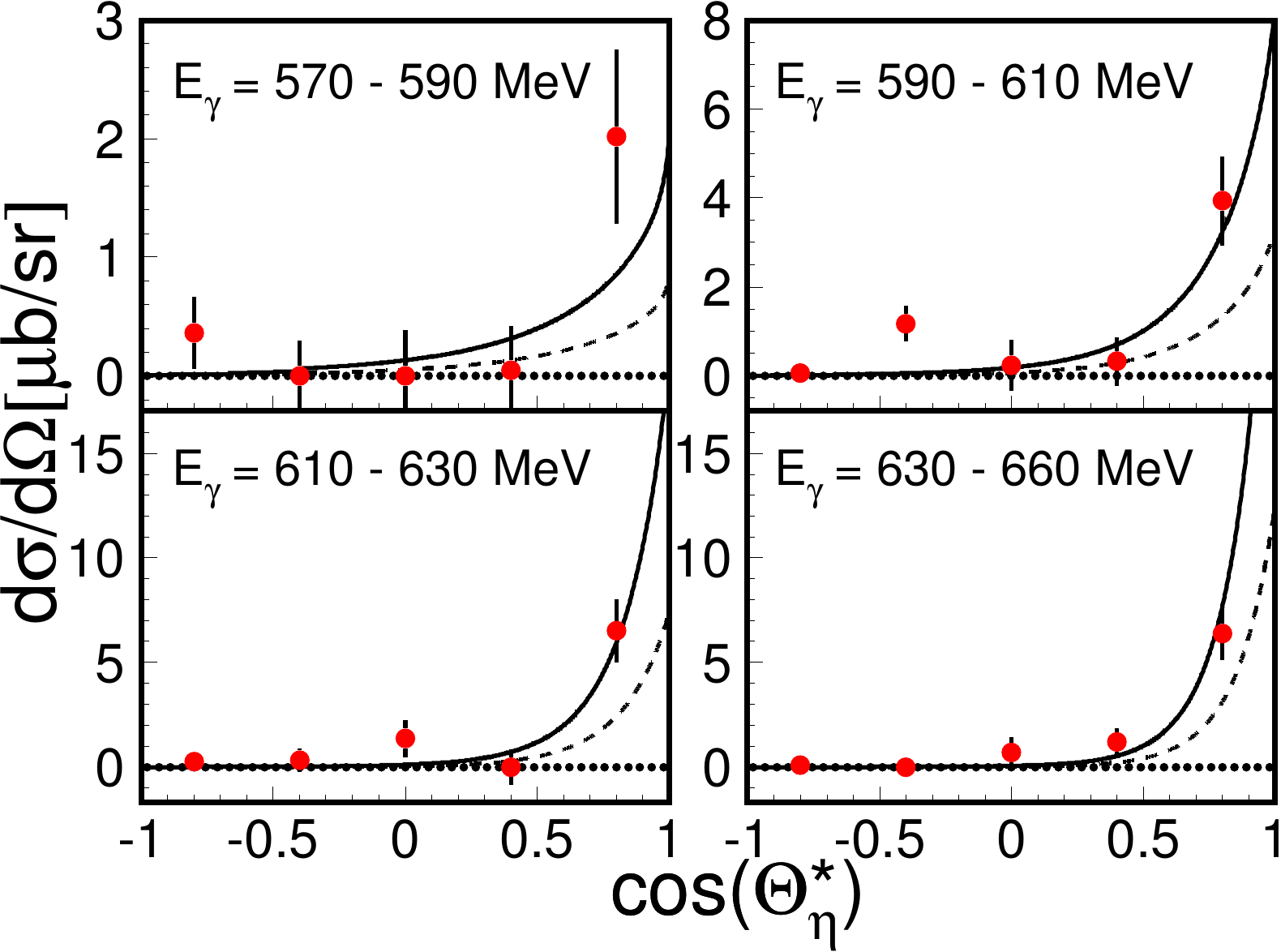,scale=0.440}
\vspace*{-0.5cm}
\begin{minipage}[t]{16.5 cm}
\caption{Angular distributions of $\gamma {\rm ^3He}\rightarrow {\rm
^3He}\,\eta$ (left hand side) and $\gamma {\rm ^7Li}\rightarrow {\rm
^7Li}\,\eta$ (right hand side) for different energy bins. For the helium
target, results from $\eta\rightarrow 2\gamma$ (red circles) and
$\eta\rightarrow 6\gamma$ (blue stars) are shown separately. The results for
the lithium target have been averaged over both decays. The curves are the
predictions from the PWIA approximation (see text). For lithium, the PWIA
results are shown for the full form factor (solid) and the $p$-wave form
factor (dotted). \label{fig:eta_coh_diff}}
\end{minipage}
\end{center}
\end{figure}

For a more quantitative discussion, both data sets have been
compared to the predictions of a simple
Plane-Wave-Impulse-Approximation (PWIA) using the formalism
outlined in \cite{Maghrbi_13,Pheron_12}. For a given incident
laboratory photon energy $E_{\gamma}$ (four-momentum
$P_{\gamma}$) and an off-shell nucleon moving with
three-momentum $\vec{p}_{N}$ (four-momentum $P_{N}$) inside the
nucleus the effective total c.m.\ energy $W=\sqrt{s^{\rm eff}}$
is obtained from
\begin{equation}
s_{\rm eff} = (P_{\gamma} + P_{N})^2.
\end{equation}
The nucleon momentum $\vec{p}_N$ is related in the
factorization approximation \cite{Drechsel_99} to the momentum
transfer $\vec{q}$ to the nucleus (note that the expressions in
reference \cite{Drechsel_99} are formulated for the c.m.\
system while in \cite{Maghrbi_13,Pheron_12} the laboratory
frame has been used)
\begin{equation}
\vec{p}_N = -\left(\frac{A-1}{2A}\right)\vec{q}\;,
\end{equation}
where $A$ is the target mass number. The amplitudes of the
elementary reactions are then evaluated at
$W(E_{\gamma},\vec{q})$. Up to this point there is analogy to
the much better studied coherent $\pi^0$ photoproduction off
(mostly $J=0$) nuclei
\cite{Drechsel_99,Krusche_02,Krusche_05a,Maghrbi_13}. In this
case the spin- and isospin-independent components of the
elementary amplitude from all nucleons must be coherently
added. However, since threshold $\eta$-production is dominated
by the $E_{0+}$ spin-flip amplitude, only the unpaired nucleon
can contribute. For $^3$He this is (dominantly) the $1s_{1/2}$
neutron and for $^{7}$Li the $1p_{3/2}$ proton. This means, of
course, that \emph{coherent} $\eta$-production does not profit
from the $A^2$ factor resulting from the coherent superposition
of scalar amplitudes. This feature, together with the larger
momentum transfers, explains the small cross sections observed.

Taking into account the phase-space factors related to the transformation
from the photon - nucleon c.m. system to the photon - nucleus c.m. system and
the nuclear form factors, the differential cross section can be written as
\begin{equation}
\frac{d\sigma_{\mbox{PWIA}}}{d\Omega}(E_{\gamma},x) =
\left(\frac{q_{\eta}^{(A)}}{k_{\gamma}^{(A)}}
\cdot\frac{k_{\gamma}^{(N)}}{q_{\eta}^{(N)}}\right)
\,\left(\frac{F_A(q^2)}{F_p(q^2)}\right)^{\!2}
\,\frac{d\sigma_{\mbox{elem}}}{d\Omega},
\label{eq:pwia}
\end{equation}
where $k_{\gamma}^{(N)}$ and $q_{\eta}^{(N)}$ are the photon
and $\eta$ three-momenta in the photon-nucleon c.m. system and
$k_{\gamma}^{(A)}$ and $q_{\eta}^{(A)}$ the same in the
photon-nucleus c.m.\ frame. We also define $x=\CT$, where
$\theta_{\eta}^{\star}$ is the polar angle of the $\eta$ meson
in the photon-nucleus c.m.\ system. The elementary cross
section $d\sigma_{\rm elem}/d\Omega$ is the (measured) $\gamma
n\rightarrow n\eta$ cross section for the $^3$He target and the
$\gamma p\rightarrow p\eta$ cross section for $^7$Li. Since
meson photoproduction probes the distribution of point-like
nucleons in the nucleus, the nuclear form factors $F_{A}$ must
be divided by the nucleon form factor $F_{p}$ (the charge form
factor of the proton is used) to account for the contribution
of the finite radii of the nucleons to the nuclear charge form
factors. For the $^3$He target the charge form factor
\cite{McCarthy_77} of this nucleus has been used.

The situation is a bit more complicated for the $^7$Li case. This nucleus has
the low lying state ($J^{\pi}=1/2^-$, 478 keV excitation energy),
corresponding to the excitation of the $1p_{3/2}$ proton to the $1p_{1/2}$
state. This contribution cannot be separated experimentally from the coherent
reaction because the missing-energy resolution is by no means good enough and
the coincident detection of a 478~keV de-excitation photon is not possible.
The form factor that should be used in Eq.~(\ref{eq:pwia}) must therefore
include contributions from both the elastic and the inelastic excitation to
this state, as given in \cite{Lichtenstadt_89}. Furthermore, since only the
$p_{3/2}$ proton contributes, there is some ambiguity whether for the elastic
part the full form factor or only the $p$-wave part is relevant. In
\cite{Maghrbi_13} the full form was used but here we compare the data also to
a model calculation based on the $p$-wave form factor.

The results of the PWIA calculations are compared to the data in
Figs.~\ref{fig:eta_coh} and \ref{fig:eta_coh_diff}. In the case of $^3$He the
rise at threshold is much less steep in the model. The ratio of data and PWIA
shown in the insert of the figure rises very strongly towards threshold (note
the logarithmic scale). The angular distributions in the model show the
expected rise to forward angles due to the form factor influence for all
energies. However, the measured distributions are almost isotropic in the
neighbourhood of the threshold and might even show a slight rise to backward
angles for the energy bin between coherent and breakup threshold. This
behaviour is interpreted as evidence for strong FSI effects.

The results for the $^7$Li target are more in line with the PWIA
approximation predictions for both the energy dependence of the total cross
section and the shape of the angular distributions. Due to the much smaller
cross section, the statistical quality of the data is of course inferior to
the $^3$He results so that the angular distributions could only be
investigated in coarser energy bins. As shown on the right hand side of
Fig.~\ref{fig:eta_coh}, the expected contribution from the excitation of the
$1/2^-$ state is substantial and the modelling using the full elastic form
factor is in better agreement with the measurement than the ansatz using the
$p$-wave form factor.

In summary, the data for $^3$He show evidence for strong FSI effects and, in
combination with the corresponding COSY data for the $dp\rightarrow {\rm
^3He}\,\eta$ reaction \cite{Mersmann_07,Smyrski_07}, make $^3$He so far the
best candidate for the observation of an $\eta$-mesic state, although there
is no final proof that the system is really quasi-bound. There were also
attempts in \cite{Pfeiffer_04,Pheron_12} to observe the decay of the
$\eta$-mesic $_{\eta}^3$He state via $\pi^0 - p$ back-to-back emission
following the capture of the $\eta$ by a nucleon into the $S_{11}$(1535)
resonance (see discussion in Sec.~\ref{ssec:virt_eta}). With this decay
channel it is possible to probe also the region below the $\eta$ threshold.
The observation of a peak-like structure at the $\eta$ threshold was
interpreted in \cite{Pfeiffer_04} as tentative evidence for this decay.
However, it was shown later \cite{Pheron_12} that this structure is an
artefact, related to the interplay of the energy and opening-angle dependence
of the $p\pi^0$ decay of nucleon resonances, which obscures any possible
signal from an $\eta$-mesic state.

The data for the $^7$Li nuclei do not show any unusual behaviour and can be
reasonably well reproduced by the PWIA approximation so that there is so far
no experimental indication of a heavier $\eta$-mesic state. Due to the
behaviour of the nuclear form factors and the spin-flip nature of the
dominant transition amplitude, most nuclei are excluded as targets for
coherent $\eta$ production. For heavier nuclei one could of course try
instead to use the breakup reaction tuned so that the recoil nucleon takes
away the momentum of the incident photon and the $\eta$ is produced (almost)
at rest in the nucleus, or to explore the coherent production of
$\pi\eta$-pairs discussed below. Such experiments have been proposed but have
not yet performed.

\subsubsection{Coherent production of $\pi^0\eta$-pairs}
As discussed in Sec.~\ref{ssec:etapi}, the production of $\eta\pi$ pairs is
dominated in the threshold region by the $\gamma N\rightarrow
D_{33}(1700)\rightarrow \eta\Delta(1232)\rightarrow \eta\pi N'$ reaction
chain. In this reaction the $\eta$ is emitted in relative $s$-wave and the
pion from the $\Delta$ resonance in relative $p$-wave. A model analysis of
the elementary reaction \cite{Fix_10} finds comparable spin-dependent and
spin-independent contributions to the amplitude, so that the coherent
production mechanism is not suppressed for spin $J=0$ nuclei as it is for
single $\eta$ production. Since, in addition, the electromagnetic excitation
of a $\Delta$ state involves only isovector photons, there are also no
cancellations. Significant cross sections have therefore been predicted for
the $\gamma A\rightarrow A\pi^0\eta$ reaction off light nuclei such as the
deuteron or $^{3,4}$He \cite{Egorov_13}. The predicted total cross sections
for $^3$He are of the same order of magnitude as for coherent $\eta$
production off this nucleus (although the elementary cross sections differs
by roughly one order of magnitude) and for $^4$He the predicted cross
sections are even larger. As a side remark, due to the coherent addition of
the amplitudes from all nucleons without cancellations, coherent cross
sections from heavier nuclei are probably substantial and might in future be
explored for detailed studies of $\eta$ interactions with nuclear matter.

\begin{figure}[htb]
\begin{center}
\epsfig{file=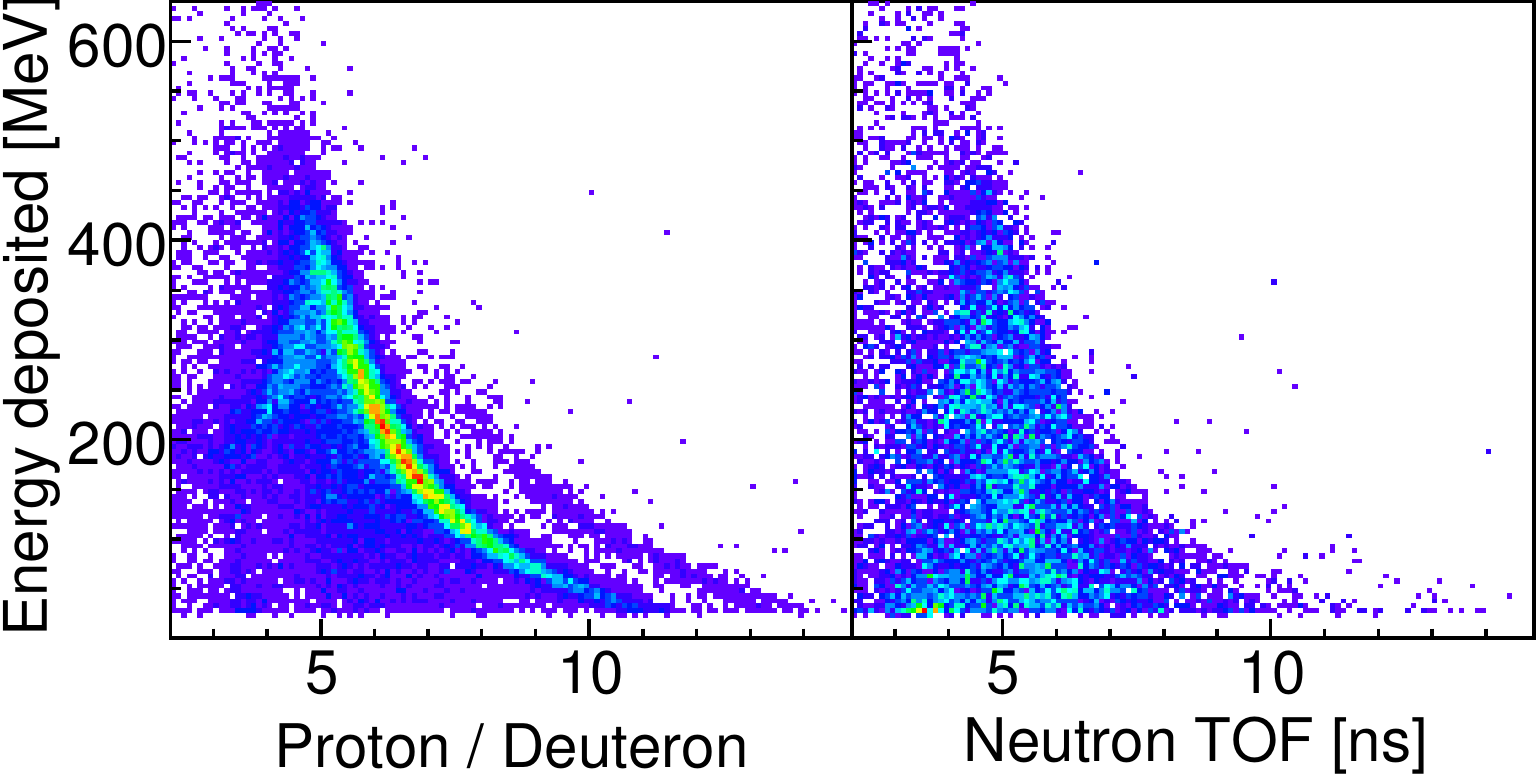,scale=0.52}
\epsfig{file=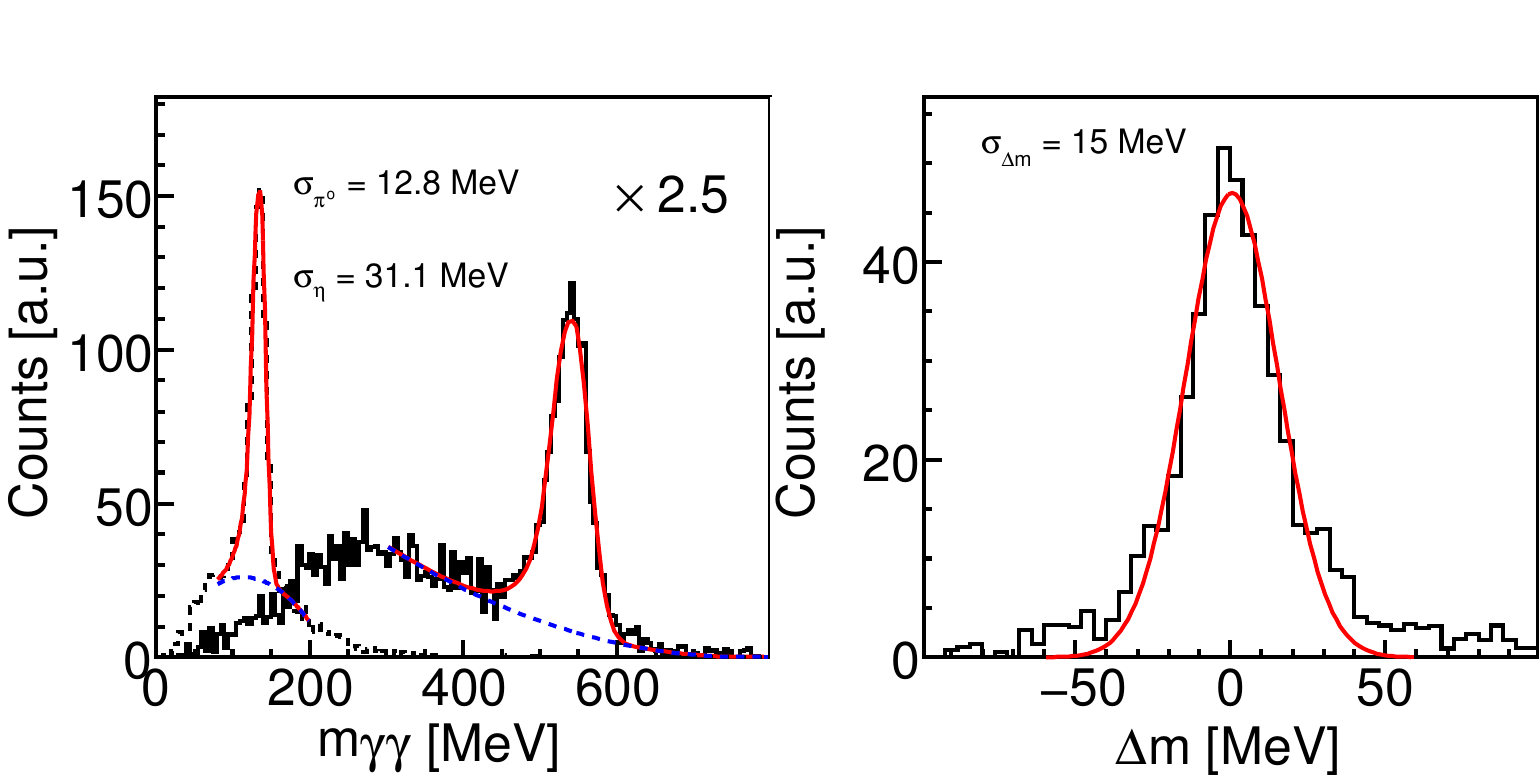,scale=0.595}
\begin{minipage}[t]{16.5 cm}
\caption{Identification of the coherent $\gamma d\rightarrow d\eta\pi^0$
reaction in the CBELSA/TAPS experiment \cite{Jaegle_09}. Left panel:
time-of-flight versus energy for charged and neutral non-photon hits in the
TAPS forward wall. For charged hits (left most plot), apart from the
dominating proton band, a pronounced deuteron band is visible. Right hand
side: invariant mass identification of $\pi^0$ and $\eta$ and missing-mass
spectrum (after cut on the deuteron band and the $\pi^0$ and the $\eta$
invariant masses). Note the small width of the missing-mass peak.
Missing-mass peaks from breakup reactions are broader, due to the Fermi
motion. \label{fig:etapi_iden}}
\end{minipage}
\end{center}
\end{figure}

\begin{figure}[htb]
\begin{center}
\epsfig{file=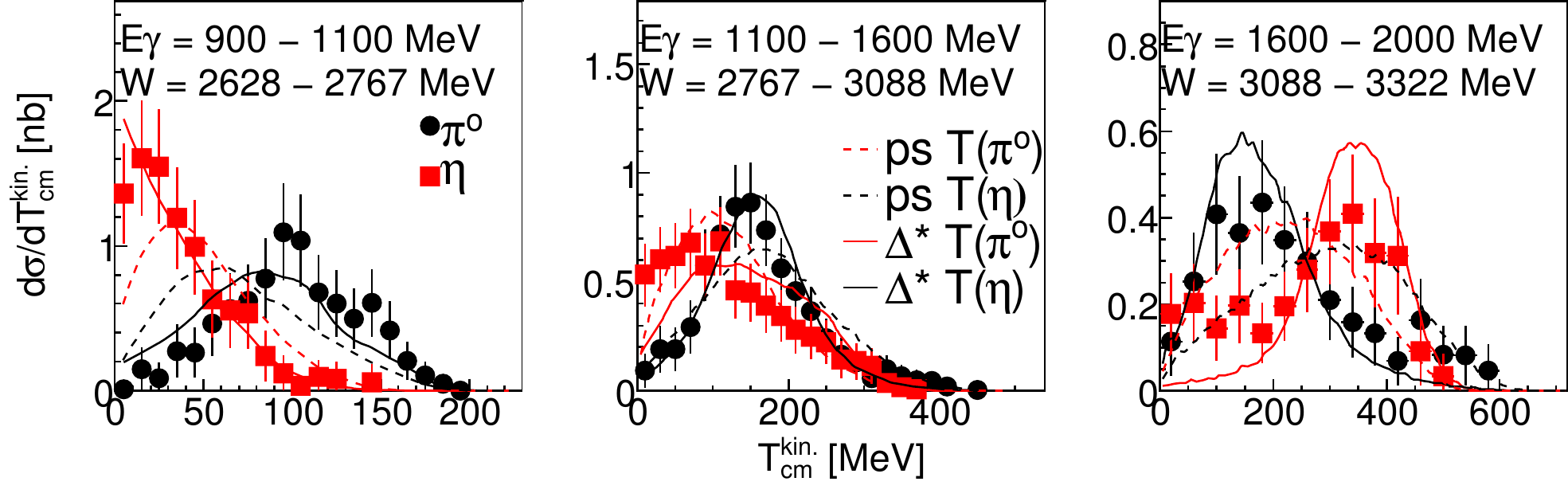,scale=0.9}
\begin{minipage}[t]{16.5 cm}
\caption{Distributions of the kinetic energies of the $\pi^0$ and $\eta$
mesons from $\gamma d\rightarrow d\pi^0\eta$ for three different ranges of
incident photon energy. (Black) circles: $\pi^0$, (red) squares $\eta$
mesons. Dashed curves: phase-space simulation, solid curves: simulation
assuming the sequential decay of a $\Delta^*$ via $\eta$ emission to the
$\Delta$(1232). Results from \cite{Jaegle_09}. \label{fig:etapi_ekin}}
\end{minipage}
\end{center}
\end{figure}

The only results available so far for coherent $\eta\pi^0$ production are
preliminary data for the deuteron target from the CBELSA/TAPS experiment at
ELSA  \cite{Jaegle_09}. The clean identification of this reaction is
demonstrated in Fig.~\ref{fig:etapi_iden}. Kinetic energy distributions of
the $\pi^0$ and $\eta$ mesons are shown in Fig.~\ref{fig:etapi_ekin}. These
data also clearly support the dominance of the
$\Delta^{\star}\rightarrow\Delta (1232)\eta\rightarrow N\pi^0\eta$ decay
chain. The kinetic energies of the $\pi^0$ mesons peak for all incident
photon energies close to the value typical for the $\Delta (1232)\rightarrow
N\pi^0$ decay ($E_{\pi}^{\rm kin}$ = 130 MeV) while the kinetic energies of
the $\eta$ mesons shift to higher values with increasing incident photon
energy. The data are in quite good agreement with the results of the model of
Egorov and Fix \cite{Fix_13}, so that their predictions for the $\gamma {\rm
^4He}\rightarrow {\rm ^4He}\,\eta\pi^0$ reaction (total cross sections up to
100 nb) are also probably realistic. With the additional degree of freedom
from the $\pi^0$ meson kinematics, where events can be selected for which the
$\eta$ is slow with respect to the $^4$He nucleus, it seems that the $\gamma
{\rm ^4He}\rightarrow {\rm ^4He}\,\eta\pi^0$ reaction may be a promising tool
for the search for $_{\eta}^4$He. A corresponding experiment proposal has
been accepted for CBall/TAPS at MAMI but the measurement has not yet been
undertaken.

\subsubsection{Coherent production of $\eta^{\prime}$ mesons}
For completeness let us make a short remark on the coherent photoproduction
of $\eta^{\prime}$ mesons off nuclei, which is so far almost completely
unexplored. The cross sections are expected to be very low, mainly because of
the large momentum transfers and the small elementary cross sections. Since,
just as for $\eta$ production, this reaction also seems to be dominated in
the threshold region by the excitation of an $S_{11}$ partial wave (although
probably not so strongly as for $\eta$ production) it will be also strongly
suppressed for spin $J=0$ targets. So far there are only a few data points
for the $\gamma d\rightarrow d\eta^{\prime}$  total cross section
\cite{Jaegle_11a}. Depending on the analysis assumptions, it seems that these
are of the order of a few nb. Model results \cite{Jaegle_11a}, based on the
$\eta^{\prime}$-MAID model, are of a similar order, but the uncertainties in
the data are still so large that no definite conclusions can be drawn.

%
%

%
%
\section{Production in pion-nucleus collisions}
\label{pions}\setcounter{equation}{0}

There have been far fewer measurements of $\eta$ production from nuclei with
low energy pion beams than with photons or nucleons and, of these, only two
are really significant.

%
%
\subsection{The $\boldsymbol{d(\pi,\eta)NN}$ reaction}
\label{pi-d}

In quark language, charge symmetry corresponds to the invariance of
interactions under the interchange of the $u$ and $d$ quarks. This symmetry
requires that the total cross sections for the interaction of $\pi^+$ and
$\pi^-$ mesons with the deuteron should be identical. This was tested many
years ago for pion laboratory energies between 70 and 370~MeV in a classical
transmission experiment~\cite{Pedroni_78}. After making corrections for
direct Coulomb effects, residual differences of up to 5\% were found between
$\sigma_{\rm tot}(\pi^+d)$ and $\sigma_{\rm tot}(\pi^-d)$. These differences
could be parameterized in terms of small amounts of charge symmetry breaking
(CSB) in the masses and widths of the $\Delta(1232)$ isobar, driven by the
$u$--$d$ mass difference.

\begin{figure}[htb]
\begin{center}
\includegraphics[width=0.5\textwidth]{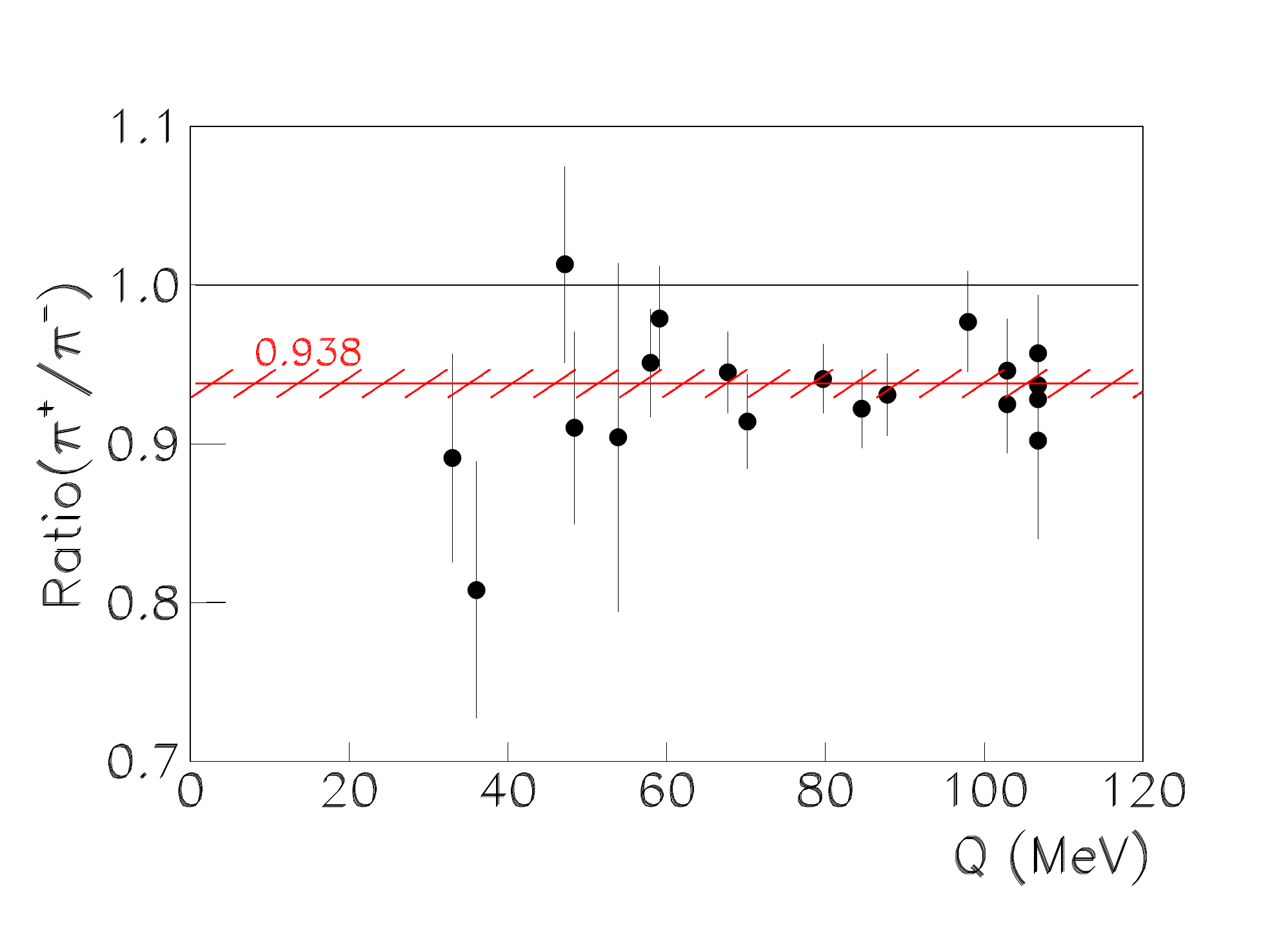}
\caption{\label{fig:Tippens} Measurements of the ratio $R=\sigma_{\rm
tot}(\pi^+d\to pp\eta)/\sigma_{\rm tot}(\pi^-d\to nn\eta)$ after adjustment
for the $n$--$p$ mass difference and initial-state Coulomb
interaction~\cite{Tippens_01}. The (red) line with hatching represents the
mean value and its uncertainty.}
\end{center}
\end{figure}

The equality of the $\pi^+d$ and $\pi^-d$ cross sections should also hold for
more exclusive reactions and this has been tested through the measurement of
the ratio $R=\sigma_{\rm tot}(\pi^+d\to pp\eta)/\sigma_{\rm tot}(\pi^-d\to
nn\eta)$ in the near-threshold region~\cite{Tippens_01}, though no attempt
was made to extract values of the individual cross sections. Only the $\eta$
meson was detected through its two-photon decay with the NaI calorimeters
that formed the central part of the spectrometer used to study the
$\pi^-{}^3\textrm{He} \to \eta\,{}^3\textrm{H}$ reaction~\cite{Peng_89}
discussed in the next section.

There are several obvious corrections that have to be made to the ratio $R$
before the results shown in Fig.~\ref{fig:Tippens} as a function of the
excess energy $Q$ can be fully assessed. These include Coulomb effects,
phase-space corrections, and the charge dependence of the $S_{11}(1535)$
mass. The residual contribution to the mean value of the ratio, $\bar{R}=
0.938\pm 0.009$, was interpreted as being due to charge-symmetry-breaking
$\eta$--$\pi^0$ mixing, with a mixing angle of
$1.5^{\circ}\pm0.4^{\circ}$~\cite{Tippens_01}. This is one contribution to
the $dd\to \pi^0\alpha$ reaction, which is the cleanest CSB test because it
does not rely on interference effects~\cite{Stepenson_03}.

%
%
\subsection{The $\boldsymbol{\pi^-{}^3\textrm{He} \to \eta\,{}^3\textrm{H}}$ reaction}

The only measurement of coherent $\eta$-production induced by a pion beam is
that of $\pi^-{}^3\textrm{He} \to \eta\,{}^3\textrm{H}$. This was first
studied at a single beam momentum for a couple of large angles by identifying
the recoil triton~\cite{Peng_87} and subsequently at five momenta in the
forward hemisphere by detecting the $\eta\to\gamma\gamma$ decay products in
the LAMPF two-photon spectrometer~\cite{Peng_89}. The incident pion
laboratory momenta between 590~MeV/$c$ and 680~MeV/$c$ corresponded to $\eta$
c.m.\ momenta between $\approx 78$~MeV/$c$ and 278~MeV/$c$. The shapes of the
angular distributions were reasonably well described by DWIA calculations
based upon $\pi^-p\to\eta n$ amplitudes dominated by the $N^*(1535)$ isobar
but these underestimated the magnitudes of the cross sections by typically a
factor of two. This deficit has been ascribed to contributions from two- and
three-nucleon mechanisms which increase the magnitudes while affecting the
angular distributions far less~\cite{Liu_92}.

\begin{figure}[htb]
\begin{center}
\includegraphics[width=0.5\textwidth]{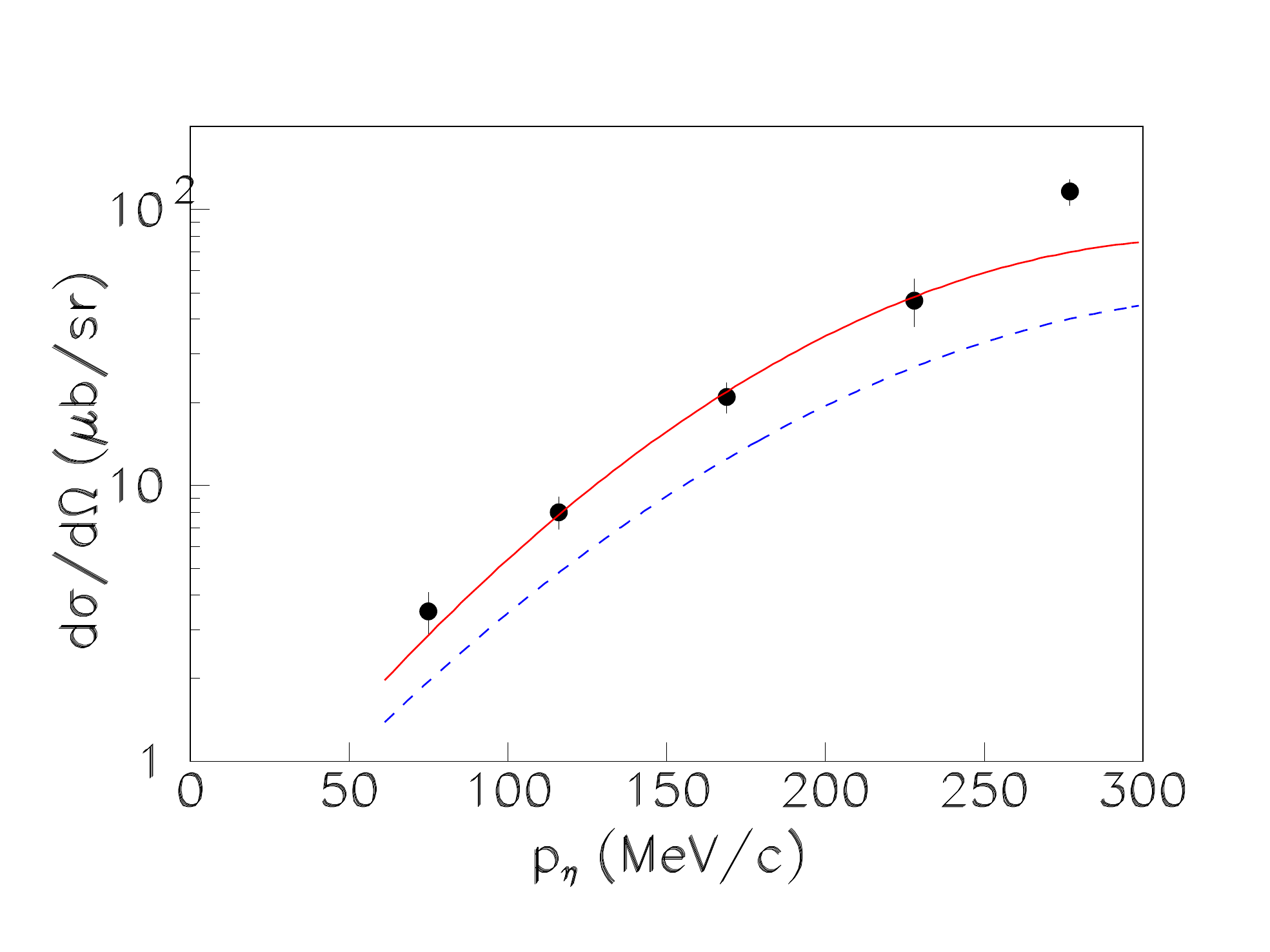}
\caption{\label{fig:Peng} Differential cross section for the
$\pi^-{}^3\textrm{He} \to \eta\,{}^3\textrm{H}$ reaction extrapolated to the
forward direction using shapes given by DWIA calculations~\cite{Peng_89}. The
(blue) dashed curve assumes only single-step contributions whereas the solid
(red) one includes also two- and three-nucleon mechanisms, though the latter
has only a minor effect~\cite{Liu_92}.}
\end{center}
\end{figure}
Figure~\ref{fig:Peng} shows the forward $\pi^-{}^3\textrm{He} \to
\eta\,{}^3\textrm{H}$ differential cross section obtained by extrapolating
the measured data using shapes given by the single-nucleon DWIA
estimates~\cite{Peng_89}. However, as shown by the dashed line, this
underestimates the absolute magnitudes by about a factor of two. Calculations
that include two- and three-nucleon terms (solid curve) lead to a much better
description~\cite{Liu_92}.

Unfortunately, the data do not extend down to the near-threshold region where
the $\eta{}^3$H final state interaction plays such an important role in other
reactions. However, the kinematics are very similar to those of the
$\gamma{}^3\textrm{He} \to \eta\,{}^3\textrm{He}$ reaction which, away from
the FSI region, are well described in impulse approximation~\cite{Pheron_12}.
Clearly, if two-step processes with intermediate virtual pions are
significant in the $(\pi^-,\eta)$ case then it would seem that this should
also be true for $(\gamma,\eta)$ as well. It might therefore be helpful if
the two reactions were analyzed simultaneously.

The LAMPF two-photon spectrometer was also used to measure the inclusive
$^{12}$C$(\pi^+,\eta)X$ reaction at 680~MeV/$c$~\cite{Peng_88} and the
limited data set has been analyzed successfully in DWIA
approaches~\cite{Kohno_90,Krippa_92}.

%
%
\section{Production in nucleon-nucleon collisions}
\label{NN}\setcounter{equation}{0}

\subsection{$\boldsymbol{\eta}$ production in $\boldsymbol{pp}$ collisions}
\label{ppeta}

\begin{figure}[htb]
\begin{center}
\includegraphics[width=0.5\textwidth]{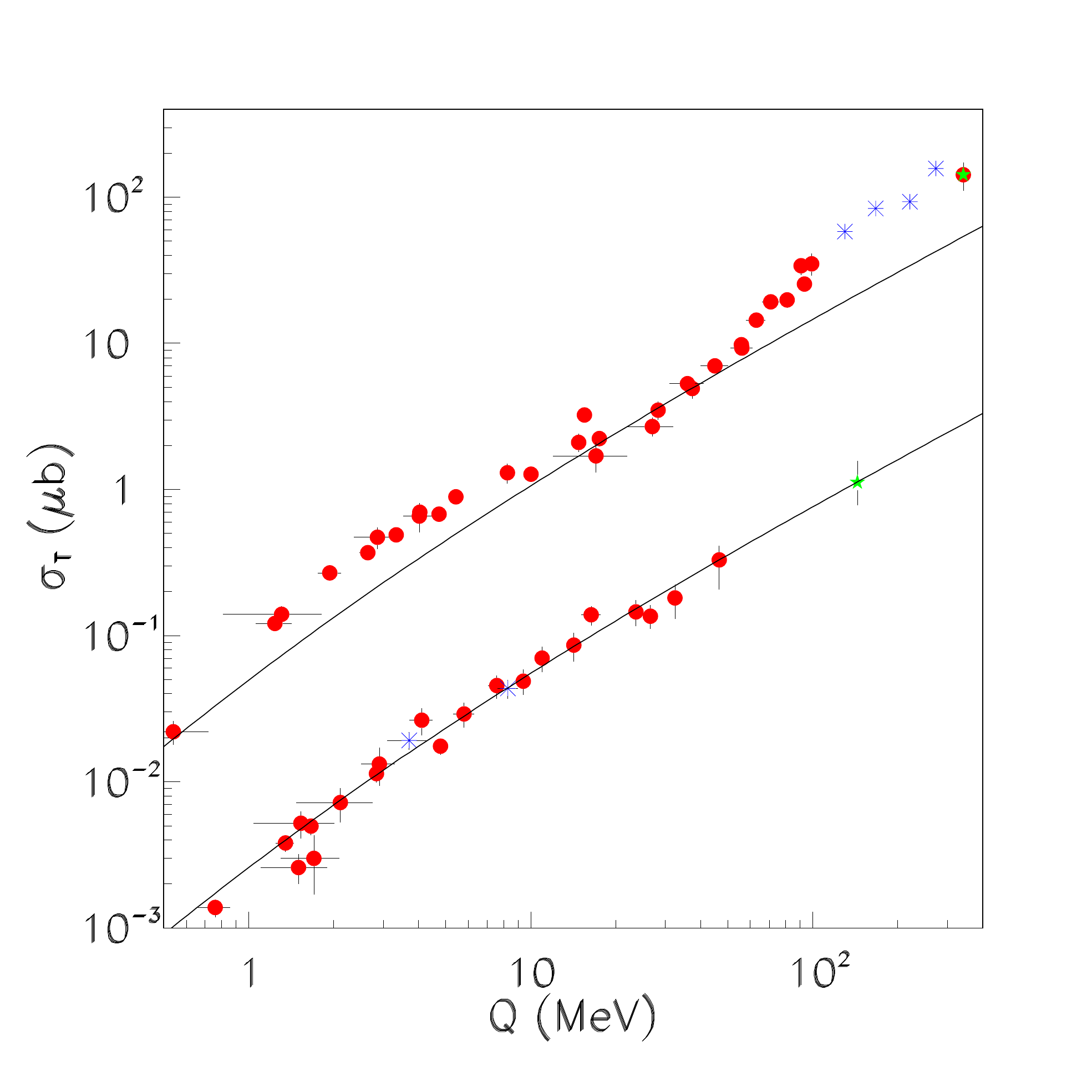}
\caption{\label{fig:eta_etaprime} Total cross sections for $pp\to pp\eta$
(upper points) and $pp\to pp\eta^{\prime}$ (lower points). The $\eta$ data
are taken from
Refs.~\cite{Chiavassa_94a,Calen_96,Bergdolt_93,Hibou_98,Smyrski_00,Moskal_04,Moskal_10}
(closed red circles), \cite{Marco_01} (blue crosses), and
\cite{Agakishiev_12} (green star) and the $\eta^{\prime}$ data from
Ref.~\cite{Bergdolt_93,Hibou_98} (blue crosses), \cite{Balestra_00} (green star), and
\cite{Moskal_98,Moskal_00,Czerwinski_14} (closed red circles). The preliminary $\eta$ point at
much higher energy~\cite{Teilab_11} is not shown. The solid curves are
arbitrarily scaled $pp$ FSI predictions of Eq.~(\ref{FW1}). }
\end{center}
\end{figure}

The total cross section for the $pp\to pp\eta$ reaction has been measured by
a variety of
groups~\cite{Chiavassa_94a,Calen_96,Bergdolt_93,Hibou_98,Smyrski_00,Moskal_04,Moskal_10,Marco_01,Agakishiev_12}
and the results are shown in Fig.~\ref{fig:eta_etaprime} as a function of the
excess energy $Q=W-2m_p-m_{\eta}$, where $W$ is the total centre-of-mass
energy. In most of these experiments the reaction was identified by measuring
the two final protons and reconstructing the $\eta$ through the missing-mass
peak in the reaction. However, in the PINOT
experiment~\cite{Chiavassa_94a,Marco_01} only the two photons from the $\eta$
decay were measured in samples of phase space so that these results could be
contaminated at the higher energies by the production of an extra
pion~\cite{Marco_01}. The exclusive HADES result at
340~MeV~\cite{Agakishiev_12}, where the $\eta$ decay products were measured
in coincidence with the two final protons, shows that this is probably not a
serious concern.

Most of the rapid rise of the total cross section with energy that is
apparent in Fig.~\ref{fig:eta_etaprime} is merely a reflection of the $Q^2$
dependence of the non-relativistic three-body phase space. However, if one
modifies this with the one-pole approximation to the $S$-wave proton-proton
final state interaction, the near-threshold energy dependence
becomes~\cite{Faldt_96}
\begin{equation}
\label{FW1}
\sigma_T(pp\to pp\eta)=C \left.\left(\frac{Q}{\varepsilon}\right)^2\right/
   \left(1+\sqrt{1+Q/\varepsilon}\right)^{2},
\end{equation}
where $C$ is constant. Since the Coulomb repulsion has here been neglected,
there is some ambiguity in the value to take for the pole position
$\varepsilon$ and the best fit to the analogous $\eta^{\prime}$ production
data~\cite{Bergdolt_93,Hibou_98,Balestra_00,Moskal_98,Moskal_00,Czerwinski_14} was achieved with $\varepsilon
= 0.75^{+0.20}_{-0.15}$~MeV~\cite{Czerwinski_14}, which is quite consistent
with the original assumptions~\cite{Faldt_96}.

Whereas all the $pp\to pp\eta^{\prime}$ total cross sections are well
described by Eq.~(\ref{FW1}) with $C\approx 0.012~\mu$b, the same is not true
for the analogous $\eta$ case. If this is normalized to the data at $Q=
20$-30~MeV, then the curve underpredicts the results at both lower and higher
values of $Q$. In the near-threshold region this is probably due to the
neglect of the strong $\eta p$ and $\eta pp$ FSI and at high energies higher
partial waves become important, as can be seen from the differential
distributions discussed later.

\begin{figure}[htb]
\begin{center}
\includegraphics[width=0.45\textwidth]{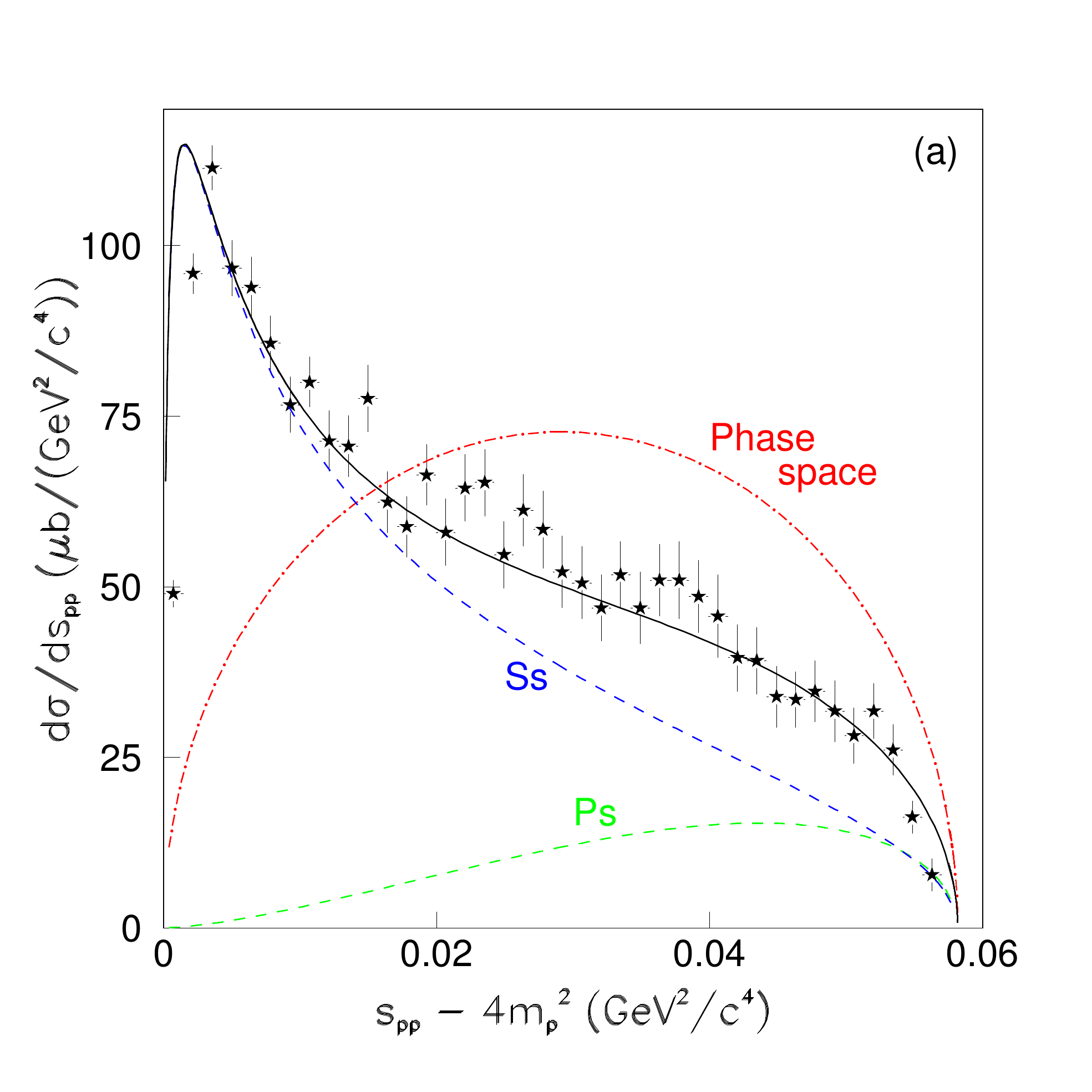}
\includegraphics[width=0.45\textwidth]{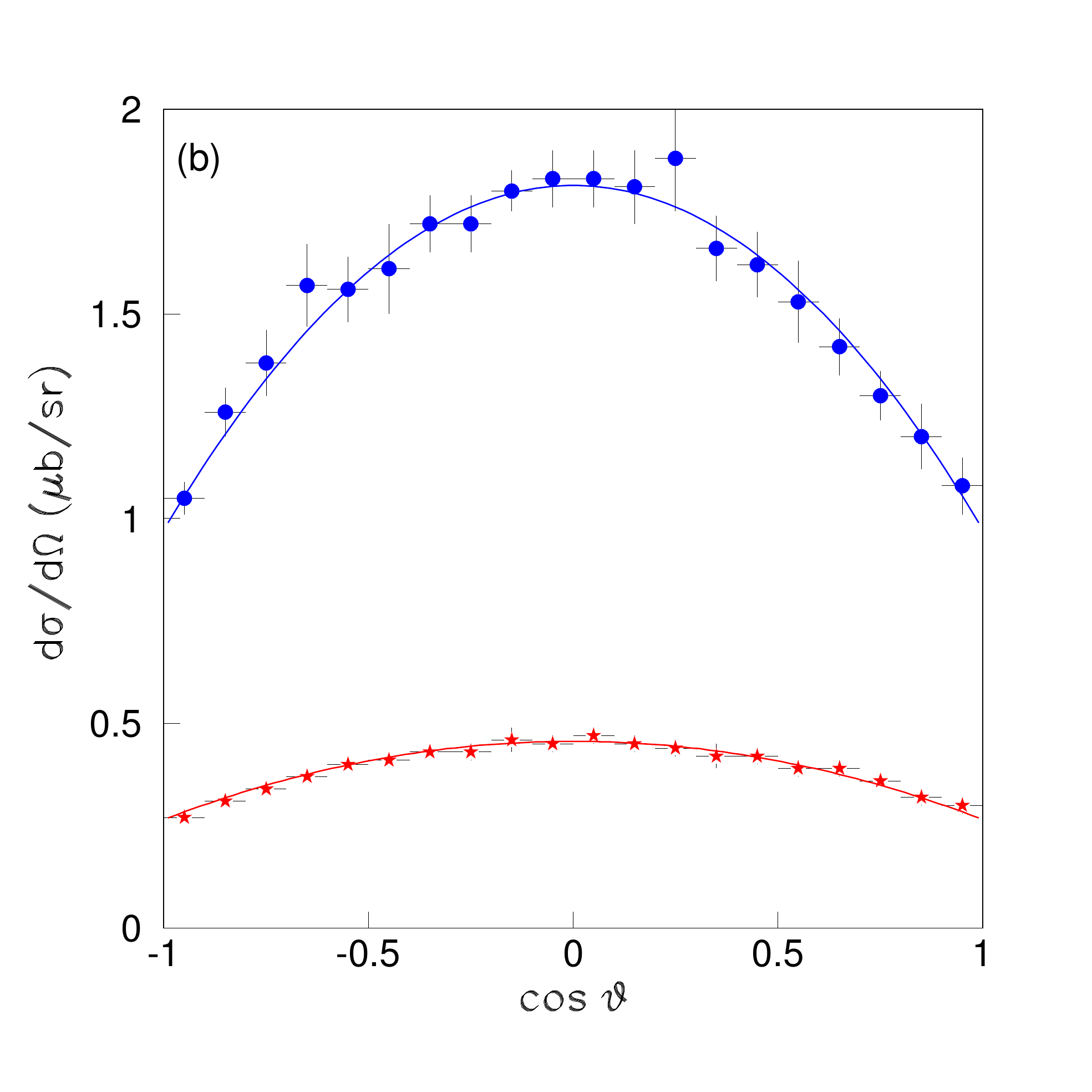}
\caption{\label{fig:pp} (a) One-dimensional distribution measured in the
$pp\to pp\eta$ reaction at $Q=15.5$~MeV~\cite{Moskal_04}; to a very good
approximation the abscissa represents $4m_pE_{pp}$. The (red) chain curve
corresponds to a phase space distribution and weighting this arbitrarily with
the $pp$ $S$-wave FSI or a $P$-wave factor gives the (blue or green) dashed
curves. The sum of $Ss$ and $Ps$ contributions (solid black curve) describes
well the shape of the data. (b) c.m.\ angular distribution of the $\eta$ in
the $pp\to pp\eta$ reaction at $Q=40$~MeV (lower red points) and 72~MeV
(upper blue points)~\cite{Petren_10}. The curves are linear fits in
$\cos^2\theta_{\eta}$.}
\end{center}
\end{figure}

More detailed information can be obtained from looking at differential
distributions of the $pp\to pp\eta$ reaction in data taken at a single
energy~\cite{Calen_99,Abdel-Bary_03}. The spectrum of the excitation energy
$E_{pp}$ in the $pp$ system is shown in Fig.~\ref{fig:pp}a at an excess
energy of $Q\approx 15.5$~MeV~\cite{Moskal_04}. What is immediately striking
here is the sharp peaking of the experimental data at very low $E_{pp}$ that
is due to the dominance of the $Ss$ wave and the very strong final state
interaction between the two protons. We are here using the notation $L\ell$
to describe the final state, where $L$ is the orbital angular momentum in the
$pp$ system and $\ell$ that of the $\eta$ relative to the diproton. There are
minor differences in the literature on how the FSI is modeled and the curve
shown in Fig.~\ref{fig:pp}a does not include Coulomb repulsion or
experimental resolution but it is clear that the model falls well below the
data at large $E_{pp}$. It was suggested that a new approach was needed for
FSI models~\cite{Deloff_04}. A more natural assumption is that at large
$E_{pp}$ there are contributions associated with $Ps$ final
waves~\cite{Nakayama_03} and the combination of $Ss$ and $Ps$ final waves
describes the data very well.

The most complete exclusive measurements of the $pp\to pp\eta$ reaction were
carried out at the CELSIUS ring at $Q=40$~MeV and 72~MeV. In addition to
measuring the two final protons, the $\eta$ was identified through either its
$3\pi^0$~\cite{Pauly_06} or $2\gamma$~\cite{Petren_10} decay. The $\eta$
angular distributions shown at the two energies in Fig.~\ref{fig:pp}b deviate
significantly from isotropy and, if these are parameterized in the form of
$d\sigma/d\Omega_{\eta} \sim 1 +\alpha\cos^2\theta_{\eta}$ then one finds
that $\alpha =-0.42$ and $-0.46$ at 40 and 72~MeV, respectively. This was
taken as evidence for the importance of $Sd$ or $Pp$ partial waves at these
energies~\cite{Petren_10}.

Other evidence for the importance of higher partial waves in
the $pp\to pp\eta$ reaction is to be found from inspection of
the Dalitz plots measured at 40 and
72~MeV~\cite{Pauly_06,Petren_10}. These show a distinct valley
when the invariant mass of one $\eta p$ pair has a similar
value to that of the other pair, viz.\ $m(\eta p_1)\approx
m(\eta p_2)$. It seems that, as the energy is raised and the
Dalitz plot opens out, the $\eta$ cannot resonate
simultaneously with both protons as an $N^*(1535)$. In terms of
the partial wave expansion, neither the $Sd$ nor $Ps$ wave can
describe such a behaviour, which requires at least a $Pp$
wave~\cite{Petren_10}. Since at least three sets of higher
partial waves are required to parameterize the 40 and 72~MeV
data, this brings into question the description of the total
cross section given in Fig.~\ref{fig:eta_etaprime} by phase
space distorted by the proton-proton final state interaction.
At low $Q$ there is some evidence for an $\eta p$ FSI but it is
hard to include this together with the $pp$ FSI in anything
other than the factorization approximation~\cite{Bernard_99}.

All the one-dimensional spectra extracted from the exclusive $pp\to pp\eta$
measurements at 40 and 72~MeV were described in terms of $Ss$, $Sd$, $Ds$,
and $Ps$ partial waves, where the only dynamics included was the $pp$
$S$-wave FSI~\cite{Petren_10}. There are, however, nine $Pp$ waves that might
potentially contribute and only one of these nine was retained in the
description. Though this choice seems adequate for the evaluation of the
acceptance of the WASA spectrometer, the authors may not have explored the
full range of ambiguity of the parameters. There could, for example, be some
conflict of the partial-wave interpretation with the limited $pp\to pp\eta$
data set reported in Ref.~\cite{Dymov_09} at small angles and low $E_{pp}$
and clearly more data would be very useful in the low $E_{pp}$ region.

By taking the amplitudes to be constant, apart from the necessary threshold
momentum factors, not only could the data at 40 and 72~MeV be reproduced, but
these parameters also describe quantitatively the $pp$ and $\eta p$
invariant-mass distributions measured at $Q=15.5$~MeV~\cite{Moskal_04}. In
particular, the fractional contribution of $Ps$ waves deduced is very similar
to that shown in Fig.~\ref{fig:pp}b.

Although there were some measurements of the proton analyzing power in the
$\pol{p}p\to pp\eta$ reaction for $Q > 320$~MeV~\cite{Balestra_04}, the
COSY-11 data at 10 and 36~MeV show very low values for $A_y$, from which it
is hard to draw any firm conclusions~\cite{Czyzykiewicz_07}. The analysis of
a more detailed COSY-WASA experiment is in progress~\cite{Hodana_11,Hodana_14}.

\subsection{$\boldsymbol{\eta}$ production in $\boldsymbol{pn}$ collisions}
\label{pneta}

The first indication that $\eta$ production is much stronger in $pn$ than in
$pp$ collisions was found from the comparison of the numbers of $\eta$ mesons
originating from $pd$ and $pp$ interactions~\cite{Chiavassa_94b,Scomparin_93}. Since only
the two photons coming from the $\eta$ decay were detected in the PINOT
spectrometer, there was no way of separating the quasi-free $pn\to d\eta$
production from the $pn \to pn\eta$ reaction. Subsequently both the two-body
$pn \to d\eta$~\cite{Calen_97,Haggstrom_97,Calen_98b} and the three-body $pn \to
pn\eta$~\cite{Calen_98a,Moskal_09} reactions were measured individually using
quasi-free production on the deuteron.

\begin{figure}[htb]
\begin{center}
\includegraphics[width=0.45\textwidth]{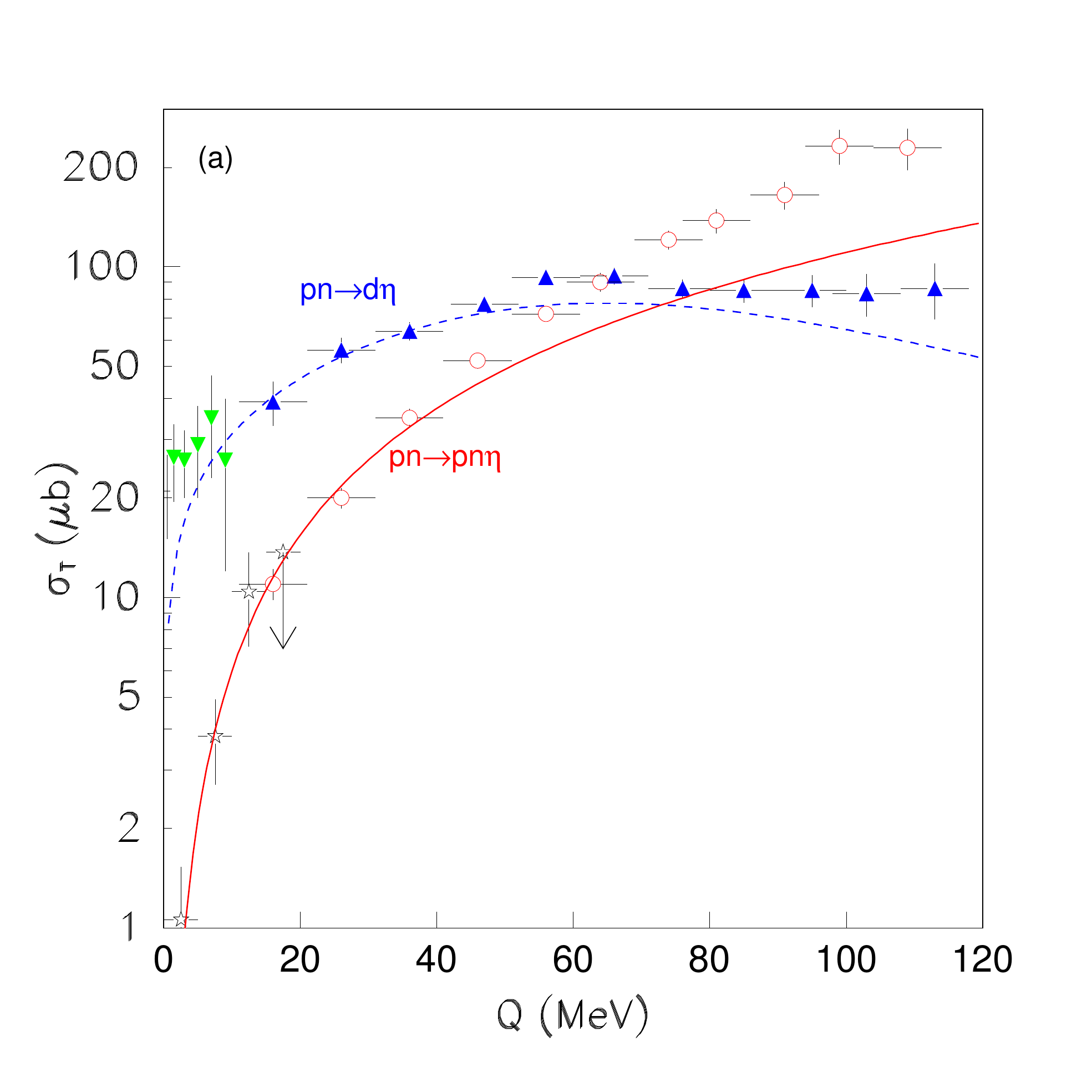}
\includegraphics[width=0.45\textwidth]{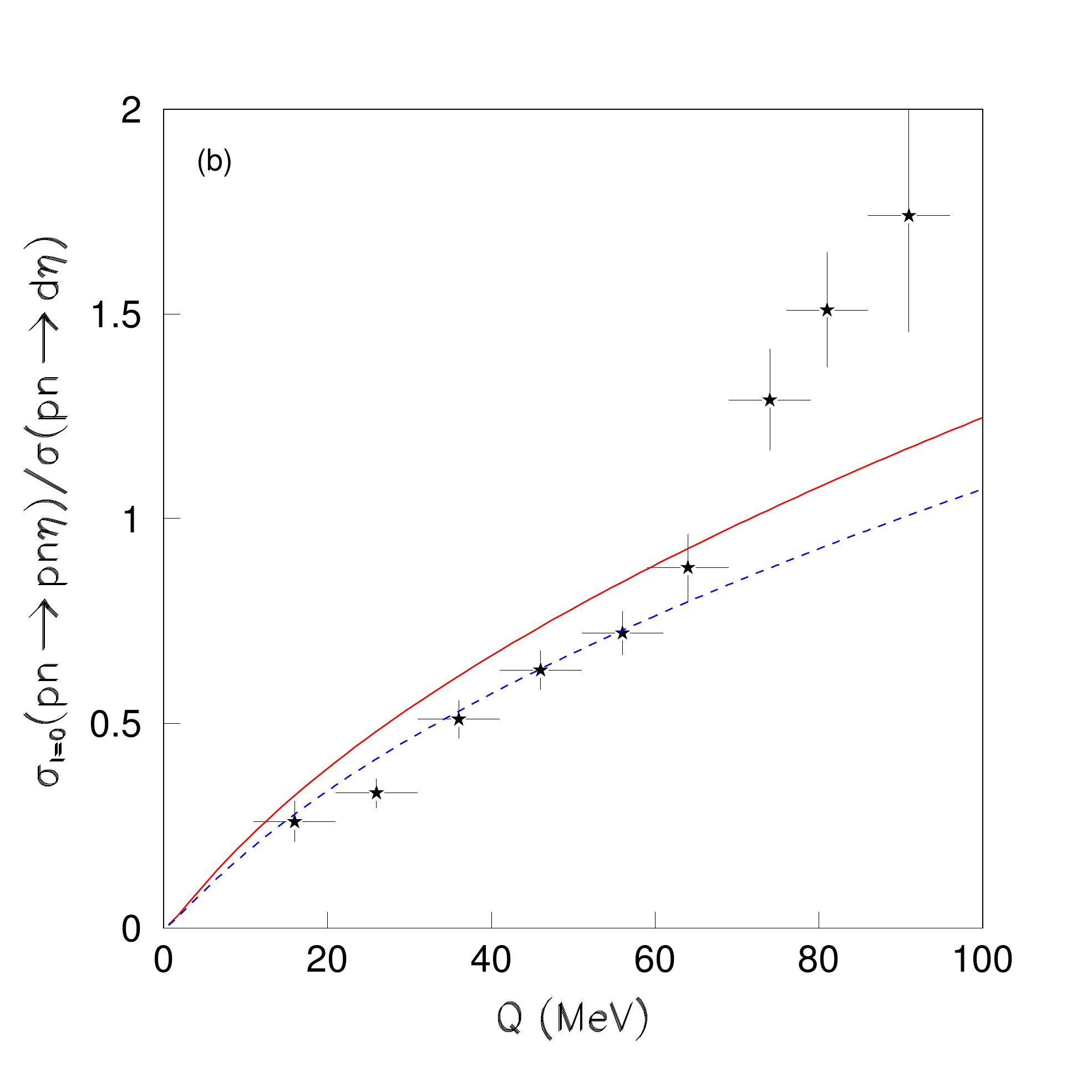}
\caption{\label{fig:pn} (a) The $pn\to d\eta$ total cross section from
Refs.~\cite{Calen_97,Haggstrom_97} (blue triangles) and \cite{Calen_98b} (green inverted
triangles) and $pn \to pn\eta$ from Refs.~\cite{Calen_98a} (red open circles)
and \cite{Moskal_09} (black open stars). The solid (red) curve represents the
shape expected from phase space modified by the $I=0$ $pn$ final-state
interaction whereas the dashed (blue) one is a simple $N^*(1535)$ resonance
shape. (b) The ratio of the isoscalar total cross sections for $pn \to
pn\eta$~\cite{Calen_98a} to $pn\to d\eta$~\cite{Calen_97,Haggstrom_97}. The solid (red)
curve is the final-state-interaction prediction of Eq.~(\ref{Master}).
Arbitrarily scaling this by a factor of 0.85 leads to the (blue) dashed
curve. }
\end{center}
\end{figure}

The comparison of the results shown in Figs.~\ref{fig:eta_etaprime} and
\ref{fig:pn}a show that, away from threshold, the total cross section for
$pn\to pn\eta$ is about 6.5 times that for $pp\to pp\eta$. Since the $pn$
channel is composed of half isospin 0 and 1, this means that
$\sigma_{I=0}(NN\to NN\eta) \approx 12\times\sigma_{I=1}(NN\to NN\eta)$. In a
one-meson-exchange model a large value for the ratio is to be expected if the
dominant exchanges are isovector mesons ($\pi$ and $\rho$) rather than
isoscalar ($\omega$ and $\eta$) but it seems that a ratio of 12 can only be
achieved if the $\pi$ and $\rho$ exchanges add constructively in the $I=0$
channel and destructively for $I=1$~\cite{Faldt_01}.

At low energies the $pn\to d\eta$ cross section is much larger than that for
$pn \to pn\eta$ due to the phase spaces being proportional to $\sqrt{Q}$ and
$Q^2$, respectively. In the $pn\to d\eta$ case, there seems to be some
evidence for the effects of the $S_{11}(1535)$ resonance because the total
cross section quickly changes from the near-threshold $\sqrt{Q}$ behaviour.
The points very close to $Q=0$, which were taken by placing counters to
detect deuterons that escaped down the CELSIUS beam pipe~\cite{Calen_98b},
show an enhancement that is discussed in Sec.~\ref{nuclei}.

In the near-threshold region final-state-interaction theory suggests that,
if one neglects relativistic corrections, there should be the simple
relation between the $pn \to pn\eta$ and $pn\to d\eta$ cross
sections~\cite{Faldt_96}:
\begin{equation}
\label{Master}
\sigma_{pn\to pn\eta}^{I=0} \approx \frac{1}{4}\left(\frac{Q}{\varepsilon}\right)^{3/2}
\left(1+\sqrt{1+Q/\varepsilon}\right)^{-2} \sigma_{pn\to d\eta},
\end{equation}
where $\varepsilon$ is the deuteron binding energy. In the derivation of
Eq.~(\ref{Master}) it is assumed that the $\eta$ production operator is of
short range and that the effects of channel coupling through the $np$ tensor
force that gives rise to the deuteron $D$ state can be neglected. In the
simplified form presented here, the $pn\to d\eta$ is taken to follow phase
space but shape corrections can be introduced~\cite{Faldt_01}. The
predictions are only relevant for the $I=0$ cross section but the small $I=1$
contribution can easily be subtracted using the data from $pp$ collisions. A
more important restriction is that Eq.~(\ref{Master}) only estimates the
cross section where the $np$ emerge in an $S$ wave so that it has to break
down at higher values of $Q$.

The rapid rise from threshold of the $pn \to pn\eta$ cross section in
Fig.~\ref{fig:pn}a is rather similar to that in the $pp\to pp\eta$ collisions
in Fig.~\ref{fig:eta_etaprime}. The shape is well described by
Eq.~(\ref{Master}) for $Q\lesssim 40$~MeV. This is examined more
quantitatively in Fig.~\ref{fig:pn}b, where the $\sigma_{pn\to pn\eta}^{I=0}
/\sigma_{pn\to d\eta}$ cross section ratio is compared to the predictions of
Eq.~(\ref{Master}). The two cross sections were measured
simultaneously~\cite{Calen_97,Haggstrom_97,Calen_98a} and so the big luminosity
uncertainty cancels but it is not clear if the small deviations between the
predictions and the data at low $Q$ are due to an oversimplified theory or to
systematic uncertainties in the event identification or the evaluation of $Q$
in the $pn \to pn\eta$ case. Reducing the predictions by 15\% gives a much
better description of the data. What is, however, very clear is that the
sharp rise above $Q=60$~MeV is likely to be due to higher partial waves in
the recoiling $pn$ system.

It has been argued~\cite{Moskal_09} that the difference in the rise of the
$pn\to pn\eta$ and $pp\to pp\eta$ total cross sections in the first 10~MeV of
excitation energy is due to the different $np$ and $pp$ final state
interactions, as described by Eqs.~(\ref{Master}) and (\ref{FW1}).

\subsection{$\boldsymbol{\eta^{\prime}}$ production in $\boldsymbol{pp}$ collisions}
\label{pnetaprime}

Only three groups have studied $\eta^{\prime}$ production near threshold
through the missing mass in the $pp\to ppX$ reaction and the existing total
cross section data~\cite{Bergdolt_93,Hibou_98,Balestra_00,Moskal_98,Moskal_00,Czerwinski_14} are compared to
those for $\eta$ production in Fig.~\ref{fig:eta_etaprime}. The first thing
to notice is that $\eta^{\prime}$ production is typically a factor $\sim 30$
weaker than that of the $\eta$ so that, for the highest energy
point~\cite{Balestra_00}, the $pp$ missing-mass criterion had to be
supplemented through an $\eta\pi\pi$ selection in order to reduce the
background. The even more striking point is that the curve of Eq.~(\ref{FW1})
describes very well all the available data. This indicates that the
low-energy interaction of the $\eta^{\prime}$ meson with nucleons is far
weaker than that of the $\eta$. This is not unexpected because there is no
strong analogue of the $N^*(1535)$ sitting close to the $\eta^{\prime}N$
threshold that would boost the interaction.

An interesting by-product of the COSY-11 $pp\to pp\eta^{\prime}$ studies is a
new and direct measurement of the natural width of the
$\eta^{\prime}$~\cite{Czerwinski_10}, $\Gamma = (0.226 \pm 0.017 \pm
0.014)$~MeV/$c^2$. This is by far the most precise value in the PDG
tabulation~\cite{Beringer_12} and can be used to normalize the decay widths
in the different channels. In order to control the systematic uncertainties,
the experiment was performed at five energies close to threshold where the
signal-to-background ratio is especially favourable; the phase space for a
four-body final state (multipion production) decreases faster than that of
the three-body $pp\eta$ as threshold is approached. For simple kinematic
reasons, the missing-mass resolution also gets better at low $Q$ such that in
the COSY-11 measurement this was $\approx 0.33$~MeV/$c^2$ at
$Q=0.8$~MeV~\cite{Czerwinski_10}, which is comparable to the line width to be
studied.

%
%
\subsection{The $\boldsymbol{pp \to pp\eta}$ reaction as a source of $\boldsymbol{\eta}$ mesons}
\label{source1}

There is typically significantly more background under a missing-mass $\eta$
peak in the $pp\to pp\eta$ case than for $pd \to {}^{3}\textrm{He}\,\eta$.
Nevertheless, because of its larger cross section, $\eta$ production in $pp$
reactions has been used by the COSY-WASA collaboration to study some of the
properties of $\eta$ decays. For example, using $1.2\times 10^5$ identified
$\eta\to 3\pi^0$ events, the deviations of the Dalitz plot from isotropy (the
$\alpha$ parameter) could be quantified~\cite{Adolph_09}. Though the result
is compatible with the current PDG compilation~\cite{Beringer_12}, some of
the experiments reported there, where the $\eta$ is produced in $\gamma p$ or
$e^+e^-$ collisions, have higher statistics and less hadronic background.

The COSY-WASA tests for the production of the $\eta^{\prime}$ meson through the
$pp\to pp\eta^{\prime}$ reaction were less promising since, in the initial runs, no
clear $\eta^{\prime}$ signal was identified~\cite{Zlomanczuk_09}. Multipion
production will certainly provide a very significant
background~\cite{Zielinski_11}.

%
%
\section{Production in proton-nucleus collisions}
\label{pA}\setcounter{equation}{0}

%
%
\subsection{Inclusive $\boldsymbol{\eta}$ production}
\label{Inclusive}

The measurement of inclusive production of the $\eta$ meson at 900~MeV and
1~GeV, i.e., well below the threshold in nucleon-nucleon collisions, was
undertaken with the PINOT spectrometer at
Saclay~\cite{Chiavassa_92,Chiavassa_93,Chiavassa_94}. This two-arm photon
detector, which had a limited energy resolution and very small geometric
acceptance, was used to study production on $^6$Li, B, C, Al, Cu, and Au
targets. Even taking into account the much stronger $\eta$ production in $pn$
compared to $pp$ collisions and a generous interpretation of the kinematics,
the folding model, where the $NN$ cross sections are convoluted with the
Fermi motion, underpredicts the measured cross section~\cite{Chiavassa_94} by
a factor of two once $\eta$ absorption is taken into
account~\cite{Vercellin_93}. There is therefore significant room for two-step
production, involving intermediate virtual pions, that seems to be crucial
for the coherent $pd \to {}^{3}\textrm{He}\,\eta$ reaction discussed in the
next section.

%
%
\subsection{The unpolarized $\boldsymbol{pd \to {}^{3\!}\textrm{He}\,\eta}$ reaction}

There have been numerous measurements of the $pd \to {}^{3}\textrm{He}\,\eta$
reaction at low energies over the last 30 years, where the $^3$He nucleus was
detected and the $\eta$ meson identified through the missing mass in the
reaction~\cite{Mersmann_07,Smyrski_07,Banaigs_73,Berthet_85,Berger_88,Kirchner_93,Mayer_96,Betigeri_00,Bilger_02,Adam_07,Rausmann_09,Adlarson_14}.
The statistical precision is generally reasonable but the systematic
uncertainties, mainly associated with the overall normalization, range from
about 7\% in the Saclay measurement~\cite{Mayer_96} up to twice that in most
of the other data, except for the large errors at COSY-11~\cite{Adam_07}.

\begin{figure}[htb]
\begin{center}
\includegraphics[width=0.7\textwidth]{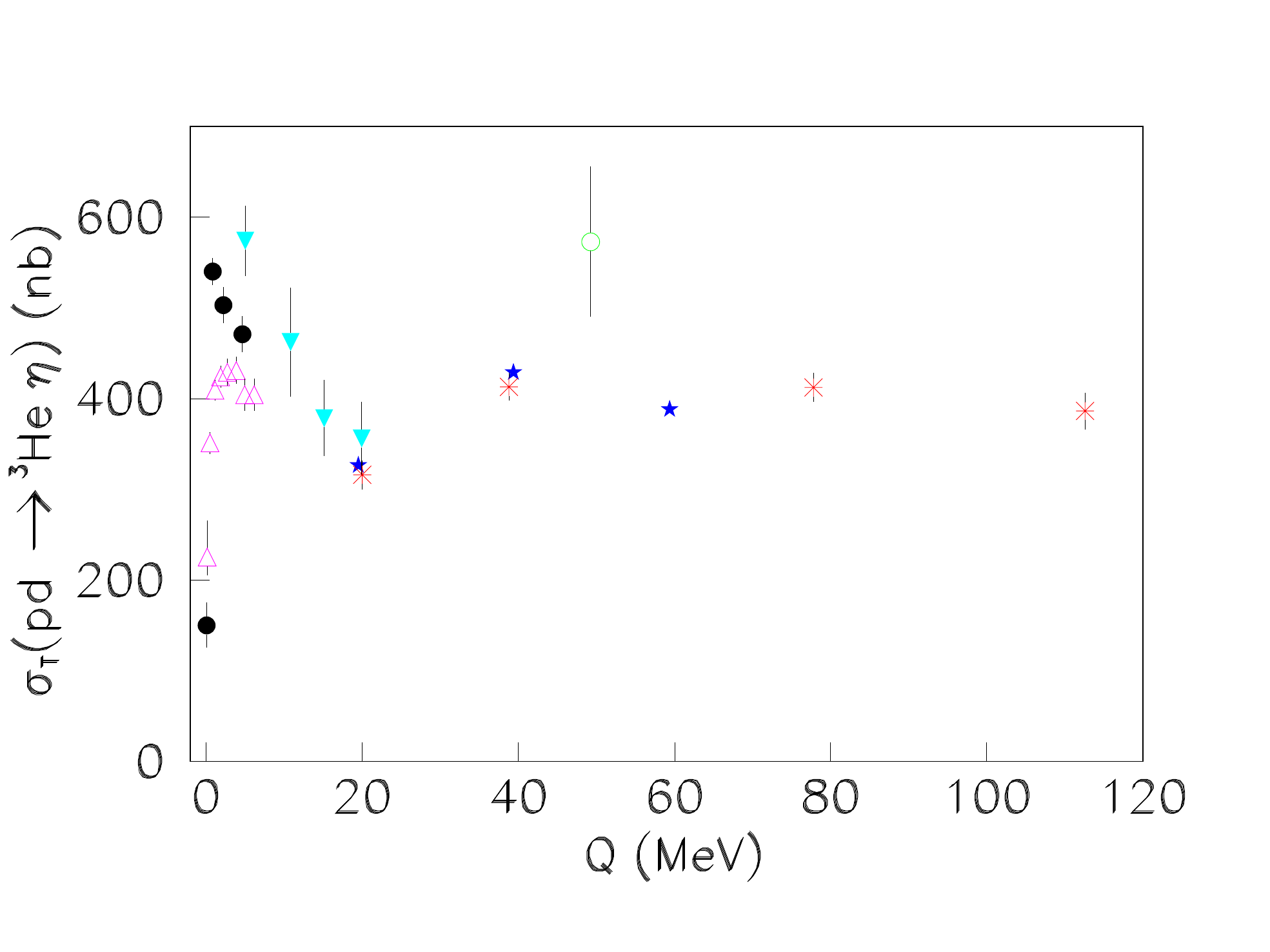}
\caption{\label{fig:sigtot3Heeta} Total cross section for the unpolarized
$pd(dp) \to {}^{3}\textrm{He}\,\eta$ reaction as a function of the excess
energy. The data are taken from Refs.~\cite{Berger_88} (black filled
circles), \cite{Mayer_96} (open magenta triangles), \cite{Betigeri_00} (open
green circle), \cite{Bilger_02} (red crosses), and \cite{Rausmann_09} (blue
stars). Systematic errors are not shown. The near-threshold data are shown in
Fig.~\ref{fig:Mersmann}.}
\end{center}
\end{figure}

\begin{figure}[thb]
\begin{center}
\includegraphics[width=0.7\textwidth]{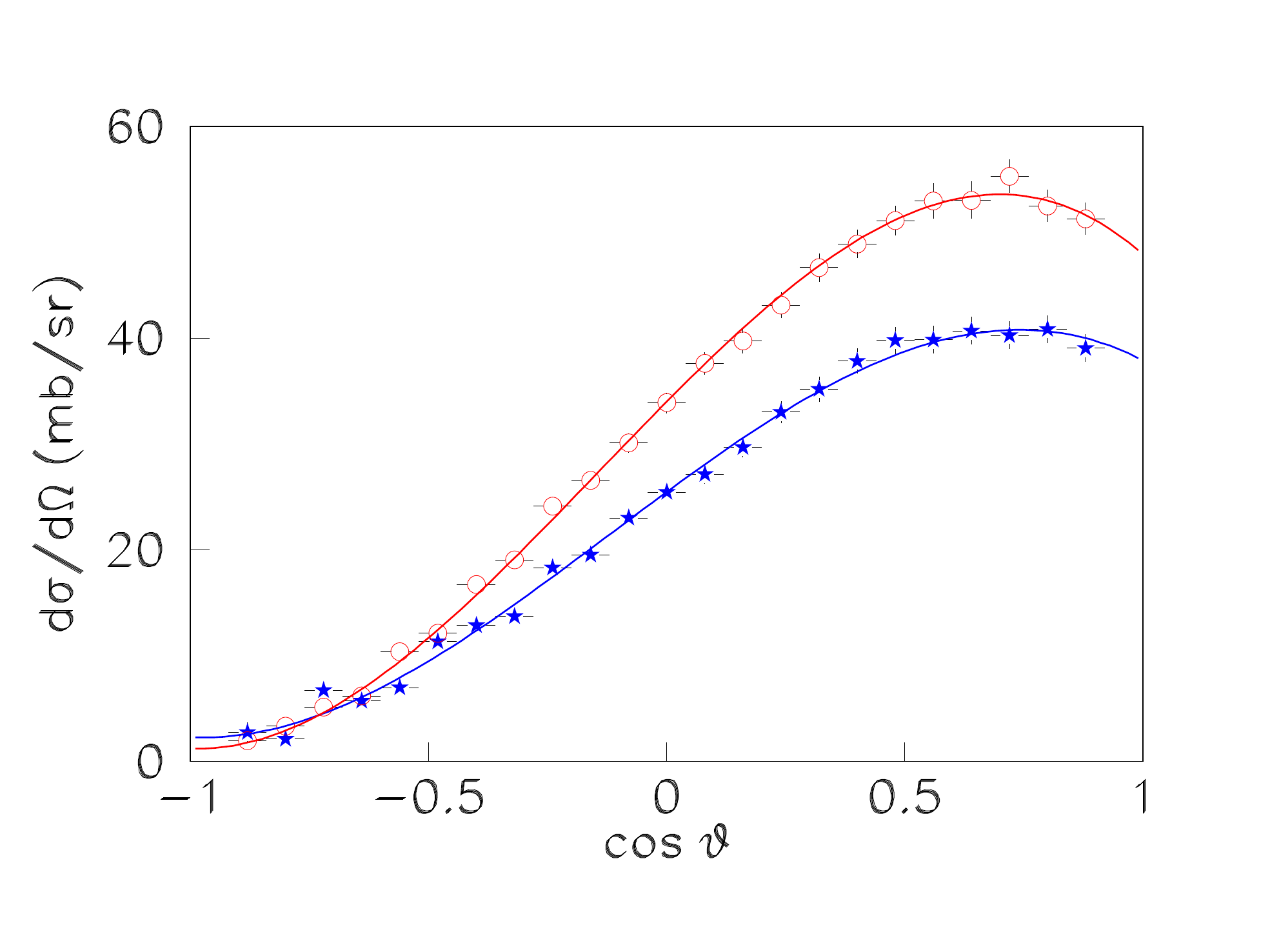}
\caption{\label{fig:Adlarson} Differential cross section for the unpolarized
$pd\to {}^{3}\textrm{He}\,\eta$ reaction  obtained at
WASA-at-COSY~\cite{Adlarson_14} at excess energies of 49 (blue filled
stars) and 60~MeV (red open circles), respectively. The curves are cubic fits
in $\cos\theta$, where $\theta$ is the c.m.\ angle between the incident
proton and the outgoing $\eta$.}
\end{center}
\end{figure}

As seen in Fig.~\ref{fig:sigtot3Heeta}, following a very rapid rise from
threshold to be discussed in the next section, the total cross section
reaches a plateau at relatively low values of the excess energy $Q$. The
threshold cross section is very similar to that of $pd
\to{}^{3}\textrm{He}\,\pi^0$, despite the much larger momentum
transfers~\cite{Nikulin_96}. Although very little structure is seen in the
total cross section for $Q\gtrsim 10$~MeV, the angular distribution is far
from flat, as illustrated by the recent data from COSY-WASA at 49 and
60~MeV~\cite{Adlarson_14} shown in Fig.~\ref{fig:Adlarson}. The data in the
backward direction (the two-nucleon transfer region) are strongly suppressed
and the cross section here falls very rapidly with incident beam
energy~\cite{Berthet_85}.

It is useful to introduce an angular asymmetry parameter $\alpha$, defined as
\begin{equation}
\label{alpha}
\alpha=\left.\frac{d\phantom{x}}{d(\cos\theta_{\eta})}
\ln\left(\frac{d\sigma}{d\Omega}\right)\right|_{\cos\theta_{\eta}=0}
\end{equation}
to describe the behaviour in the middle of the angular range. Away from the
threshold region, $\alpha$ is large and positive.

The data have some similarities to the analogous $pd \to
{}^{3}\textrm{He}\,\pi^0$ reaction, where the ratio of the forward to the
backward cross section increases fast with $Q$~\cite{Kerboul_86}. In this
case much of the effect can be understood in a single-scattering model based
upon the $pn \to d\pi^0$ with a spectator proton~\cite{Germond_90}. However,
for $\eta$ production the momentum transfers are so large ($\approx
880$~MeV/$c$ at threshold) that very high momentum components would be
required in the nuclei and the impulse approximation with reasonable values
for the $pn \to d\eta$ cross section~\cite{Calen_98b} underestimates the
production rate by almost two orders of magnitude~\cite{Germond_89}.

An alternative approach is needed and Kilian and Nann~\cite{Kilian_90}
noticed that the threshold kinematics for the $pd \to
{}^{3}\textrm{He}\,\eta$ reaction fitted well those of the sequential $pp \to
d\pi^+$ followed by $\pi^+n\to \eta p$. The relative momentum between the
proton and deuteron produced in this two-step process is very low such that
these two particles had a good chance of \emph{sticking} to produce the
observed $^3$He. Although they only made estimates within a semi-classical
Monte Carlo approach, using empirical values of the $pp \to d\pi^+$ and
$\pi^+n\to \eta p$ cross sections, they ascribed the large near-threshold
cross section to the \textit{magic kinematics}, where the intermediate pion
is essentially on-shell, which gives rise to a long-range interaction.

The near-threshold semi-classical estimates were confirmed in a
quantum-mechanical implementation of the two-step model~\cite{Faldt_95},
though there were also significant contributions where the intermediate pion
strayed from its mass shell. Although the near-threshold data could indeed be
described in terms of $pp \to d\pi^+$ followed by $\pi^+n\to \eta p$, when
this approach was tried at higher energies~\cite{Khemchandani_03} it
predicted differential cross sections that were backward peaked, in complete
contrast to the experimental data~\cite{Betigeri_00,Bilger_02,Adlarson_14}.
The kinematics become less \textit{magic} away from threshold and it is
possible that the defect in the model is associated with the neglect of the
off-shell behaviour of the $pp \to d\pi^+$ amplitude.

%
%
\subsubsection{Near-threshold data}
\label{ssec:mersmann}

\begin{figure}[htb]
\begin{center}
\includegraphics[width=0.8\textwidth]{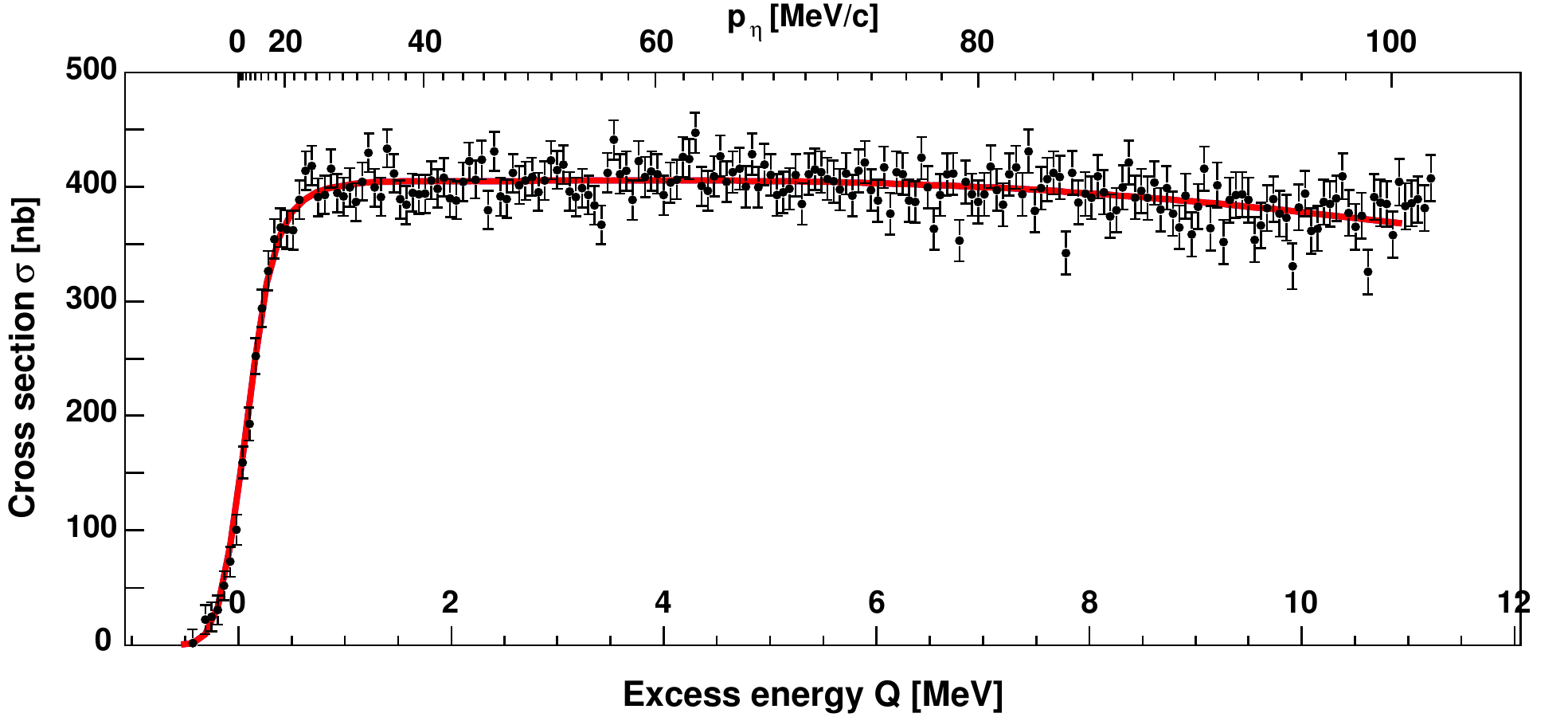}
\caption{\label{fig:Mersmann} Total cross section for the unpolarized $dp \to
{}^{3}\textrm{He}\,\eta$ total cross section measured at
COSY--ANKE~\cite{Mersmann_07} in terms of the excess energy $Q$ and $\eta$
c.m.\ momentum $p_{\eta}$. The (red) curve represents the smeared fit of
Eq.~(\ref{cross1}) with parameters given in Eq.~(\ref{mom}).}
\end{center}
\end{figure}

Of especial experimental and theoretical interest is the behaviour of the $pd
\to {}^{3\!}\textrm{He}\,\eta$  cross section in the very low $Q$ region.
Although this had been investigated at a few energies in earlier
experiments~\cite{Berger_88,Mayer_96}, more detailed measurements were
carried out by two groups at COSY~\cite{Mersmann_07,Smyrski_07} using a
deuteron beam, where the spectrometer acceptance is enhanced. The two data
sets are broadly consistent and only the ANKE results for the total cross
section are shown in Fig.~\ref{fig:Mersmann}.

What is absolutely striking here is the fact that the rise to the plateau is
achieved within the first 1~MeV. This scale is such that it must reflect the
behaviour of the $\eta{}^3$He system, i.e., be due to a final-state
interaction (FSI)~\cite{Wilkin_93}. To investigate this in greater detail,
let us evaluate the average amplitude-squared through
\begin{equation}
\label{cross1}
\overline{|f|^2} = p_d\,\sigma_T(dp \to {}^{3}\textrm{He}\,\eta)/{4\pi p_{\eta}},
\end{equation}
where $p_d$ and $p_{\eta}$ are the deuteron and $\eta$ c.m.\ momenta.

The rapid rise in the cross section shown in Fig.~\ref{fig:Mersmann}
indicates that there must be a pole in the production amplitude near $Q=0$
and so we parameterize this in the form
\begin{equation}
f_s = \frac{f_B}{(1-p_\eta/p_1)(1-p_\eta/p_2)}\,\cdot
\label{eq:fsi2}
\end{equation}
Here $p_1$ and $p_2$ could be expressed in terms of the scattering length and
effective range but the $p_i$ parameters are far less coupled in the fits.

After taking into account the momentum distribution in the circulating
deuteron beam and the smearing connected with the measuring apparatus, the
best fit was achieved with~\cite{Mersmann_07}
\begin{eqnarray}
\nonumber
p_1 &=& [(-5\pm7\,^{+2}_{-1})\pm i(19\pm2\pm 1 )]\,\textrm{MeV}/c\,\\
\label{mom}%
p_2 &=& [(106\pm5)\pm i(75\pm12\,^{+1}_{-2})]\,\textrm{MeV}/c,
\end{eqnarray}
where the first error is statistical and the second systematic. Note that
these data cannot determine the sign of the imaginary parts of the $p_i$, a
point that is critical in the discussion of $\eta$-mesic nuclei in
Sec.~\ref{nuclei}. The second pole is an effective one whose position is
unstable to changes in the data selection but $p_1$ represents a genuine
singularity which, in the excess energy plane, is situated at $Q_0 =
[(-0.36\pm0.11\pm0.04)\pm i(0.19\pm0.28\pm0.06)]$~MeV. Therefore, with zero
momentum bite and perfect apparatus, the rise in Fig.~\ref{fig:Mersmann}
would be even more precipitous with $|Q_0|\sim 0.4$~MeV.

\begin{figure}[htb]
\begin{center}
\includegraphics[width=0.57\textwidth]{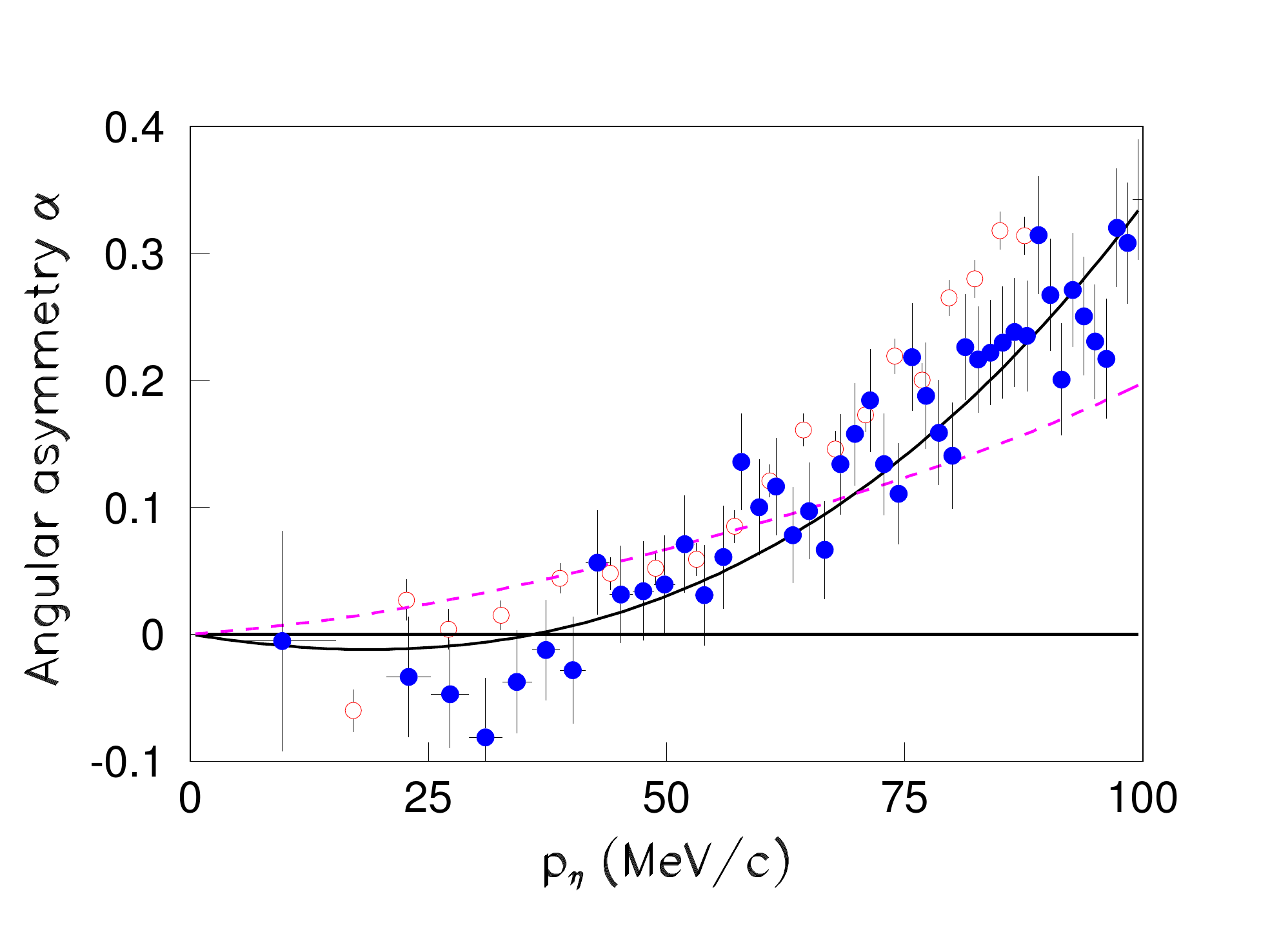}
\vspace*{-0.5cm}
\caption{\label{fig:ang_asym} Angular asymmetry parameter $\alpha$ of the $dp
\to {}^{3}\textrm{He}\,\eta$ reaction defined in Eq.~(\ref{alpha}) measured
at COSY-11~\cite{Smyrski_07} (red open symbols) and
COSY-ANKE~\cite{Mersmann_07} (blue closed symbols). Also shown are the fits
to the ANKE data with (solid black line) and without (dashed magenta line)
taking the $s$-wave phase variation into account~\cite{Wilkin_07}.}
\end{center}
\end{figure}

Further evidence for the pole hypothesis is to be found in the variation of
the angular distribution with $p_{\eta}$. There were already indications in
the Saclay data~\cite{Mayer_96} that very close to threshold the data were a
little stronger in the backward hemisphere, i.e., the parameter $\alpha$ of
Eq.~(\ref{alpha}) was slightly negative. This behaviour was confirmed in the
two COSY experiments~\cite{Mersmann_07,Smyrski_07} and their results are
shown in Fig.~\ref{fig:ang_asym}.

At low $Q$ the parameter $\alpha$ reflects the interference between $s$ and
$p$ waves and, if both the magnitude and the phase of the $s$-wave amplitude
vary fast over the first few MeV due to the pole, then this will affect the
momentum dependence of $\alpha$. The best fit in Fig.~\ref{fig:ang_asym}
assuming a constant $s$-wave phase does not allow $\alpha$ to go negative
but, if the phase variation with $p_{\eta}$ determined by Eq.~(\ref{eq:fsi2})
is taken into account then one can reproduce much better the shape of the
$\alpha$ measurements~\cite{Wilkin_07}. Therefore it seems that the pole
hypothesis in the $s$-wave production amplitude is on quite firm foundations.

In contrast to the extensive data set available for $\eta$ production, the
only published value of the $pd \to {}^{3}\textrm{He}\,\eta^{\prime}$ cross section
was obtained within the context of the threshold study of heavy meson
production~\cite{Wurzinger_96}. It is important to note here the relative
weakness of $\eta^{\prime}$ production, with the ratio of the squares of the
threshold amplitudes being a mere $|f(pd\to {}^{3}\textrm{He}\,\eta^{\prime})|^2/
|f(pd \to {}^{3}\textrm{He}\,\eta)|^2 \approx 5\times
10^{-4}$~\cite{Wurzinger_96}. In a two-step model, $\eta^{\prime}$ production
certainly seems to be anomalously weak in this channel compared to $\phi$
production~\cite{Faldt_95a}. As a consequence the signal/background ratio is
likely to be very unfavourable and a value of $\approx 1/40$ was found
experimentally at COSY-WASA at $Q\approx 64$~MeV~\cite{Zlomanczuk_09}, and
this makes it doubtful as a source of $\eta^{\prime}$ for decay studies. The
preliminary total cross section estimate of $\approx 0.6$~nb/sr corresponds
to $|f(pd\to {}^{3}\textrm{He}\,\eta^{\prime})|^2\approx 0.2$~nb/sr, which is lower
than the threshold value of $0.9\pm0.2$~nb/sr~\cite{Wurzinger_96}.

%
%
\subsection{The polarized $\boldsymbol{\pol{d}p \to {}^{3\!}\textrm{He}\,\eta}$ reaction}

The anomalous energy dependence near threshold of the $dp \to
{}^{3}\textrm{He}\,\eta$ total cross section, which jumps to its plateau
value within the first 1~MeV of excess energy
$Q$~\cite{Mersmann_07,Smyrski_07}, is evidence for a strong $s$-wave
$\eta\,^3$He final state interaction leading to a pole in the production
amplitude for $|Q|< 1$~MeV~\cite{Wilkin_93}. If this is indeed true, then the
FSI should manifest itself in broadly similar ways for different entrance
$\gamma{}^3\textrm{He} \to \eta{}^3\textrm{He}$ channels that give rise to
the same $\eta\,^3$He final state. The recent photoproduction data show a
steep increase in the first 4~MeV bin above threshold (see
Fig.~\ref{fig:eta_coh})~\cite{Pheron_12}, though the experiment could not
determine well the pole position.

In the $dp \to {}^{3}\textrm{He}\,\eta$ reaction the $s$-wave $\eta\,^3$He
final state can be accessed from either the total spin $S=\frac{3}{2}$ or the
$S=\frac{1}{2}$ initial states and the differences will influence the
deuteron tensor analyzing power $t_{20}$ in the reaction. The pure $s$-wave
FSI hypothesis would require that $t_{20}$ should remain constant, despite
the strange behaviour of the unpolarized cross section.

The tensor analyzing power of the $\pol{d}p \to {}^{3}\textrm{He}\,\eta$
total cross section has recently been measured in a missing-mass experiment
at COSY from threshold up to $Q\approx 11$~MeV~\cite{Papenbrock_14}. The
results are indeed consistent with a constant value of $t_{20}$, which offers
strong support to the FSI interpretation of the near-threshold energy
dependence. It is important to note here that the detection system is
independent of the deuteron beam polarization so that many of the systematic
effects cancel. The experiment also seems to show that the forward/backward
difference in $t_{20}$ is much smaller than for the unpolarized cross
section~\cite{Mersmann_07} and this, together with the small value found for
the vector analyzing power $it_{11}$, are both useful elements in
constraining the amplitude structure near threshold.

%
%
\subsection{The $\boldsymbol{pd \to pd\eta}$ reaction}
\label{pdeta}

\begin{figure}[htb]
\begin{center}
\includegraphics[width=0.5\textwidth]{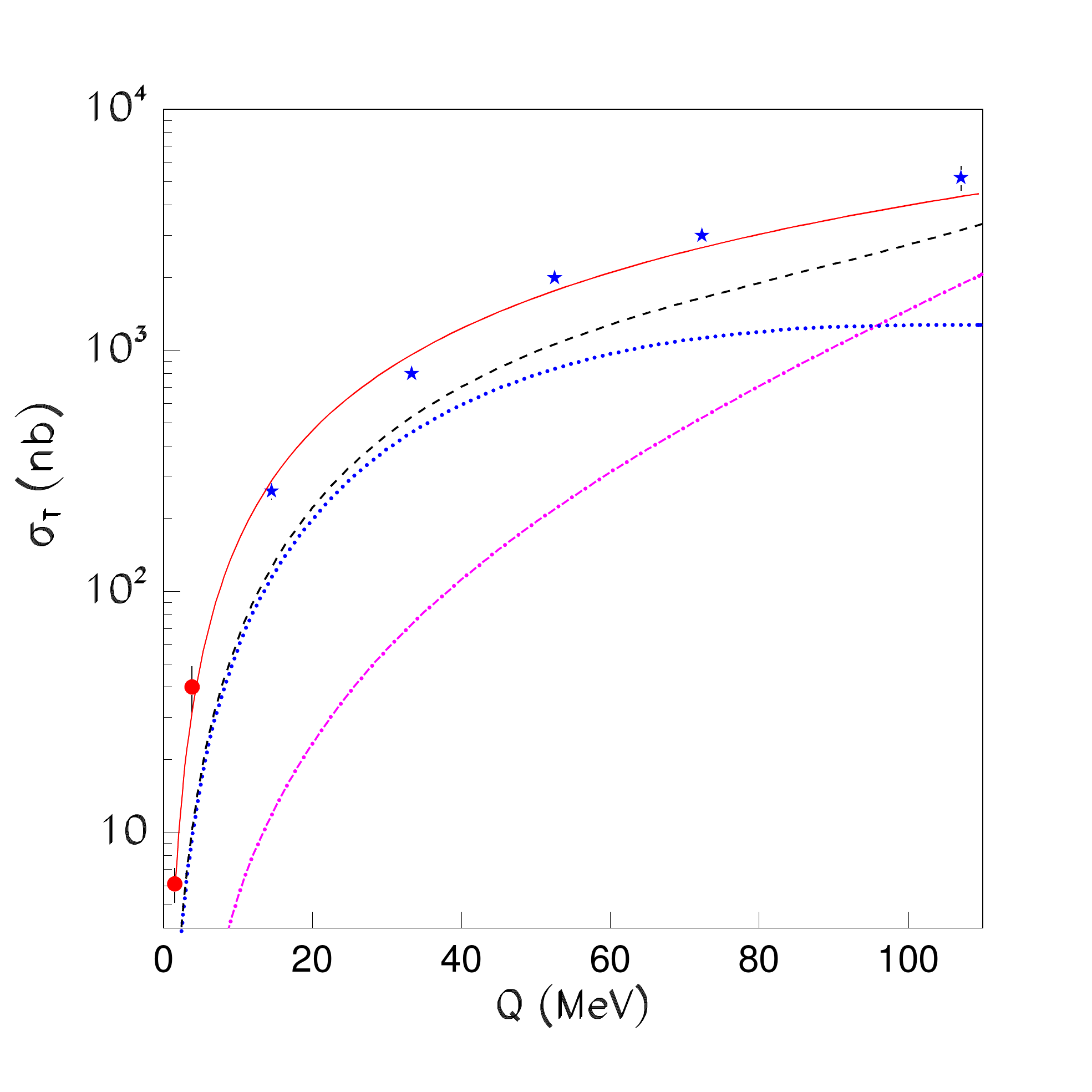}

\caption{\label{fig:Ulla-sigtot} Experimental values of the $pd \to pd\,\eta$
total cross sections from Refs.~\cite{Hibou_00} (red circles) and
\cite{Bilger_04} (blue stars). These are compared with
predictions~\cite{Tengblad_05} of the pick-up contribution (magenta chain)
and the two--step model (blue dotted), and their incoherent sum (black dashed
curve). The solid (red) curve is a phenomenological fit on the basis of
Eq.~(\ref{phenomenological}) with $C=350$~nb.}
\end{center}
\end{figure}

Since a two-step model describes much of the near-threshold $pd \to
{}^{3\!}\textrm{He}\,\eta$ data, it would be interesting to see if this
approach is equally successful for the unbound $^3$He states, i.e., for the
$pd \to pd\,\eta$ reaction. There have been two measurements of the reaction,
where the proton and deuteron were detected~\cite{Hibou_00,Bilger_04}, and
the resultant total cross sections are shown in Fig.~\ref{fig:Ulla-sigtot}.

Near threshold it is predicted~\cite{Tengblad_05} that the two-step model,
similar to that used to explain the threshold $pd \to
{}^{3\!}\textrm{He}\,\eta$ reaction, should dominate but, as $Q$ approaches
the threshold in nucleon-nucleon collisions, then the quasi-free $pn \to
d\eta$ (pick-up) contribution becomes larger. In all cases the impulse
approximation (triangle graph) is negligible. The incoherent sum of these
contributions is at least a factor of two too low but this defect is very
similar to that found for $pd \to {}^{3\!}\textrm{He}\,\eta$.

There is some evidence from the effective mass distributions measured at
$Q\approx 72$~MeV for a significant distortion in the $\eta d$ spectrum
corresponding to a large $\eta d$ scattering length~\cite{Bilger_04}. On the
other hand, the spectra do not seem to be affected by the existence of the
$^3$He bound state. Nevertheless, the data are well represented by the
phenomenological form~\cite{Faldt_96}
\begin{equation}
\label{phenomenological}
\sigma_T(pd \to pd\,\eta) = C \left.(Q/\varepsilon)^2\right/
\left(1+\sqrt{1+Q/\varepsilon}\,\right)^{\!2},
\end{equation}
where $\varepsilon = 5.5$~MeV is the $pd$ binding energy in $^3$He and the
best fit shown in the figure is achieved with $C \approx 350$~nb.

If one neglects spin-quartet production and considers only the $^3$He pole
then one expects that~\cite{Faldt_96}
\begin{equation}
\label{expect}
C =\left.\fmn{1}{4}\sqrt{\varepsilon/Q}\times\sigma_T(pd \to {}^{3\!}\textrm{He}\,\eta)\right|_{Q=0} \approx 450~\textrm{nb}.
\end{equation}

Although the extent of the agreement here may be fortuitous, it does
emphasize the close link between the $pd \to {}^{3\!}\textrm{He}$ and $pd \to
pd\,\eta$ reactions. Nevertheless, it is clearly a challenge to theory when,
as in this case, there are strong final-state interactions in both the $pd$
and $\eta\,d$ channels.

%
%
\subsection{The $\boldsymbol{\eta}$-meson mass}
\label{eta_mass}

There have been several modern measurements of the $\eta$ mass that relied on
detailed studies of the decay products of the
meson~\cite{Lai_02,Ambrosino_07,Miller_07} but there have also been two
recent ones that involved measurements in production
reactions~\cite{Nikolaev_14,Goslawski_13}, and these illustrate contrasting
approaches.

The determination of the mass of the $\eta$ meson, $m_{\eta}$, in any
two-body reaction requires the careful measurement of both the beam energy
and also the excess energy $Q$ in the reaction. Systematic effects are,
however, minimized if the data can be extrapolated to the threshold, i.e., to
$Q=0$. In the study of the $\gamma p \to \eta p$ reaction with the Crystal
Ball at MAMI~\cite{Nikolaev_14}, the photon beam energy was determined
macroscopically, starting with a precise measurement of the bending radius of
the electron beam in the first dipole magnet. A total uncertainty of
$\sigma(E_{\gamma}) = 98$~keV was found. The reaction itself was identified
cleanly through both the $\eta\to 2\gamma$ and $\eta\to 3\pi^0\to 6\gamma$
decay modes, though a very small background, originating from the target
windows, had to be subtracted. The threshold energy, $E_{\rm thr}$, of the
reaction was determined by extrapolating the arbitrarily normalized cross
section to zero, assuming pure $s$-wave dominance, viz.\ $\sigma(E_{\gamma})
\propto \sqrt{E_{\gamma} -E_{\rm thr}}$. Consistent results were obtained
from the two decay modes studied and the overall value quoted,
\[ m_{\eta}(\rm MAMI) = (547.865 \pm 0.031_{\rm stat} \pm 0.062_{\rm
syst})~\textrm{MeV}/c^2,\] agrees well with other modern
measurements~\cite{Lai_02,Ambrosino_07,Miller_07} and with the PDG
compilation~\cite{Beringer_12}. The dominant element in the 62~keV/$c^2$
systematic uncertainty originates from the determination of the beam energy.

In the $pd (dp) \to {}^3\textrm{He}\,\eta$ reaction near threshold, the meson
can be identified through the missing-mass peak in the reaction. However, in
the two cases where the beam energy was determined through the measurement of
other two-body reactions that fell within the spectrometer
acceptances~\cite{Plouin_92,Abdel-Bary_05}, the values obtained for the mass
were about 0.5~MeV/$c^2$ lower than the current PDG
average~\cite{Beringer_12}.

In a new experiment using the ANKE spectrometer at the COSY storage ring, the
internal cluster-jet target had no windows~\cite{Goslawski_13}. The momentum
of the circulating beam was determined with a relative precision of $3\times
10^{-5}$ by using a polarized deuteron beam and inducing an artificial
depolarizing resonance, which occurs at a well-defined frequency that depends
only upon the particle's speed.

The small background under the $\eta$ missing-mass peak was estimated
reliably using data obtained just below threshold. However, unlike the MAMI
$\gamma p \to \eta p$ measurement~\cite{Nikolaev_14}, the extrapolation to
threshold was not carried out using the energy dependence of the cross
section. Rather, at the 11 points measured above threshold, the excess energy
was evaluated from the size of the $^3$He momentum ellipse in the ANKE focal
plane, for which very detailed studies of the spectrometer characteristics
had to be undertaken. The result,%
\[ m_{\eta}(\rm ANKE) = (547.873 \pm 0.005_{\rm stat} \pm 0.027_{\rm
syst})~\textrm{MeV}/c^2,\]%
though agreeing with the PDG average~\cite{Beringer_12}, has the best
statistical precision of all the experiments and, arguably, the best
systematic precision also.

Although it was kinematics rather than cross section that were extrapolated
to threshold, a competitive measurement of the $\eta$ mass was only possible
due to the $\eta^3$He $s$-wave FSI such that, as shown in
Fig.~\ref{fig:Mersmann}, the cross section plateau is already reached for $Q<
1$~MeV. Thus it would be much harder to extract a reliable value for the
$\pi^0$ mass from $dp \to {}^3\textrm{He}\,\pi^0$ data because final-state
$p$-waves are very significant here, even at very low values of
$Q$~\cite{Nikulin_96}.

%
%
\subsection{The $\boldsymbol{p\,{}^{6\!}\textrm{Li}\to {}^{7\!}\textrm{Be}\,\eta}$ reaction}
\label{Li}

The inclusive cross sections for $\eta$ production reported in
Sec.~\ref{Inclusive} are generally very small and the only hope
for measuring an exclusive reaction is in the case of the
$^6$Li target. However, when the $p{}^{6}\textrm{Li}\to
{}^{7}\textrm{Be}\,\eta$ reaction was identified through the
$2\gamma$ decay of the $\eta$~\cite{Scomparin_93} the
resolution was not sufficient to decide the relative
contributions of the first four or more levels of the $^7$Be
nucleus to a c.m.\ differential cross section of
$d\sigma/d\Omega = 4.8\pm 3.8$~nb/sr at $\theta_{\eta}^{c.m.}
\approx 25^{\circ}$. In a later experiment, at a lower value of
$Q$~\cite{Budzanowski_10}, the recoiling $^7$Be nucleus was
detected and the $\eta$ identified through the missing-mass
peak in the reaction. In this case the measured
$d\sigma/d\Omega = 0.7\pm 0.3$~nb/sr must correspond to the
production of the ground-state doublet in $^7$Be because higher
states decay via the break-up of the nucleus.

The reaction has been studied in a cluster-model
approach~\cite{Al-Khalili_93}, where the initial and final
nuclear states were described in terms of their $\alpha d$ and
$\alpha \tau$ components, respectively. In this model the
process is assumed to be driven by $pd\to{}^3\textrm{He}\,\eta$
with the initial $d$ and final $\tau$ in the two nuclei. Apart
from the intrinsic crudeness of this approach, the calculations
are made much more uncertain by the limited knowledge of
cluster wave functions for such a large momentum transfer
reaction. Nevertheless, as shown explicitly in
Ref.~\cite{Budzanowski_10}, the model is compatible with both
data sets and this suggests that it is the excited $L=3$
doublet that dominates in the final state in the earlier
experiment~\cite{Scomparin_93}. The relative strengths of the
final $^7$Li states is similarly ambiguous in the
$\gamma{}^7\textrm{Li} \to \eta{}^7\textrm{Li}$
reaction~\cite{Maghrbi_13}.

%
%
\subsection{The $\boldsymbol{dd \to {}^{4\!}\textrm{He}\,\eta}$ reaction}

The third well-identified nuclear state where the $s$-wave $\eta$--nucleus
final-state interaction can be usefully investigated is that of $^4$He. Since
the final $s$-wave is forbidden in the $\gamma{}^4\textrm{He} \to
\eta{}^4\textrm{He}$ reaction, and neither the $p{}^3\textrm{H}\to
\eta^4\textrm{He}$ nor $n{}^3\textrm{He}\to \eta^4\textrm{He}$ is
experimentally \emph{appealing}, this final state has been studied in a
series of measurements of the $dd \to {}^{4\!}\textrm{He}\,\eta$
reaction~\cite{Frascaria_94,Willis_97,Wronska_05,Budzanowski_09}. It is first
important to note that the threshold cross section is about fifty times lower
than that of $pd \to {}^{3\!}\textrm{He}\,\eta$, so that its measurement
represents much more of a challenge.

Due to the identical nature of the two deuterons in the initial state, there
are strong angular momentum constraints on the amplitude structure of this
reaction, which can be written in the c.m.\ frame as:
\begin{eqnarray}
\nonumber M&=&A({\vec{\varepsilon}}_1\times{\vec{\varepsilon}}_2)
\cdot\hat{p}
+B({\vec{\varepsilon}}_1\times{\vec{\varepsilon}}_2)\cdot
\left[\hat{p}\times({\vec{p}_{\eta}}\times\hat{p})\right]
({\vec{p}_{\eta}}\cdot{\hat{p}})\\
&&\hspace{-2mm}+C\left[({\vec{\varepsilon}}_1\cdot{\hat{p}})
{\vec{\varepsilon}}_2\!\cdot\!({\vec{p}_{\eta}}\times\hat{p})
+({\vec{\varepsilon}}_2\cdot{\hat{p}})
{\vec{\varepsilon}}_1\!\cdot\!({\vec{p}_{\eta}}\times\hat{p})
\right], \label{A1}
\end{eqnarray}
where the ${\vec{\varepsilon}}_i$ are the polarization vectors of the two
deuterons, $\hat{p}$ the direction of one of the incident deuterons, and
$\vec{p}_{\eta}$ the momentum of the final $\eta$ meson. The three scalar
amplitudes $A$, $B$, and $C$ are functions of ${{p}_{\eta}}^2$, ${{p}}^2$,
and $({\vec{p}_{\eta}}\cdot{\vec{p}})^2=p_{\eta}^2p^2\cos^2\theta$, where
$\theta$ is production angle of the $\eta$ meson.

At threshold ($p_{\eta} =0$) the only non-zero contribution comes from the
$A$ amplitude and the reaction is then forbidden if the magnetic quantum
number of either deuteron is $m=0$. The Cartesian tensor analyzing power then
becomes $A_{xx}=-\half$. Although the initial SPESIV experiment was carried
out with unpolarized deuterons~\cite{Frascaria_94}, the resolution at the
SPESIII spectrometer was insufficient to identify the $\eta$ peak
cleanly~\cite{Willis_97}. By using a tensor polarized beam and assuming that
the background had only a weak analyzing power, they were able to use the
$m=0$ data to model the background and thus extract the $m=\pm 1$ cross
section. Since the $m=0$ cross section is negligible in the near-threshold
region, this allowed the unambiguous evaluation of the total cross section
and this is confirmed through the comparison of the
polarized~\cite{Willis_97} and unpolarized~\cite{Frascaria_94} cross sections
at low energies.

\begin{figure}[htb]
\begin{center}
\includegraphics[width=0.6\textwidth]{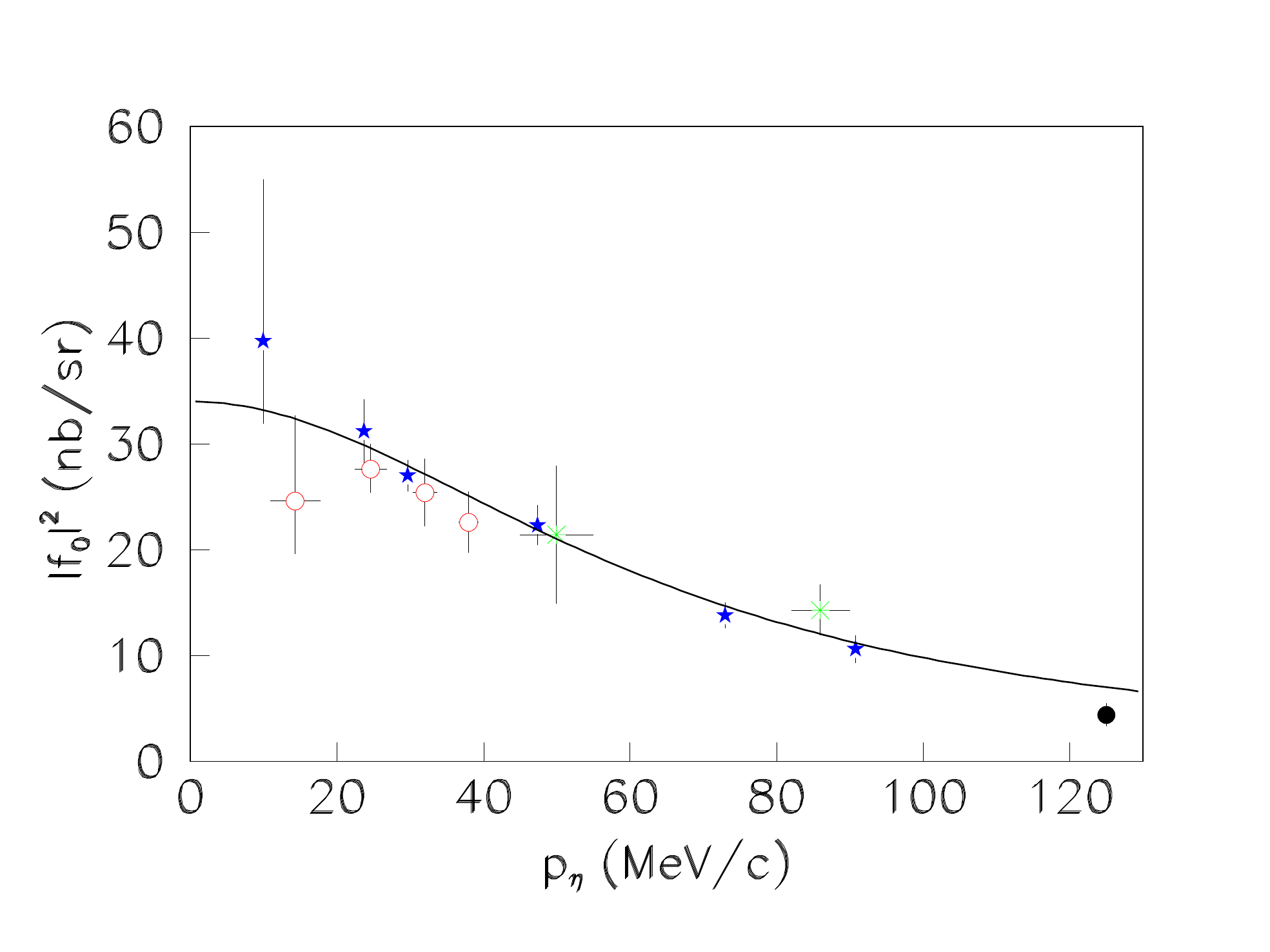}
\caption{\label{fig:dd2ae} Experimental values of the square of the $dd \to
{}^{4\!}\textrm{He}\,\eta$ $s$-wave amplitude extracted using the methodology
presented in Ref.~\cite{Budzanowski_09}. The data are taken from
Refs.~\cite{Frascaria_94} (red open circles), \cite{Willis_97} (blue stars),
\cite{Wronska_05} (green crosses), and \cite{Budzanowski_09} (black closed
circle). The curve is a scattering-length fit: $|f_0|^2 = 34/[1+
(p_{\eta}/64)^2]$~nb/sr, where $p_{\eta}$ is measured in MeV/$c$.}
\end{center}
\end{figure}

Away from threshold the angular distributions are no longer
flat~\cite{Wronska_05} but it is not possible from an unpolarized cross
section to determine whether this anisotropy arises from the square of the
$p$-wave amplitude, $C$, or from $s$--$d$ interference involving the $A$ and
$B$ amplitudes. The measurement of $A_{xx}$ away from threshold at
$Q=16.6$~MeV has a large scatter but these data have been interpreted as
suggesting that the $p$-wave amplitude $C$ is very
small~\cite{Budzanowski_09}. From the values of the fit parameters at
16.6~MeV and assuming that the $\eta$ $d$-wave amplitudes vary with threshold
factors of $p_{\eta}^2$, it is possible to attempt an extraction of the
square of the magnitude of the $s$-wave amplitude~\cite{Budzanowski_09}, and
the results of this are shown in Fig.~\ref{fig:dd2ae}.

In the scattering-length approximation, only one pole in Eq.~(\ref{eq:fsi2})
is considered and the $s$-wave production amplitude is taken in the form $f_0
= f_B/(1 - p_{\eta}/p_1)$. The best fits to the data shown in
Fig.~\ref{fig:dd2ae} are achieved with a positive value for the the real part
of $p_1$, which is forbidden by unitarity~\cite{Mersmann_07}. Putting this to
zero, the curve shown in the figure was obtained with \textit{Im}$(p_1) = \pm
(64\pm10)$~MeV/$c$ and $|f_B|^2= 34\pm 1$~nb/sr, which corresponds to a pole
in the amplitude at an excess energy $|Q_0|=
4.3\pm1.3$~MeV~\cite{Budzanowski_09}. The sign of \textit{Im}$(p_1)$ cannot
be fixed by these real $\eta$-production data and this leaves the location of
the pole in the complex $Q$ plane even more uncertain, a point that we shall
return to in the discussion of possible $\eta$-mesic nuclei. The curve does
not describe well the highest momentum point but keeping only the first pole
in Eq.~(\ref{eq:fsi2}), i.e., neglecting the effective range, is clearly a
dubious approximation at such high $p_{\eta}$.

%
%
\subsection{The $\boldsymbol{pd \to {}^{3\!}\textrm{He}\,\eta}$ reaction as a source
of $\boldsymbol{\eta}$ mesons}
\label{source2}

Since the total cross section for the $pd \to {}^{3\!}\textrm{He}\,\eta$
reaction shown in Fig.~\ref{fig:Mersmann} reaches a plateau so close to
threshold, it means that the multipion background under the $\eta$ peak at
low $Q$ is very limited~\cite{Mersmann_07}. It was already suggested at the
time of the first near-threshold measurements~\cite{Berger_88} that this
reaction could be a useful source of tagged $\eta$-mesons to study their
decay properties.

Various decays were indeed studied by the WASA collaboration at both CELSIUS
and COSY using this reaction~\cite{Bargholtz_07,Berkowski_08,Adlarson_12}.
Although the signal-to-background ratio is much more favourable than for the
$pp\to pp\eta$ reaction that was mentioned in Sec.~\ref{source1}, the $\eta$
production rate is not high enough to measure really low branching ratios.
Furthermore, when the meson is produced in $e^+e^-$ collisions or is
photoproduced there is little or no hadronic background and the reaction
tends to be much cleaner. The production rates are also high. As a
consequence, it seems that the parameters in most decay channels are
generally better determined at electron accelerators~\cite{Beringer_12}.

%
\subsection{The $\boldsymbol{pd \to {}^{3\!}\textrm{He}\,\eta\,\pi^0}$ reaction}
\label{Karin}

The only published measurement of the $pd \to {}^{3\!}\textrm{He}\,\eta\,\pi^0$
reaction~\cite{Schonning_10} was carried out at 1450~MeV using the WASA detector
at CELSIUS (Uppsala). At this energy the whole of the $D_{33}(1700)$ discussed
in Secs.~\ref{ssec:etapi} and \ref{ssec:photon_coh} could not be scanned.
Nevertheless there was some evidence for a cascade decay proceeding via a
$\Delta(1232)\eta$ intermediate state. However, when this experiment was
repeated using WASA at COSY at higher energy, there was little
clear evidence for such a decay chain~\cite{Buscher_11}.

%
%
\section{$\eta$-mesic nuclei}
\label{nuclei} \setcounter{equation}{0}

Due to the presence of the $N^*(1535)$ overlapping the $\eta p$ threshold,
the low energy $\eta p$ interaction is very strong and
attractive~\cite{Bhalerao_85}. Moreover, since the $\eta$ is an isoscalar
meson, this must also be true for the $\eta n$ system. This suggested to
Haider and Liu that the attraction might be sufficiently strong as to cause
the $\eta$ meson to bind to a nucleus~\cite{Haider_86}. The first estimates
were based upon the simple $t\rho$ optical potential and, although there is a
reasonable consensus as to what to take for the imaginary part of the $\eta
N$ scattering length $a_{\eta N}$, the same cannot be said for the real part,
where values ranging between 0.27~fm and 0.98~fm are to be found in the
literature~\cite{Haider_02}. Part of the difficulty here is that, because of
the $N^*(1535)$ resonance, the $\eta N$ scattering amplitude has a strong
energy dependence and there is uncertainty in how this should be taken into
account.

Haider and Liu~\cite{Haider_86} took a relatively low value for Re($a_{\eta
N}$) and, as a result, they only found binding for $^{12}$C and heavier
nuclei. Although there may be poles in the complex energy plane for lighter
nuclei, these would correspond to antibound (or virtual) states. Despite not
being bound, such poles might still affect nuclear reactions in a similar way
to how the antibound $^{1\!}S_0$ state may be more important than the bound
$^{3\!}S_1$ deuteron state in low energy neutron-proton scattering. It is
perhaps useful to stress here that, due to the possibility that a mesic
nucleus could decay via pion emission, described by the imaginary part of the
potential, even for strongly attractive potentials a state is at best
quasi-bound.

By increasing the size of the real part of the scattering length Re($a_{\eta
N}$) from 0.27~fm to 0.51~fm, the simplistic potential model predicts
quasi-bound $\eta$-mesic nuclei down as far as $_{\eta}^4$He~\cite{Haider_02}
but this at the expense of generating states with decay widths in the
$p$-shell nuclei of the order of 10-15~MeV. Since an $\eta$ would be expected
to stick to an excited nuclear level with about the same binding energy as
for the ground state~\cite{Haider_10}, the search for such states is clearly
made much more difficult when the decay width is larger than the nuclear
level spacing.

Several attempts have been made to refine the binding estimates but in the
$t\rho$ model one is faced with the choice of energy at which the amplitude
$t$ should be evaluated. The value is influenced by the nuclear as well as by
the $\eta$ binding energy and in a recent paper~\cite{Friedman_13} this was
estimated in a much more self-consistent way than in earlier approaches.
However, even this is not on safe grounds because we don't know how the
$N^*(1535)$ itself behaves inside a nucleus. Is it more bound or less bound
than a nucleon?

There have been two approaches to the experimental search for $\eta$-mesic
nuclei. The first involves the study of real $\eta$ production near
threshold, where one attempts to extrapolate in energy to the $\eta$-mesic
pole. The detection of an emerging $\eta$ does overcome the large multipion
background. This has been partially successful for light nuclei but it is
important to note that these above-threshold measurements can never
distinguish between a bound and an antibound state. The alternative approach
is to look directly in the bound state region, possibly trying to select
kinematic regions where the decay of the $\eta$-mesic nucleus might be
favoured. Although more direct, this approach faces more severe background
problems. We consider these two alternative approaches separately.

%
%
\subsection{Real $\boldsymbol{\eta}$ production}

The data and the analysis of the near-threshold  $dp \to
{}^{3\!}\textrm{He}\,\eta$ reaction~\cite{Mersmann_07,Smyrski_07} clearly
show that there is a pole in the production amplitude at an excitation energy
$Q_0 = [(-0.36\pm0.11\pm0.04)\pm i(0.19\pm0.28\pm0.06)]$~MeV which is most
likely a property of the $\eta^3$He system. The sign of the imaginary part of
$Q_0$, i.e., \emph{bound} or \emph{antibound}, can never be determined from
data on real $\eta$ production. Moreover, the errors on the real and
imaginary parts are strongly correlated in fits to the data. The presence of
such a pole is also reflected in the rapid rise in the total cross section of
the $\gamma{}^3\textrm{He} \to \eta\,{}^3\textrm{He}$
reaction~\cite{Pheron_12}. The angular distributions for producing the
$\eta^3$He system in $dp$ or $\gamma^3$He collisions also show anomalous
behaviour near threshold~\cite{Wilkin_07,Pheron_12} that are consistent with
the pole hypothesis.

The $s$-wave $\eta^3$He system can be accessed in the $dp \to
{}^{3\!}\textrm{He}\,\eta$ reaction from deuterons with magnetic quantum
number $|m|=1$ or $m=0$ and, in principle, these different entrance channels
could yield poles in different positions in the $Q$ plane. The low value of
the tensor analyzing power $t_{20}$ found in the initial Saclay
experiment~\cite{Berger_88} shows that the production from these two initial
states is roughly equal so that, if there were two poles, they would be
populated with about the same intensity. Fitting them with a single pole
model would then give a contribution to the width that was of the same order
as the difference in the two pole positions. However, the fits to the
data~\cite{Mersmann_07,Smyrski_07} lead to a very small value of Im($Q_0$),
which suggests that one could safely dismiss the two-pole hypothesis. A more
direct proof of this conclusion is provided by the measurement of $t_{20}$ as
a function of energy~\cite{Papenbrock_14}. One may therefore conclude that
the pole in the $dp \to {}^{3\!}\textrm{He}\,\eta$ reaction is indeed a
property of the $\eta^3$He system.

The $s$-wave $\eta^4$He system cannot be accessed in a two-body
photoproduction reaction but several experimental studies have been
undertaken in the $dd \to {}^{4\!}\textrm{He}\,\eta$ reaction. However, the
cross section is a factor of about 50 lower than that for $dp \to
{}^{3\!}\textrm{He}\,\eta$ reaction and so the results are necessarily less
precise than in the $\eta^3$He case. The exact location of the pole in the
complex $Q$ plane is far more uncertain because the data are not very
sensitive to the phase of $Q_0$ and its magnitude has also a comparatively
large error, $|Q|= 4.3\pm1.3$~MeV~\cite{Budzanowski_09}. It was argued by
Willis et al.~\cite{Willis_97} that an $\eta$ would be attracted more to
$^4$He than to $^3$He because of the extra nucleon and the smaller nuclear
radius. Hence, since $|Q_0|$ is bigger for $^4$He, it would imply that
$\eta^4$He is quasi-bound. This reasoning has been recently questioned on the
grounds that the attraction might be weakened in $^4$He due to the strong
nuclear binding~\cite{Friedman_13}.

\begin{figure}[htb]
\begin{center}
\includegraphics[width=0.65\textwidth]{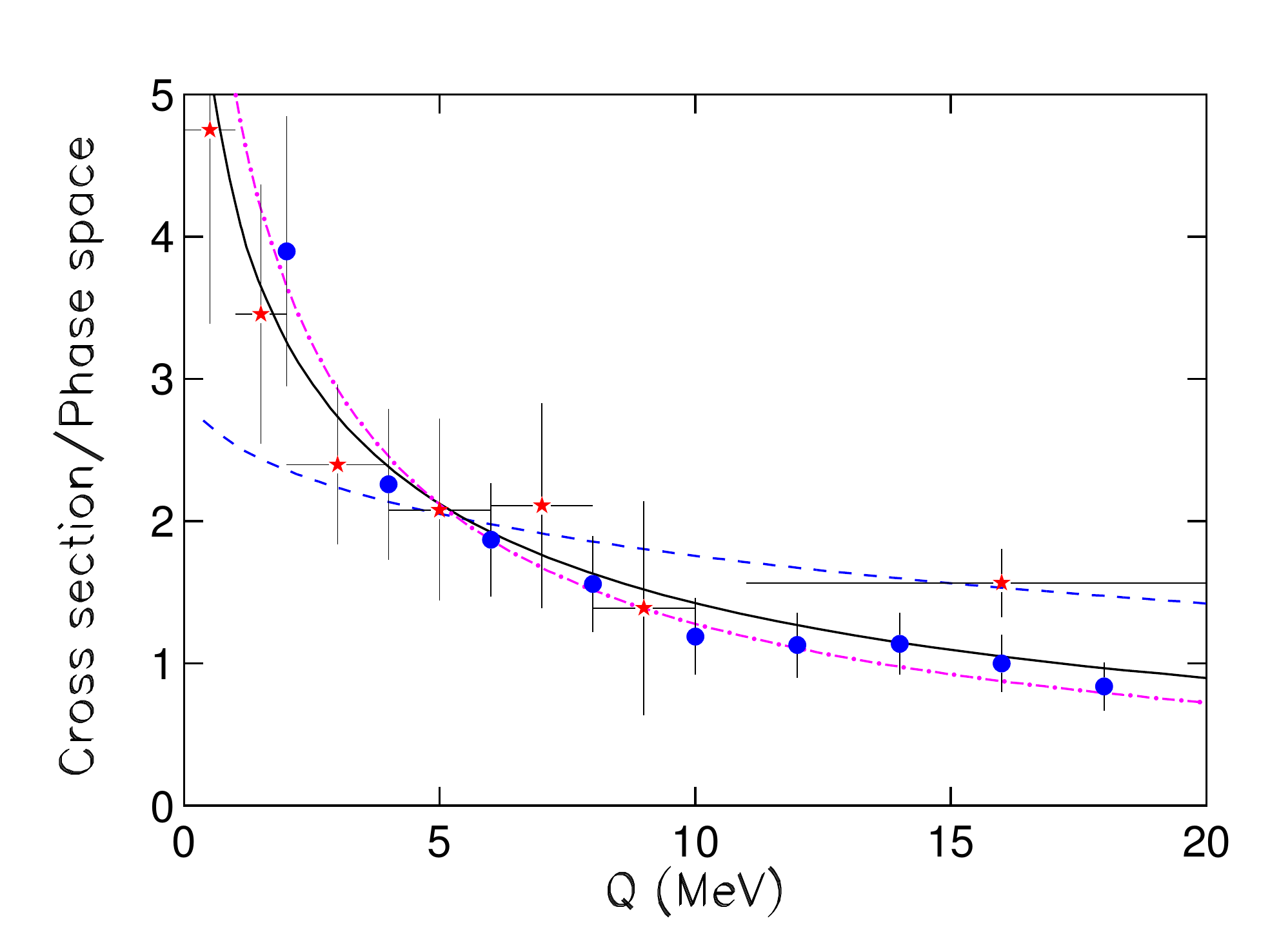}
\caption{\label{fig:etad_ratio} Ratio of the cross section for the production
of the $d\,\eta$ system to arbitrarily normalized phase space (red stars) as
a function of the energy $Q$ above the $\eta d$
threshold~\cite{Calen_97,Haggstrom_97,Calen_98b} and similarly for the $pd\to pd\,\eta$
reaction at 1032~MeV (closed blue circles)~\cite{Bilger_04}. The broken
(blue), solid (black) and chain (magenta) curves are the predictions of the
scattering length formula with input $a_{\eta d}=0.73+0.56i$, $1.64+2.99i$,
and $-4.69+1.59i$~fm, respectively~\cite{Shevchenko_98}.}
\end{center}
\end{figure}

The quasi-free $pn \to d\eta$ total cross section discussed in
Sec.~\ref{pneta} was measured in two experiments at
CELSIUS~\cite{Calen_97,Haggstrom_97,Calen_98b}. Although the cross section is larger than
that for the $pd \to {}^{3\!}\textrm{He}\,\eta$ reaction, the detection of
the two photons from the $\eta$ decay in coincidence with the fast deuteron
had a much reduced acceptance and energy resolution. The resolution was
better at lower energy when the deuteron and proton from the $dp \to dp\eta$
reaction were measured in coincidence with two photons~\cite{Bilger_04}. In
both sets of experiments there is evidence for an enhancement at low $\eta d$
invariant masses and, to make this more explicit, the data have been divided
by an arbitrarily normalized phase space in Fig.~\ref{fig:etad_ratio}. We do
not show here the near-threshold data of Ref.~\cite{Plouin_90}, where there
are uncertainties in the multipion background.

It is possible that the shape of the sub-threshold $pd\to
pd\,\eta$ ratios~\cite{Bilger_04} shown in
Fig.~\ref{fig:etad_ratio} are distorted by the reaction
mechanism such that it is not valid to divide by the three-body
phase space. This point will be resolved when the new detailed
measurements of the differential and total cross sections of
the quasi-free $np \to d\eta$ near threshold become
available~\cite{Khoukaz_12}.

Since there are only three particles in the $\eta n p$ system, the final
state interaction is then theoretically much more tractable and the
Alt-Grassberger-Sandhas equations have been resolved for various $\eta N$
inputs~\cite{Shevchenko_98}. Taking $a_{\eta N} = 0.25+0.16i$, $0.55 +
0.30i$, and $0.98+0.37i$~fm, this group found $\eta$-deuteron scattering
lengths of $a_{\eta d}=0.73+0.56i$, $1.64+2.99i$, and $-4.69+1.59i$~fm,
respectively. The enhancements expected with these scattering lengths are
also shown in Fig.~\ref{fig:etad_ratio}. The first $a_{\eta N}$ input, which
is quite close to that used by Haider and Liu~\cite{Haider_86}, does not give
a sharp enough enhancement. The other two can describe the data quite well
but, as for the $_{\eta}^3$He and $_{\eta}^4$He cases, the results cannot
distinguish between the quasi-anti-bound situation, where Im$[a_{\eta d}] >
0$, and the quasi-bound, where Im$[a_{\eta d}] < 0$.

The only other cases where the near-threshold production of the $\eta$ meson
from a nucleus was studied involves $A=7$ final states. The statistics in the
$p{}^{6\!}\textrm{Li}\to {}^{7\!}\textrm{Be}\,\eta$ measurements were
paltry~\cite{Scomparin_93b,Budzanowski_10} and, in the more extensive
$\gamma{}^7\textrm{Li} \to \eta{}^7\textrm{Li}$ data~\cite{Maghrbi_13}, the
separation of the nuclear levels is ambiguous and the lack of clean results
at very low values of $Q$ does not permit the extraction of the associated
enhancement factors.

The most straightforward (but not unique) interpretation of the data on light
nuclei is that the $\eta d$ system is unbound, the $\eta^4$He is bound, but
that the $\eta^3$He case is ambiguous. What is remarkable for $\eta^3$He is
the small value of the imaginary part of the pole position, $\textrm{Im}[Q_0]
= (0.19\pm0.28\pm0.06)$~MeV~\cite{Mersmann_07}. By adjusting the interaction
strength it is possible to get the real part to be close to zero but, in a
simple optical potential approach, one only achieves a small imaginary part
if the imaginary part of the potential is itself unreasonably
weak~\cite{Wilkin_14}.

%
%
\subsection{Virtual $\boldsymbol{\eta}$ production}
\label{ssec:virt_eta}

Since the production of real $\eta$ mesons can never distinguish between the
bound and anti-bound hypotheses, the temptation is to look directly below
threshold and search for other decay modes of an $\eta$-mesic nucleus. The
difficulty in this approach is that, without the $\eta$ trigger, the
multipion background might be hard to overcome. Inspired by the work of
Haider and Liu~\cite{Haider_86}, searches were undertaken at
Brookhaven~\cite{Chrien_88} and LAMPF~\cite{Lieb_88} using pion beams and
measuring the inclusive $(\pi^+,p)$ reaction on various nuclei with the hope
of finding a missing-mass peak corresponding to a bound $\eta$-mesic nucleus.
No conclusive signals were identified in either experiment despite, in the
LAMPF case~\cite{Lieb_88}, detecting in coincidence charged particles from
the decay of the supposed $\eta$-mesic nucleus.

A significant fraction of the decays of such an exotic nucleus might be
through the $\eta N \to \pi N$ reaction and, if one neglects the Fermi motion
in the nucleus, the pion and proton should come out back to back in the
nuclear rest frame. The searches at the Lebedev Physical Institute of the
$\gamma{}^{12}\textrm{C}\to \pi^+ n NX$ reaction  with correlated
back-to-back $\pi^+n$ pairs did not have a good enough resolution to identify
a peak coming from an $\eta$-mesic nucleus~\cite{Sokol_00}. A similar
experiment that searched for the two-nucleon decay of an $\eta$-mesic nucleus
has yet to yield conclusive results~\cite{Baskov_12} but it is possible that
this branching ratio might have been overestimated~\cite{Wilkin_14}.

The search for back-to-back $\pi^-p$ pairs was also pursued at JINR using a
deuteron beam~\cite{Afanasiev_11}. A peak was found in the $\pi^-p$ invariant
mass just below the $\eta N$ threshold but there was no indication that this
was associated with an $\eta$-mesic nucleus.

\begin{figure}[htb]
\begin{center}
\includegraphics[width=0.5\textwidth]{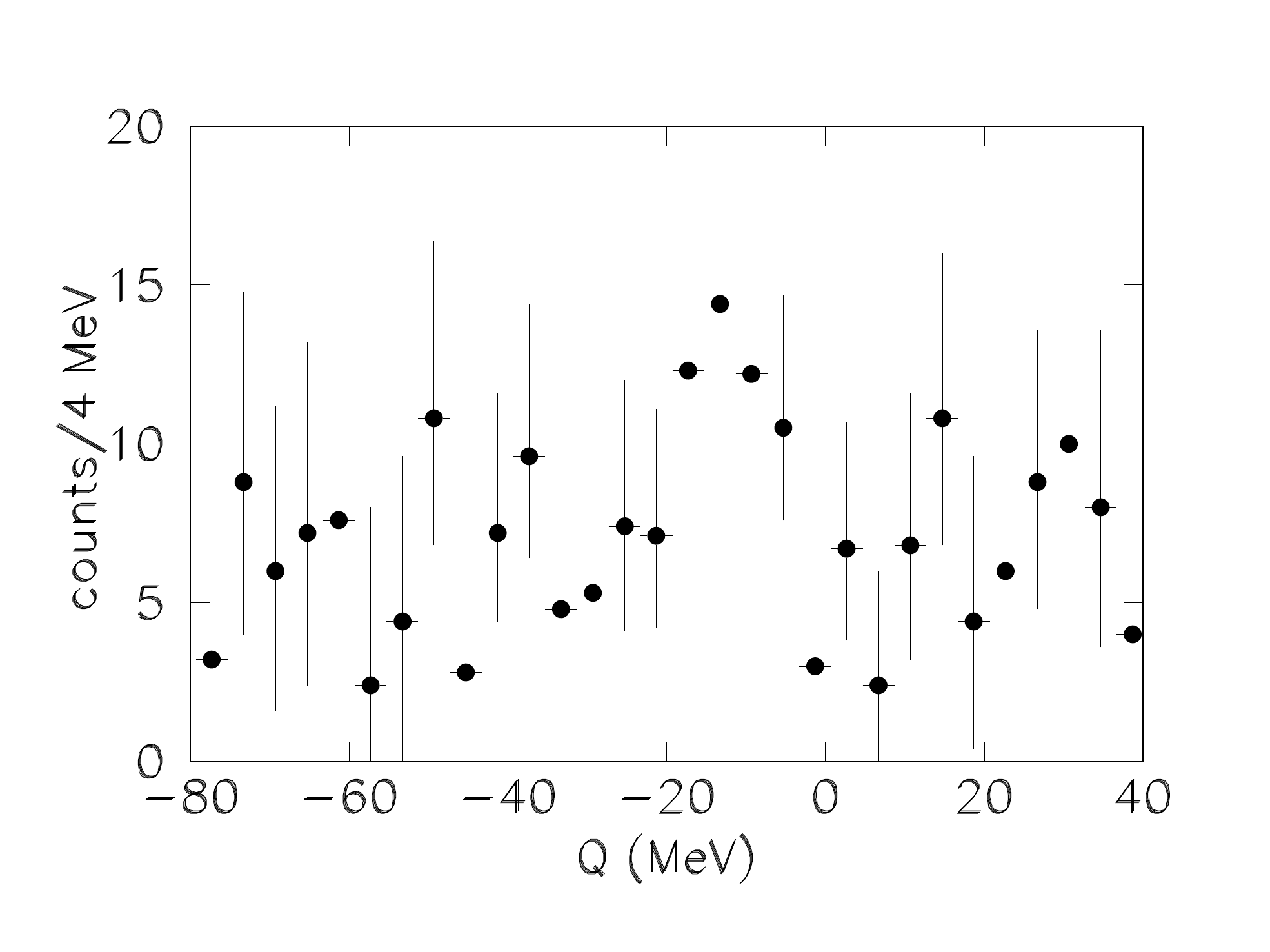}
\caption{\label{fig:Hartmut} Counts for the $p{}^{27\!}\textrm{Al} \to
{}^3\textrm{He}\, p\, \pi^- X$ reaction as a function of the excess energy in
the $\eta^{25}$Mg system~\cite{Budzanowski_09a}. }
\end{center}
\end{figure}

The strongest claim for the discovery of a peak arising from the production
of an $\eta$-mesic nucleus was made by the COSY-GEM collaboration after
measuring the $p{}^{27\!}\textrm{Al}\to{}^3\textrm{He}\, p\,\pi^-X$
reaction~\cite{Budzanowski_09a}. The kinematics were cunningly chosen such
that, for a weakly bound state, the $\eta$ was produced almost at rest so
that it had a higher chance of \textit{sticking} to the residual nucleus. As
in the Russian experiments, the emerging proton and pion were detected close
to the back-to-back region. The resulting spectrum is shown in
Fig.~\ref{fig:Hartmut} as a function of the excess energy in the
$_{\,\,\eta}^{25}$Mg system. The authors suggest that the excess of events
for $Q\approx -13$~MeV with a width FWHM of $\approx 10$~MeV might be a
signal for a $_{\,\,\eta}^{25}$Mg bound state. If this were indeed the case,
then the production cross section for this state is estimated to be $0.46\pm
0.16({\rm stat}) \pm 0.06 ({\rm syst})$~nb. If, on the other hand, it is a
statistical fluctuation, this result should be interpreted rather as an upper
limit.

The initial measurements of the $\gamma{}^3\textrm{He}\to \pi^0pX$
reaction~\cite{Pfeiffer_04} were very encouraging since, for back-to-back
$\pi^0p$ pairs, an enhancement was found in the total centre-of-mass energy
just below the $\eta^3$He threshold. Although the significance of this peak
was questioned at the time~\cite{Hanhart_05}, this did seem like \emph{prima
facie} evidence for the decay of $_{\eta}^3$He into the $\pi^0p(pn)$ channel.
As already mentioned in Sec.~\ref{ssec:photon_coh}, the later and more
detailed study of this reaction~\cite{Pheron_12} looked at events with
different ranges of $\pi^0 p$ opening angles and found that for each angular
range there were similar oscillations as a function of the c.m.\ energy but
that these were displaced with respect to each other. It therefore seems
accidental that the maximum in the back-to-back data is close to the
$\eta^3$He threshold. The newer measurements did not therefore support the
claim for the photoproduction of a quasi-bound system decaying via pion
emission.

Back-to-back $\pi p$ pairs were also measured in deuteron-deuteron
collisions, where evidence was sought for the $dd\to{} _{\eta}^{4}\textrm{He}
\to \pi^-p{}^3$He reaction~\cite{Adlarson_13}. No unambiguous signal was
found for this decay, for which the upper limit put on the total cross
section was $\approx 20$~nb, depending upon the width of the
state~\cite{Adlarson_13}. Simple estimates that start from data on the
production of real $\eta$ mesons suggest that the total cross section for $dd
\to \pi^-pX$ passing through the $_{\eta}^{4}\textrm{He}$ state might be of
the order of 30~nb~\cite{Wilkin_14}, but the $X$ here refers to all
three-nucleon states and not just the $^3$He measured at
COSY-WASA~\cite{Adlarson_13}. An upper limit of about 270~nb was found for
the $pd\to{}_{\eta}^3\textrm{He}\to \pi^- ppp$~\cite{Moskal_10a} compared to
a phenomenological estimate of about 80~nb~\cite{Wilkin_14}, though it is far
from certain that the $_{\eta}^3\textrm{He}$ pole lies in the quasi-bound
region~\cite{Mersmann_07}.

The calculations in Refs.~\cite{Wilkin_14} and \cite{Wycech_14} are crude but
a crucial point that both stress is that data on virtual $\eta$ production
should not be treated in isolation from the data on real $\eta$ production.

%
%
\subsection{Possible $\boldsymbol{\eta^{\prime}}$ nuclei}

The transparency experiments discussed in Sec~\ref{gamma-A} seem to show that
the imaginary part of the $\eta^{\prime}$-nucleus potential is quite small,
though these measurements were carried out for an $\eta^{\prime}$ momentum
$\approx$1.05~GeV/$c$~\cite{Nanova_12}. Model-dependent estimates of the
real part of the $\eta^{\prime}$-nucleus potential from measurements of
inclusive $\eta^{\prime}$ photoproduction from a $^{12}$C target suggest an
attraction of depth $\approx$37~MeV~\cite{Nanova_13}. These arguments were
behind the proposal to search for $\eta^{\prime}$ mesons bound in a
nucleus~\cite{Nanova_12p,Itashi_12}.

A counter argument would point to the COSY11 measurements of the $pp\to
pp\eta^{\prime}$ total cross section~\cite{Moskal_98,Moskal_00,Czerwinski_14}.
As is very evident
in Fig.~\ref{fig:eta_etaprime}, the near-threshold data show no sign of any
significant $\eta^{\prime}$-proton interaction and any model dependence here
is very limited. Furthermore, the production of $\eta^{\prime}$ in the $pd
\to {}^{3}\textrm{He}\,\eta^{\prime}$ reaction near threshold is about three
orders of magnitude weaker than for the $\eta$~\cite{Wurzinger_96}. As a
consequence, searches for $\eta^{\prime}$-mesic nuclei might be even less
fruitful than those for the $\eta$.

%
%
\section{Conclusions and outlook}
\setcounter{equation}{0}

We have reviewed the production of $\eta$ and $\eta^{\prime}$
mesons and $\eta\pi$ pairs with both photon and hadron beams.
Reactions on free protons and quasi-free protons and neutrons
bound in light nuclei were treated, as well as events where the
whole nucleus played a more intrinsic role. Though
electroproduction was not explicitly considered, it is
abundantly clear that, due to the limited availability of pion
beams, most of the important information in the $\eta N$ and
$\eta^{\prime}N$ sectors will come from electromagnetic probes.
This is already true with the existing data set but it will be
reinforced by the abundant data for single and double
polarization observables still under analysis or yet to be
taken. A careful analysis of these data will certainly cast
more light on the \emph{missing} $I=\frac{1}{2}$ $N^{\star}$
resonances. Furthermore, also the investigation of the decays
of this mesons in view of their transition form factors and
fundamental symmetries is shifting from hadron induced
reactions to electromagnetic production; but this would be the
topic of a different review.

The low energy $\eta N$ system, accessed with either photon or
pion beams, clearly shows evidence for the dominance of the
$S_{11}(1535)$ isobar. This sits close to the threshold and it
has roughly 50\% branching ratios into both $\eta N$ and $\pi
N$ channels. This ensures that the low energy $\eta$-nucleon
interaction is strong (and attractive). The strength is also
reflected in the energy dependence of the $pp \to pp\eta$ total
cross section and, most spectacularly, in the dependence of the
$\eta\,^3$He yields near threshold in both $\gamma\,^3$He and
$dp$ collisions. The variation of the angular shape and the
spin dependence all suggest that there is a quasi-bound or
quasi-virtual state very close by. However, none of the
experiments that detect a real $\eta$-meson emerging can ever
separate these two possibilities. On the other hand,
experiments that searched for other decay modes of the mesic
nuclei below threshold have been rather negative or, at best,
inconclusive.

Experiments to study the $\eta^{\prime}$ have been hampered by its much
smaller production cross section and the lack of such a clean decay signal to
identify the meson. It is, however, clear that there is no equivalent of the
$S_{11}(1535)$ isobar to dominate the low-energy $\eta^{\prime} N$ system.
Though there are attempts to search for $\eta^{\prime}$ bound to nuclei, such
experiments will certainly be \emph{challenging}.

Surprisingly, in the threshold region photoproduction of
$\eta\pi$ pairs is already much better understood than
production of $\eta^{\prime}$ mesons. In a sense, the $\eta\pi$
final state is more similar to $\eta$ production because it is
also clearly dominated by the excitation of one single nucleon
resonance, namely the $D_{33}(1700)$, which decays for example
via the $\Delta(1700)\rightarrow\eta\Delta(1232)\rightarrow
\eta N\pi$ chain. This is reflected in the Dalitz plots of the
reaction, angular distributions, kinetic energy spectra of the
mesons, and also the isopin dependence of the reaction.

The production of the $\eta \pi$ system might be more interesting for the
mesic nucleus hunt and the $\gamma\,^{4}\textrm{He}\to \pi^0\eta
\,^{4}\textrm{He}$ reaction might allow one to investigate the low energy
$\eta \,^{4}\textrm{He}$ system to compare with the data extracted from the
$dd\to \eta \,^{4}\textrm{He}$ reaction. However, in both these cases a real
$\eta$ meson is produced and so neither will give an unambiguous result
regarding the possible binding to the nucleus.

More data is likely to appear from proton-induced reactions. Some new data
for pion-induced reactions may also arise from, for example, the HADES
experiment at GSI. However, over the next few years it will still be
information from electron machines that is likely to dominate the database
because several of the facilities and experiments will undergo upgrades.
We look forward to this with anticipation!\\[1ex]

Many people have helped us over the years to research this material and it
would be invidious to pick any out. However, we would like to remember two
workers in the field, Ben Nefkens and Sven Kullander, who left us earlier
this year.


\clearpage
\section{Appendix A: Comparison of conflicting
{\boldmath{$\gamma n\rightarrow n\eta$}} results}
\label{app:etan}

\begin{figure}[htb]
\begin{center}
\epsfig{file=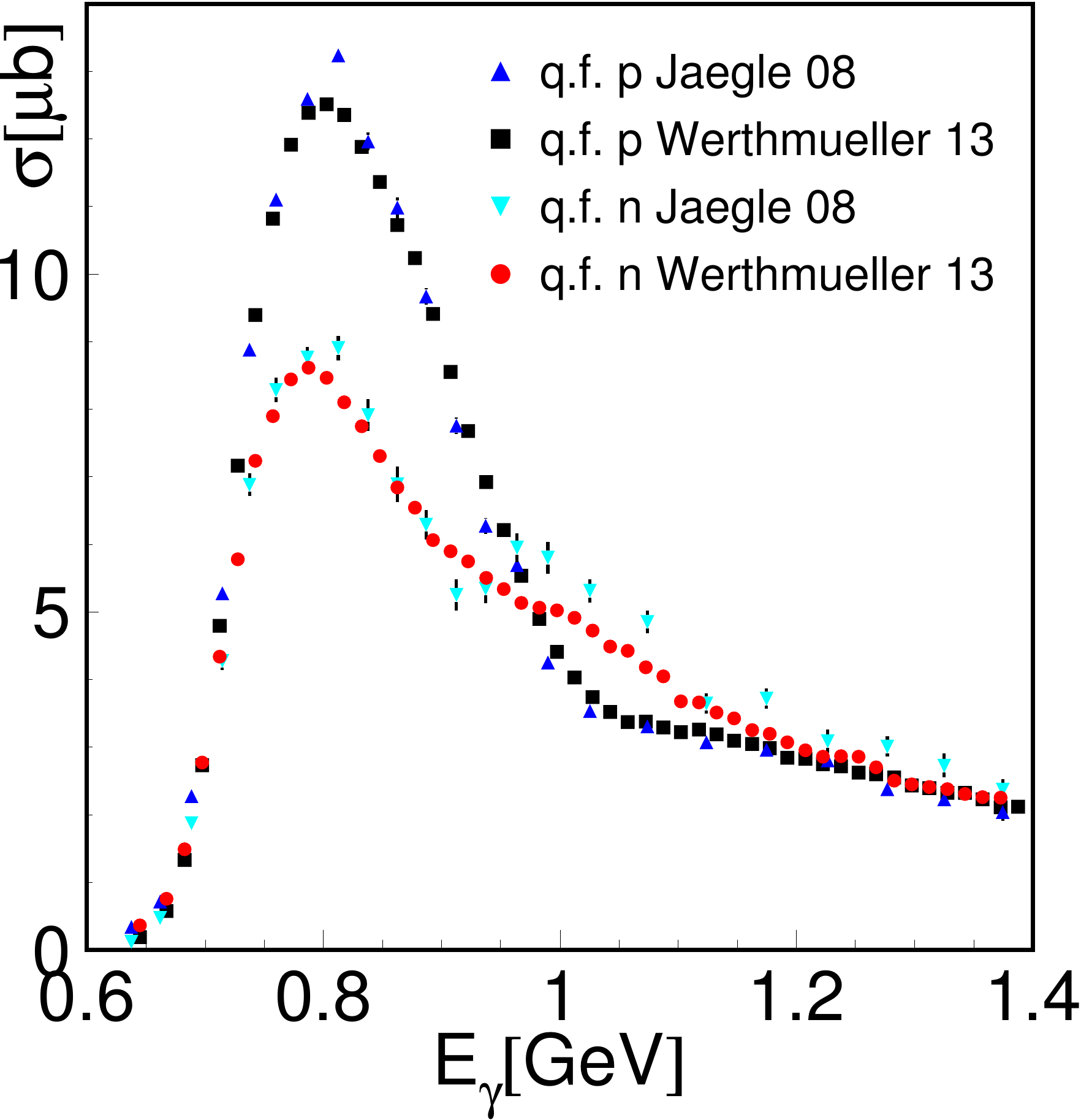,scale=0.34}
\epsfig{file=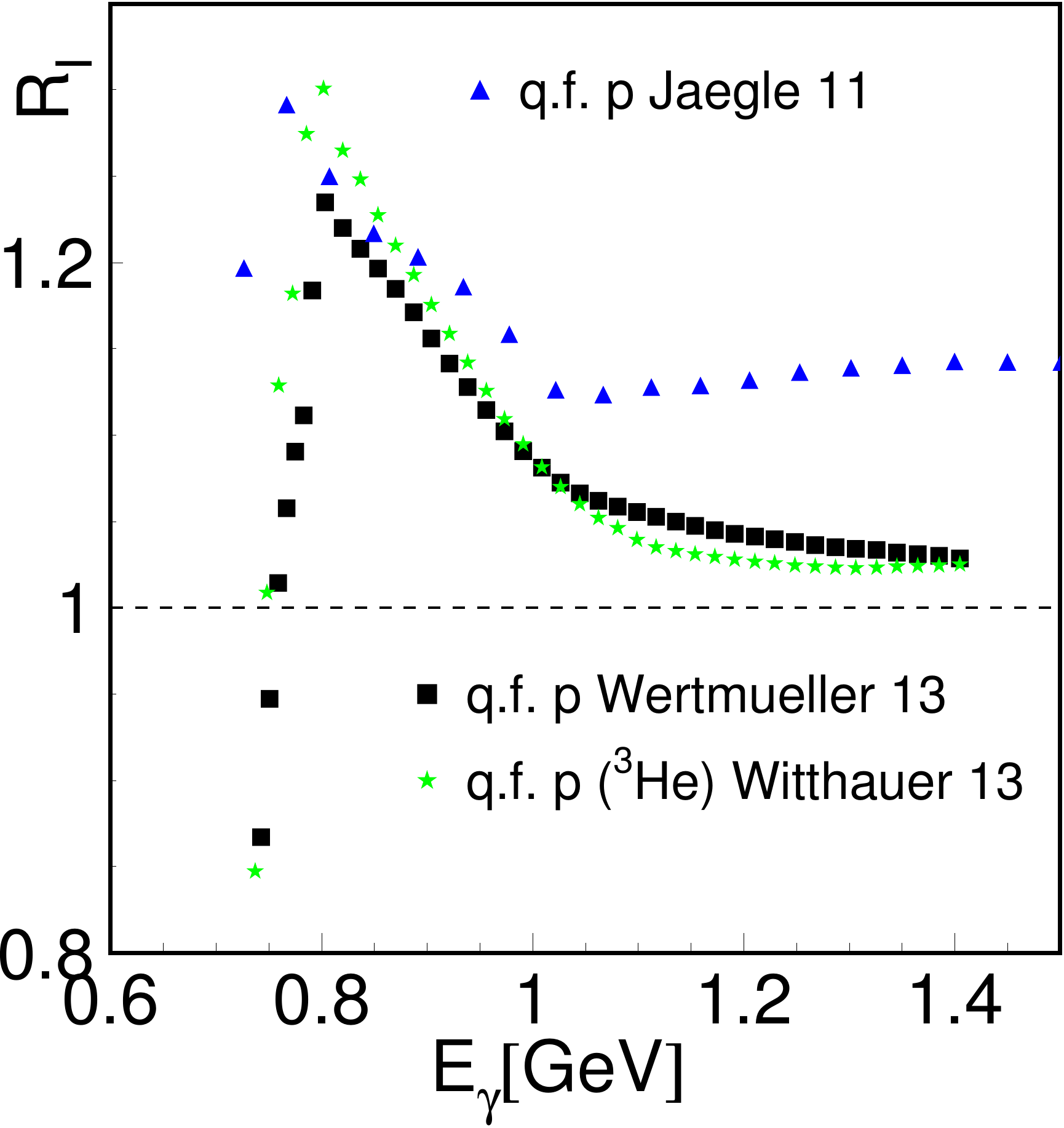,scale=0.34}
\epsfig{file=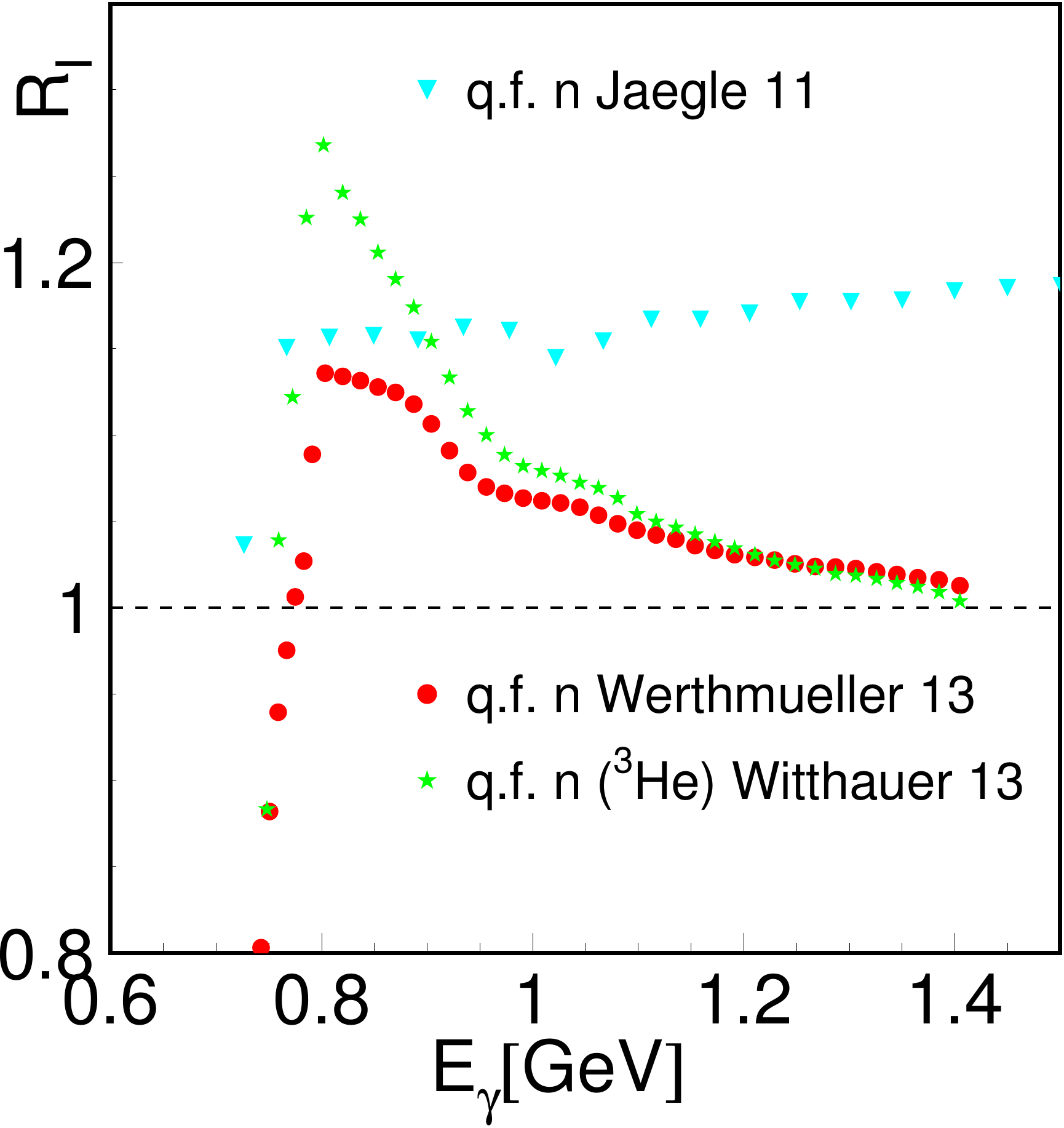,scale=0.34}
\begin{minipage}[t]{16.5 cm}
\caption{Left: comparison of quasi-free $\gamma N\rightarrow N\eta$
data for deuteron targets from ELSA \cite{Jaegle_08} and MAMI
\cite{Werthmueller_13,Werthmueller_14}. Centre: comparison of the
$R_{I}(E_{\gamma})$ factors (see text) for quasi-free protons
\cite{Jaegle_11,Werthmueller_13,Witthauer_13}. Right: the same
comparison for quasi-free neutrons. \label{fig:d_compa}}
\end{minipage}
\end{center}
\end{figure}

As mentioned in Sec.~\ref{s3ec:N_gamma_eta_free}, the kinematically
reconstructed results from the ELSA \cite{Jaegle_11} and MAMI
\cite{Werthmueller_13,Werthmueller_14} experiments for quasi-free
$\eta$-production off the deuteron differ in absolute scale. At the left hand
side of Fig.~\ref{fig:d_compa}, the quasi-free results without kinematic
reconstruction are shown as functions of the incident photon energy
$E_{\gamma}$. In contrast to the kinematically reconstructed data, they are
in excellent agreement for protons and in reasonable agreement (within
systematic uncertainties) for neutrons.

Since the kinematic reconstruction removes only the effects
from Fermi motion, it may change the shape of the observed
cross section, but the integral over energy should be conserved
(there could be small effects from the finite range of
integration but, since the cross section is small and rather
flat at the high energy limit, these effects cannot be
important). Consequently, the Fermi-smeared cross section
$\sigma (E_{\gamma})$ and the cross section $\sigma (W)$ from
kinematic reconstruction as function of $W$ should be related
by:
\begin{equation}
\int_{E_{\gamma}^{(1)}}^{E_{\gamma}^{(2)}} \sigma(E_{\gamma})\,dE_{\gamma} \approx
\int_{W^{(1)}}^{W^{(2)}} \sigma(W)\,\frac{\delta E_{\gamma}}{\delta W}\, dW
\end{equation}
for $E_{\gamma}^{(1)}$ below the $\eta$ production threshold and
$E_{\gamma}^{(2)}>>E_{\gamma}^{(1)}$. Since $W = \sqrt{2E_{\gamma}m_N
+m_N^2}$, this means that
\begin{equation}
\int_{E_{\gamma}^{(1)}}^{E_{\gamma}^{(2)}} \sigma(E_{\gamma})\,dE_{\gamma} \approx
\frac{1}{m_N}\int_{W^{(1)}}^{W^{(2)}} \sigma(W)\,W\,dW
\end{equation}
and thus
\begin{equation}
R_I\equiv \left.(1/m_N)\int \sigma (W)\,W\,dW\right/\int \sigma(E_{\gamma})\,dE_{\gamma}\rightarrow 1
\end{equation}
for $E_{\gamma}>>E_{\gamma}^{\rm thr}$. It is shown in
Fig.~\ref{fig:d_compa} that the MAMI deuteron
\cite{Werthmueller_13,Werthmueller_14} and $^3$He
\cite{Witthauer_13} experiments approximately respect this
relation for quasi-free protons and neutrons, while the ELSA
deuteron data \cite{Jaegle_08,Jaegle_11} behave differently.
This implies that there is some inconsistency between the
absolute normalization of the Fermi-smeared and the
kinematically reconstructed data for the ELSA deuteron
experiment.
%
%


\newpage
\begin{thebibliography}{999}
%
%
\bibitem{Faldt_02} G.~F\"{a}ldt, T.~Johansson, C.~Wilkin, \emph{Physica Scripta} T99 (2002) 146.
%
\bibitem{Moskal_02} P.~Moskal, M.~Wolke, A.~Khoukaz, W.~Oelert,  \emph{Prog.\ Part.\ Nucl.\ Phys.}\ 49 (2002) 1.
%
\bibitem{Kelkar_13} N.G.~Kelkar, K.P.~Khemchandani, N.J.~Upadhyay, B.K.~Jain, \emph{Rep.\ Prog.\ Phys.}\ 76 (2013) 066301.
%
\bibitem{Machner_14} H.~Machner, \emph{J.\ Phys.}\ G (\emph{in press}).
%
\bibitem{Krusche_11}         B. Krusche,                                              {\it Eur. Phys. J. Special Topics}        198     (2011)        199.
\bibitem{Aker_92}            E. Aker et al.,                                          {\it Nucl. Instr. and Meth.}          A   321     (1992)         69.
\bibitem{Gabler_94}          A.R. Gabler et al.,			                          {\it Nucl. Instr. and Meth.}          A   346     (1994)    168.
\bibitem{Hillert_06}         W. Hillert,                                              {\it Eur. Phys. J.}                   A    28     (2006)        139.
\bibitem{Mecking_03}         B.A. Mecking et al.,                                     {\it Nucl. Instr. Meth.}              A   503     (2003)        513.
\bibitem{Bartalini_05}       O. Bartalini et al.,                                     {\it Eur. Phys. J.}                   A    26     (2005)        399.
\bibitem{Starostin_01}       A. Starostin et al.,                                     {\it Phys. Rev.}	                    C    64     (2001)     055205.
\bibitem{Kaiser_08}          K.-H. Kaiser et al.,                                     {\it Nucl. Instr. Meth.}              A   593     (2008)	      159.
\bibitem{Sumihama_06}        M. Sumihama et al.,                                      {\it Phys. Rev.}                      C    73     (2006)     035214.
\bibitem{Yamazaki_05}        H. Yamazaki et al.,                                      {\it Nucl. Instr. Meth.}              A   536     (2005)         70.
\bibitem{Unverzagt_09}       M. Unverzagt et al.,                                     {\it Eur. Phys. J.}                   A    39     (2009)        169.
\bibitem{Prakhov_09}         S. Prakhov et al.,                                       {\it Phys. Rev.}                      C    79     (2009)     035204.
\bibitem{Arguar_14}          P. Aguar-Bartolom\'{e} et al.,                           {\it Phys. Rev.}                      C    89     (2014)     044608.
\bibitem{Nefkens_14}         B.M.K. Nefkens et al.,                                   {\it Phys. Rev.}                      C    90     (2014)     025206.
\bibitem{Bantes_14}          B. Bantes et al.,                                        {\it Int. J. Mod. Phys. Conf. Ser.}        26     (2014)  1460093-1.
\bibitem{Nanova_12p}         M. Nanova et al.,                                        {\it ELSA proposal: ELSA/3-2012-BGO}              (2012).
\bibitem{Crede_13}           V. Cred\'{e}, W. Roberts,                                {\it Rep. Prog. Phys.}                     76     (2013)     076301.
\bibitem{PDG_Rev}            C. Amsler, B. Krusche, T. DeGrand (Quark Model review),\\
                             in J. Beringer et al.,                                   {\it Phys. Rev.}                      D    86     (2012)     010001.
\bibitem{Beringer_12}        J.~Beringer et al.\ (Particle Data Group),               {\it Phys.\ Rev.}\ D 86 (2012) 010001 and 2013 partial update for the 2014 edition.
\bibitem{Geant4}             S. Agostinelli et al.,                                   {\it Nucl. Instr. Meth.}              A   506     (2003)        250.
\bibitem{Krusche_99}         B. Krusche et al.,                                       {\it Eur. Phys. J.}                   A     6     (1999)        309.
\bibitem{Darwish_03}         E.M. Darwish, H. Arenh{\"o}vel, M Schwamb,               {\it Eur. Phys. J.}                   A    16     (2003)        111.
\bibitem{Tarasov_11}         V.E. Tarasov et al.,                                     {\it Phys. Rev.}                      C    84     (2011)     035203.
\bibitem{Dieterle_14}        M. Dieterle et al.,                                      {\it Phys. Rev. Lett.}                    112     (2014)     142001.
\bibitem{Krusche_14}         B. Krusche,                                              {\it EPJ Web of Conf.}                     72     (2014)      00012.
\bibitem{Krusche_95}	     B. Krusche et al.,				                          {\it Phys. Rev. Lett.}		             74     (1995)       3736.
\bibitem{Renard_02}	         F. Renard et al.,				                          {\it Phys. Lett.}		                B   528     (2002)        215.
\bibitem{Dugger_02}	         M. Dugger et al.,				                          {\it Phys. Rev. Lett.}		             89     (2002)     222002.
\bibitem{Crede_05}	         V. Cred\'{e} et al.,				                      {\it Phys. Rev. Lett.}		             94     (2005)     012004.
\bibitem{Nakabayashi_06}     T. Nakabayashi et al.,  			                      {\it Phys. Rev.}		                C    74     (2006)     035202.
\bibitem{Bartalini_07}       O. Bartalini et al.,			                          {\it Eur. Phys. J.} 		            A    33     (2007)        169.
\bibitem{Williams_09}	     M. Williams et al.,			                          {\it Phys. Rev.}		                C    80     (2009)     045213.
\bibitem{Crede_09}	         V. Cred\'{e} et al.,				                      {\it Phys. Rev.}			            C    80     (2009)     055202.
\bibitem{McNicoll_10}	     E.F. McNicoll et al.,			                          {\it Phys. Rev.}		                C    82     (2010)     035208.
\bibitem{Chiang_02}          W.-T. Chiang, S.N. Yang, L. Tiator, D. Drechsel,         {\it Nucl. Phys.}                     A   700     (2002)        429.
\bibitem{Chiang_03}          W.-T. Chiang et al.,                                     {\it Phys. Rev.}                      C    68     (2003)     045202.
\bibitem{Anisovich_12a}      A.V. Anisovich et al.,                                   {\it Eur. Phys. J.}                   A    48     (2012)         15.
\bibitem{Anisovich_12}       A.V. Anisovich et al.,                                   {\it Eur. Phys. J.}                   A    48     (2012)         88; \verb=pwa.hiskp.uni-bonn.de=.
\bibitem{Jaegle_11}          I. Jaegle et al.,                                        {\it Eur. Phys. J.}                   A    47     (2011)         89.
\bibitem{Werthmueller_13}    D. Werthm\"uller et al.,                                 {\it Phys. Rev. Lett.}                    111     (2013)     232001.
\bibitem{Witthauer_13}       L. Witthauer et al.,                                     {\it Eur. Phys. J.}                   A    49     (2013)        154.
\bibitem{Anisovich_13}       V.A. Anisovich et al.,                                   {\it Eur. Phys. J.}                   A    49     (2013)         67.
\bibitem{Bartholomy_07}      O. Bartholomy et al.,			                          {\it Eur. Phys. J.} 		            A    33     (2007)        133.
\bibitem{Sumihama_09}	     M. Sumihama et al.,			                          {\it Phys. Rev.}			            C    80     (2009)  052201(R).
\bibitem{Krusche_03}         B. Krusche, S. Schadmand,                                {\it Prog. Part. Nucl. Phys.}              51     (2003)        399.
\bibitem{Werthmueller_14}    D. Werthm\"uller et al.,                                 {\it Phys. Rev.}                      C    90     (2014)     015205.
\bibitem{Krusche_95b}        B. Krusche et al.,                                       {\it Phys. Lett.}                     B   358     (1995)         40.
\bibitem{Hoffmann_97}        P. Hoffmann-Rothe et al.,                                {\it Phys. Rev. Lett.}                     78     (1997)       4697.
\bibitem{Weiss_03}           J. Wei{\ss} et al.,                                      {\it Eur. Phys. J.}                   A    16     (2003)        275.
\bibitem{Kuznetsov_07}       V. Kuznetsov et al.,                                     {\it Phys. Lett.}                     B   647     (2007)         23.
\bibitem{Miyahara_07}        F. Miyahara et al.,                                      {\it Prog. Theor. Phys. Suppl.}           168     (2007)         90.
\bibitem{Jaegle_08}          I. Jaegle et al.,                                        {\it Phys. Rev. Lett.}                    100     (2008)     252002.
\bibitem{Shklyar_13}         V.~Shklyar, H.~Lenske, U.~Mosel,                         {\it Phys. Rev.}                      C    87     (2013)     015201.
\bibitem{Saghai_01}          B. Saghai, Z. Li,                                        {\it Eur. Phys. J.}                   A    11     (2001)        217.
\bibitem{Krusche_97}         B. Krusche, N.C. Mukhopadhyay, J.-F. Zhang, M. Benmerrouche, {\it Phys. Lett.}                 B   397     (1997)        171.
\bibitem{Krusche_95c}        B. Krusche et al.,                                       {\it Z. Phys.}                        A   351     (1995)        237.
\bibitem{Nikolaev_14}        A. Nikolaev et al.,                                      {\it Eur. Phys. J.}                   A    50     (2014)         58.
\bibitem{Thompson_01}        R. Thompson et al.,                                      {\it Phys. Rev. Lett.}                     86     (2001)       1702.
\bibitem{Weiss_01}           J. Wei{\ss} et al.,                                      {\it Eur. Phys. J.}                   A    11     (2001)        371.
\bibitem{Fantini_08}         A. Fantini et al.,                                       {\it Phys. Rev.}                      C    78     (2008)     015203.
\bibitem{Bock_98}            A. Bock et al.,                                          {\it Phys. Rev. Lett.}                     81     (1998)        534.
\bibitem{Tiator_99}          L. Tiator, D. Drechsel, G. Kn{\"o}chlein, C. Bennhold,   {\it Phys. Rev.}                      C    60     (1999)     035210.
\bibitem{Akondi_14}          C.S. Akondi et al.,                                      {\it Phys. Rev. Lett.}                    113     (2014)     102001.
\bibitem{Denizli_07}         H. Denizli et al.,				                          {\it  Phys. Rev.}			            C    76     (2007)     015204.
\bibitem{Moorhouse_66}       R.G. Moorhouse,                                          {\it Phys. Rev. Lett.}                     16     (1966)        772.
\bibitem{Shklyar_07}         V. Shklyar, H. Lenske, U. Mosel,		                  {\it Phys. Lett.}		                B   650     (2007)        172.
\bibitem{Diakonov_97}        D. Diakonov, V. Petrov, M.V. Polyakov,                   {\it Z. Phys.}                        A   359     (1997)        305.
\bibitem{Polyakov_03}        M.V. Polyakov, A. Rathke,                                {\it Eur. Phys. J.}                   A    18     (2003)        691.
\bibitem{Arndt_04}           R.A. Arndt et al.,                                       {\it Phys Rev.}                       C    69     (2004)     035208.
\bibitem{Azimov_05}          Y.I. Azimov et al.,                                      {\it Eur. Phys. J.}                   A    25     (2005)        325.
\bibitem{Anisovich_13a}      A.V. Anisovich et al.,                                   {\it Phys. Lett.}                     B   719     (2013)         89.
\bibitem{Shyam_08}           R. Shyam, O. Scholten,  	            	              {\it Phys. Rev.} 		                C    78     (2008)     065201.
\bibitem{Anisovich_09}       V.A. Anisovich et al.,		            	              {\it Eur. Phys. J.}	                A    41     (2009)         13.
\bibitem{Doering_10}         M. D\"oring,  K. Nakayama,		                          {\it Phys. Lett.}		                B   683     (2010)        145.
\bibitem{Choi_06}            Ki-Seok Choi, Seung-Il Nam, Atsuhi Hosaka, Hyun-Chul Kim, {\it Phys. Lett.}                    B   636     (2006)        253.
\bibitem{Fix_07}             A. Fix, L. Tiator, M.V. Polyakov,	                      {\it Eur. Phys. J.}		            A    32     (2007)        311.
\bibitem{Shrestha_12}        M. Shrestha, D.M. Manley,                                {\it Phys. Rev.}                      C    86     (2012) 045204, {\it idem} 86 (2012) 055203.
\bibitem{Ajaka_98}           J. Ajaka et al.,                                         {\it Phys. Rev. Lett.}                     81     (1998)       1797.
\bibitem{Elsner_07}          D. Elsner et al.,		                                  {\it Eur. Phys. J.}                   A    33     (2007)        147.
\bibitem{Anisovich_14}       A.V. Anisovich et al.,                                   {\it arXiv:1402.7164 [nucl-ex]}                   (2014)
\bibitem{Capstick_92}        S. Capstick,                                             {\it Phys. Rev.}                      D    46     (1992)       2864.
\bibitem{Burkert_03}         V.D. ~Burkert et al.,                                    {\it Phys. Rev.}                      C    67     (2003)     035204.
\bibitem{Chiang_97}          W.T. Chiang, F. Tabakin,                                 {\it Phys. Rev.}                      C    55     (1997)       2054.
\bibitem{Anisovich_05}       A.V. Anisovich et al.,                                   {\it Eur. Phys. J.}                   A    25     (2005)        427.
\bibitem{Ahrens_03}          J. Ahrens et al.,                                        {\it Eur. Phys. J.}                   A    17     (2003)        241.
\bibitem{Morrison_12}        B.T. Morrison, M. Dugger, B.G.  Ritchie,                 {\it AIP Conf. Proc.}                    1432     (2012)        421.
\bibitem{Barker_75}          I.S. Barker, A. Donnachie, J.K. Storrow,                 {\it Nucl. Phys.}                     B    95     (1975)        347.
\bibitem{Nakayama_06}        K. Nakayama, H. Haberzettl,                              {\it Phys. Rev.}                      C    73     (2006)     045211.
\bibitem{ABBHHM_68}          ABBHHM collaboration,                                    {\it Phys. Rev.}                          175     (1968)       1669.
\bibitem{AHHM_76}            W. Struczinski et al.,                                   {\it Nucl. Phys.}                     B   108     (1976)         45.
\bibitem{Zhang_95}           J.-F. Zhang, N.C. Mukhopadhyay, M. Benmerrouche,         {\it Phys. Rev.}                      C    52     (1995)       1134.
\bibitem{Ploetzke_98}        R. Pl{\"o}tzke et al.,                                   {\it Phys. Lett.}                     B   444     (1998)        555.
\bibitem{Dugger_06}          M. Dugger et al.,                                        {\it Phys. Rev. Lett.}                     96     (2006)     169905.
\bibitem{Jaegle_11a}         I. Jaegle et al.,                                        {\it Eur. Phys. J.}                   A    47     (2011)         11.
\bibitem{Sibirtsev_04}       Ch. Elster, A. Sibirtsev, S. Krewald, J. Speth,          {\it AIP Conf. Proc.}                     717     (2004)        837.
\bibitem{Zhong_11}           Xian-Hui Zhong, Qiang Zhao,                              {\it Phys. Rev.}                      C    84     (2011)     065204.
\bibitem{Huang_13}           F. Huang, H. Haberzettl, K. Nakayama,                    {\it Phys. Rev.}                      C    87     (2013)     054004.
\bibitem{Afzal_14}           F.N. Afzal,                                              {\it EPJ Web of Conf.}                     73     (2014)      04005.
\bibitem{Sandri_14}          P. Levi Sandri et al.,                                   {\it arXiv:1407.6991}                             (2014).
\bibitem{Roberts_05}         W. Roberts, T. Oed,                                      {\it Phys. Rev.}                      C    71     (2005)     055201.
\bibitem{Sarantsev_08}       A.V. Sarantsev et al.,                                   {\it Phys. Lett.}                     B   659     (2008)         94.
\bibitem{Zehr_12}            F. Zehr et al.,				                          {\it Eur. Phys. J.}	                A    48	    (2012)         98.
\bibitem{Kashevarov_12}      V.L. Kashevarov et al.,			                      {\it Phys. Rev.}		                C    85	    (2012)     064610.
\bibitem{Keshelashvili_14}   I. Keshalashvili, A. K{\"a}ser,                          {\it J. Phys. Conf. Ser.}                 503     (2014)     012023.
\bibitem{Ajaka_08}           J. Ajaka et al.,                                         {\it Phys. Rev. Lett.}                    100     (2008)     052003.
\bibitem{Kashevarov_09}      V.L. Kashevarov et al.,                                  {\it Eur. Phys. J.}                   A    42     (2009)        141.
\bibitem{Gutz_14}            E. Gutz et al.,                                          {\it Eur. Phys. J.}                   A    50     (2014)         74.
\bibitem{Horn_08a}           I. Horn et al.,                                          {\it Phys. Rev. Lett.}                    101     (2008)     202002.
\bibitem{Horn_08b}           I. Horn et al.,                                          {\it Eur. Phys. J.}                   A    38     (2008)        173.
\bibitem{Gutz_08}            E. Gutz et al.,                                          {\it Eur. Phys. J.}                   A    35     (2008)        291.
\bibitem{Gutz_10}            E. Gutz et al.,                                          {\it Phys. Lett.}                     B   687     (2010)         11.
\bibitem{Kashevarov_10}      V.L. Kashevarov et al.,                                  {\it Phys. Lett.}                     B   693     (2010)        551.
\bibitem{Fix_11}             A. Fix, H. Arenh\"ovel,                                  {\it Phys. Rev.}                      C    83     (2011)     015503.
\bibitem{Strauch_05}         S. Strauch et al.,                                       {\it Phys. Rev. Lett.}                     95     (2005)     162003.
\bibitem{Krambrich_09}       D. Krambrich et al.,                                     {\it Phys. Rev. Lett.}                    103     (2009)     052002.
\bibitem{Oberle_13}          M. Oberle et al.,                                        {\it Phys. Lett.}                     B   721     (2013)        237.
\bibitem{Oberle_14}          M. Oberle et al.,                                        {\it Eur. Phys. J.}                   A    50     (2014)         54.
\bibitem{Fix_10}             A. Fix, V.L. Kashevarov, A. Lee, M. Ostrick,             {\it Phys. Rev.}                      C    82     (2010)     035207.
\bibitem{Doering_06a}        M. D\"oring, E. Oset, D. Strottman,                      {\it Phys. Lett.}                     B   639     (2006)         59.
\bibitem{Doering_06b}        M. D\"oring, E. Oset, D. Strottman,                      {\it Phys. Rev.}                      C    73     (2006)     045209.
\bibitem{Fix_13}             A. Fix,                                                  {\it private communication}                       (2013)
\bibitem{Krusche_14b}        B. Krusche,                                              {\it Acta Phys. Pol.}                 B    45     (2014)        639.
%
\bibitem{Bulos_64} F.~Bulos et al., \emph{Phys.\ Rev.\ Lett.}\ 13 (1964) 486.
%
\bibitem{Deinet_69} W.~Deinet et al., \emph{Nucl.\ Phys.}\ B 11 (1969) 495.
%
\bibitem{Richards_70} W.B.~Richards et al., \emph{Phys.\ Rev.}\ D 1 (1970) 10.
%
\bibitem{Binnie_73} D.M.~Binnie et al., \emph{Phys.\ Rev.}\ D 8 (1973) 2789.
%
\bibitem{Debeham_75} N.C.~Debeham et al., \emph{Phys.\ Rev.}\ D 12 (1975) 2545.
%
\bibitem{Feltesse_75} J.~Feltesse et al., \emph{Nucl.\ Phys.}\ B 93 (1975) 242.
%
\bibitem{Brown_79} R.M.~Brown et al., \emph{Nucl.\ Phys.}\ B 153 (1979) 89.
%
\bibitem{Baker_79} R.D.~Baker et al., \emph{Nucl.\ Phys.}\ B 156 (1979) 93.
%
\bibitem{Prakhov_05} S.~Prakhov et al., \emph{Phys.\ Rev.}\ C 72 (2005) 015203.
%
\bibitem{Bayadilov_12} D.E.~Bayadilov et al., \emph{Phys.\ Atom.\ Nuclei} 75 (2012) 923.
%
\bibitem{Wilkin_93} C.~Wilkin, \emph{Phys.\ Rev.}\ C 47 (1993) R938.
%
\bibitem{Landolt_88} H.H.~Landolt, R.~B{\"o}rnstein, \emph{New Series, Springer} vol I/12a   (1988).
%
\bibitem{Rader_72} R.K.~Rader et al.,  \emph{Phys. Rev.}  D 6 (1972) 3059.
%
\bibitem{Starostin_05} A.~Starostin et al., \emph{Phys.\ Rev.}\ C 72 (2005) 015205.
%
\bibitem{Arndt_06} R.A.~Arndt, W.J.~Briscoe, I.I.~Strakovsky, R.L.~Workman, \emph{Phys.\ Rev.}\ C 74 (2006) 045205; \verb=http://gwdac.phys.gwu.edu=.
%
\bibitem{Leupold_10}         S. Leupold, V. Metag, U. Mosel,                          {\it Int. J. Mod. Phys.}              E    19     (2010)        147.
\bibitem{Krusche_05}         B. Krusche,                                              {\it Prog. Part. Nucl. Phys.}              55     (2005)         46.
\bibitem{Drechsel_99}        D. Drechsel, L. Tiator, S.S. Kamalov, S.N. Yang,         {\it Nucl. Phys.}                     A   660     (1999)        423.
\bibitem{Krusche_02}         B. Krusche et al.,                                       {\it Phys. Lett.}                     B   526     (2002)        287.
\bibitem{Krusche_05a}        B. Krusche,                                              {\it Eur. Phys. J.}                   A    26     (2005)          7.
\bibitem{Maghrbi_13}         Y. Maghrbi et al.,                                       {\it Eur. Phys. J.}                   A    49     (2013)         38.
\bibitem{Tarbert_14}         C.M. Tarbert et al.,                                     {\it Phys. Rev. Lett.}                    112     (2014)     242502.
\bibitem{Tarbert_08}         C.M. Tarbert et al.,                                     {\it Phys. Rev. Lett.}                    100     (2008)     132301.
\bibitem{Krusche_04}         B. Krusche et al.,                                       {\it Eur. Phys. J.}                   A    22     (2004)	      277.
\bibitem{Roebig_96}          M. R\"obig-Landau et al.,                                {\it Phys. Lett.}                     B   373     (1996)	       45.
\bibitem{Mertens_08}         T. Mertens et al.,                                       {\it Eur. Phys. J.}                   A    38     (2008)        195.
\bibitem{Nanova_12}          M. Nanova et al.,                                        {\it Phys. Lett.}                     B   710     (2012)        600.
\bibitem{Kotulla_08}         M. Kotulla et al.,                                       {\it Phys. Rev. Lett.}                    100     (2008)     192302.
\bibitem{Bianchi_96}         N. Bianchi et al.,                                       {\it Phys. Rev.}                      C    54     (1996)       1688.
\bibitem{Geissel_02}         H. Geissel, et al.,                                      {\it Phys. Rev. Lett.}                     88     (2002)     122301.
\bibitem{Ishikawa_05}        T. Ishikawa et al.,                                      {\it Phys. Lett.}                     B   608     (2005)        215.
\bibitem{Yorita_00}          T. Yorita et al.,                                        {\it Phys. Lett.}                     B   476     (2000)        226.
\bibitem{Kinoshita_06}       T. Kinoshita et al.,                                     {\it Phys. Lett.}                     B   639     (2006)        429.
\bibitem{Buss_12}            O. Buss et al.,                                          {\it Phys. Reports}                       512     (2012)          1.
\bibitem{Hombach_95}         A. Hombach, A. Engel, S. Teis, U. Mosel,                 {\it Z. Phys.}                        A   352     (1995)        223.
\bibitem{Effenberger_97}     M. Effenberger, A. Hombach, S. Teis, U. Mosel,           {\it Nucl. Phys.}                     A   614     (1997)        501.
\bibitem{Lehr_00}            J. Lehr, M. Effenberger, U. Mosel,                       {\it Nucl. Phys.}                     A   671     (2000)        503.
\bibitem{Lehr_01}            J. Lehr, U. Mosel,                                       {\it Phys. Rev.}                      C    64     (2001)     042202.
\bibitem{Hernandez_92}       E. Hern\'{a}ndez, E. Oset,                               {\it Z. Phys.}                        A   341     (1992)        201.
\bibitem{Magas_05}           V.K. Magas, L. Roca, E. Oset,                            {\it Phys. Rev.}                      C    71     (2005)     065202.
\bibitem{Kaskulov_07}        M. Kaskulov, E. Hernandez, E. Oset,                      {\it Eur. Phys. J.}                   A    31     (2007)        245.
\bibitem{Nanova_13}          M. Nanova et al.,                                        {\it Phys. Lett.}                     B   727     (2013)        417.
\bibitem{Paryev_13}          E.Ya. Paryev,                                            {\it J. Phys.}                        G    40     (2013)     025201.
\bibitem{Nasseripour_07}     R. Nasseripour et al.,                                   {\it Phys. Rev. Lett.}                     99     (2007)     262302.
\bibitem{Wood_08}            M.H. Wood et al.,                                        {\it Phys. Rev.}                      C    78     (2008)     015201.
\bibitem{Thiel_13}           M. Thiel et al.,                                         {\it Eur. Phys. J.}                   A    49     (2013)        132.
\bibitem{Weil_13}            J. Weil, U. Mosel, V. Metag,                             {\it Phys. Lett.}                     B   723     (2013)        120.
\bibitem{Jido_12}            D. Jido, H. Nagahiro, S. Hirenzaki,                      {\it Phys. Rev.}                      C    85     (2012)     032201.
\bibitem{Nagahiro_05}        H. Nagahiro et al.,                                      {\it Phys. Rev. Lett.}                     94     (2005)     232503.
\bibitem{Oset_11}            E. Oset, A. Ramos,                                       {\it Phys. Lett.}                     B   704     (2011)        334.
\bibitem{Hejny_99}           V. Hejny et al.,			                              {\it Eur. Phys. J.}                   A     6     (1999)         83.
\bibitem{Hejny_02}           V. Hejny et al.,			                              {\it Eur. Phys. J.}                   A    13     (2002)        493.
\bibitem{Pfeiffer_04}        M. Pfeiffer et al.,                                      {\it Phys. Rev. Lett.}                     92     (2004)     252001; \textit{idem} 94 (2005) 049102.
\bibitem{Pheron_12}          F. Pheron et al.,                                        {\it Phys. Lett.}                     B   709     (2012)         21.
\bibitem{Mersmann_07}        T. Mersmann et al.,                                      {\it Phys. Rev. Lett.}                     98     (2007)     242301.
\bibitem{Smyrski_07}         J. Smyrski et al.,                                       {\it Phys. Lett.}                     B   649     (2007)        258.
\bibitem{McCarthy_77}        J. S. McCarthy, I. Sick, R. R. Whitney,                  {\it Phys. Rev.}                      C    15     (1977)       1396.
\bibitem{Lichtenstadt_89}    L. Lichtenstadt et al.,                                  {\it Phys. Lett.}                     B   219     (1989)        394.
\bibitem{Egorov_13}          M. Egorov, A. Fix,                                       {\it Phys. Rev.}                      C    88     (2013)     054611.
\bibitem{Jaegle_09}          I. Jaegle,                                               {\it Chinese Phys.}                   C    33     (2009)       1340.
%
\bibitem{Pedroni_78} E.~Pedroni et al., \emph{Nucl.\ Phys.}\ A 300 (1978) 321.
%
\bibitem{Tippens_01} W.B.~Tippens et al., \emph{Phys.\ Rev.}\ D 63 (2001) 052001.
%
\bibitem{Peng_89} J.C.~Peng et al., \emph{Phys.\ Rev.\ Lett.}\ 63 (1989) 2353.
%
\bibitem{Stepenson_03} E.J.~Stephenson et al., \emph{Phys.\ Rev.\ Lett.}\ 91 (2003) 142302.
%
\bibitem{Peng_87} J.C.~Peng et al., \emph{Phys.\ Rev.\ Lett.}\ 58 (1987) 2027.
%
\bibitem{Liu_92} L.C.~Liu, \emph{Phys.\ Lett.}\ B 288 (1992) 18.
%
\bibitem{Peng_88} J.C.~Peng, in \textit{Production and decay of light mesons}, Ed.\ P.~Fleury (World Scientific, 1988) p.~102.
%
\bibitem{Kohno_90} M.~Kohno, H.~Tanabe, \emph{Nucl.\ Phys.}\ A 519 (1990) 755.
%
\bibitem{Krippa_92} B.V.~Krippa, J.T.~Londergan, \emph{Phys.\ Lett.}\ B 286 (1992) 216.
%
\bibitem{Chiavassa_94a} E.~Chiavassa et al., \emph{Phys.\ Lett.}\ B {322} (1994) 270. 
%
\bibitem{Calen_96} H.~Cal\'en et al., \emph{Phys.\ Lett.}\ B {366} (1996) 39. 
%
\bibitem{Bergdolt_93} A.M.~Bergdolt, et~al., \emph{Phys.\ Rev.}\ D 48 (1993) R2969;
%
\bibitem{Hibou_98} F.~Hibou et al., \emph{Phys.\ Lett.}\ B {438} (1998) 41. 
%
\bibitem{Smyrski_00} J.~Smyrski et al., \emph{Phys.\ Lett.}\ B  {474} (2000) 182. 
%
\bibitem{Moskal_04} P.~Moskal et al., \emph{Phys.\ Rev.}\ C {69} (2004) 025203. 
%
\bibitem{Moskal_10} P.~Moskal et al., \emph{Eur.\ Phys.\ J.}\ A {43} (2010) 131. 
%
\bibitem{Marco_01} N.~de Marco, \textit{private communication} (2001). 
%
\bibitem{Agakishiev_12} G.~Agakishiev et al., \emph{Eur.\ Phys.\ J.}\ A {48} (2012) 74. 
%
\bibitem{Balestra_00} F.~Balestra et al., \emph{Phys.\ Lett.}\ B 491 (2000) 29. 
%
\bibitem{Moskal_98} P.~Moskal et al., \emph{Phys. Rev.\ Lett.}\ 80 (1998) 3202.
%
\bibitem{Moskal_00} P.~Moskal et al., \emph{Phys.\ Lett.}\ B 474 (2000) 416.
%
\bibitem{Czerwinski_14} E.~Czerwi\'{n}ski, P.~Moskal, M.~Silarski, \emph{Acta Phys. Polon.} B 45 (2014) 739.
%
\bibitem{Teilab_11} K.~Teilab, \emph{Int.\ J.\ Mod.\ Phys.}\ A {26} (2011) 694. 
%
\bibitem{Faldt_96} G.~F{\"a}ldt, C.~Wilkin, \emph{Phys.\ Lett.}\ B 382 (1996) 209.
%
\bibitem{Petren_10} H.~Petr\'{e}n et al., \emph{Phys.\ Rev.}\ C {82} (2010) 055206. 
%
\bibitem{Calen_99} H.~Cal\'en et al., \emph{Phys.\ Lett.}\ B {458} (1999) 190. 
%
\bibitem{Abdel-Bary_03} M.~Abdel-Bary et al., \emph{Eur.\ Phys.\ J.}\ A {16} (2003) 127. 
%
\bibitem{Deloff_04} A.~Deloff, \emph{Phys.\ Rev.}\ C 69 (2004) 035206.
%
\bibitem{Nakayama_03} K.~Nakayama, J.~Haidenbauer, C.~Hanhart, J.~Speth, \emph{Phys.\ Rev.}\ C 68 (2003) 045201.
%
\bibitem{Pauly_06} C.~Pauly, Ph.D. thesis, Hamburg University, 2006.
%
\bibitem{Bernard_99} V.~Bernard, N.~Kaiser, U.-G.~Mei{\ss}ner, \emph{Eur.\ Phys.\ J.}\ A 4 (1999) 259.
%
\bibitem{Dymov_09} S.~Dymov et al., \emph{Phys.\ Rev.\ Lett}.\  102 (2009) 192301.
%
\bibitem{Balestra_04} F.~Balestra et al., \emph{Phys.\ Rev.}\ C 69 (2004) 064003.
%
\bibitem{Czyzykiewicz_07} R.~Czy\.{z}ykiewicz et al., \emph{Phys.\ Rev.\ Lett.}\ 98 (2007) 122003.
%
\bibitem{Hodana_11} P.~Moskal, M.~Hodana, \emph{J.\ Phys.\ Conf.\ Ser.}\ 295 (2011) 012080.
%
\bibitem{Hodana_14} M.~Hodana, P.~Moskal, I.~Ozerianska, M.J.~Zieli\'{n}ski, \emph{Acta Phys.\ Pol.}\ B 45 (2014) 697.
%
\bibitem{Chiavassa_94b} E.~Chiavassa et al., \emph{Phys.\ Lett.}\ B  {337} (1994) 192.
%
\bibitem{Scomparin_93} E.~Scomparin, PhD thesis, University of Turin (1993). 
%
\bibitem{Calen_97} H.~Cal\'en et al., \emph{Phys.\ Rev.\ Lett.}\ {79} (1997) 2642.
%
\bibitem{Haggstrom_97} S.~H\"aggstr\"om, PhD thesis, \textit{Acta Universitatis Upsaliensis} {13} (1997). 
%
\bibitem{Calen_98b} H.~Cal\'{e}n et al., \emph{Phys.\ Rev.\ Lett.}\ {80} (1998) 2069. 
%
\bibitem{Calen_98a} H.~Cal\'en et al., \emph{Phys.\ Rev.}\ C {58} (1998) 2667.
%
\bibitem{Moskal_09} P.~Moskal et al., \emph{Phys.\ Rev.}\ C 79 (2009) 015208. 
%
\bibitem{Faldt_01} G.~F\"{a}ldt, C.~Wilkin, \emph{Physica Scripta} 64 (2001) 427.
%
%
\bibitem{Czerwinski_10} E.~Czerwi\'{n}ski et al., \emph{Phys.\ Rev.\ Lett.}\ {105} (2010) 122001. 
%
\bibitem{Adolph_09} C.~Adolph et al., \emph{Phys.\ Lett.}\ B 677 (2009) 24.
%
\bibitem{Zlomanczuk_09} J.~Z{\l}oma\'nczuk, PrimeNet Workshop, Bonn (2009);\\
\verb=http://www.itkp.uni-bonn.de/~kubis/PrimeNet/Program.html=.
%
\bibitem{Zielinski_11} M.J.~Zieli\'{n}ski, P.~Moskal, A.~Kupsc,  \emph{Eur.\ Phys.\ J.}\ A 47 (2011) 93.
%
\bibitem{Chiavassa_92} E.~Chiavassa et al., \emph{Zeit.\ Phys.}\ A 342 (1992) 107.
%
\bibitem{Chiavassa_93} E.~Chiavassa et al., \emph{Zeit.\ Phys.}\ A 344 (1993) 345.
%
\bibitem{Chiavassa_94} E.~Chiavassa et al., \emph{Nuovo Cim.}\ 107 A (1994) 1195.
%
\bibitem{Vercellin_93} E.~Vercellin et al., \emph{Nuovo Cim.}\ 106 A (1993) 861.
%
\bibitem{Banaigs_73} J.~Banaigs et al., \emph{Phys.\ Lett.}\ 45B (1973) 394.
%
\bibitem{Berthet_85} P.~Berthet et al., \emph{Nucl.\ Phys.}\ A 443 (1985) 589.
%
\bibitem{Berger_88} J.~Berger et al., \emph{Phys.\ Rev.\ Lett.}\ 61 (1988) 919.
%
\bibitem{Kirchner_93} T.~Kirchner, PhD thesis, Universit\'e de Paris-Sud, (1993).
%
\bibitem{Mayer_96} B.~Mayer et al., \emph{Phys.\ Rev.}\ C 53 (1996) 2068.
%
\bibitem{Betigeri_00} M.~Betigeri et al., \emph{Phys.\ Lett.}\ B 472 (2000) 267.
%
\bibitem{Bilger_02} R.~Bilger et al., \emph{Phys.\ Rev.}\ C 65 (2002) 044608.
%
\bibitem{Adam_07} H.-H.~Adam et al., \emph{Phys.\ Rev.}\ C 75 (2007) 014004.
%
\bibitem{Rausmann_09} T.~Rausmann et al., \emph{Phys.\ Rev.}\ C 80 (2009) 017001.
%
\bibitem{Adlarson_14} P.~Adlarson et al., \emph{Eur.\ Phys.\ J.}\ A 50 (2014) 100.
%
\bibitem{Nikulin_96} V.N.~Nikulin et al., \emph{Phys.\ Rev.}\ C 54 (1996) 1732.
%
\bibitem{Kerboul_86} C.~Kerboul et al., \emph{Phys.\ Lett.}\ B 181 (1986) 28.
%
\bibitem{Germond_90} J.-F.~Germond, C.~Wilkin, \emph{J.\ Phys.}\ G 16 (1990) 381.
%
\bibitem{Germond_89} J.-F.~Germond, C.~Wilkin, \emph{J.\ Phys.}\ G 15 (1989) 437.
%
\bibitem{Kilian_90} K.~Kilian, H.~Nann, \emph{AIP Conf.\ Proc.}\ 221 (1991) 185.
%
\bibitem{Faldt_95} G.~F\"{a}ldt, C.~Wilkin, \emph{Nucl.\ Phys.}\ A 587 (1995) 769.
%
\bibitem{Khemchandani_03} K.P.~Khemchandani, N.G.~Kelkar, B.K.~Jain, \emph{Phys.\ Rev.}\ C 68 (2003) 064610.
%
\bibitem{Wilkin_07} C.~Wilkin et al., \emph{Phys.\ Lett.}\ B 654 (2007) 92.
%
\bibitem{Wurzinger_96} R.~Wurzinger et al., \emph{Phys.\ Lett.}\ B 374 (1996) 283.
%
\bibitem{Faldt_95a} G.~F\"{a}ldt, C.Wilkin, \emph{Phys.\ Lett.}\ B 354 (1995) 20.
%
\bibitem{Papenbrock_14} M.~Papenbrock et al., \emph{Phys.\ Lett.}\ B 734 (2014) 333.
%
\bibitem{Hibou_00} F.~Hibou et al., \emph{Eur.\ Phys.\ J.}\ A 7 (2000) 537.
%
\bibitem{Bilger_04} R.~Bilger et al., \emph{Phys.\ Rev.}\ C 69 (2004) 014003.
%
\bibitem{Tengblad_05} U.~Tengblad, G.~F\"{a}ldt, C.~Wilkin, \emph{Eur.\ Phys.\ J.}\ A 25 (2005) 267.
%
\bibitem{Lai_02} A.~Lai et al., \emph{Phys.\ Lett.}\ B 533 (2002) 196.
%
\bibitem{Ambrosino_07} F.~Ambrosino et al., \emph{J.\ High Energy Phys.}\ 12 (2007) 73.
%
\bibitem{Miller_07} D.H.~Miller et al., \emph{Phys.\ Rev.\  Lett.}\ 99 (2007) 122002.
%
\bibitem{Goslawski_13} P.~Goslawski et al., \emph{Phys.\ Rev.}\ D 85 (2012) 112011.
%
\bibitem{Plouin_92} F.~Plouin et al., \emph{Phys.\ Lett.}\ B 276 (1992) 526.
%
\bibitem{Abdel-Bary_05} M.~Abdel-Bary et al.,\emph{ Phys.\ Lett.}\ B 619 (2005)  281.
%
\bibitem{Scomparin_93b} E.~Scomparin et al., \emph{J.\ Phys.}\ G 19 (1993) L51.
%
\bibitem{Budzanowski_10} A.~Budzanowski et al., \emph{Phys.\ Rev.}\ C 82 (2010) 041001 (R).
%
\bibitem{Al-Khalili_93} J.S.~Al-Khalili, M.B.~Barbaro, C.~Wilkin, \emph{J.\ Phys.}\ G 19 (1993) 403.
%
\bibitem{Frascaria_94} R.~Frascaria, et~al., \emph{Phys.\ Rev.}\  C 50 (1994) 537 (R).
%
\bibitem{Willis_97} N.~Willis, et~al., \emph{Phys.\ Lett.}\ B 406 (1997) 14.
%
\bibitem{Wronska_05} A.~Wro\'{n}ska et al., \emph{Eur.\ Phys.}\ J.\ A 26 (2005) 421.
%
\bibitem{Budzanowski_09} A.~Budzanowski et al., \emph{Nucl.\ Phys.}\ A 821 (2009) 193.
%
\bibitem{Bargholtz_07} C.~Bargholtz et al., \emph{Phys.\ Lett.}\ B 644 (2007) 299.
%
\bibitem{Berkowski_08} M.~Ber{\l}owski et al., \emph{Phys.\ Rev.}\ D 77 (2008) 032004.
%
\bibitem{Adlarson_12} P.~Adlarson et al., \emph{Phys.\ Lett.}\ B 707 (2012) 243.
%
\bibitem{Schonning_10} K.~Sch\"{o}nning et al., \emph{Phys.\ Lett.}\ B 685 (2010) 33.
%
\bibitem{Buscher_11} M.~B\"{u}scher, P.~Fedorets, \emph{private communication} (2011).
%
\bibitem{Bhalerao_85} R.S.~Bhalerao, L.C.~Liu, \emph{Phys.\ Rev.\ Lett.}\ 54 (1985) 865.
%
\bibitem{Haider_86} Q.~Haider, L.C.~Liu, \emph{Phys.\ Lett.}\ B 172 (1986) 257, \textit{idem} 174 (1986) 465E.
%
\bibitem{Haider_02} Q.~Haider, L.C.~Liu, \emph{Phys.\ Rev.}\ C 66 (2002) 045208.
%
\bibitem{Haider_10} Q.~Haider, \textit{private communication} (2010).
%
\bibitem{Friedman_13} E.~Friedman, A.~Gal, J.~Mare\v{s}, \emph{Phys.\ Lett.}\ B 725 (2013) 334.
%
\bibitem{Plouin_90} F.~Plouin, P.~Fleury, C.~Wilkin, \emph{Phys.\ Rev.\ Lett.}\ 65 (1990) 690.
%
\bibitem{Shevchenko_98} N.V.~Shevchenko et al., \emph{Phys.\ Rev.}\ C 58 (1998) 3055 (R).
%
\bibitem{Khoukaz_12} A.~Khoukaz, D.~Schroer, COSY proposal \#211 (2012),\\ \verb=http://collaborations.fz-juelich.de/ikp/anke/proposal/Proposal_211_pndeta.pdf=.
%
\bibitem{Wilkin_14} C.~Wilkin, \emph{Acta Phys.\ Pol.}\ B 45 (2014) 603.
%
\bibitem{Chrien_88} R.E.~Chrien et al., \emph{Phys.\ Rev.\ Lett.}\ 60 (1988) 2595.
%
\bibitem{Lieb_88} J.B.~Lieb, in \emph{Proc.\ Int.\ Conf.\ Nucl.\ Phys.}, Sao Paulo, Brazil (1988).
%
\bibitem{Sokol_00} G.A.~Sokol et al., \emph{Part\ Nucl.\ Lett.}\ 102 (2000) 71.
%
\bibitem{Baskov_12} V.A.~Baskov et al., \emph{PoS (Baldin ISHEPP XXI)} (2012) 102.
%
\bibitem{Afanasiev_11} S.V.~Afanasiev et al., \emph{Nucl.\ Phys.\ B (Proc.\ Suppl.)} 219-220 (2011) 255.
%
\bibitem{Budzanowski_09a} A.~Budzanowski et al., \emph{Phys.\ Rev.}\ C 79 (2009) 012201 (R).
%
\bibitem{Hanhart_05} C.~Hanhart, \emph{Phys.\ Rev.\ Lett.}\ 94 (2005) 049101.
%
\bibitem{Adlarson_13} P.~Adlarson et al., \emph{Phys.\ Rev.}\ C 87 (2013) 035204.
%
\bibitem{Moskal_10a} P.~Moskal, J.~Smyrski, \emph{Acta Phys.\ Pol.}\ B 41 (2010) 2281.
%
\bibitem{Wycech_14} S.~Wycech, W.~Krzemie{\'n}, \emph{Acta Phys.\ Pol.}\ B 45 (2014) 745.
%
\bibitem{Itashi_12} K.~Itahashi et al., \emph{Prog.\ Theor.\ Phys.}\ 128 (2012) 601.
%
\end{thebibliography}
\end{document}